%% file: main.tex
\documentclass[runningheads]{llncs}

\usepackage[paperwidth=155mm,
            paperheight=235mm,
            left=16mm,
            right=16mm,
            top=17mm,
            bottom=25mm
           ]{geometry}

\usepackage[]{sosy-paper}
\usepackage{wrapfig}
\usepackage[super]{nth}

\newcommand{\inv}{Inv} 
\newcommand{\start}{\lambda} 
\newcommand{\injecttoi}{\textsubscript{\textbf{i}}\injectdynamic}
\newcommand{\injecttof}{\textsubscript{\textbf{f}}\injectdynamic}

\definetool{\spacer}{Spacer}
\definetool{\ikos}{Ikos}
\definetool{\cpacheckerabbrv}{CPA}

\newcommand{\solved}{{\color{green}{\cmark}}\xspace}
\newcommand{\timeout}{--\xspace}

\title{Augmenting Interpolation-Based Model Checking with Auxiliary Invariants}
\subtitle{(Extended Version)}
\titlerunning{Augmenting IMC with Auxiliary Invariants (Extended Version)}

\begin{document}

\author{Dirk Beyer\orcidID{0000-0003-4832-7662}
    \and
    Po-Chun Chien\orcidID{0000-0001-5139-5178}
    \and
    Nian-Ze Lee\orcidID{0000-0002-8096-5595}}

\authorrunning{D.~Beyer, P.-C.~Chien, and N.-Z.~Lee}
\institute{LMU Munich, Munich, Germany}

\blfootnote{A conference version~\cite{IMCDF} of this manuscript is published at SPIN 2024.}

\newcommand{\mypaperkeywords}{
    Software model checking \and
    Program invariants \and
    Invariant injection \and
    Craig interpolation \and
    Data-flow analysis \and
    SMT \and
    SAT
}

\maketitle

\input{evaluation/tex/data-commands}

\begin{abstract}
    \input{abstract}
\end{abstract}

\input{introduction}
\input{related-work}
\input{background}
\input{approach}
\input{implementation}
\input{evaluation}
\input{conclusion}

\subsubsection{Data-Availability Statement}
To enhance the verifiability and transparency of the paper,
all relevant materials, including used tools, benchmark tasks, and raw experimental data,
are available in a supplemental reproduction package~\cite{IMCDF-artifact-SPIN24-submission}.
More information is available at \url{https://www.sosy-lab.org/research/imc-df/}.

\subsubsection{Funding Statement}
This project was funded in part by the Deutsche Forschungsgemeinschaft (DFG)
-- \href{http://gepris.dfg.de/gepris/projekt/378803395}{378803395} (ConVeY).

\bibliography{bib/artifacts, bib/dbeyer, bib/svcomp, bib/sw, bib/svcomp-artifacts}

\end{document}

%% file: evaluation/tex/data-commands.tex
\providecommand\StoreBenchExecResult[7]{\expandafter\newcommand\csname#1#2#3#4#5#6\endcsname{#7}}%
\StoreBenchExecResult{ReducedEvalSlDfS}{ImcHardSafe}{Total}{}{Count}{}{870}%
\StoreBenchExecResult{ReducedEvalSlDfS}{ImcHardSafe}{Total}{}{Cputime}{}{300655.012786016}%
\StoreBenchExecResult{ReducedEvalSlDfS}{ImcHardSafe}{Total}{}{Cputime}{Avg}{345.5804744666850574712643678}%
\StoreBenchExecResult{ReducedEvalSlDfS}{ImcHardSafe}{Total}{}{Cputime}{Median}{27.1679739745}%
\StoreBenchExecResult{ReducedEvalSlDfS}{ImcHardSafe}{Total}{}{Cputime}{Min}{4.244829399}%
\StoreBenchExecResult{ReducedEvalSlDfS}{ImcHardSafe}{Total}{}{Cputime}{Max}{913.342992329}%
\StoreBenchExecResult{ReducedEvalSlDfS}{ImcHardSafe}{Total}{}{Cputime}{Stdev}{415.7696730122742251930677500}%
\StoreBenchExecResult{ReducedEvalSlDfS}{ImcHardSafe}{Total}{}{Walltime}{}{291982.6438434834129888}%
\StoreBenchExecResult{ReducedEvalSlDfS}{ImcHardSafe}{Total}{}{Walltime}{Avg}{335.6122343028544976882758621}%
\StoreBenchExecResult{ReducedEvalSlDfS}{ImcHardSafe}{Total}{}{Walltime}{Median}{15.1187923440011215}%
\StoreBenchExecResult{ReducedEvalSlDfS}{ImcHardSafe}{Total}{}{Walltime}{Min}{1.5823553358204663}%
\StoreBenchExecResult{ReducedEvalSlDfS}{ImcHardSafe}{Total}{}{Walltime}{Max}{906.2711261240765}%
\StoreBenchExecResult{ReducedEvalSlDfS}{ImcHardSafe}{Total}{}{Walltime}{Stdev}{411.9616860843426465762985796}%
\StoreBenchExecResult{ReducedEvalSlDfS}{ImcHardSafe}{Correct}{}{Count}{}{534}%
\StoreBenchExecResult{ReducedEvalSlDfS}{ImcHardSafe}{Correct}{}{Cputime}{}{29969.754082084}%
\StoreBenchExecResult{ReducedEvalSlDfS}{ImcHardSafe}{Correct}{}{Cputime}{Avg}{56.12313498517602996254681648}%
\StoreBenchExecResult{ReducedEvalSlDfS}{ImcHardSafe}{Correct}{}{Cputime}{Median}{10.4909821335}%
\StoreBenchExecResult{ReducedEvalSlDfS}{ImcHardSafe}{Correct}{}{Cputime}{Min}{4.244829399}%
\StoreBenchExecResult{ReducedEvalSlDfS}{ImcHardSafe}{Correct}{}{Cputime}{Max}{864.226558973}%
\StoreBenchExecResult{ReducedEvalSlDfS}{ImcHardSafe}{Correct}{}{Cputime}{Stdev}{130.8632543775968979837164436}%
\StoreBenchExecResult{ReducedEvalSlDfS}{ImcHardSafe}{Correct}{}{Walltime}{}{25731.8120062216184775}%
\StoreBenchExecResult{ReducedEvalSlDfS}{ImcHardSafe}{Correct}{}{Walltime}{Avg}{48.18691386932887355337078652}%
\StoreBenchExecResult{ReducedEvalSlDfS}{ImcHardSafe}{Correct}{}{Walltime}{Median}{4.3605187805369495}%
\StoreBenchExecResult{ReducedEvalSlDfS}{ImcHardSafe}{Correct}{}{Walltime}{Min}{1.5823553358204663}%
\StoreBenchExecResult{ReducedEvalSlDfS}{ImcHardSafe}{Correct}{}{Walltime}{Max}{853.9676453699358}%
\StoreBenchExecResult{ReducedEvalSlDfS}{ImcHardSafe}{Correct}{}{Walltime}{Stdev}{127.1558985613344855249871413}%
\StoreBenchExecResult{ReducedEvalSlDfS}{ImcHardSafe}{Correct}{True}{Count}{}{534}%
\StoreBenchExecResult{ReducedEvalSlDfS}{ImcHardSafe}{Correct}{True}{Cputime}{}{29969.754082084}%
\StoreBenchExecResult{ReducedEvalSlDfS}{ImcHardSafe}{Correct}{True}{Cputime}{Avg}{56.12313498517602996254681648}%
\StoreBenchExecResult{ReducedEvalSlDfS}{ImcHardSafe}{Correct}{True}{Cputime}{Median}{10.4909821335}%
\StoreBenchExecResult{ReducedEvalSlDfS}{ImcHardSafe}{Correct}{True}{Cputime}{Min}{4.244829399}%
\StoreBenchExecResult{ReducedEvalSlDfS}{ImcHardSafe}{Correct}{True}{Cputime}{Max}{864.226558973}%
\StoreBenchExecResult{ReducedEvalSlDfS}{ImcHardSafe}{Correct}{True}{Cputime}{Stdev}{130.8632543775968979837164436}%
\StoreBenchExecResult{ReducedEvalSlDfS}{ImcHardSafe}{Correct}{True}{Walltime}{}{25731.8120062216184775}%
\StoreBenchExecResult{ReducedEvalSlDfS}{ImcHardSafe}{Correct}{True}{Walltime}{Avg}{48.18691386932887355337078652}%
\StoreBenchExecResult{ReducedEvalSlDfS}{ImcHardSafe}{Correct}{True}{Walltime}{Median}{4.3605187805369495}%
\StoreBenchExecResult{ReducedEvalSlDfS}{ImcHardSafe}{Correct}{True}{Walltime}{Min}{1.5823553358204663}%
\StoreBenchExecResult{ReducedEvalSlDfS}{ImcHardSafe}{Correct}{True}{Walltime}{Max}{853.9676453699358}%
\StoreBenchExecResult{ReducedEvalSlDfS}{ImcHardSafe}{Correct}{True}{Walltime}{Stdev}{127.1558985613344855249871413}%

\StoreBenchExecResult{ReducedEvalSlDfS}{ImcHardSafe}{Error}{}{Count}{}{336}%
\StoreBenchExecResult{ReducedEvalSlDfS}{ImcHardSafe}{Error}{}{Cputime}{}{270685.258703932}%
\StoreBenchExecResult{ReducedEvalSlDfS}{ImcHardSafe}{Error}{}{Cputime}{Avg}{805.6108889997976190476190476}%
\StoreBenchExecResult{ReducedEvalSlDfS}{ImcHardSafe}{Error}{}{Cputime}{Median}{902.0707596905}%
\StoreBenchExecResult{ReducedEvalSlDfS}{ImcHardSafe}{Error}{}{Cputime}{Min}{4.409323771}%
\StoreBenchExecResult{ReducedEvalSlDfS}{ImcHardSafe}{Error}{}{Cputime}{Max}{913.342992329}%
\StoreBenchExecResult{ReducedEvalSlDfS}{ImcHardSafe}{Error}{}{Cputime}{Stdev}{274.9388871498509733635371654}%
\StoreBenchExecResult{ReducedEvalSlDfS}{ImcHardSafe}{Error}{}{Walltime}{}{266250.8318372617945113}%
\StoreBenchExecResult{ReducedEvalSlDfS}{ImcHardSafe}{Error}{}{Walltime}{Avg}{792.4131899918505789026785714}%
\StoreBenchExecResult{ReducedEvalSlDfS}{ImcHardSafe}{Error}{}{Walltime}{Median}{888.61855757853485}%
\StoreBenchExecResult{ReducedEvalSlDfS}{ImcHardSafe}{Error}{}{Walltime}{Min}{1.7914597410708666}%
\StoreBenchExecResult{ReducedEvalSlDfS}{ImcHardSafe}{Error}{}{Walltime}{Max}{906.2711261240765}%
\StoreBenchExecResult{ReducedEvalSlDfS}{ImcHardSafe}{Error}{}{Walltime}{Stdev}{271.6139495837103277595000264}%
\StoreBenchExecResult{ReducedEvalSlDfS}{ImcHardSafe}{Error}{Error}{Count}{}{38}%
\StoreBenchExecResult{ReducedEvalSlDfS}{ImcHardSafe}{Error}{Error}{Cputime}{}{1555.597390537}%
\StoreBenchExecResult{ReducedEvalSlDfS}{ImcHardSafe}{Error}{Error}{Cputime}{Avg}{40.93677343518421052631578947}%
\StoreBenchExecResult{ReducedEvalSlDfS}{ImcHardSafe}{Error}{Error}{Cputime}{Median}{9.9339096745}%
\StoreBenchExecResult{ReducedEvalSlDfS}{ImcHardSafe}{Error}{Error}{Cputime}{Min}{4.409323771}%
\StoreBenchExecResult{ReducedEvalSlDfS}{ImcHardSafe}{Error}{Error}{Cputime}{Max}{548.373546117}%
\StoreBenchExecResult{ReducedEvalSlDfS}{ImcHardSafe}{Error}{Error}{Cputime}{Stdev}{94.23070954576991667281082107}%
\StoreBenchExecResult{ReducedEvalSlDfS}{ImcHardSafe}{Error}{Error}{Walltime}{}{1417.4675342177506113}%
\StoreBenchExecResult{ReducedEvalSlDfS}{ImcHardSafe}{Error}{Error}{Walltime}{Avg}{37.30177721625659503421052632}%
\StoreBenchExecResult{ReducedEvalSlDfS}{ImcHardSafe}{Error}{Error}{Walltime}{Median}{5.8008855393854905}%
\StoreBenchExecResult{ReducedEvalSlDfS}{ImcHardSafe}{Error}{Error}{Walltime}{Min}{1.7914597410708666}%
\StoreBenchExecResult{ReducedEvalSlDfS}{ImcHardSafe}{Error}{Error}{Walltime}{Max}{544.9657988150138}%
\StoreBenchExecResult{ReducedEvalSlDfS}{ImcHardSafe}{Error}{Error}{Walltime}{Stdev}{94.34067085453063476191626269}%
\StoreBenchExecResult{ReducedEvalSlDfS}{ImcHardSafe}{Error}{SegmentationFault}{Count}{}{1}%
\StoreBenchExecResult{ReducedEvalSlDfS}{ImcHardSafe}{Error}{SegmentationFault}{Cputime}{}{829.112224541}%
\StoreBenchExecResult{ReducedEvalSlDfS}{ImcHardSafe}{Error}{SegmentationFault}{Cputime}{Avg}{829.112224541}%
\StoreBenchExecResult{ReducedEvalSlDfS}{ImcHardSafe}{Error}{SegmentationFault}{Cputime}{Median}{829.112224541}%
\StoreBenchExecResult{ReducedEvalSlDfS}{ImcHardSafe}{Error}{SegmentationFault}{Cputime}{Min}{829.112224541}%
\StoreBenchExecResult{ReducedEvalSlDfS}{ImcHardSafe}{Error}{SegmentationFault}{Cputime}{Max}{829.112224541}%
\StoreBenchExecResult{ReducedEvalSlDfS}{ImcHardSafe}{Error}{SegmentationFault}{Cputime}{Stdev}{0E-14}%
\StoreBenchExecResult{ReducedEvalSlDfS}{ImcHardSafe}{Error}{SegmentationFault}{Walltime}{}{822.4779165070504}%
\StoreBenchExecResult{ReducedEvalSlDfS}{ImcHardSafe}{Error}{SegmentationFault}{Walltime}{Avg}{822.4779165070504}%
\StoreBenchExecResult{ReducedEvalSlDfS}{ImcHardSafe}{Error}{SegmentationFault}{Walltime}{Median}{822.4779165070504}%
\StoreBenchExecResult{ReducedEvalSlDfS}{ImcHardSafe}{Error}{SegmentationFault}{Walltime}{Min}{822.4779165070504}%
\StoreBenchExecResult{ReducedEvalSlDfS}{ImcHardSafe}{Error}{SegmentationFault}{Walltime}{Max}{822.4779165070504}%
\StoreBenchExecResult{ReducedEvalSlDfS}{ImcHardSafe}{Error}{SegmentationFault}{Walltime}{Stdev}{0E-14}%
\StoreBenchExecResult{ReducedEvalSlDfS}{ImcHardSafe}{Error}{Timeout}{Count}{}{297}%
\StoreBenchExecResult{ReducedEvalSlDfS}{ImcHardSafe}{Error}{Timeout}{Cputime}{}{268300.549088854}%
\StoreBenchExecResult{ReducedEvalSlDfS}{ImcHardSafe}{Error}{Timeout}{Cputime}{Avg}{903.3688521510235690235690236}%
\StoreBenchExecResult{ReducedEvalSlDfS}{ImcHardSafe}{Error}{Timeout}{Cputime}{Median}{902.172749168}%
\StoreBenchExecResult{ReducedEvalSlDfS}{ImcHardSafe}{Error}{Timeout}{Cputime}{Min}{901.355122778}%
\StoreBenchExecResult{ReducedEvalSlDfS}{ImcHardSafe}{Error}{Timeout}{Cputime}{Max}{913.342992329}%
\StoreBenchExecResult{ReducedEvalSlDfS}{ImcHardSafe}{Error}{Timeout}{Cputime}{Stdev}{3.078114049980596698323377590}%
\StoreBenchExecResult{ReducedEvalSlDfS}{ImcHardSafe}{Error}{Timeout}{Walltime}{}{264010.8863865369935}%
\StoreBenchExecResult{ReducedEvalSlDfS}{ImcHardSafe}{Error}{Timeout}{Walltime}{Avg}{888.9255433890134461279461279}%
\StoreBenchExecResult{ReducedEvalSlDfS}{ImcHardSafe}{Error}{Timeout}{Walltime}{Median}{889.50945615815}%
\StoreBenchExecResult{ReducedEvalSlDfS}{ImcHardSafe}{Error}{Timeout}{Walltime}{Min}{871.6039576940238}%
\StoreBenchExecResult{ReducedEvalSlDfS}{ImcHardSafe}{Error}{Timeout}{Walltime}{Max}{906.2711261240765}%
\StoreBenchExecResult{ReducedEvalSlDfS}{ImcHardSafe}{Error}{Timeout}{Walltime}{Stdev}{7.156369104823719523263748330}%
\ifdefined\ReducedEvalSlDfSImcHardSafeTotalCount\else\edef\ReducedEvalSlDfSImcHardSafeTotalCount{0}\fi
\ifdefined\ReducedEvalSlDfSImcHardSafeCorrectCount\else\edef\ReducedEvalSlDfSImcHardSafeCorrectCount{0}\fi
\ifdefined\ReducedEvalSlDfSImcHardSafeCorrectTrueCount\else\edef\ReducedEvalSlDfSImcHardSafeCorrectTrueCount{0}\fi
\ifdefined\ReducedEvalSlDfSImcHardSafeCorrectFalseCount\else\edef\ReducedEvalSlDfSImcHardSafeCorrectFalseCount{0}\fi
\ifdefined\ReducedEvalSlDfSImcHardSafeWrongTrueCount\else\edef\ReducedEvalSlDfSImcHardSafeWrongTrueCount{0}\fi
\ifdefined\ReducedEvalSlDfSImcHardSafeWrongFalseCount\else\edef\ReducedEvalSlDfSImcHardSafeWrongFalseCount{0}\fi
\ifdefined\ReducedEvalSlDfSImcHardSafeErrorTimeoutCount\else\edef\ReducedEvalSlDfSImcHardSafeErrorTimeoutCount{0}\fi
\ifdefined\ReducedEvalSlDfSImcHardSafeErrorOutOfMemoryCount\else\edef\ReducedEvalSlDfSImcHardSafeErrorOutOfMemoryCount{0}\fi
\ifdefined\ReducedEvalSlDfSImcHardSafeCorrectCputime\else\edef\ReducedEvalSlDfSImcHardSafeCorrectCputime{0}\fi
\ifdefined\ReducedEvalSlDfSImcHardSafeCorrectCputimeAvg\else\edef\ReducedEvalSlDfSImcHardSafeCorrectCputimeAvg{None}\fi
\ifdefined\ReducedEvalSlDfSImcHardSafeCorrectWalltime\else\edef\ReducedEvalSlDfSImcHardSafeCorrectWalltime{0}\fi
\ifdefined\ReducedEvalSlDfSImcHardSafeCorrectWalltimeAvg\else\edef\ReducedEvalSlDfSImcHardSafeCorrectWalltimeAvg{None}\fi
\edef\ReducedEvalSlDfSImcHardSafeErrorOtherInconclusiveCount{\the\numexpr \ReducedEvalSlDfSImcHardSafeTotalCount - \ReducedEvalSlDfSImcHardSafeCorrectCount - \ReducedEvalSlDfSImcHardSafeWrongTrueCount - \ReducedEvalSlDfSImcHardSafeWrongFalseCount - \ReducedEvalSlDfSImcHardSafeErrorTimeoutCount - \ReducedEvalSlDfSImcHardSafeErrorOutOfMemoryCount \relax}
\providecommand\StoreBenchExecResult[7]{\expandafter\newcommand\csname#1#2#3#4#5#6\endcsname{#7}}%
\StoreBenchExecResult{ReducedEvalSlDfS}{ImcIgndfiHardSafe}{Total}{}{Count}{}{870}%
\StoreBenchExecResult{ReducedEvalSlDfS}{ImcIgndfiHardSafe}{Total}{}{Cputime}{}{309303.419607832}%
\StoreBenchExecResult{ReducedEvalSlDfS}{ImcIgndfiHardSafe}{Total}{}{Cputime}{Avg}{355.5211719630252873563218391}%
\StoreBenchExecResult{ReducedEvalSlDfS}{ImcIgndfiHardSafe}{Total}{}{Cputime}{Median}{77.4893762715}%
\StoreBenchExecResult{ReducedEvalSlDfS}{ImcIgndfiHardSafe}{Total}{}{Cputime}{Min}{4.264615926}%
\StoreBenchExecResult{ReducedEvalSlDfS}{ImcIgndfiHardSafe}{Total}{}{Cputime}{Max}{913.163518055}%
\StoreBenchExecResult{ReducedEvalSlDfS}{ImcIgndfiHardSafe}{Total}{}{Cputime}{Stdev}{404.9774748279143765099713063}%
\StoreBenchExecResult{ReducedEvalSlDfS}{ImcIgndfiHardSafe}{Total}{}{Walltime}{}{264867.4108252616131332}%
\StoreBenchExecResult{ReducedEvalSlDfS}{ImcIgndfiHardSafe}{Total}{}{Walltime}{Avg}{304.4452997991512794634482759}%
\StoreBenchExecResult{ReducedEvalSlDfS}{ImcIgndfiHardSafe}{Total}{}{Walltime}{Median}{28.55570562905632}%
\StoreBenchExecResult{ReducedEvalSlDfS}{ImcIgndfiHardSafe}{Total}{}{Walltime}{Min}{1.6341450288891792}%
\StoreBenchExecResult{ReducedEvalSlDfS}{ImcIgndfiHardSafe}{Total}{}{Walltime}{Max}{891.9258381379768}%
\StoreBenchExecResult{ReducedEvalSlDfS}{ImcIgndfiHardSafe}{Total}{}{Walltime}{Stdev}{368.9925792296065299142514261}%
\StoreBenchExecResult{ReducedEvalSlDfS}{ImcIgndfiHardSafe}{Correct}{}{Count}{}{544}%
\StoreBenchExecResult{ReducedEvalSlDfS}{ImcIgndfiHardSafe}{Correct}{}{Cputime}{}{43934.712608342}%
\StoreBenchExecResult{ReducedEvalSlDfS}{ImcIgndfiHardSafe}{Correct}{}{Cputime}{Avg}{80.76233935356985294117647059}%
\StoreBenchExecResult{ReducedEvalSlDfS}{ImcIgndfiHardSafe}{Correct}{}{Cputime}{Median}{14.818945497}%
\StoreBenchExecResult{ReducedEvalSlDfS}{ImcIgndfiHardSafe}{Correct}{}{Cputime}{Min}{4.264615926}%
\StoreBenchExecResult{ReducedEvalSlDfS}{ImcIgndfiHardSafe}{Correct}{}{Cputime}{Max}{887.278560398}%
\StoreBenchExecResult{ReducedEvalSlDfS}{ImcIgndfiHardSafe}{Correct}{}{Cputime}{Stdev}{157.8846595319528594670621148}%
\StoreBenchExecResult{ReducedEvalSlDfS}{ImcIgndfiHardSafe}{Correct}{}{Walltime}{}{28336.6891551057340462}%
\StoreBenchExecResult{ReducedEvalSlDfS}{ImcIgndfiHardSafe}{Correct}{}{Walltime}{Avg}{52.08950212335612876139705882}%
\StoreBenchExecResult{ReducedEvalSlDfS}{ImcIgndfiHardSafe}{Correct}{}{Walltime}{Median}{5.3046461464837195}%
\StoreBenchExecResult{ReducedEvalSlDfS}{ImcIgndfiHardSafe}{Correct}{}{Walltime}{Min}{1.6341450288891792}%
\StoreBenchExecResult{ReducedEvalSlDfS}{ImcIgndfiHardSafe}{Correct}{}{Walltime}{Max}{788.5136345371138}%
\StoreBenchExecResult{ReducedEvalSlDfS}{ImcIgndfiHardSafe}{Correct}{}{Walltime}{Stdev}{131.5311616757876054953217169}%
\StoreBenchExecResult{ReducedEvalSlDfS}{ImcIgndfiHardSafe}{Correct}{True}{Count}{}{544}%
\StoreBenchExecResult{ReducedEvalSlDfS}{ImcIgndfiHardSafe}{Correct}{True}{Cputime}{}{43934.712608342}%
\StoreBenchExecResult{ReducedEvalSlDfS}{ImcIgndfiHardSafe}{Correct}{True}{Cputime}{Avg}{80.76233935356985294117647059}%
\StoreBenchExecResult{ReducedEvalSlDfS}{ImcIgndfiHardSafe}{Correct}{True}{Cputime}{Median}{14.818945497}%
\StoreBenchExecResult{ReducedEvalSlDfS}{ImcIgndfiHardSafe}{Correct}{True}{Cputime}{Min}{4.264615926}%
\StoreBenchExecResult{ReducedEvalSlDfS}{ImcIgndfiHardSafe}{Correct}{True}{Cputime}{Max}{887.278560398}%
\StoreBenchExecResult{ReducedEvalSlDfS}{ImcIgndfiHardSafe}{Correct}{True}{Cputime}{Stdev}{157.8846595319528594670621148}%
\StoreBenchExecResult{ReducedEvalSlDfS}{ImcIgndfiHardSafe}{Correct}{True}{Walltime}{}{28336.6891551057340462}%
\StoreBenchExecResult{ReducedEvalSlDfS}{ImcIgndfiHardSafe}{Correct}{True}{Walltime}{Avg}{52.08950212335612876139705882}%
\StoreBenchExecResult{ReducedEvalSlDfS}{ImcIgndfiHardSafe}{Correct}{True}{Walltime}{Median}{5.3046461464837195}%
\StoreBenchExecResult{ReducedEvalSlDfS}{ImcIgndfiHardSafe}{Correct}{True}{Walltime}{Min}{1.6341450288891792}%
\StoreBenchExecResult{ReducedEvalSlDfS}{ImcIgndfiHardSafe}{Correct}{True}{Walltime}{Max}{788.5136345371138}%
\StoreBenchExecResult{ReducedEvalSlDfS}{ImcIgndfiHardSafe}{Correct}{True}{Walltime}{Stdev}{131.5311616757876054953217169}%

\StoreBenchExecResult{ReducedEvalSlDfS}{ImcIgndfiHardSafe}{Error}{}{Count}{}{326}%
\StoreBenchExecResult{ReducedEvalSlDfS}{ImcIgndfiHardSafe}{Error}{}{Cputime}{}{265368.706999490}%
\StoreBenchExecResult{ReducedEvalSlDfS}{ImcIgndfiHardSafe}{Error}{}{Cputime}{Avg}{814.0144386487423312883435583}%
\StoreBenchExecResult{ReducedEvalSlDfS}{ImcIgndfiHardSafe}{Error}{}{Cputime}{Median}{901.9240303895}%
\StoreBenchExecResult{ReducedEvalSlDfS}{ImcIgndfiHardSafe}{Error}{}{Cputime}{Min}{11.584360526}%
\StoreBenchExecResult{ReducedEvalSlDfS}{ImcIgndfiHardSafe}{Error}{}{Cputime}{Max}{913.163518055}%
\StoreBenchExecResult{ReducedEvalSlDfS}{ImcIgndfiHardSafe}{Error}{}{Cputime}{Stdev}{244.7420465150263852720549357}%
\StoreBenchExecResult{ReducedEvalSlDfS}{ImcIgndfiHardSafe}{Error}{}{Walltime}{}{236530.721670155879087}%
\StoreBenchExecResult{ReducedEvalSlDfS}{ImcIgndfiHardSafe}{Error}{}{Walltime}{Avg}{725.5543609513984021073619632}%
\StoreBenchExecResult{ReducedEvalSlDfS}{ImcIgndfiHardSafe}{Error}{}{Walltime}{Median}{804.56101240194405}%
\StoreBenchExecResult{ReducedEvalSlDfS}{ImcIgndfiHardSafe}{Error}{}{Walltime}{Min}{6.804908343125135}%
\StoreBenchExecResult{ReducedEvalSlDfS}{ImcIgndfiHardSafe}{Error}{}{Walltime}{Max}{891.9258381379768}%
\StoreBenchExecResult{ReducedEvalSlDfS}{ImcIgndfiHardSafe}{Error}{}{Walltime}{Stdev}{225.5841784823372628832699072}%
\StoreBenchExecResult{ReducedEvalSlDfS}{ImcIgndfiHardSafe}{Error}{Error}{Count}{}{38}%
\StoreBenchExecResult{ReducedEvalSlDfS}{ImcIgndfiHardSafe}{Error}{Error}{Cputime}{}{5635.882824445}%
\StoreBenchExecResult{ReducedEvalSlDfS}{ImcIgndfiHardSafe}{Error}{Error}{Cputime}{Avg}{148.3127059064473684210526316}%
\StoreBenchExecResult{ReducedEvalSlDfS}{ImcIgndfiHardSafe}{Error}{Error}{Cputime}{Median}{153.704890649}%
\StoreBenchExecResult{ReducedEvalSlDfS}{ImcIgndfiHardSafe}{Error}{Error}{Cputime}{Min}{11.584360526}%
\StoreBenchExecResult{ReducedEvalSlDfS}{ImcIgndfiHardSafe}{Error}{Error}{Cputime}{Max}{634.966942936}%
\StoreBenchExecResult{ReducedEvalSlDfS}{ImcIgndfiHardSafe}{Error}{Error}{Cputime}{Stdev}{96.78110833424672954374417821}%
\StoreBenchExecResult{ReducedEvalSlDfS}{ImcIgndfiHardSafe}{Error}{Error}{Walltime}{}{4242.370571961160787}%
\StoreBenchExecResult{ReducedEvalSlDfS}{ImcIgndfiHardSafe}{Error}{Error}{Walltime}{Avg}{111.6413308410831786052631579}%
\StoreBenchExecResult{ReducedEvalSlDfS}{ImcIgndfiHardSafe}{Error}{Error}{Walltime}{Median}{123.997571681975385}%
\StoreBenchExecResult{ReducedEvalSlDfS}{ImcIgndfiHardSafe}{Error}{Error}{Walltime}{Min}{6.804908343125135}%
\StoreBenchExecResult{ReducedEvalSlDfS}{ImcIgndfiHardSafe}{Error}{Error}{Walltime}{Max}{548.886447101133}%
\StoreBenchExecResult{ReducedEvalSlDfS}{ImcIgndfiHardSafe}{Error}{Error}{Walltime}{Stdev}{86.54235479818014852592115561}%
\StoreBenchExecResult{ReducedEvalSlDfS}{ImcIgndfiHardSafe}{Error}{SegmentationFault}{Count}{}{1}%
\StoreBenchExecResult{ReducedEvalSlDfS}{ImcIgndfiHardSafe}{Error}{SegmentationFault}{Cputime}{}{575.70592272}%
\StoreBenchExecResult{ReducedEvalSlDfS}{ImcIgndfiHardSafe}{Error}{SegmentationFault}{Cputime}{Avg}{575.70592272}%
\StoreBenchExecResult{ReducedEvalSlDfS}{ImcIgndfiHardSafe}{Error}{SegmentationFault}{Cputime}{Median}{575.70592272}%
\StoreBenchExecResult{ReducedEvalSlDfS}{ImcIgndfiHardSafe}{Error}{SegmentationFault}{Cputime}{Min}{575.70592272}%
\StoreBenchExecResult{ReducedEvalSlDfS}{ImcIgndfiHardSafe}{Error}{SegmentationFault}{Cputime}{Max}{575.70592272}%
\StoreBenchExecResult{ReducedEvalSlDfS}{ImcIgndfiHardSafe}{Error}{SegmentationFault}{Cputime}{Stdev}{0E-14}%
\StoreBenchExecResult{ReducedEvalSlDfS}{ImcIgndfiHardSafe}{Error}{SegmentationFault}{Walltime}{}{570.0703720431775}%
\StoreBenchExecResult{ReducedEvalSlDfS}{ImcIgndfiHardSafe}{Error}{SegmentationFault}{Walltime}{Avg}{570.0703720431775}%
\StoreBenchExecResult{ReducedEvalSlDfS}{ImcIgndfiHardSafe}{Error}{SegmentationFault}{Walltime}{Median}{570.0703720431775}%
\StoreBenchExecResult{ReducedEvalSlDfS}{ImcIgndfiHardSafe}{Error}{SegmentationFault}{Walltime}{Min}{570.0703720431775}%
\StoreBenchExecResult{ReducedEvalSlDfS}{ImcIgndfiHardSafe}{Error}{SegmentationFault}{Walltime}{Max}{570.0703720431775}%
\StoreBenchExecResult{ReducedEvalSlDfS}{ImcIgndfiHardSafe}{Error}{SegmentationFault}{Walltime}{Stdev}{0E-14}%
\StoreBenchExecResult{ReducedEvalSlDfS}{ImcIgndfiHardSafe}{Error}{Timeout}{Count}{}{287}%
\StoreBenchExecResult{ReducedEvalSlDfS}{ImcIgndfiHardSafe}{Error}{Timeout}{Cputime}{}{259157.118252325}%
\StoreBenchExecResult{ReducedEvalSlDfS}{ImcIgndfiHardSafe}{Error}{Timeout}{Cputime}{Avg}{902.9864747467770034843205575}%
\StoreBenchExecResult{ReducedEvalSlDfS}{ImcIgndfiHardSafe}{Error}{Timeout}{Cputime}{Median}{902.004197987}%
\StoreBenchExecResult{ReducedEvalSlDfS}{ImcIgndfiHardSafe}{Error}{Timeout}{Cputime}{Min}{901.133121685}%
\StoreBenchExecResult{ReducedEvalSlDfS}{ImcIgndfiHardSafe}{Error}{Timeout}{Cputime}{Max}{913.163518055}%
\StoreBenchExecResult{ReducedEvalSlDfS}{ImcIgndfiHardSafe}{Error}{Timeout}{Cputime}{Stdev}{2.838361287366368748485279314}%
\StoreBenchExecResult{ReducedEvalSlDfS}{ImcIgndfiHardSafe}{Error}{Timeout}{Walltime}{}{231718.2807261515408}%
\StoreBenchExecResult{ReducedEvalSlDfS}{ImcIgndfiHardSafe}{Error}{Timeout}{Walltime}{Avg}{807.3807690806673895470383275}%
\StoreBenchExecResult{ReducedEvalSlDfS}{ImcIgndfiHardSafe}{Error}{Timeout}{Walltime}{Median}{805.3098062351346}%
\StoreBenchExecResult{ReducedEvalSlDfS}{ImcIgndfiHardSafe}{Error}{Timeout}{Walltime}{Min}{793.4792728479952}%
\StoreBenchExecResult{ReducedEvalSlDfS}{ImcIgndfiHardSafe}{Error}{Timeout}{Walltime}{Max}{891.9258381379768}%
\StoreBenchExecResult{ReducedEvalSlDfS}{ImcIgndfiHardSafe}{Error}{Timeout}{Walltime}{Stdev}{11.41015583901474103978136358}%
\ifdefined\ReducedEvalSlDfSImcIgndfiHardSafeTotalCount\else\edef\ReducedEvalSlDfSImcIgndfiHardSafeTotalCount{0}\fi
\ifdefined\ReducedEvalSlDfSImcIgndfiHardSafeCorrectCount\else\edef\ReducedEvalSlDfSImcIgndfiHardSafeCorrectCount{0}\fi
\ifdefined\ReducedEvalSlDfSImcIgndfiHardSafeCorrectTrueCount\else\edef\ReducedEvalSlDfSImcIgndfiHardSafeCorrectTrueCount{0}\fi
\ifdefined\ReducedEvalSlDfSImcIgndfiHardSafeCorrectFalseCount\else\edef\ReducedEvalSlDfSImcIgndfiHardSafeCorrectFalseCount{0}\fi
\ifdefined\ReducedEvalSlDfSImcIgndfiHardSafeWrongTrueCount\else\edef\ReducedEvalSlDfSImcIgndfiHardSafeWrongTrueCount{0}\fi
\ifdefined\ReducedEvalSlDfSImcIgndfiHardSafeWrongFalseCount\else\edef\ReducedEvalSlDfSImcIgndfiHardSafeWrongFalseCount{0}\fi
\ifdefined\ReducedEvalSlDfSImcIgndfiHardSafeErrorTimeoutCount\else\edef\ReducedEvalSlDfSImcIgndfiHardSafeErrorTimeoutCount{0}\fi
\ifdefined\ReducedEvalSlDfSImcIgndfiHardSafeErrorOutOfMemoryCount\else\edef\ReducedEvalSlDfSImcIgndfiHardSafeErrorOutOfMemoryCount{0}\fi
\ifdefined\ReducedEvalSlDfSImcIgndfiHardSafeCorrectCputime\else\edef\ReducedEvalSlDfSImcIgndfiHardSafeCorrectCputime{0}\fi
\ifdefined\ReducedEvalSlDfSImcIgndfiHardSafeCorrectCputimeAvg\else\edef\ReducedEvalSlDfSImcIgndfiHardSafeCorrectCputimeAvg{None}\fi
\ifdefined\ReducedEvalSlDfSImcIgndfiHardSafeCorrectWalltime\else\edef\ReducedEvalSlDfSImcIgndfiHardSafeCorrectWalltime{0}\fi
\ifdefined\ReducedEvalSlDfSImcIgndfiHardSafeCorrectWalltimeAvg\else\edef\ReducedEvalSlDfSImcIgndfiHardSafeCorrectWalltimeAvg{None}\fi
\edef\ReducedEvalSlDfSImcIgndfiHardSafeErrorOtherInconclusiveCount{\the\numexpr \ReducedEvalSlDfSImcIgndfiHardSafeTotalCount - \ReducedEvalSlDfSImcIgndfiHardSafeCorrectCount - \ReducedEvalSlDfSImcIgndfiHardSafeWrongTrueCount - \ReducedEvalSlDfSImcIgndfiHardSafeWrongFalseCount - \ReducedEvalSlDfSImcIgndfiHardSafeErrorTimeoutCount - \ReducedEvalSlDfSImcIgndfiHardSafeErrorOutOfMemoryCount \relax}
\providecommand\StoreBenchExecResult[7]{\expandafter\newcommand\csname#1#2#3#4#5#6\endcsname{#7}}%
\StoreBenchExecResult{ReducedEvalSlDfS}{ImcIgndffHardSafe}{Total}{}{Count}{}{870}%
\StoreBenchExecResult{ReducedEvalSlDfS}{ImcIgndffHardSafe}{Total}{}{Cputime}{}{314816.319374998}%
\StoreBenchExecResult{ReducedEvalSlDfS}{ImcIgndffHardSafe}{Total}{}{Cputime}{Avg}{361.8578383620666666666666667}%
\StoreBenchExecResult{ReducedEvalSlDfS}{ImcIgndffHardSafe}{Total}{}{Cputime}{Median}{90.7977464765}%
\StoreBenchExecResult{ReducedEvalSlDfS}{ImcIgndffHardSafe}{Total}{}{Cputime}{Min}{4.389563423}%
\StoreBenchExecResult{ReducedEvalSlDfS}{ImcIgndffHardSafe}{Total}{}{Cputime}{Max}{960.844313854}%
\StoreBenchExecResult{ReducedEvalSlDfS}{ImcIgndffHardSafe}{Total}{}{Cputime}{Stdev}{406.5686745237305120621645820}%
\StoreBenchExecResult{ReducedEvalSlDfS}{ImcIgndffHardSafe}{Total}{}{Walltime}{}{270077.5419443259011914}%
\StoreBenchExecResult{ReducedEvalSlDfS}{ImcIgndffHardSafe}{Total}{}{Walltime}{Avg}{310.4339562578458634383908046}%
\StoreBenchExecResult{ReducedEvalSlDfS}{ImcIgndffHardSafe}{Total}{}{Walltime}{Median}{33.4916003005346275}%
\StoreBenchExecResult{ReducedEvalSlDfS}{ImcIgndffHardSafe}{Total}{}{Walltime}{Min}{1.6207367829047143}%
\StoreBenchExecResult{ReducedEvalSlDfS}{ImcIgndffHardSafe}{Total}{}{Walltime}{Max}{898.3665967129637}%
\StoreBenchExecResult{ReducedEvalSlDfS}{ImcIgndffHardSafe}{Total}{}{Walltime}{Stdev}{370.8956363130227859871072612}%
\StoreBenchExecResult{ReducedEvalSlDfS}{ImcIgndffHardSafe}{Correct}{}{Count}{}{536}%
\StoreBenchExecResult{ReducedEvalSlDfS}{ImcIgndffHardSafe}{Correct}{}{Cputime}{}{42135.127288814}%
\StoreBenchExecResult{ReducedEvalSlDfS}{ImcIgndffHardSafe}{Correct}{}{Cputime}{Avg}{78.61031210599626865671641791}%
\StoreBenchExecResult{ReducedEvalSlDfS}{ImcIgndffHardSafe}{Correct}{}{Cputime}{Median}{15.104333281}%
\StoreBenchExecResult{ReducedEvalSlDfS}{ImcIgndffHardSafe}{Correct}{}{Cputime}{Min}{4.389563423}%
\StoreBenchExecResult{ReducedEvalSlDfS}{ImcIgndffHardSafe}{Correct}{}{Cputime}{Max}{886.738056765}%
\StoreBenchExecResult{ReducedEvalSlDfS}{ImcIgndffHardSafe}{Correct}{}{Cputime}{Stdev}{150.9258400024836893127959003}%
\StoreBenchExecResult{ReducedEvalSlDfS}{ImcIgndffHardSafe}{Correct}{}{Walltime}{}{26840.8293929656500329}%
\StoreBenchExecResult{ReducedEvalSlDfS}{ImcIgndffHardSafe}{Correct}{}{Walltime}{Avg}{50.07617424060755603152985075}%
\StoreBenchExecResult{ReducedEvalSlDfS}{ImcIgndffHardSafe}{Correct}{}{Walltime}{Median}{5.4172254455043005}%
\StoreBenchExecResult{ReducedEvalSlDfS}{ImcIgndffHardSafe}{Correct}{}{Walltime}{Min}{1.6207367829047143}%
\StoreBenchExecResult{ReducedEvalSlDfS}{ImcIgndffHardSafe}{Correct}{}{Walltime}{Max}{790.9702519162092}%
\StoreBenchExecResult{ReducedEvalSlDfS}{ImcIgndffHardSafe}{Correct}{}{Walltime}{Stdev}{124.8341274320198760295557334}%
\StoreBenchExecResult{ReducedEvalSlDfS}{ImcIgndffHardSafe}{Correct}{True}{Count}{}{536}%
\StoreBenchExecResult{ReducedEvalSlDfS}{ImcIgndffHardSafe}{Correct}{True}{Cputime}{}{42135.127288814}%
\StoreBenchExecResult{ReducedEvalSlDfS}{ImcIgndffHardSafe}{Correct}{True}{Cputime}{Avg}{78.61031210599626865671641791}%
\StoreBenchExecResult{ReducedEvalSlDfS}{ImcIgndffHardSafe}{Correct}{True}{Cputime}{Median}{15.104333281}%
\StoreBenchExecResult{ReducedEvalSlDfS}{ImcIgndffHardSafe}{Correct}{True}{Cputime}{Min}{4.389563423}%
\StoreBenchExecResult{ReducedEvalSlDfS}{ImcIgndffHardSafe}{Correct}{True}{Cputime}{Max}{886.738056765}%
\StoreBenchExecResult{ReducedEvalSlDfS}{ImcIgndffHardSafe}{Correct}{True}{Cputime}{Stdev}{150.9258400024836893127959003}%
\StoreBenchExecResult{ReducedEvalSlDfS}{ImcIgndffHardSafe}{Correct}{True}{Walltime}{}{26840.8293929656500329}%
\StoreBenchExecResult{ReducedEvalSlDfS}{ImcIgndffHardSafe}{Correct}{True}{Walltime}{Avg}{50.07617424060755603152985075}%
\StoreBenchExecResult{ReducedEvalSlDfS}{ImcIgndffHardSafe}{Correct}{True}{Walltime}{Median}{5.4172254455043005}%
\StoreBenchExecResult{ReducedEvalSlDfS}{ImcIgndffHardSafe}{Correct}{True}{Walltime}{Min}{1.6207367829047143}%
\StoreBenchExecResult{ReducedEvalSlDfS}{ImcIgndffHardSafe}{Correct}{True}{Walltime}{Max}{790.9702519162092}%
\StoreBenchExecResult{ReducedEvalSlDfS}{ImcIgndffHardSafe}{Correct}{True}{Walltime}{Stdev}{124.8341274320198760295557334}%

\StoreBenchExecResult{ReducedEvalSlDfS}{ImcIgndffHardSafe}{Error}{}{Count}{}{334}%
\StoreBenchExecResult{ReducedEvalSlDfS}{ImcIgndffHardSafe}{Error}{}{Cputime}{}{272681.192086184}%
\StoreBenchExecResult{ReducedEvalSlDfS}{ImcIgndffHardSafe}{Error}{}{Cputime}{Avg}{816.4107547490538922155688623}%
\StoreBenchExecResult{ReducedEvalSlDfS}{ImcIgndffHardSafe}{Error}{}{Cputime}{Median}{901.829625269}%
\StoreBenchExecResult{ReducedEvalSlDfS}{ImcIgndffHardSafe}{Error}{}{Cputime}{Min}{12.334949564}%
\StoreBenchExecResult{ReducedEvalSlDfS}{ImcIgndffHardSafe}{Error}{}{Cputime}{Max}{960.844313854}%
\StoreBenchExecResult{ReducedEvalSlDfS}{ImcIgndffHardSafe}{Error}{}{Cputime}{Stdev}{242.1624184397172213727280205}%
\StoreBenchExecResult{ReducedEvalSlDfS}{ImcIgndffHardSafe}{Error}{}{Walltime}{}{243236.7125513602511585}%
\StoreBenchExecResult{ReducedEvalSlDfS}{ImcIgndffHardSafe}{Error}{}{Walltime}{Avg}{728.2536303932941651452095808}%
\StoreBenchExecResult{ReducedEvalSlDfS}{ImcIgndffHardSafe}{Error}{}{Walltime}{Median}{804.74800015904475}%
\StoreBenchExecResult{ReducedEvalSlDfS}{ImcIgndffHardSafe}{Error}{}{Walltime}{Min}{6.9568020009901375}%
\StoreBenchExecResult{ReducedEvalSlDfS}{ImcIgndffHardSafe}{Error}{}{Walltime}{Max}{898.3665967129637}%
\StoreBenchExecResult{ReducedEvalSlDfS}{ImcIgndffHardSafe}{Error}{}{Walltime}{Stdev}{223.5174412837703122715940745}%
\StoreBenchExecResult{ReducedEvalSlDfS}{ImcIgndffHardSafe}{Error}{Error}{Count}{}{38}%
\StoreBenchExecResult{ReducedEvalSlDfS}{ImcIgndffHardSafe}{Error}{Error}{Cputime}{}{5629.088669122}%
\StoreBenchExecResult{ReducedEvalSlDfS}{ImcIgndffHardSafe}{Error}{Error}{Cputime}{Avg}{148.1339123453157894736842105}%
\StoreBenchExecResult{ReducedEvalSlDfS}{ImcIgndffHardSafe}{Error}{Error}{Cputime}{Median}{153.713725871}%
\StoreBenchExecResult{ReducedEvalSlDfS}{ImcIgndffHardSafe}{Error}{Error}{Cputime}{Min}{12.334949564}%
\StoreBenchExecResult{ReducedEvalSlDfS}{ImcIgndffHardSafe}{Error}{Error}{Cputime}{Max}{640.015438062}%
\StoreBenchExecResult{ReducedEvalSlDfS}{ImcIgndffHardSafe}{Error}{Error}{Cputime}{Stdev}{96.92139864983600617525012158}%
\StoreBenchExecResult{ReducedEvalSlDfS}{ImcIgndffHardSafe}{Error}{Error}{Walltime}{}{4233.4550810465589585}%
\StoreBenchExecResult{ReducedEvalSlDfS}{ImcIgndffHardSafe}{Error}{Error}{Walltime}{Avg}{111.4067126591199725921052632}%
\StoreBenchExecResult{ReducedEvalSlDfS}{ImcIgndffHardSafe}{Error}{Error}{Walltime}{Median}{124.97714194294531}%
\StoreBenchExecResult{ReducedEvalSlDfS}{ImcIgndffHardSafe}{Error}{Error}{Walltime}{Min}{6.9568020009901375}%
\StoreBenchExecResult{ReducedEvalSlDfS}{ImcIgndffHardSafe}{Error}{Error}{Walltime}{Max}{552.121082775062}%
\StoreBenchExecResult{ReducedEvalSlDfS}{ImcIgndffHardSafe}{Error}{Error}{Walltime}{Stdev}{86.05062289530439781831468229}%
\StoreBenchExecResult{ReducedEvalSlDfS}{ImcIgndffHardSafe}{Error}{SegmentationFault}{Count}{}{1}%
\StoreBenchExecResult{ReducedEvalSlDfS}{ImcIgndffHardSafe}{Error}{SegmentationFault}{Cputime}{}{629.188798755}%
\StoreBenchExecResult{ReducedEvalSlDfS}{ImcIgndffHardSafe}{Error}{SegmentationFault}{Cputime}{Avg}{629.188798755}%
\StoreBenchExecResult{ReducedEvalSlDfS}{ImcIgndffHardSafe}{Error}{SegmentationFault}{Cputime}{Median}{629.188798755}%
\StoreBenchExecResult{ReducedEvalSlDfS}{ImcIgndffHardSafe}{Error}{SegmentationFault}{Cputime}{Min}{629.188798755}%
\StoreBenchExecResult{ReducedEvalSlDfS}{ImcIgndffHardSafe}{Error}{SegmentationFault}{Cputime}{Max}{629.188798755}%
\StoreBenchExecResult{ReducedEvalSlDfS}{ImcIgndffHardSafe}{Error}{SegmentationFault}{Cputime}{Stdev}{0E-14}%
\StoreBenchExecResult{ReducedEvalSlDfS}{ImcIgndffHardSafe}{Error}{SegmentationFault}{Walltime}{}{623.2380462910514}%
\StoreBenchExecResult{ReducedEvalSlDfS}{ImcIgndffHardSafe}{Error}{SegmentationFault}{Walltime}{Avg}{623.2380462910514}%
\StoreBenchExecResult{ReducedEvalSlDfS}{ImcIgndffHardSafe}{Error}{SegmentationFault}{Walltime}{Median}{623.2380462910514}%
\StoreBenchExecResult{ReducedEvalSlDfS}{ImcIgndffHardSafe}{Error}{SegmentationFault}{Walltime}{Min}{623.2380462910514}%
\StoreBenchExecResult{ReducedEvalSlDfS}{ImcIgndffHardSafe}{Error}{SegmentationFault}{Walltime}{Max}{623.2380462910514}%
\StoreBenchExecResult{ReducedEvalSlDfS}{ImcIgndffHardSafe}{Error}{SegmentationFault}{Walltime}{Stdev}{0E-14}%
\StoreBenchExecResult{ReducedEvalSlDfS}{ImcIgndffHardSafe}{Error}{Timeout}{Count}{}{295}%
\StoreBenchExecResult{ReducedEvalSlDfS}{ImcIgndffHardSafe}{Error}{Timeout}{Cputime}{}{266422.914618307}%
\StoreBenchExecResult{ReducedEvalSlDfS}{ImcIgndffHardSafe}{Error}{Timeout}{Cputime}{Avg}{903.1285241298542372881355932}%
\StoreBenchExecResult{ReducedEvalSlDfS}{ImcIgndffHardSafe}{Error}{Timeout}{Cputime}{Median}{901.918224984}%
\StoreBenchExecResult{ReducedEvalSlDfS}{ImcIgndffHardSafe}{Error}{Timeout}{Cputime}{Min}{901.167930677}%
\StoreBenchExecResult{ReducedEvalSlDfS}{ImcIgndffHardSafe}{Error}{Timeout}{Cputime}{Max}{960.844313854}%
\StoreBenchExecResult{ReducedEvalSlDfS}{ImcIgndffHardSafe}{Error}{Timeout}{Cputime}{Stdev}{4.382024871749907634325978894}%
\StoreBenchExecResult{ReducedEvalSlDfS}{ImcIgndffHardSafe}{Error}{Timeout}{Walltime}{}{238380.0194240226408}%
\StoreBenchExecResult{ReducedEvalSlDfS}{ImcIgndffHardSafe}{Error}{Timeout}{Walltime}{Avg}{808.0678624543140366101694915}%
\StoreBenchExecResult{ReducedEvalSlDfS}{ImcIgndffHardSafe}{Error}{Timeout}{Walltime}{Median}{805.3656359971501}%
\StoreBenchExecResult{ReducedEvalSlDfS}{ImcIgndffHardSafe}{Error}{Timeout}{Walltime}{Min}{790.2945519289933}%
\StoreBenchExecResult{ReducedEvalSlDfS}{ImcIgndffHardSafe}{Error}{Timeout}{Walltime}{Max}{898.3665967129637}%
\StoreBenchExecResult{ReducedEvalSlDfS}{ImcIgndffHardSafe}{Error}{Timeout}{Walltime}{Stdev}{13.77812295860502932172094289}%
\ifdefined\ReducedEvalSlDfSImcIgndffHardSafeTotalCount\else\edef\ReducedEvalSlDfSImcIgndffHardSafeTotalCount{0}\fi
\ifdefined\ReducedEvalSlDfSImcIgndffHardSafeCorrectCount\else\edef\ReducedEvalSlDfSImcIgndffHardSafeCorrectCount{0}\fi
\ifdefined\ReducedEvalSlDfSImcIgndffHardSafeCorrectTrueCount\else\edef\ReducedEvalSlDfSImcIgndffHardSafeCorrectTrueCount{0}\fi
\ifdefined\ReducedEvalSlDfSImcIgndffHardSafeCorrectFalseCount\else\edef\ReducedEvalSlDfSImcIgndffHardSafeCorrectFalseCount{0}\fi
\ifdefined\ReducedEvalSlDfSImcIgndffHardSafeWrongTrueCount\else\edef\ReducedEvalSlDfSImcIgndffHardSafeWrongTrueCount{0}\fi
\ifdefined\ReducedEvalSlDfSImcIgndffHardSafeWrongFalseCount\else\edef\ReducedEvalSlDfSImcIgndffHardSafeWrongFalseCount{0}\fi
\ifdefined\ReducedEvalSlDfSImcIgndffHardSafeErrorTimeoutCount\else\edef\ReducedEvalSlDfSImcIgndffHardSafeErrorTimeoutCount{0}\fi
\ifdefined\ReducedEvalSlDfSImcIgndffHardSafeErrorOutOfMemoryCount\else\edef\ReducedEvalSlDfSImcIgndffHardSafeErrorOutOfMemoryCount{0}\fi
\ifdefined\ReducedEvalSlDfSImcIgndffHardSafeCorrectCputime\else\edef\ReducedEvalSlDfSImcIgndffHardSafeCorrectCputime{0}\fi
\ifdefined\ReducedEvalSlDfSImcIgndffHardSafeCorrectCputimeAvg\else\edef\ReducedEvalSlDfSImcIgndffHardSafeCorrectCputimeAvg{None}\fi
\ifdefined\ReducedEvalSlDfSImcIgndffHardSafeCorrectWalltime\else\edef\ReducedEvalSlDfSImcIgndffHardSafeCorrectWalltime{0}\fi
\ifdefined\ReducedEvalSlDfSImcIgndffHardSafeCorrectWalltimeAvg\else\edef\ReducedEvalSlDfSImcIgndffHardSafeCorrectWalltimeAvg{None}\fi
\edef\ReducedEvalSlDfSImcIgndffHardSafeErrorOtherInconclusiveCount{\the\numexpr \ReducedEvalSlDfSImcIgndffHardSafeTotalCount - \ReducedEvalSlDfSImcIgndffHardSafeCorrectCount - \ReducedEvalSlDfSImcIgndffHardSafeWrongTrueCount - \ReducedEvalSlDfSImcIgndffHardSafeWrongFalseCount - \ReducedEvalSlDfSImcIgndffHardSafeErrorTimeoutCount - \ReducedEvalSlDfSImcIgndffHardSafeErrorOutOfMemoryCount \relax}
\providecommand\StoreBenchExecResult[7]{\expandafter\newcommand\csname#1#2#3#4#5#6\endcsname{#7}}%
\StoreBenchExecResult{EvalSlDfS}{ImcHardSafe}{Total}{}{Count}{}{1623}%
\StoreBenchExecResult{EvalSlDfS}{ImcHardSafe}{Total}{}{Cputime}{}{623635.253742403}%
\StoreBenchExecResult{EvalSlDfS}{ImcHardSafe}{Total}{}{Cputime}{Avg}{384.2484619484922982131854590}%
\StoreBenchExecResult{EvalSlDfS}{ImcHardSafe}{Total}{}{Cputime}{Median}{54.502199301}%
\StoreBenchExecResult{EvalSlDfS}{ImcHardSafe}{Total}{}{Cputime}{Min}{3.893722656}%
\StoreBenchExecResult{EvalSlDfS}{ImcHardSafe}{Total}{}{Cputime}{Max}{936.266226017}%
\StoreBenchExecResult{EvalSlDfS}{ImcHardSafe}{Total}{}{Cputime}{Stdev}{420.0109407345892945929934348}%
\StoreBenchExecResult{EvalSlDfS}{ImcHardSafe}{Total}{}{Walltime}{}{591464.3737796372739864}%
\StoreBenchExecResult{EvalSlDfS}{ImcHardSafe}{Total}{}{Walltime}{Avg}{364.4266012197395403489833641}%
\StoreBenchExecResult{EvalSlDfS}{ImcHardSafe}{Total}{}{Walltime}{Median}{42.995444857981056}%
\StoreBenchExecResult{EvalSlDfS}{ImcHardSafe}{Total}{}{Walltime}{Min}{1.5823553358204663}%
\StoreBenchExecResult{EvalSlDfS}{ImcHardSafe}{Total}{}{Walltime}{Max}{906.2711261240765}%
\StoreBenchExecResult{EvalSlDfS}{ImcHardSafe}{Total}{}{Walltime}{Stdev}{412.2453049459989973109160718}%
\StoreBenchExecResult{EvalSlDfS}{ImcHardSafe}{Correct}{}{Count}{}{861}%
\StoreBenchExecResult{EvalSlDfS}{ImcHardSafe}{Correct}{}{Cputime}{}{40966.360523549}%
\StoreBenchExecResult{EvalSlDfS}{ImcHardSafe}{Correct}{}{Cputime}{Avg}{47.57997737926713124274099884}%
\StoreBenchExecResult{EvalSlDfS}{ImcHardSafe}{Correct}{}{Cputime}{Median}{11.099334399}%
\StoreBenchExecResult{EvalSlDfS}{ImcHardSafe}{Correct}{}{Cputime}{Min}{3.893722656}%
\StoreBenchExecResult{EvalSlDfS}{ImcHardSafe}{Correct}{}{Cputime}{Max}{864.226558973}%
\StoreBenchExecResult{EvalSlDfS}{ImcHardSafe}{Correct}{}{Cputime}{Stdev}{112.1072572903452216464938689}%
\StoreBenchExecResult{EvalSlDfS}{ImcHardSafe}{Correct}{}{Walltime}{}{32687.9022667715325789}%
\StoreBenchExecResult{EvalSlDfS}{ImcHardSafe}{Correct}{}{Walltime}{Avg}{37.96504328312605409860627178}%
\StoreBenchExecResult{EvalSlDfS}{ImcHardSafe}{Correct}{}{Walltime}{Median}{4.556960269808769}%
\StoreBenchExecResult{EvalSlDfS}{ImcHardSafe}{Correct}{}{Walltime}{Min}{1.5823553358204663}%
\StoreBenchExecResult{EvalSlDfS}{ImcHardSafe}{Correct}{}{Walltime}{Max}{853.9676453699358}%
\StoreBenchExecResult{EvalSlDfS}{ImcHardSafe}{Correct}{}{Walltime}{Stdev}{108.9820334837581903464217846}%
\StoreBenchExecResult{EvalSlDfS}{ImcHardSafe}{Correct}{True}{Count}{}{861}%
\StoreBenchExecResult{EvalSlDfS}{ImcHardSafe}{Correct}{True}{Cputime}{}{40966.360523549}%
\StoreBenchExecResult{EvalSlDfS}{ImcHardSafe}{Correct}{True}{Cputime}{Avg}{47.57997737926713124274099884}%
\StoreBenchExecResult{EvalSlDfS}{ImcHardSafe}{Correct}{True}{Cputime}{Median}{11.099334399}%
\StoreBenchExecResult{EvalSlDfS}{ImcHardSafe}{Correct}{True}{Cputime}{Min}{3.893722656}%
\StoreBenchExecResult{EvalSlDfS}{ImcHardSafe}{Correct}{True}{Cputime}{Max}{864.226558973}%
\StoreBenchExecResult{EvalSlDfS}{ImcHardSafe}{Correct}{True}{Cputime}{Stdev}{112.1072572903452216464938689}%
\StoreBenchExecResult{EvalSlDfS}{ImcHardSafe}{Correct}{True}{Walltime}{}{32687.9022667715325789}%
\StoreBenchExecResult{EvalSlDfS}{ImcHardSafe}{Correct}{True}{Walltime}{Avg}{37.96504328312605409860627178}%
\StoreBenchExecResult{EvalSlDfS}{ImcHardSafe}{Correct}{True}{Walltime}{Median}{4.556960269808769}%
\StoreBenchExecResult{EvalSlDfS}{ImcHardSafe}{Correct}{True}{Walltime}{Min}{1.5823553358204663}%
\StoreBenchExecResult{EvalSlDfS}{ImcHardSafe}{Correct}{True}{Walltime}{Max}{853.9676453699358}%
\StoreBenchExecResult{EvalSlDfS}{ImcHardSafe}{Correct}{True}{Walltime}{Stdev}{108.9820334837581903464217846}%

\StoreBenchExecResult{EvalSlDfS}{ImcHardSafe}{Error}{}{Count}{}{762}%
\StoreBenchExecResult{EvalSlDfS}{ImcHardSafe}{Error}{}{Cputime}{}{582668.893218854}%
\StoreBenchExecResult{EvalSlDfS}{ImcHardSafe}{Error}{}{Cputime}{Avg}{764.6573401822230971128608924}%
\StoreBenchExecResult{EvalSlDfS}{ImcHardSafe}{Error}{}{Cputime}{Median}{901.8237785455}%
\StoreBenchExecResult{EvalSlDfS}{ImcHardSafe}{Error}{}{Cputime}{Min}{4.409323771}%
\StoreBenchExecResult{EvalSlDfS}{ImcHardSafe}{Error}{}{Cputime}{Max}{936.266226017}%
\StoreBenchExecResult{EvalSlDfS}{ImcHardSafe}{Error}{}{Cputime}{Stdev}{297.9164268438606275121668814}%
\StoreBenchExecResult{EvalSlDfS}{ImcHardSafe}{Error}{}{Walltime}{}{558776.4715128657414075}%
\StoreBenchExecResult{EvalSlDfS}{ImcHardSafe}{Error}{}{Walltime}{Avg}{733.3024560536295818996062992}%
\StoreBenchExecResult{EvalSlDfS}{ImcHardSafe}{Error}{}{Walltime}{Median}{885.4109070175327}%
\StoreBenchExecResult{EvalSlDfS}{ImcHardSafe}{Error}{}{Walltime}{Min}{1.7712977770715952}%
\StoreBenchExecResult{EvalSlDfS}{ImcHardSafe}{Error}{}{Walltime}{Max}{906.2711261240765}%
\StoreBenchExecResult{EvalSlDfS}{ImcHardSafe}{Error}{}{Walltime}{Stdev}{303.4116912303244820900066199}%
\StoreBenchExecResult{EvalSlDfS}{ImcHardSafe}{Error}{Error}{Count}{}{95}%
\StoreBenchExecResult{EvalSlDfS}{ImcHardSafe}{Error}{Error}{Cputime}{}{2184.220121717}%
\StoreBenchExecResult{EvalSlDfS}{ImcHardSafe}{Error}{Error}{Cputime}{Avg}{22.99179075491578947368421053}%
\StoreBenchExecResult{EvalSlDfS}{ImcHardSafe}{Error}{Error}{Cputime}{Median}{8.208014288}%
\StoreBenchExecResult{EvalSlDfS}{ImcHardSafe}{Error}{Error}{Cputime}{Min}{4.409323771}%
\StoreBenchExecResult{EvalSlDfS}{ImcHardSafe}{Error}{Error}{Cputime}{Max}{548.373546117}%
\StoreBenchExecResult{EvalSlDfS}{ImcHardSafe}{Error}{Error}{Cputime}{Stdev}{61.91641428551672529040337702}%
\StoreBenchExecResult{EvalSlDfS}{ImcHardSafe}{Error}{Error}{Walltime}{}{1756.0222281333990005}%
\StoreBenchExecResult{EvalSlDfS}{ImcHardSafe}{Error}{Error}{Walltime}{Avg}{18.4844445066673579}%
\StoreBenchExecResult{EvalSlDfS}{ImcHardSafe}{Error}{Error}{Walltime}{Median}{2.838396789971739}%
\StoreBenchExecResult{EvalSlDfS}{ImcHardSafe}{Error}{Error}{Walltime}{Min}{1.7712977770715952}%
\StoreBenchExecResult{EvalSlDfS}{ImcHardSafe}{Error}{Error}{Walltime}{Max}{544.9657988150138}%
\StoreBenchExecResult{EvalSlDfS}{ImcHardSafe}{Error}{Error}{Walltime}{Stdev}{62.09682565264074990306448363}%
\StoreBenchExecResult{EvalSlDfS}{ImcHardSafe}{Error}{OutOfMemory}{Count}{}{89}%
\StoreBenchExecResult{EvalSlDfS}{ImcHardSafe}{Error}{OutOfMemory}{Cputime}{}{60169.019080696}%
\StoreBenchExecResult{EvalSlDfS}{ImcHardSafe}{Error}{OutOfMemory}{Cputime}{Avg}{676.0563941651235955056179775}%
\StoreBenchExecResult{EvalSlDfS}{ImcHardSafe}{Error}{OutOfMemory}{Cputime}{Median}{692.912840715}%
\StoreBenchExecResult{EvalSlDfS}{ImcHardSafe}{Error}{OutOfMemory}{Cputime}{Min}{299.766382408}%
\StoreBenchExecResult{EvalSlDfS}{ImcHardSafe}{Error}{OutOfMemory}{Cputime}{Max}{898.821722975}%
\StoreBenchExecResult{EvalSlDfS}{ImcHardSafe}{Error}{OutOfMemory}{Cputime}{Stdev}{155.2745936117885592626777614}%
\StoreBenchExecResult{EvalSlDfS}{ImcHardSafe}{Error}{OutOfMemory}{Walltime}{}{47603.62796989851660}%
\StoreBenchExecResult{EvalSlDfS}{ImcHardSafe}{Error}{OutOfMemory}{Walltime}{Avg}{534.8722243808822089887640449}%
\StoreBenchExecResult{EvalSlDfS}{ImcHardSafe}{Error}{OutOfMemory}{Walltime}{Median}{546.8767519900575}%
\StoreBenchExecResult{EvalSlDfS}{ImcHardSafe}{Error}{OutOfMemory}{Walltime}{Min}{219.80549752991647}%
\StoreBenchExecResult{EvalSlDfS}{ImcHardSafe}{Error}{OutOfMemory}{Walltime}{Max}{819.5512235269416}%
\StoreBenchExecResult{EvalSlDfS}{ImcHardSafe}{Error}{OutOfMemory}{Walltime}{Stdev}{188.6793913133499693387281059}%
\StoreBenchExecResult{EvalSlDfS}{ImcHardSafe}{Error}{OutOfNativeMemory}{Count}{}{2}%
\StoreBenchExecResult{EvalSlDfS}{ImcHardSafe}{Error}{OutOfNativeMemory}{Cputime}{}{51.406086997}%
\StoreBenchExecResult{EvalSlDfS}{ImcHardSafe}{Error}{OutOfNativeMemory}{Cputime}{Avg}{25.7030434985}%
\StoreBenchExecResult{EvalSlDfS}{ImcHardSafe}{Error}{OutOfNativeMemory}{Cputime}{Median}{25.7030434985}%
\StoreBenchExecResult{EvalSlDfS}{ImcHardSafe}{Error}{OutOfNativeMemory}{Cputime}{Min}{22.465384977}%
\StoreBenchExecResult{EvalSlDfS}{ImcHardSafe}{Error}{OutOfNativeMemory}{Cputime}{Max}{28.94070202}%
\StoreBenchExecResult{EvalSlDfS}{ImcHardSafe}{Error}{OutOfNativeMemory}{Cputime}{Stdev}{3.2376585215000}%
\StoreBenchExecResult{EvalSlDfS}{ImcHardSafe}{Error}{OutOfNativeMemory}{Walltime}{}{34.874126139795407}%
\StoreBenchExecResult{EvalSlDfS}{ImcHardSafe}{Error}{OutOfNativeMemory}{Walltime}{Avg}{17.4370630698977035}%
\StoreBenchExecResult{EvalSlDfS}{ImcHardSafe}{Error}{OutOfNativeMemory}{Walltime}{Median}{17.4370630698977035}%
\StoreBenchExecResult{EvalSlDfS}{ImcHardSafe}{Error}{OutOfNativeMemory}{Walltime}{Min}{14.434731234796345}%
\StoreBenchExecResult{EvalSlDfS}{ImcHardSafe}{Error}{OutOfNativeMemory}{Walltime}{Max}{20.439394904999062}%
\StoreBenchExecResult{EvalSlDfS}{ImcHardSafe}{Error}{OutOfNativeMemory}{Walltime}{Stdev}{3.002331835101358500000000000}%
\StoreBenchExecResult{EvalSlDfS}{ImcHardSafe}{Error}{SegmentationFault}{Count}{}{1}%
\StoreBenchExecResult{EvalSlDfS}{ImcHardSafe}{Error}{SegmentationFault}{Cputime}{}{829.112224541}%
\StoreBenchExecResult{EvalSlDfS}{ImcHardSafe}{Error}{SegmentationFault}{Cputime}{Avg}{829.112224541}%
\StoreBenchExecResult{EvalSlDfS}{ImcHardSafe}{Error}{SegmentationFault}{Cputime}{Median}{829.112224541}%
\StoreBenchExecResult{EvalSlDfS}{ImcHardSafe}{Error}{SegmentationFault}{Cputime}{Min}{829.112224541}%
\StoreBenchExecResult{EvalSlDfS}{ImcHardSafe}{Error}{SegmentationFault}{Cputime}{Max}{829.112224541}%
\StoreBenchExecResult{EvalSlDfS}{ImcHardSafe}{Error}{SegmentationFault}{Cputime}{Stdev}{0E-14}%
\StoreBenchExecResult{EvalSlDfS}{ImcHardSafe}{Error}{SegmentationFault}{Walltime}{}{822.4779165070504}%
\StoreBenchExecResult{EvalSlDfS}{ImcHardSafe}{Error}{SegmentationFault}{Walltime}{Avg}{822.4779165070504}%
\StoreBenchExecResult{EvalSlDfS}{ImcHardSafe}{Error}{SegmentationFault}{Walltime}{Median}{822.4779165070504}%
\StoreBenchExecResult{EvalSlDfS}{ImcHardSafe}{Error}{SegmentationFault}{Walltime}{Min}{822.4779165070504}%
\StoreBenchExecResult{EvalSlDfS}{ImcHardSafe}{Error}{SegmentationFault}{Walltime}{Max}{822.4779165070504}%
\StoreBenchExecResult{EvalSlDfS}{ImcHardSafe}{Error}{SegmentationFault}{Walltime}{Stdev}{0E-14}%
\StoreBenchExecResult{EvalSlDfS}{ImcHardSafe}{Error}{Timeout}{Count}{}{575}%
\StoreBenchExecResult{EvalSlDfS}{ImcHardSafe}{Error}{Timeout}{Cputime}{}{519435.135704903}%
\StoreBenchExecResult{EvalSlDfS}{ImcHardSafe}{Error}{Timeout}{Cputime}{Avg}{903.3654533998313043478260870}%
\StoreBenchExecResult{EvalSlDfS}{ImcHardSafe}{Error}{Timeout}{Cputime}{Median}{902.076231343}%
\StoreBenchExecResult{EvalSlDfS}{ImcHardSafe}{Error}{Timeout}{Cputime}{Min}{900.124635784}%
\StoreBenchExecResult{EvalSlDfS}{ImcHardSafe}{Error}{Timeout}{Cputime}{Max}{936.266226017}%
\StoreBenchExecResult{EvalSlDfS}{ImcHardSafe}{Error}{Timeout}{Cputime}{Stdev}{3.321943028896274263014591528}%
\StoreBenchExecResult{EvalSlDfS}{ImcHardSafe}{Error}{Timeout}{Walltime}{}{508559.4692721869800}%
\StoreBenchExecResult{EvalSlDfS}{ImcHardSafe}{Error}{Timeout}{Walltime}{Avg}{884.4512509081512695652173913}%
\StoreBenchExecResult{EvalSlDfS}{ImcHardSafe}{Error}{Timeout}{Walltime}{Median}{890.3328494820744}%
\StoreBenchExecResult{EvalSlDfS}{ImcHardSafe}{Error}{Timeout}{Walltime}{Min}{451.8912322949618}%
\StoreBenchExecResult{EvalSlDfS}{ImcHardSafe}{Error}{Timeout}{Walltime}{Max}{906.2711261240765}%
\StoreBenchExecResult{EvalSlDfS}{ImcHardSafe}{Error}{Timeout}{Walltime}{Stdev}{26.33125314052430896169829240}%
\ifdefined\EvalSlDfSImcHardSafeTotalCount\else\edef\EvalSlDfSImcHardSafeTotalCount{0}\fi
\ifdefined\EvalSlDfSImcHardSafeCorrectCount\else\edef\EvalSlDfSImcHardSafeCorrectCount{0}\fi
\ifdefined\EvalSlDfSImcHardSafeCorrectTrueCount\else\edef\EvalSlDfSImcHardSafeCorrectTrueCount{0}\fi
\ifdefined\EvalSlDfSImcHardSafeCorrectFalseCount\else\edef\EvalSlDfSImcHardSafeCorrectFalseCount{0}\fi
\ifdefined\EvalSlDfSImcHardSafeWrongTrueCount\else\edef\EvalSlDfSImcHardSafeWrongTrueCount{0}\fi
\ifdefined\EvalSlDfSImcHardSafeWrongFalseCount\else\edef\EvalSlDfSImcHardSafeWrongFalseCount{0}\fi
\ifdefined\EvalSlDfSImcHardSafeErrorTimeoutCount\else\edef\EvalSlDfSImcHardSafeErrorTimeoutCount{0}\fi
\ifdefined\EvalSlDfSImcHardSafeErrorOutOfMemoryCount\else\edef\EvalSlDfSImcHardSafeErrorOutOfMemoryCount{0}\fi
\ifdefined\EvalSlDfSImcHardSafeCorrectCputime\else\edef\EvalSlDfSImcHardSafeCorrectCputime{0}\fi
\ifdefined\EvalSlDfSImcHardSafeCorrectCputimeAvg\else\edef\EvalSlDfSImcHardSafeCorrectCputimeAvg{None}\fi
\ifdefined\EvalSlDfSImcHardSafeCorrectWalltime\else\edef\EvalSlDfSImcHardSafeCorrectWalltime{0}\fi
\ifdefined\EvalSlDfSImcHardSafeCorrectWalltimeAvg\else\edef\EvalSlDfSImcHardSafeCorrectWalltimeAvg{None}\fi
\edef\EvalSlDfSImcHardSafeErrorOtherInconclusiveCount{\the\numexpr \EvalSlDfSImcHardSafeTotalCount - \EvalSlDfSImcHardSafeCorrectCount - \EvalSlDfSImcHardSafeWrongTrueCount - \EvalSlDfSImcHardSafeWrongFalseCount - \EvalSlDfSImcHardSafeErrorTimeoutCount - \EvalSlDfSImcHardSafeErrorOutOfMemoryCount \relax}
\providecommand\StoreBenchExecResult[7]{\expandafter\newcommand\csname#1#2#3#4#5#6\endcsname{#7}}%
\StoreBenchExecResult{EvalSlDfS}{ImcIgndfiHardSafe}{Total}{}{Count}{}{1623}%
\StoreBenchExecResult{EvalSlDfS}{ImcIgndfiHardSafe}{Total}{}{Cputime}{}{621960.101180177}%
\StoreBenchExecResult{EvalSlDfS}{ImcIgndfiHardSafe}{Total}{}{Cputime}{Avg}{383.2163285152045594577942083}%
\StoreBenchExecResult{EvalSlDfS}{ImcIgndfiHardSafe}{Total}{}{Cputime}{Median}{153.452999399}%
\StoreBenchExecResult{EvalSlDfS}{ImcIgndfiHardSafe}{Total}{}{Cputime}{Min}{4.264615926}%
\StoreBenchExecResult{EvalSlDfS}{ImcIgndfiHardSafe}{Total}{}{Cputime}{Max}{936.754802521}%
\StoreBenchExecResult{EvalSlDfS}{ImcIgndfiHardSafe}{Total}{}{Cputime}{Stdev}{404.0270110562732604481016363}%
\StoreBenchExecResult{EvalSlDfS}{ImcIgndfiHardSafe}{Total}{}{Walltime}{}{527486.6745337841574952}%
\StoreBenchExecResult{EvalSlDfS}{ImcIgndfiHardSafe}{Total}{}{Walltime}{Avg}{325.0071931816291789865680838}%
\StoreBenchExecResult{EvalSlDfS}{ImcIgndfiHardSafe}{Total}{}{Walltime}{Median}{58.7128456721548}%
\StoreBenchExecResult{EvalSlDfS}{ImcIgndfiHardSafe}{Total}{}{Walltime}{Min}{1.6150803179480135}%
\StoreBenchExecResult{EvalSlDfS}{ImcIgndfiHardSafe}{Total}{}{Walltime}{Max}{898.3241822579876}%
\StoreBenchExecResult{EvalSlDfS}{ImcIgndfiHardSafe}{Total}{}{Walltime}{Stdev}{373.2110516352093448062181141}%
\StoreBenchExecResult{EvalSlDfS}{ImcIgndfiHardSafe}{Correct}{}{Count}{}{871}%
\StoreBenchExecResult{EvalSlDfS}{ImcIgndfiHardSafe}{Correct}{}{Cputime}{}{61852.107907284}%
\StoreBenchExecResult{EvalSlDfS}{ImcIgndfiHardSafe}{Correct}{}{Cputime}{Avg}{71.01275305084270952927669346}%
\StoreBenchExecResult{EvalSlDfS}{ImcIgndfiHardSafe}{Correct}{}{Cputime}{Median}{16.300551097}%
\StoreBenchExecResult{EvalSlDfS}{ImcIgndfiHardSafe}{Correct}{}{Cputime}{Min}{4.264615926}%
\StoreBenchExecResult{EvalSlDfS}{ImcIgndfiHardSafe}{Correct}{}{Cputime}{Max}{887.278560398}%
\StoreBenchExecResult{EvalSlDfS}{ImcIgndfiHardSafe}{Correct}{}{Cputime}{Stdev}{134.7584092416860238467751425}%
\StoreBenchExecResult{EvalSlDfS}{ImcIgndfiHardSafe}{Correct}{}{Walltime}{}{35681.7231501184867982}%
\StoreBenchExecResult{EvalSlDfS}{ImcIgndfiHardSafe}{Correct}{}{Walltime}{Avg}{40.96638708394774603696900115}%
\StoreBenchExecResult{EvalSlDfS}{ImcIgndfiHardSafe}{Correct}{}{Walltime}{Median}{5.7562073490116745}%
\StoreBenchExecResult{EvalSlDfS}{ImcIgndfiHardSafe}{Correct}{}{Walltime}{Min}{1.6150803179480135}%
\StoreBenchExecResult{EvalSlDfS}{ImcIgndfiHardSafe}{Correct}{}{Walltime}{Max}{788.5136345371138}%
\StoreBenchExecResult{EvalSlDfS}{ImcIgndfiHardSafe}{Correct}{}{Walltime}{Stdev}{110.9663584652824841163509773}%
\StoreBenchExecResult{EvalSlDfS}{ImcIgndfiHardSafe}{Correct}{True}{Count}{}{871}%
\StoreBenchExecResult{EvalSlDfS}{ImcIgndfiHardSafe}{Correct}{True}{Cputime}{}{61852.107907284}%
\StoreBenchExecResult{EvalSlDfS}{ImcIgndfiHardSafe}{Correct}{True}{Cputime}{Avg}{71.01275305084270952927669346}%
\StoreBenchExecResult{EvalSlDfS}{ImcIgndfiHardSafe}{Correct}{True}{Cputime}{Median}{16.300551097}%
\StoreBenchExecResult{EvalSlDfS}{ImcIgndfiHardSafe}{Correct}{True}{Cputime}{Min}{4.264615926}%
\StoreBenchExecResult{EvalSlDfS}{ImcIgndfiHardSafe}{Correct}{True}{Cputime}{Max}{887.278560398}%
\StoreBenchExecResult{EvalSlDfS}{ImcIgndfiHardSafe}{Correct}{True}{Cputime}{Stdev}{134.7584092416860238467751425}%
\StoreBenchExecResult{EvalSlDfS}{ImcIgndfiHardSafe}{Correct}{True}{Walltime}{}{35681.7231501184867982}%
\StoreBenchExecResult{EvalSlDfS}{ImcIgndfiHardSafe}{Correct}{True}{Walltime}{Avg}{40.96638708394774603696900115}%
\StoreBenchExecResult{EvalSlDfS}{ImcIgndfiHardSafe}{Correct}{True}{Walltime}{Median}{5.7562073490116745}%
\StoreBenchExecResult{EvalSlDfS}{ImcIgndfiHardSafe}{Correct}{True}{Walltime}{Min}{1.6150803179480135}%
\StoreBenchExecResult{EvalSlDfS}{ImcIgndfiHardSafe}{Correct}{True}{Walltime}{Max}{788.5136345371138}%
\StoreBenchExecResult{EvalSlDfS}{ImcIgndfiHardSafe}{Correct}{True}{Walltime}{Stdev}{110.9663584652824841163509773}%

\StoreBenchExecResult{EvalSlDfS}{ImcIgndfiHardSafe}{Error}{}{Count}{}{752}%
\StoreBenchExecResult{EvalSlDfS}{ImcIgndfiHardSafe}{Error}{}{Cputime}{}{560107.993272893}%
\StoreBenchExecResult{EvalSlDfS}{ImcIgndfiHardSafe}{Error}{}{Cputime}{Avg}{744.8244591394853723404255319}%
\StoreBenchExecResult{EvalSlDfS}{ImcIgndfiHardSafe}{Error}{}{Cputime}{Median}{901.6776979085}%
\StoreBenchExecResult{EvalSlDfS}{ImcIgndfiHardSafe}{Error}{}{Cputime}{Min}{4.890311028}%
\StoreBenchExecResult{EvalSlDfS}{ImcIgndfiHardSafe}{Error}{}{Cputime}{Max}{936.754802521}%
\StoreBenchExecResult{EvalSlDfS}{ImcIgndfiHardSafe}{Error}{}{Cputime}{Stdev}{296.0031289289186176523275292}%
\StoreBenchExecResult{EvalSlDfS}{ImcIgndfiHardSafe}{Error}{}{Walltime}{}{491804.9513836656706970}%
\StoreBenchExecResult{EvalSlDfS}{ImcIgndfiHardSafe}{Error}{}{Walltime}{Avg}{653.9959459889171152885638298}%
\StoreBenchExecResult{EvalSlDfS}{ImcIgndfiHardSafe}{Error}{}{Walltime}{Median}{803.16415317507925}%
\StoreBenchExecResult{EvalSlDfS}{ImcIgndfiHardSafe}{Error}{}{Walltime}{Min}{1.8038417058996856}%
\StoreBenchExecResult{EvalSlDfS}{ImcIgndfiHardSafe}{Error}{}{Walltime}{Max}{898.3241822579876}%
\StoreBenchExecResult{EvalSlDfS}{ImcIgndfiHardSafe}{Error}{}{Walltime}{Stdev}{290.9852407781219473544768253}%
\StoreBenchExecResult{EvalSlDfS}{ImcIgndfiHardSafe}{Error}{Error}{Count}{}{95}%
\StoreBenchExecResult{EvalSlDfS}{ImcIgndfiHardSafe}{Error}{Error}{Cputime}{}{6955.628980194}%
\StoreBenchExecResult{EvalSlDfS}{ImcIgndfiHardSafe}{Error}{Error}{Cputime}{Avg}{73.21714715993684210526315789}%
\StoreBenchExecResult{EvalSlDfS}{ImcIgndfiHardSafe}{Error}{Error}{Cputime}{Median}{25.268025112}%
\StoreBenchExecResult{EvalSlDfS}{ImcIgndfiHardSafe}{Error}{Error}{Cputime}{Min}{4.890311028}%
\StoreBenchExecResult{EvalSlDfS}{ImcIgndfiHardSafe}{Error}{Error}{Cputime}{Max}{634.966942936}%
\StoreBenchExecResult{EvalSlDfS}{ImcIgndfiHardSafe}{Error}{Error}{Cputime}{Stdev}{90.14260288807897302586071811}%
\StoreBenchExecResult{EvalSlDfS}{ImcIgndfiHardSafe}{Error}{Error}{Walltime}{}{4939.6969694169238620}%
\StoreBenchExecResult{EvalSlDfS}{ImcIgndfiHardSafe}{Error}{Error}{Walltime}{Avg}{51.99681020438867223157894737}%
\StoreBenchExecResult{EvalSlDfS}{ImcIgndfiHardSafe}{Error}{Error}{Walltime}{Median}{10.747336158994585}%
\StoreBenchExecResult{EvalSlDfS}{ImcIgndfiHardSafe}{Error}{Error}{Walltime}{Min}{1.8038417058996856}%
\StoreBenchExecResult{EvalSlDfS}{ImcIgndfiHardSafe}{Error}{Error}{Walltime}{Max}{548.886447101133}%
\StoreBenchExecResult{EvalSlDfS}{ImcIgndfiHardSafe}{Error}{Error}{Walltime}{Stdev}{76.03884879467201063514236316}%
\StoreBenchExecResult{EvalSlDfS}{ImcIgndfiHardSafe}{Error}{Exception}{Count}{}{2}%
\StoreBenchExecResult{EvalSlDfS}{ImcIgndfiHardSafe}{Error}{Exception}{Cputime}{}{220.979790127}%
\StoreBenchExecResult{EvalSlDfS}{ImcIgndfiHardSafe}{Error}{Exception}{Cputime}{Avg}{110.4898950635}%
\StoreBenchExecResult{EvalSlDfS}{ImcIgndfiHardSafe}{Error}{Exception}{Cputime}{Median}{110.4898950635}%
\StoreBenchExecResult{EvalSlDfS}{ImcIgndfiHardSafe}{Error}{Exception}{Cputime}{Min}{46.040358444}%
\StoreBenchExecResult{EvalSlDfS}{ImcIgndfiHardSafe}{Error}{Exception}{Cputime}{Max}{174.939431683}%
\StoreBenchExecResult{EvalSlDfS}{ImcIgndfiHardSafe}{Error}{Exception}{Cputime}{Stdev}{64.449536619500}%
\StoreBenchExecResult{EvalSlDfS}{ImcIgndfiHardSafe}{Error}{Exception}{Walltime}{}{159.99651051918045}%
\StoreBenchExecResult{EvalSlDfS}{ImcIgndfiHardSafe}{Error}{Exception}{Walltime}{Avg}{79.998255259590225}%
\StoreBenchExecResult{EvalSlDfS}{ImcIgndfiHardSafe}{Error}{Exception}{Walltime}{Median}{79.998255259590225}%
\StoreBenchExecResult{EvalSlDfS}{ImcIgndfiHardSafe}{Error}{Exception}{Walltime}{Min}{12.61084886197932}%
\StoreBenchExecResult{EvalSlDfS}{ImcIgndfiHardSafe}{Error}{Exception}{Walltime}{Max}{147.38566165720113}%
\StoreBenchExecResult{EvalSlDfS}{ImcIgndfiHardSafe}{Error}{Exception}{Walltime}{Stdev}{67.38740639761090500000000000}%
\StoreBenchExecResult{EvalSlDfS}{ImcIgndfiHardSafe}{Error}{OutOfMemory}{Count}{}{110}%
\StoreBenchExecResult{EvalSlDfS}{ImcIgndfiHardSafe}{Error}{OutOfMemory}{Cputime}{}{62887.186652512}%
\StoreBenchExecResult{EvalSlDfS}{ImcIgndfiHardSafe}{Error}{OutOfMemory}{Cputime}{Avg}{571.7016968410181818181818182}%
\StoreBenchExecResult{EvalSlDfS}{ImcIgndfiHardSafe}{Error}{OutOfMemory}{Cputime}{Median}{508.584332852}%
\StoreBenchExecResult{EvalSlDfS}{ImcIgndfiHardSafe}{Error}{OutOfMemory}{Cputime}{Min}{285.633690449}%
\StoreBenchExecResult{EvalSlDfS}{ImcIgndfiHardSafe}{Error}{OutOfMemory}{Cputime}{Max}{879.752389359}%
\StoreBenchExecResult{EvalSlDfS}{ImcIgndfiHardSafe}{Error}{OutOfMemory}{Cputime}{Stdev}{179.1119437779375238801902620}%
\StoreBenchExecResult{EvalSlDfS}{ImcIgndfiHardSafe}{Error}{OutOfMemory}{Walltime}{}{41549.62496188515794}%
\StoreBenchExecResult{EvalSlDfS}{ImcIgndfiHardSafe}{Error}{OutOfMemory}{Walltime}{Avg}{377.7238632898650721818181818}%
\StoreBenchExecResult{EvalSlDfS}{ImcIgndfiHardSafe}{Error}{OutOfMemory}{Walltime}{Median}{325.584877796121875}%
\StoreBenchExecResult{EvalSlDfS}{ImcIgndfiHardSafe}{Error}{OutOfMemory}{Walltime}{Min}{165.82995136105455}%
\StoreBenchExecResult{EvalSlDfS}{ImcIgndfiHardSafe}{Error}{OutOfMemory}{Walltime}{Max}{739.3721200518776}%
\StoreBenchExecResult{EvalSlDfS}{ImcIgndfiHardSafe}{Error}{OutOfMemory}{Walltime}{Stdev}{149.5144052566973001180678090}%
\StoreBenchExecResult{EvalSlDfS}{ImcIgndfiHardSafe}{Error}{OutOfNativeMemory}{Count}{}{2}%
\StoreBenchExecResult{EvalSlDfS}{ImcIgndfiHardSafe}{Error}{OutOfNativeMemory}{Cputime}{}{50.467133237}%
\StoreBenchExecResult{EvalSlDfS}{ImcIgndfiHardSafe}{Error}{OutOfNativeMemory}{Cputime}{Avg}{25.2335666185}%
\StoreBenchExecResult{EvalSlDfS}{ImcIgndfiHardSafe}{Error}{OutOfNativeMemory}{Cputime}{Median}{25.2335666185}%
\StoreBenchExecResult{EvalSlDfS}{ImcIgndfiHardSafe}{Error}{OutOfNativeMemory}{Cputime}{Min}{22.978721878}%
\StoreBenchExecResult{EvalSlDfS}{ImcIgndfiHardSafe}{Error}{OutOfNativeMemory}{Cputime}{Max}{27.488411359}%
\StoreBenchExecResult{EvalSlDfS}{ImcIgndfiHardSafe}{Error}{OutOfNativeMemory}{Cputime}{Stdev}{2.2548447405000}%
\StoreBenchExecResult{EvalSlDfS}{ImcIgndfiHardSafe}{Error}{OutOfNativeMemory}{Walltime}{}{31.223575392737985}%
\StoreBenchExecResult{EvalSlDfS}{ImcIgndfiHardSafe}{Error}{OutOfNativeMemory}{Walltime}{Avg}{15.6117876963689925}%
\StoreBenchExecResult{EvalSlDfS}{ImcIgndfiHardSafe}{Error}{OutOfNativeMemory}{Walltime}{Median}{15.6117876963689925}%
\StoreBenchExecResult{EvalSlDfS}{ImcIgndfiHardSafe}{Error}{OutOfNativeMemory}{Walltime}{Min}{14.158378108870238}%
\StoreBenchExecResult{EvalSlDfS}{ImcIgndfiHardSafe}{Error}{OutOfNativeMemory}{Walltime}{Max}{17.065197283867747}%
\StoreBenchExecResult{EvalSlDfS}{ImcIgndfiHardSafe}{Error}{OutOfNativeMemory}{Walltime}{Stdev}{1.453409587498754500000000000}%
\StoreBenchExecResult{EvalSlDfS}{ImcIgndfiHardSafe}{Error}{SegmentationFault}{Count}{}{1}%
\StoreBenchExecResult{EvalSlDfS}{ImcIgndfiHardSafe}{Error}{SegmentationFault}{Cputime}{}{575.70592272}%
\StoreBenchExecResult{EvalSlDfS}{ImcIgndfiHardSafe}{Error}{SegmentationFault}{Cputime}{Avg}{575.70592272}%
\StoreBenchExecResult{EvalSlDfS}{ImcIgndfiHardSafe}{Error}{SegmentationFault}{Cputime}{Median}{575.70592272}%
\StoreBenchExecResult{EvalSlDfS}{ImcIgndfiHardSafe}{Error}{SegmentationFault}{Cputime}{Min}{575.70592272}%
\StoreBenchExecResult{EvalSlDfS}{ImcIgndfiHardSafe}{Error}{SegmentationFault}{Cputime}{Max}{575.70592272}%
\StoreBenchExecResult{EvalSlDfS}{ImcIgndfiHardSafe}{Error}{SegmentationFault}{Cputime}{Stdev}{0E-14}%
\StoreBenchExecResult{EvalSlDfS}{ImcIgndfiHardSafe}{Error}{SegmentationFault}{Walltime}{}{570.0703720431775}%
\StoreBenchExecResult{EvalSlDfS}{ImcIgndfiHardSafe}{Error}{SegmentationFault}{Walltime}{Avg}{570.0703720431775}%
\StoreBenchExecResult{EvalSlDfS}{ImcIgndfiHardSafe}{Error}{SegmentationFault}{Walltime}{Median}{570.0703720431775}%
\StoreBenchExecResult{EvalSlDfS}{ImcIgndfiHardSafe}{Error}{SegmentationFault}{Walltime}{Min}{570.0703720431775}%
\StoreBenchExecResult{EvalSlDfS}{ImcIgndfiHardSafe}{Error}{SegmentationFault}{Walltime}{Max}{570.0703720431775}%
\StoreBenchExecResult{EvalSlDfS}{ImcIgndfiHardSafe}{Error}{SegmentationFault}{Walltime}{Stdev}{0E-14}%
\StoreBenchExecResult{EvalSlDfS}{ImcIgndfiHardSafe}{Error}{Timeout}{Count}{}{542}%
\StoreBenchExecResult{EvalSlDfS}{ImcIgndfiHardSafe}{Error}{Timeout}{Cputime}{}{489418.024794103}%
\StoreBenchExecResult{EvalSlDfS}{ImcIgndfiHardSafe}{Error}{Timeout}{Cputime}{Avg}{902.9852855979760147601476015}%
\StoreBenchExecResult{EvalSlDfS}{ImcIgndfiHardSafe}{Error}{Timeout}{Cputime}{Median}{901.905484295}%
\StoreBenchExecResult{EvalSlDfS}{ImcIgndfiHardSafe}{Error}{Timeout}{Cputime}{Min}{901.053377178}%
\StoreBenchExecResult{EvalSlDfS}{ImcIgndfiHardSafe}{Error}{Timeout}{Cputime}{Max}{936.754802521}%
\StoreBenchExecResult{EvalSlDfS}{ImcIgndfiHardSafe}{Error}{Timeout}{Cputime}{Stdev}{3.096227848334105650189761177}%
\StoreBenchExecResult{EvalSlDfS}{ImcIgndfiHardSafe}{Error}{Timeout}{Walltime}{}{444554.33899440849296}%
\StoreBenchExecResult{EvalSlDfS}{ImcIgndfiHardSafe}{Error}{Timeout}{Walltime}{Avg}{820.2109575542592121033210332}%
\StoreBenchExecResult{EvalSlDfS}{ImcIgndfiHardSafe}{Error}{Timeout}{Walltime}{Median}{806.5669892249862}%
\StoreBenchExecResult{EvalSlDfS}{ImcIgndfiHardSafe}{Error}{Timeout}{Walltime}{Min}{387.01273836614564}%
\StoreBenchExecResult{EvalSlDfS}{ImcIgndfiHardSafe}{Error}{Timeout}{Walltime}{Max}{898.3241822579876}%
\StoreBenchExecResult{EvalSlDfS}{ImcIgndfiHardSafe}{Error}{Timeout}{Walltime}{Stdev}{50.40404426479683899592296240}%
\ifdefined\EvalSlDfSImcIgndfiHardSafeTotalCount\else\edef\EvalSlDfSImcIgndfiHardSafeTotalCount{0}\fi
\ifdefined\EvalSlDfSImcIgndfiHardSafeCorrectCount\else\edef\EvalSlDfSImcIgndfiHardSafeCorrectCount{0}\fi
\ifdefined\EvalSlDfSImcIgndfiHardSafeCorrectTrueCount\else\edef\EvalSlDfSImcIgndfiHardSafeCorrectTrueCount{0}\fi
\ifdefined\EvalSlDfSImcIgndfiHardSafeCorrectFalseCount\else\edef\EvalSlDfSImcIgndfiHardSafeCorrectFalseCount{0}\fi
\ifdefined\EvalSlDfSImcIgndfiHardSafeWrongTrueCount\else\edef\EvalSlDfSImcIgndfiHardSafeWrongTrueCount{0}\fi
\ifdefined\EvalSlDfSImcIgndfiHardSafeWrongFalseCount\else\edef\EvalSlDfSImcIgndfiHardSafeWrongFalseCount{0}\fi
\ifdefined\EvalSlDfSImcIgndfiHardSafeErrorTimeoutCount\else\edef\EvalSlDfSImcIgndfiHardSafeErrorTimeoutCount{0}\fi
\ifdefined\EvalSlDfSImcIgndfiHardSafeErrorOutOfMemoryCount\else\edef\EvalSlDfSImcIgndfiHardSafeErrorOutOfMemoryCount{0}\fi
\ifdefined\EvalSlDfSImcIgndfiHardSafeCorrectCputime\else\edef\EvalSlDfSImcIgndfiHardSafeCorrectCputime{0}\fi
\ifdefined\EvalSlDfSImcIgndfiHardSafeCorrectCputimeAvg\else\edef\EvalSlDfSImcIgndfiHardSafeCorrectCputimeAvg{None}\fi
\ifdefined\EvalSlDfSImcIgndfiHardSafeCorrectWalltime\else\edef\EvalSlDfSImcIgndfiHardSafeCorrectWalltime{0}\fi
\ifdefined\EvalSlDfSImcIgndfiHardSafeCorrectWalltimeAvg\else\edef\EvalSlDfSImcIgndfiHardSafeCorrectWalltimeAvg{None}\fi
\edef\EvalSlDfSImcIgndfiHardSafeErrorOtherInconclusiveCount{\the\numexpr \EvalSlDfSImcIgndfiHardSafeTotalCount - \EvalSlDfSImcIgndfiHardSafeCorrectCount - \EvalSlDfSImcIgndfiHardSafeWrongTrueCount - \EvalSlDfSImcIgndfiHardSafeWrongFalseCount - \EvalSlDfSImcIgndfiHardSafeErrorTimeoutCount - \EvalSlDfSImcIgndfiHardSafeErrorOutOfMemoryCount \relax}
\providecommand\StoreBenchExecResult[7]{\expandafter\newcommand\csname#1#2#3#4#5#6\endcsname{#7}}%
\StoreBenchExecResult{EvalSlDfS}{ImcIgndffHardSafe}{Total}{}{Count}{}{1623}%
\StoreBenchExecResult{EvalSlDfS}{ImcIgndffHardSafe}{Total}{}{Cputime}{}{623815.650820764}%
\StoreBenchExecResult{EvalSlDfS}{ImcIgndffHardSafe}{Total}{}{Cputime}{Avg}{384.3596123356524953789279113}%
\StoreBenchExecResult{EvalSlDfS}{ImcIgndffHardSafe}{Total}{}{Cputime}{Median}{153.6376579}%
\StoreBenchExecResult{EvalSlDfS}{ImcIgndffHardSafe}{Total}{}{Cputime}{Min}{4.389563423}%
\StoreBenchExecResult{EvalSlDfS}{ImcIgndffHardSafe}{Total}{}{Cputime}{Max}{960.844313854}%
\StoreBenchExecResult{EvalSlDfS}{ImcIgndffHardSafe}{Total}{}{Cputime}{Stdev}{402.5398204265345802023131683}%
\StoreBenchExecResult{EvalSlDfS}{ImcIgndffHardSafe}{Total}{}{Walltime}{}{529021.6198876225387696}%
\StoreBenchExecResult{EvalSlDfS}{ImcIgndffHardSafe}{Total}{}{Walltime}{Avg}{325.9529389326078489030191004}%
\StoreBenchExecResult{EvalSlDfS}{ImcIgndffHardSafe}{Total}{}{Walltime}{Median}{61.45514691597782}%
\StoreBenchExecResult{EvalSlDfS}{ImcIgndffHardSafe}{Total}{}{Walltime}{Min}{1.6207367829047143}%
\StoreBenchExecResult{EvalSlDfS}{ImcIgndffHardSafe}{Total}{}{Walltime}{Max}{898.3665967129637}%
\StoreBenchExecResult{EvalSlDfS}{ImcIgndffHardSafe}{Total}{}{Walltime}{Stdev}{372.4728817405750670475603778}%
\StoreBenchExecResult{EvalSlDfS}{ImcIgndffHardSafe}{Correct}{}{Count}{}{861}%
\StoreBenchExecResult{EvalSlDfS}{ImcIgndffHardSafe}{Correct}{}{Cputime}{}{60636.461671619}%
\StoreBenchExecResult{EvalSlDfS}{ImcIgndffHardSafe}{Correct}{}{Cputime}{Avg}{70.42562331198490127758420441}%
\StoreBenchExecResult{EvalSlDfS}{ImcIgndffHardSafe}{Correct}{}{Cputime}{Median}{16.490194401}%
\StoreBenchExecResult{EvalSlDfS}{ImcIgndffHardSafe}{Correct}{}{Cputime}{Min}{4.389563423}%
\StoreBenchExecResult{EvalSlDfS}{ImcIgndffHardSafe}{Correct}{}{Cputime}{Max}{898.545311845}%
\StoreBenchExecResult{EvalSlDfS}{ImcIgndffHardSafe}{Correct}{}{Cputime}{Stdev}{131.9448684821684376725783843}%
\StoreBenchExecResult{EvalSlDfS}{ImcIgndffHardSafe}{Correct}{}{Walltime}{}{34685.1921361668499549}%
\StoreBenchExecResult{EvalSlDfS}{ImcIgndffHardSafe}{Correct}{}{Walltime}{Avg}{40.28477600019378624262485482}%
\StoreBenchExecResult{EvalSlDfS}{ImcIgndffHardSafe}{Correct}{}{Walltime}{Median}{5.800977509003133}%
\StoreBenchExecResult{EvalSlDfS}{ImcIgndffHardSafe}{Correct}{}{Walltime}{Min}{1.6207367829047143}%
\StoreBenchExecResult{EvalSlDfS}{ImcIgndffHardSafe}{Correct}{}{Walltime}{Max}{807.4744842050131}%
\StoreBenchExecResult{EvalSlDfS}{ImcIgndffHardSafe}{Correct}{}{Walltime}{Stdev}{108.2944419252977317822000669}%
\StoreBenchExecResult{EvalSlDfS}{ImcIgndffHardSafe}{Correct}{True}{Count}{}{861}%
\StoreBenchExecResult{EvalSlDfS}{ImcIgndffHardSafe}{Correct}{True}{Cputime}{}{60636.461671619}%
\StoreBenchExecResult{EvalSlDfS}{ImcIgndffHardSafe}{Correct}{True}{Cputime}{Avg}{70.42562331198490127758420441}%
\StoreBenchExecResult{EvalSlDfS}{ImcIgndffHardSafe}{Correct}{True}{Cputime}{Median}{16.490194401}%
\StoreBenchExecResult{EvalSlDfS}{ImcIgndffHardSafe}{Correct}{True}{Cputime}{Min}{4.389563423}%
\StoreBenchExecResult{EvalSlDfS}{ImcIgndffHardSafe}{Correct}{True}{Cputime}{Max}{898.545311845}%
\StoreBenchExecResult{EvalSlDfS}{ImcIgndffHardSafe}{Correct}{True}{Cputime}{Stdev}{131.9448684821684376725783843}%
\StoreBenchExecResult{EvalSlDfS}{ImcIgndffHardSafe}{Correct}{True}{Walltime}{}{34685.1921361668499549}%
\StoreBenchExecResult{EvalSlDfS}{ImcIgndffHardSafe}{Correct}{True}{Walltime}{Avg}{40.28477600019378624262485482}%
\StoreBenchExecResult{EvalSlDfS}{ImcIgndffHardSafe}{Correct}{True}{Walltime}{Median}{5.800977509003133}%
\StoreBenchExecResult{EvalSlDfS}{ImcIgndffHardSafe}{Correct}{True}{Walltime}{Min}{1.6207367829047143}%
\StoreBenchExecResult{EvalSlDfS}{ImcIgndffHardSafe}{Correct}{True}{Walltime}{Max}{807.4744842050131}%
\StoreBenchExecResult{EvalSlDfS}{ImcIgndffHardSafe}{Correct}{True}{Walltime}{Stdev}{108.2944419252977317822000669}%

\StoreBenchExecResult{EvalSlDfS}{ImcIgndffHardSafe}{Error}{}{Count}{}{762}%
\StoreBenchExecResult{EvalSlDfS}{ImcIgndffHardSafe}{Error}{}{Cputime}{}{563179.189149145}%
\StoreBenchExecResult{EvalSlDfS}{ImcIgndffHardSafe}{Error}{}{Cputime}{Avg}{739.0803007206627296587926509}%
\StoreBenchExecResult{EvalSlDfS}{ImcIgndffHardSafe}{Error}{}{Cputime}{Median}{901.647590858}%
\StoreBenchExecResult{EvalSlDfS}{ImcIgndffHardSafe}{Error}{}{Cputime}{Min}{5.910879265}%
\StoreBenchExecResult{EvalSlDfS}{ImcIgndffHardSafe}{Error}{}{Cputime}{Max}{960.844313854}%
\StoreBenchExecResult{EvalSlDfS}{ImcIgndffHardSafe}{Error}{}{Cputime}{Stdev}{297.1058427736401663653891887}%
\StoreBenchExecResult{EvalSlDfS}{ImcIgndffHardSafe}{Error}{}{Walltime}{}{494336.4277514556888147}%
\StoreBenchExecResult{EvalSlDfS}{ImcIgndffHardSafe}{Error}{}{Walltime}{Avg}{648.7354694900993291531496063}%
\StoreBenchExecResult{EvalSlDfS}{ImcIgndffHardSafe}{Error}{}{Walltime}{Median}{803.3862628475763}%
\StoreBenchExecResult{EvalSlDfS}{ImcIgndffHardSafe}{Error}{}{Walltime}{Min}{1.997819446027279}%
\StoreBenchExecResult{EvalSlDfS}{ImcIgndffHardSafe}{Error}{}{Walltime}{Max}{898.3665967129637}%
\StoreBenchExecResult{EvalSlDfS}{ImcIgndffHardSafe}{Error}{}{Walltime}{Stdev}{292.9987045743662388852113921}%
\StoreBenchExecResult{EvalSlDfS}{ImcIgndffHardSafe}{Error}{Error}{Count}{}{95}%
\StoreBenchExecResult{EvalSlDfS}{ImcIgndffHardSafe}{Error}{Error}{Cputime}{}{6959.199193577}%
\StoreBenchExecResult{EvalSlDfS}{ImcIgndffHardSafe}{Error}{Error}{Cputime}{Avg}{73.25472835344210526315789474}%
\StoreBenchExecResult{EvalSlDfS}{ImcIgndffHardSafe}{Error}{Error}{Cputime}{Median}{26.654241179}%
\StoreBenchExecResult{EvalSlDfS}{ImcIgndffHardSafe}{Error}{Error}{Cputime}{Min}{5.910879265}%
\StoreBenchExecResult{EvalSlDfS}{ImcIgndffHardSafe}{Error}{Error}{Cputime}{Max}{640.015438062}%
\StoreBenchExecResult{EvalSlDfS}{ImcIgndffHardSafe}{Error}{Error}{Cputime}{Stdev}{90.26311239937641689022927636}%
\StoreBenchExecResult{EvalSlDfS}{ImcIgndffHardSafe}{Error}{Error}{Walltime}{}{4912.3958494486287437}%
\StoreBenchExecResult{EvalSlDfS}{ImcIgndffHardSafe}{Error}{Error}{Walltime}{Avg}{51.70942999419609203894736842}%
\StoreBenchExecResult{EvalSlDfS}{ImcIgndffHardSafe}{Error}{Error}{Walltime}{Median}{11.359909028047696}%
\StoreBenchExecResult{EvalSlDfS}{ImcIgndffHardSafe}{Error}{Error}{Walltime}{Min}{1.997819446027279}%
\StoreBenchExecResult{EvalSlDfS}{ImcIgndffHardSafe}{Error}{Error}{Walltime}{Max}{552.121082775062}%
\StoreBenchExecResult{EvalSlDfS}{ImcIgndffHardSafe}{Error}{Error}{Walltime}{Stdev}{75.70178231308046515170572144}%
\StoreBenchExecResult{EvalSlDfS}{ImcIgndffHardSafe}{Error}{Exception}{Count}{}{6}%
\StoreBenchExecResult{EvalSlDfS}{ImcIgndffHardSafe}{Error}{Exception}{Cputime}{}{1743.023941437}%
\StoreBenchExecResult{EvalSlDfS}{ImcIgndffHardSafe}{Error}{Exception}{Cputime}{Avg}{290.5039902395}%
\StoreBenchExecResult{EvalSlDfS}{ImcIgndffHardSafe}{Error}{Exception}{Cputime}{Median}{83.943433781}%
\StoreBenchExecResult{EvalSlDfS}{ImcIgndffHardSafe}{Error}{Exception}{Cputime}{Min}{30.442761466}%
\StoreBenchExecResult{EvalSlDfS}{ImcIgndffHardSafe}{Error}{Exception}{Cputime}{Max}{807.813435666}%
\StoreBenchExecResult{EvalSlDfS}{ImcIgndffHardSafe}{Error}{Exception}{Cputime}{Stdev}{328.5654841537445972185956222}%
\StoreBenchExecResult{EvalSlDfS}{ImcIgndffHardSafe}{Error}{Exception}{Walltime}{}{1405.065314560430158}%
\StoreBenchExecResult{EvalSlDfS}{ImcIgndffHardSafe}{Error}{Exception}{Walltime}{Avg}{234.1775524267383596666666667}%
\StoreBenchExecResult{EvalSlDfS}{ImcIgndffHardSafe}{Error}{Exception}{Walltime}{Median}{33.771831851452589}%
\StoreBenchExecResult{EvalSlDfS}{ImcIgndffHardSafe}{Error}{Exception}{Walltime}{Min}{10.49653640203178}%
\StoreBenchExecResult{EvalSlDfS}{ImcIgndffHardSafe}{Error}{Exception}{Walltime}{Max}{715.7052715891041}%
\StoreBenchExecResult{EvalSlDfS}{ImcIgndffHardSafe}{Error}{Exception}{Walltime}{Stdev}{301.6087437955067170936337657}%
\StoreBenchExecResult{EvalSlDfS}{ImcIgndffHardSafe}{Error}{OutOfMemory}{Count}{}{113}%
\StoreBenchExecResult{EvalSlDfS}{ImcIgndffHardSafe}{Error}{OutOfMemory}{Cputime}{}{61675.578978554}%
\StoreBenchExecResult{EvalSlDfS}{ImcIgndffHardSafe}{Error}{OutOfMemory}{Cputime}{Avg}{545.8015838810088495575221239}%
\StoreBenchExecResult{EvalSlDfS}{ImcIgndffHardSafe}{Error}{OutOfMemory}{Cputime}{Median}{497.283625869}%
\StoreBenchExecResult{EvalSlDfS}{ImcIgndffHardSafe}{Error}{OutOfMemory}{Cputime}{Min}{292.461037721}%
\StoreBenchExecResult{EvalSlDfS}{ImcIgndffHardSafe}{Error}{OutOfMemory}{Cputime}{Max}{869.041417936}%
\StoreBenchExecResult{EvalSlDfS}{ImcIgndffHardSafe}{Error}{OutOfMemory}{Cputime}{Stdev}{152.0466497006579609370584103}%
\StoreBenchExecResult{EvalSlDfS}{ImcIgndffHardSafe}{Error}{OutOfMemory}{Walltime}{}{40939.31031058402727}%
\StoreBenchExecResult{EvalSlDfS}{ImcIgndffHardSafe}{Error}{OutOfMemory}{Walltime}{Avg}{362.2947815095931616814159292}%
\StoreBenchExecResult{EvalSlDfS}{ImcIgndffHardSafe}{Error}{OutOfMemory}{Walltime}{Median}{325.99138604220934}%
\StoreBenchExecResult{EvalSlDfS}{ImcIgndffHardSafe}{Error}{OutOfMemory}{Walltime}{Min}{178.3885821159929}%
\StoreBenchExecResult{EvalSlDfS}{ImcIgndffHardSafe}{Error}{OutOfMemory}{Walltime}{Max}{726.6958846680354}%
\StoreBenchExecResult{EvalSlDfS}{ImcIgndffHardSafe}{Error}{OutOfMemory}{Walltime}{Stdev}{119.8516793073265118901074500}%
\StoreBenchExecResult{EvalSlDfS}{ImcIgndffHardSafe}{Error}{OutOfNativeMemory}{Count}{}{2}%
\StoreBenchExecResult{EvalSlDfS}{ImcIgndffHardSafe}{Error}{OutOfNativeMemory}{Cputime}{}{46.780725960}%
\StoreBenchExecResult{EvalSlDfS}{ImcIgndffHardSafe}{Error}{OutOfNativeMemory}{Cputime}{Avg}{23.390362980}%
\StoreBenchExecResult{EvalSlDfS}{ImcIgndffHardSafe}{Error}{OutOfNativeMemory}{Cputime}{Median}{23.390362980}%
\StoreBenchExecResult{EvalSlDfS}{ImcIgndffHardSafe}{Error}{OutOfNativeMemory}{Cputime}{Min}{20.345002841}%
\StoreBenchExecResult{EvalSlDfS}{ImcIgndffHardSafe}{Error}{OutOfNativeMemory}{Cputime}{Max}{26.435723119}%
\StoreBenchExecResult{EvalSlDfS}{ImcIgndffHardSafe}{Error}{OutOfNativeMemory}{Cputime}{Stdev}{3.0453601390000}%
\StoreBenchExecResult{EvalSlDfS}{ImcIgndffHardSafe}{Error}{OutOfNativeMemory}{Walltime}{}{29.318209047196433}%
\StoreBenchExecResult{EvalSlDfS}{ImcIgndffHardSafe}{Error}{OutOfNativeMemory}{Walltime}{Avg}{14.6591045235982165}%
\StoreBenchExecResult{EvalSlDfS}{ImcIgndffHardSafe}{Error}{OutOfNativeMemory}{Walltime}{Median}{14.6591045235982165}%
\StoreBenchExecResult{EvalSlDfS}{ImcIgndffHardSafe}{Error}{OutOfNativeMemory}{Walltime}{Min}{12.18005455005914}%
\StoreBenchExecResult{EvalSlDfS}{ImcIgndffHardSafe}{Error}{OutOfNativeMemory}{Walltime}{Max}{17.138154497137293}%
\StoreBenchExecResult{EvalSlDfS}{ImcIgndffHardSafe}{Error}{OutOfNativeMemory}{Walltime}{Stdev}{2.479049973539076500000000000}%
\StoreBenchExecResult{EvalSlDfS}{ImcIgndffHardSafe}{Error}{SegmentationFault}{Count}{}{1}%
\StoreBenchExecResult{EvalSlDfS}{ImcIgndffHardSafe}{Error}{SegmentationFault}{Cputime}{}{629.188798755}%
\StoreBenchExecResult{EvalSlDfS}{ImcIgndffHardSafe}{Error}{SegmentationFault}{Cputime}{Avg}{629.188798755}%
\StoreBenchExecResult{EvalSlDfS}{ImcIgndffHardSafe}{Error}{SegmentationFault}{Cputime}{Median}{629.188798755}%
\StoreBenchExecResult{EvalSlDfS}{ImcIgndffHardSafe}{Error}{SegmentationFault}{Cputime}{Min}{629.188798755}%
\StoreBenchExecResult{EvalSlDfS}{ImcIgndffHardSafe}{Error}{SegmentationFault}{Cputime}{Max}{629.188798755}%
\StoreBenchExecResult{EvalSlDfS}{ImcIgndffHardSafe}{Error}{SegmentationFault}{Cputime}{Stdev}{0E-14}%
\StoreBenchExecResult{EvalSlDfS}{ImcIgndffHardSafe}{Error}{SegmentationFault}{Walltime}{}{623.2380462910514}%
\StoreBenchExecResult{EvalSlDfS}{ImcIgndffHardSafe}{Error}{SegmentationFault}{Walltime}{Avg}{623.2380462910514}%
\StoreBenchExecResult{EvalSlDfS}{ImcIgndffHardSafe}{Error}{SegmentationFault}{Walltime}{Median}{623.2380462910514}%
\StoreBenchExecResult{EvalSlDfS}{ImcIgndffHardSafe}{Error}{SegmentationFault}{Walltime}{Min}{623.2380462910514}%
\StoreBenchExecResult{EvalSlDfS}{ImcIgndffHardSafe}{Error}{SegmentationFault}{Walltime}{Max}{623.2380462910514}%
\StoreBenchExecResult{EvalSlDfS}{ImcIgndffHardSafe}{Error}{SegmentationFault}{Walltime}{Stdev}{0E-14}%
\StoreBenchExecResult{EvalSlDfS}{ImcIgndffHardSafe}{Error}{Timeout}{Count}{}{545}%
\StoreBenchExecResult{EvalSlDfS}{ImcIgndffHardSafe}{Error}{Timeout}{Cputime}{}{492125.417510862}%
\StoreBenchExecResult{EvalSlDfS}{ImcIgndffHardSafe}{Error}{Timeout}{Cputime}{Avg}{902.9824174511229357798165138}%
\StoreBenchExecResult{EvalSlDfS}{ImcIgndffHardSafe}{Error}{Timeout}{Cputime}{Median}{901.865317422}%
\StoreBenchExecResult{EvalSlDfS}{ImcIgndffHardSafe}{Error}{Timeout}{Cputime}{Min}{901.046999356}%
\StoreBenchExecResult{EvalSlDfS}{ImcIgndffHardSafe}{Error}{Timeout}{Cputime}{Max}{960.844313854}%
\StoreBenchExecResult{EvalSlDfS}{ImcIgndffHardSafe}{Error}{Timeout}{Cputime}{Stdev}{3.674195155440456588084654942}%
\StoreBenchExecResult{EvalSlDfS}{ImcIgndffHardSafe}{Error}{Timeout}{Walltime}{}{446427.10002152435481}%
\StoreBenchExecResult{EvalSlDfS}{ImcIgndffHardSafe}{Error}{Timeout}{Walltime}{Avg}{819.1322936174758803853211009}%
\StoreBenchExecResult{EvalSlDfS}{ImcIgndffHardSafe}{Error}{Timeout}{Walltime}{Median}{806.9942094781436}%
\StoreBenchExecResult{EvalSlDfS}{ImcIgndffHardSafe}{Error}{Timeout}{Walltime}{Min}{331.9345284518786}%
\StoreBenchExecResult{EvalSlDfS}{ImcIgndffHardSafe}{Error}{Timeout}{Walltime}{Max}{898.3665967129637}%
\StoreBenchExecResult{EvalSlDfS}{ImcIgndffHardSafe}{Error}{Timeout}{Walltime}{Stdev}{59.19416565358766566387871721}%
\ifdefined\EvalSlDfSImcIgndffHardSafeTotalCount\else\edef\EvalSlDfSImcIgndffHardSafeTotalCount{0}\fi
\ifdefined\EvalSlDfSImcIgndffHardSafeCorrectCount\else\edef\EvalSlDfSImcIgndffHardSafeCorrectCount{0}\fi
\ifdefined\EvalSlDfSImcIgndffHardSafeCorrectTrueCount\else\edef\EvalSlDfSImcIgndffHardSafeCorrectTrueCount{0}\fi
\ifdefined\EvalSlDfSImcIgndffHardSafeCorrectFalseCount\else\edef\EvalSlDfSImcIgndffHardSafeCorrectFalseCount{0}\fi
\ifdefined\EvalSlDfSImcIgndffHardSafeWrongTrueCount\else\edef\EvalSlDfSImcIgndffHardSafeWrongTrueCount{0}\fi
\ifdefined\EvalSlDfSImcIgndffHardSafeWrongFalseCount\else\edef\EvalSlDfSImcIgndffHardSafeWrongFalseCount{0}\fi
\ifdefined\EvalSlDfSImcIgndffHardSafeErrorTimeoutCount\else\edef\EvalSlDfSImcIgndffHardSafeErrorTimeoutCount{0}\fi
\ifdefined\EvalSlDfSImcIgndffHardSafeErrorOutOfMemoryCount\else\edef\EvalSlDfSImcIgndffHardSafeErrorOutOfMemoryCount{0}\fi
\ifdefined\EvalSlDfSImcIgndffHardSafeCorrectCputime\else\edef\EvalSlDfSImcIgndffHardSafeCorrectCputime{0}\fi
\ifdefined\EvalSlDfSImcIgndffHardSafeCorrectCputimeAvg\else\edef\EvalSlDfSImcIgndffHardSafeCorrectCputimeAvg{None}\fi
\ifdefined\EvalSlDfSImcIgndffHardSafeCorrectWalltime\else\edef\EvalSlDfSImcIgndffHardSafeCorrectWalltime{0}\fi
\ifdefined\EvalSlDfSImcIgndffHardSafeCorrectWalltimeAvg\else\edef\EvalSlDfSImcIgndffHardSafeCorrectWalltimeAvg{None}\fi
\edef\EvalSlDfSImcIgndffHardSafeErrorOtherInconclusiveCount{\the\numexpr \EvalSlDfSImcIgndffHardSafeTotalCount - \EvalSlDfSImcIgndffHardSafeCorrectCount - \EvalSlDfSImcIgndffHardSafeWrongTrueCount - \EvalSlDfSImcIgndffHardSafeWrongFalseCount - \EvalSlDfSImcIgndffHardSafeErrorTimeoutCount - \EvalSlDfSImcIgndffHardSafeErrorOutOfMemoryCount \relax}
\providecommand\StoreBenchExecResult[7]{\expandafter\newcommand\csname#1#2#3#4#5#6\endcsname{#7}}%
\StoreBenchExecResult{EvalSlDfS}{KiIgndfHardSafe}{Total}{}{Count}{}{1623}%
\StoreBenchExecResult{EvalSlDfS}{KiIgndfHardSafe}{Total}{}{Cputime}{}{805512.433306398}%
\StoreBenchExecResult{EvalSlDfS}{KiIgndfHardSafe}{Total}{}{Cputime}{Avg}{496.3108030230425138632162662}%
\StoreBenchExecResult{EvalSlDfS}{KiIgndfHardSafe}{Total}{}{Cputime}{Median}{579.189004198}%
\StoreBenchExecResult{EvalSlDfS}{KiIgndfHardSafe}{Total}{}{Cputime}{Min}{4.340483952}%
\StoreBenchExecResult{EvalSlDfS}{KiIgndfHardSafe}{Total}{}{Cputime}{Max}{913.612097539}%
\StoreBenchExecResult{EvalSlDfS}{KiIgndfHardSafe}{Total}{}{Cputime}{Stdev}{393.8272603741783528724520825}%
\StoreBenchExecResult{EvalSlDfS}{KiIgndfHardSafe}{Total}{}{Walltime}{}{702653.0435650919111027}%
\StoreBenchExecResult{EvalSlDfS}{KiIgndfHardSafe}{Total}{}{Walltime}{Avg}{432.9347156901367289603820086}%
\StoreBenchExecResult{EvalSlDfS}{KiIgndfHardSafe}{Total}{}{Walltime}{Median}{382.72340535302646}%
\StoreBenchExecResult{EvalSlDfS}{KiIgndfHardSafe}{Total}{}{Walltime}{Min}{1.6930798878893256}%
\StoreBenchExecResult{EvalSlDfS}{KiIgndfHardSafe}{Total}{}{Walltime}{Max}{898.0256430990994}%
\StoreBenchExecResult{EvalSlDfS}{KiIgndfHardSafe}{Total}{}{Walltime}{Stdev}{386.2097394056667035174611769}%
\StoreBenchExecResult{EvalSlDfS}{KiIgndfHardSafe}{Correct}{}{Count}{}{631}%
\StoreBenchExecResult{EvalSlDfS}{KiIgndfHardSafe}{Correct}{}{Cputime}{}{56775.422464576}%
\StoreBenchExecResult{EvalSlDfS}{KiIgndfHardSafe}{Correct}{}{Cputime}{Avg}{89.97689772516006339144215531}%
\StoreBenchExecResult{EvalSlDfS}{KiIgndfHardSafe}{Correct}{}{Cputime}{Median}{47.823091281}%
\StoreBenchExecResult{EvalSlDfS}{KiIgndfHardSafe}{Correct}{}{Cputime}{Min}{4.340483952}%
\StoreBenchExecResult{EvalSlDfS}{KiIgndfHardSafe}{Correct}{}{Cputime}{Max}{869.27901315}%
\StoreBenchExecResult{EvalSlDfS}{KiIgndfHardSafe}{Correct}{}{Cputime}{Stdev}{123.2033152971355241484552010}%
\StoreBenchExecResult{EvalSlDfS}{KiIgndfHardSafe}{Correct}{}{Walltime}{}{28724.0737446353303807}%
\StoreBenchExecResult{EvalSlDfS}{KiIgndfHardSafe}{Correct}{}{Walltime}{Avg}{45.52151148119703705340729002}%
\StoreBenchExecResult{EvalSlDfS}{KiIgndfHardSafe}{Correct}{}{Walltime}{Median}{12.883415329968557}%
\StoreBenchExecResult{EvalSlDfS}{KiIgndfHardSafe}{Correct}{}{Walltime}{Min}{1.6930798878893256}%
\StoreBenchExecResult{EvalSlDfS}{KiIgndfHardSafe}{Correct}{}{Walltime}{Max}{767.7303217090666}%
\StoreBenchExecResult{EvalSlDfS}{KiIgndfHardSafe}{Correct}{}{Walltime}{Stdev}{98.55729178095425510224753841}%
\StoreBenchExecResult{EvalSlDfS}{KiIgndfHardSafe}{Correct}{True}{Count}{}{631}%
\StoreBenchExecResult{EvalSlDfS}{KiIgndfHardSafe}{Correct}{True}{Cputime}{}{56775.422464576}%
\StoreBenchExecResult{EvalSlDfS}{KiIgndfHardSafe}{Correct}{True}{Cputime}{Avg}{89.97689772516006339144215531}%
\StoreBenchExecResult{EvalSlDfS}{KiIgndfHardSafe}{Correct}{True}{Cputime}{Median}{47.823091281}%
\StoreBenchExecResult{EvalSlDfS}{KiIgndfHardSafe}{Correct}{True}{Cputime}{Min}{4.340483952}%
\StoreBenchExecResult{EvalSlDfS}{KiIgndfHardSafe}{Correct}{True}{Cputime}{Max}{869.27901315}%
\StoreBenchExecResult{EvalSlDfS}{KiIgndfHardSafe}{Correct}{True}{Cputime}{Stdev}{123.2033152971355241484552010}%
\StoreBenchExecResult{EvalSlDfS}{KiIgndfHardSafe}{Correct}{True}{Walltime}{}{28724.0737446353303807}%
\StoreBenchExecResult{EvalSlDfS}{KiIgndfHardSafe}{Correct}{True}{Walltime}{Avg}{45.52151148119703705340729002}%
\StoreBenchExecResult{EvalSlDfS}{KiIgndfHardSafe}{Correct}{True}{Walltime}{Median}{12.883415329968557}%
\StoreBenchExecResult{EvalSlDfS}{KiIgndfHardSafe}{Correct}{True}{Walltime}{Min}{1.6930798878893256}%
\StoreBenchExecResult{EvalSlDfS}{KiIgndfHardSafe}{Correct}{True}{Walltime}{Max}{767.7303217090666}%
\StoreBenchExecResult{EvalSlDfS}{KiIgndfHardSafe}{Correct}{True}{Walltime}{Stdev}{98.55729178095425510224753841}%

\StoreBenchExecResult{EvalSlDfS}{KiIgndfHardSafe}{Error}{}{Count}{}{992}%
\StoreBenchExecResult{EvalSlDfS}{KiIgndfHardSafe}{Error}{}{Cputime}{}{748737.010841822}%
\StoreBenchExecResult{EvalSlDfS}{KiIgndfHardSafe}{Error}{}{Cputime}{Avg}{754.7752125421592741935483871}%
\StoreBenchExecResult{EvalSlDfS}{KiIgndfHardSafe}{Error}{}{Cputime}{Median}{901.9469235975}%
\StoreBenchExecResult{EvalSlDfS}{KiIgndfHardSafe}{Error}{}{Cputime}{Min}{5.640087163}%
\StoreBenchExecResult{EvalSlDfS}{KiIgndfHardSafe}{Error}{}{Cputime}{Max}{913.612097539}%
\StoreBenchExecResult{EvalSlDfS}{KiIgndfHardSafe}{Error}{}{Cputime}{Stdev}{268.8406269064438609220579998}%
\StoreBenchExecResult{EvalSlDfS}{KiIgndfHardSafe}{Error}{}{Walltime}{}{673928.9698204565807220}%
\StoreBenchExecResult{EvalSlDfS}{KiIgndfHardSafe}{Error}{}{Walltime}{Avg}{679.3638808673957466955645161}%
\StoreBenchExecResult{EvalSlDfS}{KiIgndfHardSafe}{Error}{}{Walltime}{Median}{811.69320455903655}%
\StoreBenchExecResult{EvalSlDfS}{KiIgndfHardSafe}{Error}{}{Walltime}{Min}{1.9460978580173105}%
\StoreBenchExecResult{EvalSlDfS}{KiIgndfHardSafe}{Error}{}{Walltime}{Max}{898.0256430990994}%
\StoreBenchExecResult{EvalSlDfS}{KiIgndfHardSafe}{Error}{}{Walltime}{Stdev}{285.7616839232830949100171829}%
\StoreBenchExecResult{EvalSlDfS}{KiIgndfHardSafe}{Error}{Error}{Count}{}{46}%
\StoreBenchExecResult{EvalSlDfS}{KiIgndfHardSafe}{Error}{Error}{Cputime}{}{493.792274237}%
\StoreBenchExecResult{EvalSlDfS}{KiIgndfHardSafe}{Error}{Error}{Cputime}{Avg}{10.73461465732608695652173913}%
\StoreBenchExecResult{EvalSlDfS}{KiIgndfHardSafe}{Error}{Error}{Cputime}{Median}{9.2851304905}%
\StoreBenchExecResult{EvalSlDfS}{KiIgndfHardSafe}{Error}{Error}{Cputime}{Min}{5.640087163}%
\StoreBenchExecResult{EvalSlDfS}{KiIgndfHardSafe}{Error}{Error}{Cputime}{Max}{30.091284439}%
\StoreBenchExecResult{EvalSlDfS}{KiIgndfHardSafe}{Error}{Error}{Cputime}{Stdev}{4.522060413940556055935759300}%
\StoreBenchExecResult{EvalSlDfS}{KiIgndfHardSafe}{Error}{Error}{Walltime}{}{164.8060932073276490}%
\StoreBenchExecResult{EvalSlDfS}{KiIgndfHardSafe}{Error}{Error}{Walltime}{Avg}{3.582741156681035847826086957}%
\StoreBenchExecResult{EvalSlDfS}{KiIgndfHardSafe}{Error}{Error}{Walltime}{Median}{2.97280241106636815}%
\StoreBenchExecResult{EvalSlDfS}{KiIgndfHardSafe}{Error}{Error}{Walltime}{Min}{1.9460978580173105}%
\StoreBenchExecResult{EvalSlDfS}{KiIgndfHardSafe}{Error}{Error}{Walltime}{Max}{12.572210507001728}%
\StoreBenchExecResult{EvalSlDfS}{KiIgndfHardSafe}{Error}{Error}{Walltime}{Stdev}{1.950974011605352132409749537}%
\StoreBenchExecResult{EvalSlDfS}{KiIgndfHardSafe}{Error}{Exception}{Count}{}{2}%
\StoreBenchExecResult{EvalSlDfS}{KiIgndfHardSafe}{Error}{Exception}{Cputime}{}{88.211240487}%
\StoreBenchExecResult{EvalSlDfS}{KiIgndfHardSafe}{Error}{Exception}{Cputime}{Avg}{44.1056202435}%
\StoreBenchExecResult{EvalSlDfS}{KiIgndfHardSafe}{Error}{Exception}{Cputime}{Median}{44.1056202435}%
\StoreBenchExecResult{EvalSlDfS}{KiIgndfHardSafe}{Error}{Exception}{Cputime}{Min}{19.52778297}%
\StoreBenchExecResult{EvalSlDfS}{KiIgndfHardSafe}{Error}{Exception}{Cputime}{Max}{68.683457517}%
\StoreBenchExecResult{EvalSlDfS}{KiIgndfHardSafe}{Error}{Exception}{Cputime}{Stdev}{24.577837273500}%
\StoreBenchExecResult{EvalSlDfS}{KiIgndfHardSafe}{Error}{Exception}{Walltime}{}{24.319282640935853}%
\StoreBenchExecResult{EvalSlDfS}{KiIgndfHardSafe}{Error}{Exception}{Walltime}{Avg}{12.1596413204679265}%
\StoreBenchExecResult{EvalSlDfS}{KiIgndfHardSafe}{Error}{Exception}{Walltime}{Median}{12.1596413204679265}%
\StoreBenchExecResult{EvalSlDfS}{KiIgndfHardSafe}{Error}{Exception}{Walltime}{Min}{5.658640498993918}%
\StoreBenchExecResult{EvalSlDfS}{KiIgndfHardSafe}{Error}{Exception}{Walltime}{Max}{18.660642141941935}%
\StoreBenchExecResult{EvalSlDfS}{KiIgndfHardSafe}{Error}{Exception}{Walltime}{Stdev}{6.501000821474008500000000000}%
\StoreBenchExecResult{EvalSlDfS}{KiIgndfHardSafe}{Error}{OutOfMemory}{Count}{}{237}%
\StoreBenchExecResult{EvalSlDfS}{KiIgndfHardSafe}{Error}{OutOfMemory}{Cputime}{}{109222.735476261}%
\StoreBenchExecResult{EvalSlDfS}{KiIgndfHardSafe}{Error}{OutOfMemory}{Cputime}{Avg}{460.8554239504683544303797468}%
\StoreBenchExecResult{EvalSlDfS}{KiIgndfHardSafe}{Error}{OutOfMemory}{Cputime}{Median}{451.556631065}%
\StoreBenchExecResult{EvalSlDfS}{KiIgndfHardSafe}{Error}{OutOfMemory}{Cputime}{Min}{186.300403209}%
\StoreBenchExecResult{EvalSlDfS}{KiIgndfHardSafe}{Error}{OutOfMemory}{Cputime}{Max}{895.98419357}%
\StoreBenchExecResult{EvalSlDfS}{KiIgndfHardSafe}{Error}{OutOfMemory}{Cputime}{Stdev}{195.4599735349624510714124824}%
\StoreBenchExecResult{EvalSlDfS}{KiIgndfHardSafe}{Error}{OutOfMemory}{Walltime}{}{74954.66343123721922}%
\StoreBenchExecResult{EvalSlDfS}{KiIgndfHardSafe}{Error}{OutOfMemory}{Walltime}{Avg}{316.2644026634481823628691983}%
\StoreBenchExecResult{EvalSlDfS}{KiIgndfHardSafe}{Error}{OutOfMemory}{Walltime}{Median}{271.73661891696975}%
\StoreBenchExecResult{EvalSlDfS}{KiIgndfHardSafe}{Error}{OutOfMemory}{Walltime}{Min}{87.98165804683231}%
\StoreBenchExecResult{EvalSlDfS}{KiIgndfHardSafe}{Error}{OutOfMemory}{Walltime}{Max}{798.8651493170764}%
\StoreBenchExecResult{EvalSlDfS}{KiIgndfHardSafe}{Error}{OutOfMemory}{Walltime}{Stdev}{170.0374393107617026754958322}%
\StoreBenchExecResult{EvalSlDfS}{KiIgndfHardSafe}{Error}{Timeout}{Count}{}{707}%
\StoreBenchExecResult{EvalSlDfS}{KiIgndfHardSafe}{Error}{Timeout}{Cputime}{}{638932.271850837}%
\StoreBenchExecResult{EvalSlDfS}{KiIgndfHardSafe}{Error}{Timeout}{Cputime}{Avg}{903.7231567904342291371994342}%
\StoreBenchExecResult{EvalSlDfS}{KiIgndfHardSafe}{Error}{Timeout}{Cputime}{Median}{902.411463625}%
\StoreBenchExecResult{EvalSlDfS}{KiIgndfHardSafe}{Error}{Timeout}{Cputime}{Min}{900.992192044}%
\StoreBenchExecResult{EvalSlDfS}{KiIgndfHardSafe}{Error}{Timeout}{Cputime}{Max}{913.612097539}%
\StoreBenchExecResult{EvalSlDfS}{KiIgndfHardSafe}{Error}{Timeout}{Cputime}{Stdev}{2.816043400350343209618748503}%
\StoreBenchExecResult{EvalSlDfS}{KiIgndfHardSafe}{Error}{Timeout}{Walltime}{}{598785.1810133710980}%
\StoreBenchExecResult{EvalSlDfS}{KiIgndfHardSafe}{Error}{Timeout}{Walltime}{Avg}{846.9380212353197991513437058}%
\StoreBenchExecResult{EvalSlDfS}{KiIgndfHardSafe}{Error}{Timeout}{Walltime}{Median}{862.5694193840027}%
\StoreBenchExecResult{EvalSlDfS}{KiIgndfHardSafe}{Error}{Timeout}{Walltime}{Min}{747.8283931100741}%
\StoreBenchExecResult{EvalSlDfS}{KiIgndfHardSafe}{Error}{Timeout}{Walltime}{Max}{898.0256430990994}%
\StoreBenchExecResult{EvalSlDfS}{KiIgndfHardSafe}{Error}{Timeout}{Walltime}{Stdev}{40.44608502862143752991965158}%
\ifdefined\EvalSlDfSKiIgndfHardSafeTotalCount\else\edef\EvalSlDfSKiIgndfHardSafeTotalCount{0}\fi
\ifdefined\EvalSlDfSKiIgndfHardSafeCorrectCount\else\edef\EvalSlDfSKiIgndfHardSafeCorrectCount{0}\fi
\ifdefined\EvalSlDfSKiIgndfHardSafeCorrectTrueCount\else\edef\EvalSlDfSKiIgndfHardSafeCorrectTrueCount{0}\fi
\ifdefined\EvalSlDfSKiIgndfHardSafeCorrectFalseCount\else\edef\EvalSlDfSKiIgndfHardSafeCorrectFalseCount{0}\fi
\ifdefined\EvalSlDfSKiIgndfHardSafeWrongTrueCount\else\edef\EvalSlDfSKiIgndfHardSafeWrongTrueCount{0}\fi
\ifdefined\EvalSlDfSKiIgndfHardSafeWrongFalseCount\else\edef\EvalSlDfSKiIgndfHardSafeWrongFalseCount{0}\fi
\ifdefined\EvalSlDfSKiIgndfHardSafeErrorTimeoutCount\else\edef\EvalSlDfSKiIgndfHardSafeErrorTimeoutCount{0}\fi
\ifdefined\EvalSlDfSKiIgndfHardSafeErrorOutOfMemoryCount\else\edef\EvalSlDfSKiIgndfHardSafeErrorOutOfMemoryCount{0}\fi
\ifdefined\EvalSlDfSKiIgndfHardSafeCorrectCputime\else\edef\EvalSlDfSKiIgndfHardSafeCorrectCputime{0}\fi
\ifdefined\EvalSlDfSKiIgndfHardSafeCorrectCputimeAvg\else\edef\EvalSlDfSKiIgndfHardSafeCorrectCputimeAvg{None}\fi
\ifdefined\EvalSlDfSKiIgndfHardSafeCorrectWalltime\else\edef\EvalSlDfSKiIgndfHardSafeCorrectWalltime{0}\fi
\ifdefined\EvalSlDfSKiIgndfHardSafeCorrectWalltimeAvg\else\edef\EvalSlDfSKiIgndfHardSafeCorrectWalltimeAvg{None}\fi
\edef\EvalSlDfSKiIgndfHardSafeErrorOtherInconclusiveCount{\the\numexpr \EvalSlDfSKiIgndfHardSafeTotalCount - \EvalSlDfSKiIgndfHardSafeCorrectCount - \EvalSlDfSKiIgndfHardSafeWrongTrueCount - \EvalSlDfSKiIgndfHardSafeWrongFalseCount - \EvalSlDfSKiIgndfHardSafeErrorTimeoutCount - \EvalSlDfSKiIgndfHardSafeErrorOutOfMemoryCount \relax}
\providecommand\StoreBenchExecResult[7]{\expandafter\newcommand\csname#1#2#3#4#5#6\endcsname{#7}}%
\StoreBenchExecResult{EvalSlDfS}{ImpactHardSafe}{Total}{}{Count}{}{1623}%
\StoreBenchExecResult{EvalSlDfS}{ImpactHardSafe}{Total}{}{Cputime}{}{625675.806832697}%
\StoreBenchExecResult{EvalSlDfS}{ImpactHardSafe}{Total}{}{Cputime}{Avg}{385.5057343393080714725816389}%
\StoreBenchExecResult{EvalSlDfS}{ImpactHardSafe}{Total}{}{Cputime}{Median}{83.122919743}%
\StoreBenchExecResult{EvalSlDfS}{ImpactHardSafe}{Total}{}{Cputime}{Min}{3.667642453}%
\StoreBenchExecResult{EvalSlDfS}{ImpactHardSafe}{Total}{}{Cputime}{Max}{937.452378711}%
\StoreBenchExecResult{EvalSlDfS}{ImpactHardSafe}{Total}{}{Cputime}{Stdev}{422.6342445123661034027545785}%
\StoreBenchExecResult{EvalSlDfS}{ImpactHardSafe}{Total}{}{Walltime}{}{601208.3073872148527365}%
\StoreBenchExecResult{EvalSlDfS}{ImpactHardSafe}{Total}{}{Walltime}{Avg}{370.4302571701878328629081947}%
\StoreBenchExecResult{EvalSlDfS}{ImpactHardSafe}{Total}{}{Walltime}{Median}{66.59498809603974}%
\StoreBenchExecResult{EvalSlDfS}{ImpactHardSafe}{Total}{}{Walltime}{Min}{1.4115563579834998}%
\StoreBenchExecResult{EvalSlDfS}{ImpactHardSafe}{Total}{}{Walltime}{Max}{909.8247127579525}%
\StoreBenchExecResult{EvalSlDfS}{ImpactHardSafe}{Total}{}{Walltime}{Stdev}{415.8289502357941722762987904}%
\StoreBenchExecResult{EvalSlDfS}{ImpactHardSafe}{Correct}{}{Count}{}{801}%
\StoreBenchExecResult{EvalSlDfS}{ImpactHardSafe}{Correct}{}{Cputime}{}{35512.777931906}%
\StoreBenchExecResult{EvalSlDfS}{ImpactHardSafe}{Correct}{}{Cputime}{Avg}{44.33555297366541822721598002}%
\StoreBenchExecResult{EvalSlDfS}{ImpactHardSafe}{Correct}{}{Cputime}{Median}{9.350704616}%
\StoreBenchExecResult{EvalSlDfS}{ImpactHardSafe}{Correct}{}{Cputime}{Min}{3.667642453}%
\StoreBenchExecResult{EvalSlDfS}{ImpactHardSafe}{Correct}{}{Cputime}{Max}{877.092239784}%
\StoreBenchExecResult{EvalSlDfS}{ImpactHardSafe}{Correct}{}{Cputime}{Stdev}{105.6738410810966334577687457}%
\StoreBenchExecResult{EvalSlDfS}{ImpactHardSafe}{Correct}{}{Walltime}{}{27995.2619197696913031}%
\StoreBenchExecResult{EvalSlDfS}{ImpactHardSafe}{Correct}{}{Walltime}{Avg}{34.95038941294593171423220974}%
\StoreBenchExecResult{EvalSlDfS}{ImpactHardSafe}{Correct}{}{Walltime}{Median}{3.252674941904843}%
\StoreBenchExecResult{EvalSlDfS}{ImpactHardSafe}{Correct}{}{Walltime}{Min}{1.472306028008461}%
\StoreBenchExecResult{EvalSlDfS}{ImpactHardSafe}{Correct}{}{Walltime}{Max}{860.8184966859408}%
\StoreBenchExecResult{EvalSlDfS}{ImpactHardSafe}{Correct}{}{Walltime}{Stdev}{101.8449939355819747182418895}%
\StoreBenchExecResult{EvalSlDfS}{ImpactHardSafe}{Correct}{True}{Count}{}{801}%
\StoreBenchExecResult{EvalSlDfS}{ImpactHardSafe}{Correct}{True}{Cputime}{}{35512.777931906}%
\StoreBenchExecResult{EvalSlDfS}{ImpactHardSafe}{Correct}{True}{Cputime}{Avg}{44.33555297366541822721598002}%
\StoreBenchExecResult{EvalSlDfS}{ImpactHardSafe}{Correct}{True}{Cputime}{Median}{9.350704616}%
\StoreBenchExecResult{EvalSlDfS}{ImpactHardSafe}{Correct}{True}{Cputime}{Min}{3.667642453}%
\StoreBenchExecResult{EvalSlDfS}{ImpactHardSafe}{Correct}{True}{Cputime}{Max}{877.092239784}%
\StoreBenchExecResult{EvalSlDfS}{ImpactHardSafe}{Correct}{True}{Cputime}{Stdev}{105.6738410810966334577687457}%
\StoreBenchExecResult{EvalSlDfS}{ImpactHardSafe}{Correct}{True}{Walltime}{}{27995.2619197696913031}%
\StoreBenchExecResult{EvalSlDfS}{ImpactHardSafe}{Correct}{True}{Walltime}{Avg}{34.95038941294593171423220974}%
\StoreBenchExecResult{EvalSlDfS}{ImpactHardSafe}{Correct}{True}{Walltime}{Median}{3.252674941904843}%
\StoreBenchExecResult{EvalSlDfS}{ImpactHardSafe}{Correct}{True}{Walltime}{Min}{1.472306028008461}%
\StoreBenchExecResult{EvalSlDfS}{ImpactHardSafe}{Correct}{True}{Walltime}{Max}{860.8184966859408}%
\StoreBenchExecResult{EvalSlDfS}{ImpactHardSafe}{Correct}{True}{Walltime}{Stdev}{101.8449939355819747182418895}%

\StoreBenchExecResult{EvalSlDfS}{ImpactHardSafe}{Error}{}{Count}{}{822}%
\StoreBenchExecResult{EvalSlDfS}{ImpactHardSafe}{Error}{}{Cputime}{}{590163.028900791}%
\StoreBenchExecResult{EvalSlDfS}{ImpactHardSafe}{Error}{}{Cputime}{Avg}{717.9598891737116788321167883}%
\StoreBenchExecResult{EvalSlDfS}{ImpactHardSafe}{Error}{}{Cputime}{Median}{902.010961956}%
\StoreBenchExecResult{EvalSlDfS}{ImpactHardSafe}{Error}{}{Cputime}{Min}{3.718918754}%
\StoreBenchExecResult{EvalSlDfS}{ImpactHardSafe}{Error}{}{Cputime}{Max}{937.452378711}%
\StoreBenchExecResult{EvalSlDfS}{ImpactHardSafe}{Error}{}{Cputime}{Stdev}{343.2859586469269266229151093}%
\StoreBenchExecResult{EvalSlDfS}{ImpactHardSafe}{Error}{}{Walltime}{}{573213.0454674451614334}%
\StoreBenchExecResult{EvalSlDfS}{ImpactHardSafe}{Error}{}{Walltime}{Avg}{697.3394713715877876318734793}%
\StoreBenchExecResult{EvalSlDfS}{ImpactHardSafe}{Error}{}{Walltime}{Median}{888.33448113547635}%
\StoreBenchExecResult{EvalSlDfS}{ImpactHardSafe}{Error}{}{Walltime}{Min}{1.4115563579834998}%
\StoreBenchExecResult{EvalSlDfS}{ImpactHardSafe}{Error}{}{Walltime}{Max}{909.8247127579525}%
\StoreBenchExecResult{EvalSlDfS}{ImpactHardSafe}{Error}{}{Walltime}{Stdev}{338.7645485329195387492829669}%
\StoreBenchExecResult{EvalSlDfS}{ImpactHardSafe}{Error}{Error}{Count}{}{115}%
\StoreBenchExecResult{EvalSlDfS}{ImpactHardSafe}{Error}{Error}{Cputime}{}{3212.319413170}%
\StoreBenchExecResult{EvalSlDfS}{ImpactHardSafe}{Error}{Error}{Cputime}{Avg}{27.93321228843478260869565217}%
\StoreBenchExecResult{EvalSlDfS}{ImpactHardSafe}{Error}{Error}{Cputime}{Median}{7.330223879}%
\StoreBenchExecResult{EvalSlDfS}{ImpactHardSafe}{Error}{Error}{Cputime}{Min}{3.718918754}%
\StoreBenchExecResult{EvalSlDfS}{ImpactHardSafe}{Error}{Error}{Cputime}{Max}{854.874453185}%
\StoreBenchExecResult{EvalSlDfS}{ImpactHardSafe}{Error}{Error}{Cputime}{Stdev}{92.09252259700966671826367008}%
\StoreBenchExecResult{EvalSlDfS}{ImpactHardSafe}{Error}{Error}{Walltime}{}{2607.8426847066730130}%
\StoreBenchExecResult{EvalSlDfS}{ImpactHardSafe}{Error}{Error}{Walltime}{Avg}{22.67689291049280880869565217}%
\StoreBenchExecResult{EvalSlDfS}{ImpactHardSafe}{Error}{Error}{Walltime}{Median}{2.603470346191898}%
\StoreBenchExecResult{EvalSlDfS}{ImpactHardSafe}{Error}{Error}{Walltime}{Min}{1.4115563579834998}%
\StoreBenchExecResult{EvalSlDfS}{ImpactHardSafe}{Error}{Error}{Walltime}{Max}{844.2773178759962}%
\StoreBenchExecResult{EvalSlDfS}{ImpactHardSafe}{Error}{Error}{Walltime}{Stdev}{91.42938780517330488695124720}%
\StoreBenchExecResult{EvalSlDfS}{ImpactHardSafe}{Error}{Exception}{Count}{}{11}%
\StoreBenchExecResult{EvalSlDfS}{ImpactHardSafe}{Error}{Exception}{Cputime}{}{259.874609054}%
\StoreBenchExecResult{EvalSlDfS}{ImpactHardSafe}{Error}{Exception}{Cputime}{Avg}{23.62496445945454545454545455}%
\StoreBenchExecResult{EvalSlDfS}{ImpactHardSafe}{Error}{Exception}{Cputime}{Median}{17.350155321}%
\StoreBenchExecResult{EvalSlDfS}{ImpactHardSafe}{Error}{Exception}{Cputime}{Min}{6.719567262}%
\StoreBenchExecResult{EvalSlDfS}{ImpactHardSafe}{Error}{Exception}{Cputime}{Max}{53.275712396}%
\StoreBenchExecResult{EvalSlDfS}{ImpactHardSafe}{Error}{Exception}{Cputime}{Stdev}{15.11163129345403760088303460}%
\StoreBenchExecResult{EvalSlDfS}{ImpactHardSafe}{Error}{Exception}{Walltime}{}{115.5171431771013884}%
\StoreBenchExecResult{EvalSlDfS}{ImpactHardSafe}{Error}{Exception}{Walltime}{Avg}{10.50155847064558076363636364}%
\StoreBenchExecResult{EvalSlDfS}{ImpactHardSafe}{Error}{Exception}{Walltime}{Median}{6.9386294670403}%
\StoreBenchExecResult{EvalSlDfS}{ImpactHardSafe}{Error}{Exception}{Walltime}{Min}{2.465361797949299}%
\StoreBenchExecResult{EvalSlDfS}{ImpactHardSafe}{Error}{Exception}{Walltime}{Max}{27.692742268089205}%
\StoreBenchExecResult{EvalSlDfS}{ImpactHardSafe}{Error}{Exception}{Walltime}{Stdev}{8.098720256280821297096248964}%
\StoreBenchExecResult{EvalSlDfS}{ImpactHardSafe}{Error}{OutOfMemory}{Count}{}{75}%
\StoreBenchExecResult{EvalSlDfS}{ImpactHardSafe}{Error}{OutOfMemory}{Cputime}{}{27353.864941681}%
\StoreBenchExecResult{EvalSlDfS}{ImpactHardSafe}{Error}{OutOfMemory}{Cputime}{Avg}{364.7181992224133333333333333}%
\StoreBenchExecResult{EvalSlDfS}{ImpactHardSafe}{Error}{OutOfMemory}{Cputime}{Median}{276.503043325}%
\StoreBenchExecResult{EvalSlDfS}{ImpactHardSafe}{Error}{OutOfMemory}{Cputime}{Min}{96.778708281}%
\StoreBenchExecResult{EvalSlDfS}{ImpactHardSafe}{Error}{OutOfMemory}{Cputime}{Max}{894.945118973}%
\StoreBenchExecResult{EvalSlDfS}{ImpactHardSafe}{Error}{OutOfMemory}{Cputime}{Stdev}{210.5914738907465689358640346}%
\StoreBenchExecResult{EvalSlDfS}{ImpactHardSafe}{Error}{OutOfMemory}{Walltime}{}{25262.49850797862750}%
\StoreBenchExecResult{EvalSlDfS}{ImpactHardSafe}{Error}{OutOfMemory}{Walltime}{Avg}{336.8333134397150333333333333}%
\StoreBenchExecResult{EvalSlDfS}{ImpactHardSafe}{Error}{OutOfMemory}{Walltime}{Median}{253.50899201305583}%
\StoreBenchExecResult{EvalSlDfS}{ImpactHardSafe}{Error}{OutOfMemory}{Walltime}{Min}{89.82783197704703}%
\StoreBenchExecResult{EvalSlDfS}{ImpactHardSafe}{Error}{OutOfMemory}{Walltime}{Max}{837.8989532811102}%
\StoreBenchExecResult{EvalSlDfS}{ImpactHardSafe}{Error}{OutOfMemory}{Walltime}{Stdev}{196.1591819621856127836882820}%
\StoreBenchExecResult{EvalSlDfS}{ImpactHardSafe}{Error}{OutOfNativeMemory}{Count}{}{2}%
\StoreBenchExecResult{EvalSlDfS}{ImpactHardSafe}{Error}{OutOfNativeMemory}{Cputime}{}{18.182446655}%
\StoreBenchExecResult{EvalSlDfS}{ImpactHardSafe}{Error}{OutOfNativeMemory}{Cputime}{Avg}{9.0912233275}%
\StoreBenchExecResult{EvalSlDfS}{ImpactHardSafe}{Error}{OutOfNativeMemory}{Cputime}{Median}{9.0912233275}%
\StoreBenchExecResult{EvalSlDfS}{ImpactHardSafe}{Error}{OutOfNativeMemory}{Cputime}{Min}{7.707902978}%
\StoreBenchExecResult{EvalSlDfS}{ImpactHardSafe}{Error}{OutOfNativeMemory}{Cputime}{Max}{10.474543677}%
\StoreBenchExecResult{EvalSlDfS}{ImpactHardSafe}{Error}{OutOfNativeMemory}{Cputime}{Stdev}{1.38332034950000}%
\StoreBenchExecResult{EvalSlDfS}{ImpactHardSafe}{Error}{OutOfNativeMemory}{Walltime}{}{6.940640599932521}%
\StoreBenchExecResult{EvalSlDfS}{ImpactHardSafe}{Error}{OutOfNativeMemory}{Walltime}{Avg}{3.4703202999662605}%
\StoreBenchExecResult{EvalSlDfS}{ImpactHardSafe}{Error}{OutOfNativeMemory}{Walltime}{Median}{3.4703202999662605}%
\StoreBenchExecResult{EvalSlDfS}{ImpactHardSafe}{Error}{OutOfNativeMemory}{Walltime}{Min}{2.597530666971579}%
\StoreBenchExecResult{EvalSlDfS}{ImpactHardSafe}{Error}{OutOfNativeMemory}{Walltime}{Max}{4.343109932960942}%
\StoreBenchExecResult{EvalSlDfS}{ImpactHardSafe}{Error}{OutOfNativeMemory}{Walltime}{Stdev}{0.8727896329946815000000000000}%
\StoreBenchExecResult{EvalSlDfS}{ImpactHardSafe}{Error}{SegmentationFault}{Count}{}{1}%
\StoreBenchExecResult{EvalSlDfS}{ImpactHardSafe}{Error}{SegmentationFault}{Cputime}{}{7.767350061}%
\StoreBenchExecResult{EvalSlDfS}{ImpactHardSafe}{Error}{SegmentationFault}{Cputime}{Avg}{7.767350061}%
\StoreBenchExecResult{EvalSlDfS}{ImpactHardSafe}{Error}{SegmentationFault}{Cputime}{Median}{7.767350061}%
\StoreBenchExecResult{EvalSlDfS}{ImpactHardSafe}{Error}{SegmentationFault}{Cputime}{Min}{7.767350061}%
\StoreBenchExecResult{EvalSlDfS}{ImpactHardSafe}{Error}{SegmentationFault}{Cputime}{Max}{7.767350061}%
\StoreBenchExecResult{EvalSlDfS}{ImpactHardSafe}{Error}{SegmentationFault}{Cputime}{Stdev}{0E-14}%
\StoreBenchExecResult{EvalSlDfS}{ImpactHardSafe}{Error}{SegmentationFault}{Walltime}{}{4.293340474134311}%
\StoreBenchExecResult{EvalSlDfS}{ImpactHardSafe}{Error}{SegmentationFault}{Walltime}{Avg}{4.293340474134311}%
\StoreBenchExecResult{EvalSlDfS}{ImpactHardSafe}{Error}{SegmentationFault}{Walltime}{Median}{4.293340474134311}%
\StoreBenchExecResult{EvalSlDfS}{ImpactHardSafe}{Error}{SegmentationFault}{Walltime}{Min}{4.293340474134311}%
\StoreBenchExecResult{EvalSlDfS}{ImpactHardSafe}{Error}{SegmentationFault}{Walltime}{Max}{4.293340474134311}%
\StoreBenchExecResult{EvalSlDfS}{ImpactHardSafe}{Error}{SegmentationFault}{Walltime}{Stdev}{0E-15}%
\StoreBenchExecResult{EvalSlDfS}{ImpactHardSafe}{Error}{Timeout}{Count}{}{618}%
\StoreBenchExecResult{EvalSlDfS}{ImpactHardSafe}{Error}{Timeout}{Cputime}{}{559311.020140170}%
\StoreBenchExecResult{EvalSlDfS}{ImpactHardSafe}{Error}{Timeout}{Cputime}{Avg}{905.0340131717961165048543689}%
\StoreBenchExecResult{EvalSlDfS}{ImpactHardSafe}{Error}{Timeout}{Cputime}{Median}{903.801759968}%
\StoreBenchExecResult{EvalSlDfS}{ImpactHardSafe}{Error}{Timeout}{Cputime}{Min}{901.124683105}%
\StoreBenchExecResult{EvalSlDfS}{ImpactHardSafe}{Error}{Timeout}{Cputime}{Max}{937.452378711}%
\StoreBenchExecResult{EvalSlDfS}{ImpactHardSafe}{Error}{Timeout}{Cputime}{Stdev}{4.050752163115742841305157892}%
\StoreBenchExecResult{EvalSlDfS}{ImpactHardSafe}{Error}{Timeout}{Walltime}{}{545215.9531505086927}%
\StoreBenchExecResult{EvalSlDfS}{ImpactHardSafe}{Error}{Timeout}{Walltime}{Avg}{882.2264614085901176375404531}%
\StoreBenchExecResult{EvalSlDfS}{ImpactHardSafe}{Error}{Timeout}{Walltime}{Median}{893.3046599810477}%
\StoreBenchExecResult{EvalSlDfS}{ImpactHardSafe}{Error}{Timeout}{Walltime}{Min}{536.1248709810898}%
\StoreBenchExecResult{EvalSlDfS}{ImpactHardSafe}{Error}{Timeout}{Walltime}{Max}{909.8247127579525}%
\StoreBenchExecResult{EvalSlDfS}{ImpactHardSafe}{Error}{Timeout}{Walltime}{Stdev}{32.09147362505997872086881999}%
\ifdefined\EvalSlDfSImpactHardSafeTotalCount\else\edef\EvalSlDfSImpactHardSafeTotalCount{0}\fi
\ifdefined\EvalSlDfSImpactHardSafeCorrectCount\else\edef\EvalSlDfSImpactHardSafeCorrectCount{0}\fi
\ifdefined\EvalSlDfSImpactHardSafeCorrectTrueCount\else\edef\EvalSlDfSImpactHardSafeCorrectTrueCount{0}\fi
\ifdefined\EvalSlDfSImpactHardSafeCorrectFalseCount\else\edef\EvalSlDfSImpactHardSafeCorrectFalseCount{0}\fi
\ifdefined\EvalSlDfSImpactHardSafeWrongTrueCount\else\edef\EvalSlDfSImpactHardSafeWrongTrueCount{0}\fi
\ifdefined\EvalSlDfSImpactHardSafeWrongFalseCount\else\edef\EvalSlDfSImpactHardSafeWrongFalseCount{0}\fi
\ifdefined\EvalSlDfSImpactHardSafeErrorTimeoutCount\else\edef\EvalSlDfSImpactHardSafeErrorTimeoutCount{0}\fi
\ifdefined\EvalSlDfSImpactHardSafeErrorOutOfMemoryCount\else\edef\EvalSlDfSImpactHardSafeErrorOutOfMemoryCount{0}\fi
\ifdefined\EvalSlDfSImpactHardSafeCorrectCputime\else\edef\EvalSlDfSImpactHardSafeCorrectCputime{0}\fi
\ifdefined\EvalSlDfSImpactHardSafeCorrectCputimeAvg\else\edef\EvalSlDfSImpactHardSafeCorrectCputimeAvg{None}\fi
\ifdefined\EvalSlDfSImpactHardSafeCorrectWalltime\else\edef\EvalSlDfSImpactHardSafeCorrectWalltime{0}\fi
\ifdefined\EvalSlDfSImpactHardSafeCorrectWalltimeAvg\else\edef\EvalSlDfSImpactHardSafeCorrectWalltimeAvg{None}\fi
\edef\EvalSlDfSImpactHardSafeErrorOtherInconclusiveCount{\the\numexpr \EvalSlDfSImpactHardSafeTotalCount - \EvalSlDfSImpactHardSafeCorrectCount - \EvalSlDfSImpactHardSafeWrongTrueCount - \EvalSlDfSImpactHardSafeWrongFalseCount - \EvalSlDfSImpactHardSafeErrorTimeoutCount - \EvalSlDfSImpactHardSafeErrorOutOfMemoryCount \relax}
\providecommand\StoreBenchExecResult[7]{\expandafter\newcommand\csname#1#2#3#4#5#6\endcsname{#7}}%
\StoreBenchExecResult{EvalSlDfS}{PredabsHardSafe}{Total}{}{Count}{}{1623}%
\StoreBenchExecResult{EvalSlDfS}{PredabsHardSafe}{Total}{}{Cputime}{}{698172.936011471}%
\StoreBenchExecResult{EvalSlDfS}{PredabsHardSafe}{Total}{}{Cputime}{Avg}{430.1743290274004929143561306}%
\StoreBenchExecResult{EvalSlDfS}{PredabsHardSafe}{Total}{}{Cputime}{Median}{126.31787407}%
\StoreBenchExecResult{EvalSlDfS}{PredabsHardSafe}{Total}{}{Cputime}{Min}{3.83685884}%
\StoreBenchExecResult{EvalSlDfS}{PredabsHardSafe}{Total}{}{Cputime}{Max}{935.255565302}%
\StoreBenchExecResult{EvalSlDfS}{PredabsHardSafe}{Total}{}{Cputime}{Stdev}{431.6485778676166661828955487}%
\StoreBenchExecResult{EvalSlDfS}{PredabsHardSafe}{Total}{}{Walltime}{}{666416.6618090423292261}%
\StoreBenchExecResult{EvalSlDfS}{PredabsHardSafe}{Total}{}{Walltime}{Avg}{410.6079247129034684079482440}%
\StoreBenchExecResult{EvalSlDfS}{PredabsHardSafe}{Total}{}{Walltime}{Median}{96.26409202907234}%
\StoreBenchExecResult{EvalSlDfS}{PredabsHardSafe}{Total}{}{Walltime}{Min}{1.5104425649624318}%
\StoreBenchExecResult{EvalSlDfS}{PredabsHardSafe}{Total}{}{Walltime}{Max}{908.897384821903}%
\StoreBenchExecResult{EvalSlDfS}{PredabsHardSafe}{Total}{}{Walltime}{Stdev}{421.5092299885446562460959936}%
\StoreBenchExecResult{EvalSlDfS}{PredabsHardSafe}{Correct}{}{Count}{}{780}%
\StoreBenchExecResult{EvalSlDfS}{PredabsHardSafe}{Correct}{}{Cputime}{}{42267.117379623}%
\StoreBenchExecResult{EvalSlDfS}{PredabsHardSafe}{Correct}{}{Cputime}{Avg}{54.18861202515769230769230769}%
\StoreBenchExecResult{EvalSlDfS}{PredabsHardSafe}{Correct}{}{Cputime}{Median}{10.303212481}%
\StoreBenchExecResult{EvalSlDfS}{PredabsHardSafe}{Correct}{}{Cputime}{Min}{3.83685884}%
\StoreBenchExecResult{EvalSlDfS}{PredabsHardSafe}{Correct}{}{Cputime}{Max}{867.060171817}%
\StoreBenchExecResult{EvalSlDfS}{PredabsHardSafe}{Correct}{}{Cputime}{Stdev}{120.5837287983612518279333221}%
\StoreBenchExecResult{EvalSlDfS}{PredabsHardSafe}{Correct}{}{Walltime}{}{34244.0497224796565856}%
\StoreBenchExecResult{EvalSlDfS}{PredabsHardSafe}{Correct}{}{Walltime}{Avg}{43.90262784933289305846153846}%
\StoreBenchExecResult{EvalSlDfS}{PredabsHardSafe}{Correct}{}{Walltime}{Median}{3.68548357754480095}%
\StoreBenchExecResult{EvalSlDfS}{PredabsHardSafe}{Correct}{}{Walltime}{Min}{1.5104425649624318}%
\StoreBenchExecResult{EvalSlDfS}{PredabsHardSafe}{Correct}{}{Walltime}{Max}{835.0918126839679}%
\StoreBenchExecResult{EvalSlDfS}{PredabsHardSafe}{Correct}{}{Walltime}{Stdev}{117.1834928985434656281659052}%
\StoreBenchExecResult{EvalSlDfS}{PredabsHardSafe}{Correct}{True}{Count}{}{780}%
\StoreBenchExecResult{EvalSlDfS}{PredabsHardSafe}{Correct}{True}{Cputime}{}{42267.117379623}%
\StoreBenchExecResult{EvalSlDfS}{PredabsHardSafe}{Correct}{True}{Cputime}{Avg}{54.18861202515769230769230769}%
\StoreBenchExecResult{EvalSlDfS}{PredabsHardSafe}{Correct}{True}{Cputime}{Median}{10.303212481}%
\StoreBenchExecResult{EvalSlDfS}{PredabsHardSafe}{Correct}{True}{Cputime}{Min}{3.83685884}%
\StoreBenchExecResult{EvalSlDfS}{PredabsHardSafe}{Correct}{True}{Cputime}{Max}{867.060171817}%
\StoreBenchExecResult{EvalSlDfS}{PredabsHardSafe}{Correct}{True}{Cputime}{Stdev}{120.5837287983612518279333221}%
\StoreBenchExecResult{EvalSlDfS}{PredabsHardSafe}{Correct}{True}{Walltime}{}{34244.0497224796565856}%
\StoreBenchExecResult{EvalSlDfS}{PredabsHardSafe}{Correct}{True}{Walltime}{Avg}{43.90262784933289305846153846}%
\StoreBenchExecResult{EvalSlDfS}{PredabsHardSafe}{Correct}{True}{Walltime}{Median}{3.68548357754480095}%
\StoreBenchExecResult{EvalSlDfS}{PredabsHardSafe}{Correct}{True}{Walltime}{Min}{1.5104425649624318}%
\StoreBenchExecResult{EvalSlDfS}{PredabsHardSafe}{Correct}{True}{Walltime}{Max}{835.0918126839679}%
\StoreBenchExecResult{EvalSlDfS}{PredabsHardSafe}{Correct}{True}{Walltime}{Stdev}{117.1834928985434656281659052}%

\StoreBenchExecResult{EvalSlDfS}{PredabsHardSafe}{Error}{}{Count}{}{843}%
\StoreBenchExecResult{EvalSlDfS}{PredabsHardSafe}{Error}{}{Cputime}{}{655905.818631848}%
\StoreBenchExecResult{EvalSlDfS}{PredabsHardSafe}{Error}{}{Cputime}{Avg}{778.0614693141731909845788849}%
\StoreBenchExecResult{EvalSlDfS}{PredabsHardSafe}{Error}{}{Cputime}{Median}{902.30590276}%
\StoreBenchExecResult{EvalSlDfS}{PredabsHardSafe}{Error}{}{Cputime}{Min}{3.845252283}%
\StoreBenchExecResult{EvalSlDfS}{PredabsHardSafe}{Error}{}{Cputime}{Max}{935.255565302}%
\StoreBenchExecResult{EvalSlDfS}{PredabsHardSafe}{Error}{}{Cputime}{Stdev}{305.6744019079400068581703514}%
\StoreBenchExecResult{EvalSlDfS}{PredabsHardSafe}{Error}{}{Walltime}{}{632172.6120865626726405}%
\StoreBenchExecResult{EvalSlDfS}{PredabsHardSafe}{Error}{}{Walltime}{Avg}{749.9081993909402996921708185}%
\StoreBenchExecResult{EvalSlDfS}{PredabsHardSafe}{Error}{}{Walltime}{Median}{885.9045387019869}%
\StoreBenchExecResult{EvalSlDfS}{PredabsHardSafe}{Error}{}{Walltime}{Min}{1.5583017419558018}%
\StoreBenchExecResult{EvalSlDfS}{PredabsHardSafe}{Error}{}{Walltime}{Max}{908.897384821903}%
\StoreBenchExecResult{EvalSlDfS}{PredabsHardSafe}{Error}{}{Walltime}{Stdev}{299.6808432115865926093468268}%
\StoreBenchExecResult{EvalSlDfS}{PredabsHardSafe}{Error}{Error}{Count}{}{118}%
\StoreBenchExecResult{EvalSlDfS}{PredabsHardSafe}{Error}{Error}{Cputime}{}{4525.869151006}%
\StoreBenchExecResult{EvalSlDfS}{PredabsHardSafe}{Error}{Error}{Cputime}{Avg}{38.35482331361016949152542373}%
\StoreBenchExecResult{EvalSlDfS}{PredabsHardSafe}{Error}{Error}{Cputime}{Median}{8.0809586145}%
\StoreBenchExecResult{EvalSlDfS}{PredabsHardSafe}{Error}{Error}{Cputime}{Min}{3.845252283}%
\StoreBenchExecResult{EvalSlDfS}{PredabsHardSafe}{Error}{Error}{Cputime}{Max}{773.592629881}%
\StoreBenchExecResult{EvalSlDfS}{PredabsHardSafe}{Error}{Error}{Cputime}{Stdev}{106.1160467078988795342908502}%
\StoreBenchExecResult{EvalSlDfS}{PredabsHardSafe}{Error}{Error}{Walltime}{}{3814.0359228048473039}%
\StoreBenchExecResult{EvalSlDfS}{PredabsHardSafe}{Error}{Error}{Walltime}{Avg}{32.32233832885463816864406780}%
\StoreBenchExecResult{EvalSlDfS}{PredabsHardSafe}{Error}{Error}{Walltime}{Median}{2.8217260305536911}%
\StoreBenchExecResult{EvalSlDfS}{PredabsHardSafe}{Error}{Error}{Walltime}{Min}{1.5583017419558018}%
\StoreBenchExecResult{EvalSlDfS}{PredabsHardSafe}{Error}{Error}{Walltime}{Max}{765.3318508500233}%
\StoreBenchExecResult{EvalSlDfS}{PredabsHardSafe}{Error}{Error}{Walltime}{Stdev}{105.0603282030256204666092568}%
\StoreBenchExecResult{EvalSlDfS}{PredabsHardSafe}{Error}{OutOfMemory}{Count}{}{11}%
\StoreBenchExecResult{EvalSlDfS}{PredabsHardSafe}{Error}{OutOfMemory}{Cputime}{}{7510.680548906}%
\StoreBenchExecResult{EvalSlDfS}{PredabsHardSafe}{Error}{OutOfMemory}{Cputime}{Avg}{682.7891408096363636363636364}%
\StoreBenchExecResult{EvalSlDfS}{PredabsHardSafe}{Error}{OutOfMemory}{Cputime}{Median}{662.86969673}%
\StoreBenchExecResult{EvalSlDfS}{PredabsHardSafe}{Error}{OutOfMemory}{Cputime}{Min}{450.849853889}%
\StoreBenchExecResult{EvalSlDfS}{PredabsHardSafe}{Error}{OutOfMemory}{Cputime}{Max}{866.147408546}%
\StoreBenchExecResult{EvalSlDfS}{PredabsHardSafe}{Error}{OutOfMemory}{Cputime}{Stdev}{148.2072901358198403976273631}%
\StoreBenchExecResult{EvalSlDfS}{PredabsHardSafe}{Error}{OutOfMemory}{Walltime}{}{6364.93902181205355}%
\StoreBenchExecResult{EvalSlDfS}{PredabsHardSafe}{Error}{OutOfMemory}{Walltime}{Avg}{578.6308201647321409090909091}%
\StoreBenchExecResult{EvalSlDfS}{PredabsHardSafe}{Error}{OutOfMemory}{Walltime}{Median}{548.8872601580806}%
\StoreBenchExecResult{EvalSlDfS}{PredabsHardSafe}{Error}{OutOfMemory}{Walltime}{Min}{357.43307993910275}%
\StoreBenchExecResult{EvalSlDfS}{PredabsHardSafe}{Error}{OutOfMemory}{Walltime}{Max}{758.2471495999489}%
\StoreBenchExecResult{EvalSlDfS}{PredabsHardSafe}{Error}{OutOfMemory}{Walltime}{Stdev}{144.3525423159530994514347951}%
\StoreBenchExecResult{EvalSlDfS}{PredabsHardSafe}{Error}{OutOfNativeMemory}{Count}{}{2}%
\StoreBenchExecResult{EvalSlDfS}{PredabsHardSafe}{Error}{OutOfNativeMemory}{Cputime}{}{21.014669087}%
\StoreBenchExecResult{EvalSlDfS}{PredabsHardSafe}{Error}{OutOfNativeMemory}{Cputime}{Avg}{10.5073345435}%
\StoreBenchExecResult{EvalSlDfS}{PredabsHardSafe}{Error}{OutOfNativeMemory}{Cputime}{Median}{10.5073345435}%
\StoreBenchExecResult{EvalSlDfS}{PredabsHardSafe}{Error}{OutOfNativeMemory}{Cputime}{Min}{10.292220588}%
\StoreBenchExecResult{EvalSlDfS}{PredabsHardSafe}{Error}{OutOfNativeMemory}{Cputime}{Max}{10.722448499}%
\StoreBenchExecResult{EvalSlDfS}{PredabsHardSafe}{Error}{OutOfNativeMemory}{Cputime}{Stdev}{0.21511395550000}%
\StoreBenchExecResult{EvalSlDfS}{PredabsHardSafe}{Error}{OutOfNativeMemory}{Walltime}{}{9.7531176027841866}%
\StoreBenchExecResult{EvalSlDfS}{PredabsHardSafe}{Error}{OutOfNativeMemory}{Walltime}{Avg}{4.8765588013920933}%
\StoreBenchExecResult{EvalSlDfS}{PredabsHardSafe}{Error}{OutOfNativeMemory}{Walltime}{Median}{4.8765588013920933}%
\StoreBenchExecResult{EvalSlDfS}{PredabsHardSafe}{Error}{OutOfNativeMemory}{Walltime}{Min}{4.8364225819241256}%
\StoreBenchExecResult{EvalSlDfS}{PredabsHardSafe}{Error}{OutOfNativeMemory}{Walltime}{Max}{4.916695020860061}%
\StoreBenchExecResult{EvalSlDfS}{PredabsHardSafe}{Error}{OutOfNativeMemory}{Walltime}{Stdev}{0.04013621946796770000000000000}%
\StoreBenchExecResult{EvalSlDfS}{PredabsHardSafe}{Error}{Timeout}{Count}{}{712}%
\StoreBenchExecResult{EvalSlDfS}{PredabsHardSafe}{Error}{Timeout}{Cputime}{}{643848.254262849}%
\StoreBenchExecResult{EvalSlDfS}{PredabsHardSafe}{Error}{Timeout}{Cputime}{Avg}{904.2812559871474719101123596}%
\StoreBenchExecResult{EvalSlDfS}{PredabsHardSafe}{Error}{Timeout}{Cputime}{Median}{903.0127982455}%
\StoreBenchExecResult{EvalSlDfS}{PredabsHardSafe}{Error}{Timeout}{Cputime}{Min}{901.129047127}%
\StoreBenchExecResult{EvalSlDfS}{PredabsHardSafe}{Error}{Timeout}{Cputime}{Max}{935.255565302}%
\StoreBenchExecResult{EvalSlDfS}{PredabsHardSafe}{Error}{Timeout}{Cputime}{Stdev}{3.747676624004133769631899653}%
\StoreBenchExecResult{EvalSlDfS}{PredabsHardSafe}{Error}{Timeout}{Walltime}{}{621983.8840243429876}%
\StoreBenchExecResult{EvalSlDfS}{PredabsHardSafe}{Error}{Timeout}{Walltime}{Avg}{873.5728708207064432584269663}%
\StoreBenchExecResult{EvalSlDfS}{PredabsHardSafe}{Error}{Timeout}{Walltime}{Median}{887.6327370353974}%
\StoreBenchExecResult{EvalSlDfS}{PredabsHardSafe}{Error}{Timeout}{Walltime}{Min}{530.7303732892033}%
\StoreBenchExecResult{EvalSlDfS}{PredabsHardSafe}{Error}{Timeout}{Walltime}{Max}{908.897384821903}%
\StoreBenchExecResult{EvalSlDfS}{PredabsHardSafe}{Error}{Timeout}{Walltime}{Stdev}{39.19661677171333336639456033}%
\ifdefined\EvalSlDfSPredabsHardSafeTotalCount\else\edef\EvalSlDfSPredabsHardSafeTotalCount{0}\fi
\ifdefined\EvalSlDfSPredabsHardSafeCorrectCount\else\edef\EvalSlDfSPredabsHardSafeCorrectCount{0}\fi
\ifdefined\EvalSlDfSPredabsHardSafeCorrectTrueCount\else\edef\EvalSlDfSPredabsHardSafeCorrectTrueCount{0}\fi
\ifdefined\EvalSlDfSPredabsHardSafeCorrectFalseCount\else\edef\EvalSlDfSPredabsHardSafeCorrectFalseCount{0}\fi
\ifdefined\EvalSlDfSPredabsHardSafeWrongTrueCount\else\edef\EvalSlDfSPredabsHardSafeWrongTrueCount{0}\fi
\ifdefined\EvalSlDfSPredabsHardSafeWrongFalseCount\else\edef\EvalSlDfSPredabsHardSafeWrongFalseCount{0}\fi
\ifdefined\EvalSlDfSPredabsHardSafeErrorTimeoutCount\else\edef\EvalSlDfSPredabsHardSafeErrorTimeoutCount{0}\fi
\ifdefined\EvalSlDfSPredabsHardSafeErrorOutOfMemoryCount\else\edef\EvalSlDfSPredabsHardSafeErrorOutOfMemoryCount{0}\fi
\ifdefined\EvalSlDfSPredabsHardSafeCorrectCputime\else\edef\EvalSlDfSPredabsHardSafeCorrectCputime{0}\fi
\ifdefined\EvalSlDfSPredabsHardSafeCorrectCputimeAvg\else\edef\EvalSlDfSPredabsHardSafeCorrectCputimeAvg{None}\fi
\ifdefined\EvalSlDfSPredabsHardSafeCorrectWalltime\else\edef\EvalSlDfSPredabsHardSafeCorrectWalltime{0}\fi
\ifdefined\EvalSlDfSPredabsHardSafeCorrectWalltimeAvg\else\edef\EvalSlDfSPredabsHardSafeCorrectWalltimeAvg{None}\fi
\edef\EvalSlDfSPredabsHardSafeErrorOtherInconclusiveCount{\the\numexpr \EvalSlDfSPredabsHardSafeTotalCount - \EvalSlDfSPredabsHardSafeCorrectCount - \EvalSlDfSPredabsHardSafeWrongTrueCount - \EvalSlDfSPredabsHardSafeWrongFalseCount - \EvalSlDfSPredabsHardSafeErrorTimeoutCount - \EvalSlDfSPredabsHardSafeErrorOutOfMemoryCount \relax}
\providecommand\StoreBenchExecResult[7]{\expandafter\newcommand\csname#1#2#3#4#5#6\endcsname{#7}}%
\StoreBenchExecResult{Ls}{DefaultHardSafe}{Total}{}{Count}{}{1623}%
\StoreBenchExecResult{Ls}{DefaultHardSafe}{Total}{}{Cputime}{}{719932.890804972}%
\StoreBenchExecResult{Ls}{DefaultHardSafe}{Total}{}{Cputime}{Avg}{443.5815716604879852125693161}%
\StoreBenchExecResult{Ls}{DefaultHardSafe}{Total}{}{Cputime}{Median}{175.645951556}%
\StoreBenchExecResult{Ls}{DefaultHardSafe}{Total}{}{Cputime}{Min}{0.058135813}%
\StoreBenchExecResult{Ls}{DefaultHardSafe}{Total}{}{Cputime}{Max}{961.759953167}%
\StoreBenchExecResult{Ls}{DefaultHardSafe}{Total}{}{Cputime}{Stdev}{451.0300220927029897039867058}%
\StoreBenchExecResult{Ls}{DefaultHardSafe}{Total}{}{Walltime}{}{719925.12175116525026377}%
\StoreBenchExecResult{Ls}{DefaultHardSafe}{Total}{}{Walltime}{Avg}{443.5767848127943624545717807}%
\StoreBenchExecResult{Ls}{DefaultHardSafe}{Total}{}{Walltime}{Median}{174.66767428978346}%
\StoreBenchExecResult{Ls}{DefaultHardSafe}{Total}{}{Walltime}{Min}{0.05626319604925811}%
\StoreBenchExecResult{Ls}{DefaultHardSafe}{Total}{}{Walltime}{Max}{962.1028034440242}%
\StoreBenchExecResult{Ls}{DefaultHardSafe}{Total}{}{Walltime}{Stdev}{451.0834743638158949255995844}%
\StoreBenchExecResult{Ls}{DefaultHardSafe}{Correct}{}{Count}{}{556}%
\StoreBenchExecResult{Ls}{DefaultHardSafe}{Correct}{}{Cputime}{}{59817.903917362}%
\StoreBenchExecResult{Ls}{DefaultHardSafe}{Correct}{}{Cputime}{Avg}{107.5861581247517985611510791}%
\StoreBenchExecResult{Ls}{DefaultHardSafe}{Correct}{}{Cputime}{Median}{9.619554029}%
\StoreBenchExecResult{Ls}{DefaultHardSafe}{Correct}{}{Cputime}{Min}{0.067917895}%
\StoreBenchExecResult{Ls}{DefaultHardSafe}{Correct}{}{Cputime}{Max}{716.375306938}%
\StoreBenchExecResult{Ls}{DefaultHardSafe}{Correct}{}{Cputime}{Stdev}{178.6913341459215198091882361}%
\StoreBenchExecResult{Ls}{DefaultHardSafe}{Correct}{}{Walltime}{}{59829.09684327757024599}%
\StoreBenchExecResult{Ls}{DefaultHardSafe}{Correct}{}{Walltime}{Avg}{107.6062892864704500827158273}%
\StoreBenchExecResult{Ls}{DefaultHardSafe}{Correct}{}{Walltime}{Median}{9.6240481265122075}%
\StoreBenchExecResult{Ls}{DefaultHardSafe}{Correct}{}{Walltime}{Min}{0.06595819001086056}%
\StoreBenchExecResult{Ls}{DefaultHardSafe}{Correct}{}{Walltime}{Max}{716.3847211168613}%
\StoreBenchExecResult{Ls}{DefaultHardSafe}{Correct}{}{Walltime}{Stdev}{178.7289274845079925849354047}%
\StoreBenchExecResult{Ls}{DefaultHardSafe}{Correct}{True}{Count}{}{556}%
\StoreBenchExecResult{Ls}{DefaultHardSafe}{Correct}{True}{Cputime}{}{59817.903917362}%
\StoreBenchExecResult{Ls}{DefaultHardSafe}{Correct}{True}{Cputime}{Avg}{107.5861581247517985611510791}%
\StoreBenchExecResult{Ls}{DefaultHardSafe}{Correct}{True}{Cputime}{Median}{9.619554029}%
\StoreBenchExecResult{Ls}{DefaultHardSafe}{Correct}{True}{Cputime}{Min}{0.067917895}%
\StoreBenchExecResult{Ls}{DefaultHardSafe}{Correct}{True}{Cputime}{Max}{716.375306938}%
\StoreBenchExecResult{Ls}{DefaultHardSafe}{Correct}{True}{Cputime}{Stdev}{178.6913341459215198091882361}%
\StoreBenchExecResult{Ls}{DefaultHardSafe}{Correct}{True}{Walltime}{}{59829.09684327757024599}%
\StoreBenchExecResult{Ls}{DefaultHardSafe}{Correct}{True}{Walltime}{Avg}{107.6062892864704500827158273}%
\StoreBenchExecResult{Ls}{DefaultHardSafe}{Correct}{True}{Walltime}{Median}{9.6240481265122075}%
\StoreBenchExecResult{Ls}{DefaultHardSafe}{Correct}{True}{Walltime}{Min}{0.06595819001086056}%
\StoreBenchExecResult{Ls}{DefaultHardSafe}{Correct}{True}{Walltime}{Max}{716.3847211168613}%
\StoreBenchExecResult{Ls}{DefaultHardSafe}{Correct}{True}{Walltime}{Stdev}{178.7289274845079925849354047}%

\StoreBenchExecResult{Ls}{DefaultHardSafe}{Error}{}{Count}{}{739}%
\StoreBenchExecResult{Ls}{DefaultHardSafe}{Error}{}{Cputime}{}{659278.441365491}%
\StoreBenchExecResult{Ls}{DefaultHardSafe}{Error}{}{Cputime}{Avg}{892.1223834445074424898511502}%
\StoreBenchExecResult{Ls}{DefaultHardSafe}{Error}{}{Cputime}{Median}{960.985898694}%
\StoreBenchExecResult{Ls}{DefaultHardSafe}{Error}{}{Cputime}{Min}{48.567998614}%
\StoreBenchExecResult{Ls}{DefaultHardSafe}{Error}{}{Cputime}{Max}{961.759953167}%
\StoreBenchExecResult{Ls}{DefaultHardSafe}{Error}{}{Cputime}{Stdev}{224.1319140916381391110515076}%
\StoreBenchExecResult{Ls}{DefaultHardSafe}{Error}{}{Walltime}{}{659260.072617943165385}%
\StoreBenchExecResult{Ls}{DefaultHardSafe}{Error}{}{Walltime}{Avg}{892.0975272231977880717185386}%
\StoreBenchExecResult{Ls}{DefaultHardSafe}{Error}{}{Walltime}{Median}{961.0507924461272}%
\StoreBenchExecResult{Ls}{DefaultHardSafe}{Error}{}{Walltime}{Min}{47.612525469157845}%
\StoreBenchExecResult{Ls}{DefaultHardSafe}{Error}{}{Walltime}{Max}{962.1028034440242}%
\StoreBenchExecResult{Ls}{DefaultHardSafe}{Error}{}{Walltime}{Stdev}{224.4162480450697828206497533}%
\StoreBenchExecResult{Ls}{DefaultHardSafe}{Error}{OutOfMemory}{Count}{}{64}%
\StoreBenchExecResult{Ls}{DefaultHardSafe}{Error}{OutOfMemory}{Cputime}{}{10555.554316269}%
\StoreBenchExecResult{Ls}{DefaultHardSafe}{Error}{OutOfMemory}{Cputime}{Avg}{164.930536191703125}%
\StoreBenchExecResult{Ls}{DefaultHardSafe}{Error}{OutOfMemory}{Cputime}{Median}{174.2138363245}%
\StoreBenchExecResult{Ls}{DefaultHardSafe}{Error}{OutOfMemory}{Cputime}{Min}{48.567998614}%
\StoreBenchExecResult{Ls}{DefaultHardSafe}{Error}{OutOfMemory}{Cputime}{Max}{206.941452422}%
\StoreBenchExecResult{Ls}{DefaultHardSafe}{Error}{OutOfMemory}{Cputime}{Stdev}{33.33701620315738919084020530}%
\StoreBenchExecResult{Ls}{DefaultHardSafe}{Error}{OutOfMemory}{Walltime}{}{10494.676643047714585}%
\StoreBenchExecResult{Ls}{DefaultHardSafe}{Error}{OutOfMemory}{Walltime}{Avg}{163.979322547620540390625}%
\StoreBenchExecResult{Ls}{DefaultHardSafe}{Error}{OutOfMemory}{Walltime}{Median}{173.266255603521125}%
\StoreBenchExecResult{Ls}{DefaultHardSafe}{Error}{OutOfMemory}{Walltime}{Min}{47.612525469157845}%
\StoreBenchExecResult{Ls}{DefaultHardSafe}{Error}{OutOfMemory}{Walltime}{Max}{205.86021886602975}%
\StoreBenchExecResult{Ls}{DefaultHardSafe}{Error}{OutOfMemory}{Walltime}{Stdev}{33.28873226379080794596598653}%
\StoreBenchExecResult{Ls}{DefaultHardSafe}{Error}{Timeout}{Count}{}{675}%
\StoreBenchExecResult{Ls}{DefaultHardSafe}{Error}{Timeout}{Cputime}{}{648722.887049222}%
\StoreBenchExecResult{Ls}{DefaultHardSafe}{Error}{Timeout}{Cputime}{Avg}{961.0709437766251851851851852}%
\StoreBenchExecResult{Ls}{DefaultHardSafe}{Error}{Timeout}{Cputime}{Median}{960.989120093}%
\StoreBenchExecResult{Ls}{DefaultHardSafe}{Error}{Timeout}{Cputime}{Min}{960.940005152}%
\StoreBenchExecResult{Ls}{DefaultHardSafe}{Error}{Timeout}{Cputime}{Max}{961.759953167}%
\StoreBenchExecResult{Ls}{DefaultHardSafe}{Error}{Timeout}{Cputime}{Stdev}{0.2000801946045754328163022414}%
\StoreBenchExecResult{Ls}{DefaultHardSafe}{Error}{Timeout}{Walltime}{}{648765.3959748954508}%
\StoreBenchExecResult{Ls}{DefaultHardSafe}{Error}{Timeout}{Walltime}{Avg}{961.1339199628080752592592593}%
\StoreBenchExecResult{Ls}{DefaultHardSafe}{Error}{Timeout}{Walltime}{Median}{961.073038879782}%
\StoreBenchExecResult{Ls}{DefaultHardSafe}{Error}{Timeout}{Walltime}{Min}{960.9507412961684}%
\StoreBenchExecResult{Ls}{DefaultHardSafe}{Error}{Timeout}{Walltime}{Max}{962.1028034440242}%
\StoreBenchExecResult{Ls}{DefaultHardSafe}{Error}{Timeout}{Walltime}{Stdev}{0.1789549292913334400367941806}%
\StoreBenchExecResult{Ls}{DefaultHardSafe}{Unknown}{}{Count}{}{328}%
\StoreBenchExecResult{Ls}{DefaultHardSafe}{Unknown}{}{Cputime}{}{836.545522119}%
\StoreBenchExecResult{Ls}{DefaultHardSafe}{Unknown}{}{Cputime}{Avg}{2.550443664996951219512195122}%
\StoreBenchExecResult{Ls}{DefaultHardSafe}{Unknown}{}{Cputime}{Median}{0.2923550955}%
\StoreBenchExecResult{Ls}{DefaultHardSafe}{Unknown}{}{Cputime}{Min}{0.058135813}%
\StoreBenchExecResult{Ls}{DefaultHardSafe}{Unknown}{}{Cputime}{Max}{120.48928875}%
\StoreBenchExecResult{Ls}{DefaultHardSafe}{Unknown}{}{Cputime}{Stdev}{11.15471967573265335189299111}%
\StoreBenchExecResult{Ls}{DefaultHardSafe}{Unknown}{}{Walltime}{}{835.95228994451463278}%
\StoreBenchExecResult{Ls}{DefaultHardSafe}{Unknown}{}{Walltime}{Avg}{2.548635030318642173109756098}%
\StoreBenchExecResult{Ls}{DefaultHardSafe}{Unknown}{}{Walltime}{Median}{0.2905340545112267}%
\StoreBenchExecResult{Ls}{DefaultHardSafe}{Unknown}{}{Walltime}{Min}{0.05626319604925811}%
\StoreBenchExecResult{Ls}{DefaultHardSafe}{Unknown}{}{Walltime}{Max}{120.54215457406826}%
\StoreBenchExecResult{Ls}{DefaultHardSafe}{Unknown}{}{Walltime}{Stdev}{11.15690973872125450481307865}%
\StoreBenchExecResult{Ls}{DefaultHardSafe}{Unknown}{Unknown}{Count}{}{328}%
\StoreBenchExecResult{Ls}{DefaultHardSafe}{Unknown}{Unknown}{Cputime}{}{836.545522119}%
\StoreBenchExecResult{Ls}{DefaultHardSafe}{Unknown}{Unknown}{Cputime}{Avg}{2.550443664996951219512195122}%
\StoreBenchExecResult{Ls}{DefaultHardSafe}{Unknown}{Unknown}{Cputime}{Median}{0.2923550955}%
\StoreBenchExecResult{Ls}{DefaultHardSafe}{Unknown}{Unknown}{Cputime}{Min}{0.058135813}%
\StoreBenchExecResult{Ls}{DefaultHardSafe}{Unknown}{Unknown}{Cputime}{Max}{120.48928875}%
\StoreBenchExecResult{Ls}{DefaultHardSafe}{Unknown}{Unknown}{Cputime}{Stdev}{11.15471967573265335189299111}%
\StoreBenchExecResult{Ls}{DefaultHardSafe}{Unknown}{Unknown}{Walltime}{}{835.95228994451463278}%
\StoreBenchExecResult{Ls}{DefaultHardSafe}{Unknown}{Unknown}{Walltime}{Avg}{2.548635030318642173109756098}%
\StoreBenchExecResult{Ls}{DefaultHardSafe}{Unknown}{Unknown}{Walltime}{Median}{0.2905340545112267}%
\StoreBenchExecResult{Ls}{DefaultHardSafe}{Unknown}{Unknown}{Walltime}{Min}{0.05626319604925811}%
\StoreBenchExecResult{Ls}{DefaultHardSafe}{Unknown}{Unknown}{Walltime}{Max}{120.54215457406826}%
\StoreBenchExecResult{Ls}{DefaultHardSafe}{Unknown}{Unknown}{Walltime}{Stdev}{11.15690973872125450481307865}%
\ifdefined\LsDefaultHardSafeTotalCount\else\edef\LsDefaultHardSafeTotalCount{0}\fi
\ifdefined\LsDefaultHardSafeCorrectCount\else\edef\LsDefaultHardSafeCorrectCount{0}\fi
\ifdefined\LsDefaultHardSafeCorrectTrueCount\else\edef\LsDefaultHardSafeCorrectTrueCount{0}\fi
\ifdefined\LsDefaultHardSafeCorrectFalseCount\else\edef\LsDefaultHardSafeCorrectFalseCount{0}\fi
\ifdefined\LsDefaultHardSafeWrongTrueCount\else\edef\LsDefaultHardSafeWrongTrueCount{0}\fi
\ifdefined\LsDefaultHardSafeWrongFalseCount\else\edef\LsDefaultHardSafeWrongFalseCount{0}\fi
\ifdefined\LsDefaultHardSafeErrorTimeoutCount\else\edef\LsDefaultHardSafeErrorTimeoutCount{0}\fi
\ifdefined\LsDefaultHardSafeErrorOutOfMemoryCount\else\edef\LsDefaultHardSafeErrorOutOfMemoryCount{0}\fi
\ifdefined\LsDefaultHardSafeCorrectCputime\else\edef\LsDefaultHardSafeCorrectCputime{0}\fi
\ifdefined\LsDefaultHardSafeCorrectCputimeAvg\else\edef\LsDefaultHardSafeCorrectCputimeAvg{None}\fi
\ifdefined\LsDefaultHardSafeCorrectWalltime\else\edef\LsDefaultHardSafeCorrectWalltime{0}\fi
\ifdefined\LsDefaultHardSafeCorrectWalltimeAvg\else\edef\LsDefaultHardSafeCorrectWalltimeAvg{None}\fi
\edef\LsDefaultHardSafeErrorOtherInconclusiveCount{\the\numexpr \LsDefaultHardSafeTotalCount - \LsDefaultHardSafeCorrectCount - \LsDefaultHardSafeWrongTrueCount - \LsDefaultHardSafeWrongFalseCount - \LsDefaultHardSafeErrorTimeoutCount - \LsDefaultHardSafeErrorOutOfMemoryCount \relax}
\providecommand\StoreBenchExecResult[7]{\expandafter\newcommand\csname#1#2#3#4#5#6\endcsname{#7}}%
\StoreBenchExecResult{Symbiotic}{SvcompHardSafe}{Total}{}{Count}{}{1623}%
\StoreBenchExecResult{Symbiotic}{SvcompHardSafe}{Total}{}{Cputime}{}{962564.209624467}%
\StoreBenchExecResult{Symbiotic}{SvcompHardSafe}{Total}{}{Cputime}{Avg}{593.0771470267818853974121996}%
\StoreBenchExecResult{Symbiotic}{SvcompHardSafe}{Total}{}{Cputime}{Median}{747.145650466}%
\StoreBenchExecResult{Symbiotic}{SvcompHardSafe}{Total}{}{Cputime}{Min}{0.349282553}%
\StoreBenchExecResult{Symbiotic}{SvcompHardSafe}{Total}{}{Cputime}{Max}{962.052878053}%
\StoreBenchExecResult{Symbiotic}{SvcompHardSafe}{Total}{}{Cputime}{Stdev}{388.0080710115454913841572131}%
\StoreBenchExecResult{Symbiotic}{SvcompHardSafe}{Total}{}{Walltime}{}{962518.44204927491826754}%
\StoreBenchExecResult{Symbiotic}{SvcompHardSafe}{Total}{}{Walltime}{Avg}{593.0489476582100543854220579}%
\StoreBenchExecResult{Symbiotic}{SvcompHardSafe}{Total}{}{Walltime}{Median}{747.6349683480803}%
\StoreBenchExecResult{Symbiotic}{SvcompHardSafe}{Total}{}{Walltime}{Min}{0.36273218505084515}%
\StoreBenchExecResult{Symbiotic}{SvcompHardSafe}{Total}{}{Walltime}{Max}{961.8926863400266}%
\StoreBenchExecResult{Symbiotic}{SvcompHardSafe}{Total}{}{Walltime}{Stdev}{388.0455976387177955935638473}%
\StoreBenchExecResult{Symbiotic}{SvcompHardSafe}{Correct}{}{Count}{}{319}%
\StoreBenchExecResult{Symbiotic}{SvcompHardSafe}{Correct}{}{Cputime}{}{26715.419290801}%
\StoreBenchExecResult{Symbiotic}{SvcompHardSafe}{Correct}{}{Cputime}{Avg}{83.74739589592789968652037618}%
\StoreBenchExecResult{Symbiotic}{SvcompHardSafe}{Correct}{}{Cputime}{Median}{0.489191758}%
\StoreBenchExecResult{Symbiotic}{SvcompHardSafe}{Correct}{}{Cputime}{Min}{0.349282553}%
\StoreBenchExecResult{Symbiotic}{SvcompHardSafe}{Correct}{}{Cputime}{Max}{687.00283368}%
\StoreBenchExecResult{Symbiotic}{SvcompHardSafe}{Correct}{}{Cputime}{Stdev}{153.0256082401796875631039533}%
\StoreBenchExecResult{Symbiotic}{SvcompHardSafe}{Correct}{}{Walltime}{}{26739.79898214130664644}%
\StoreBenchExecResult{Symbiotic}{SvcompHardSafe}{Correct}{}{Walltime}{Avg}{83.82382126063105531799373041}%
\StoreBenchExecResult{Symbiotic}{SvcompHardSafe}{Correct}{}{Walltime}{Median}{0.544163707876578}%
\StoreBenchExecResult{Symbiotic}{SvcompHardSafe}{Correct}{}{Walltime}{Min}{0.36273218505084515}%
\StoreBenchExecResult{Symbiotic}{SvcompHardSafe}{Correct}{}{Walltime}{Max}{686.6393246790394}%
\StoreBenchExecResult{Symbiotic}{SvcompHardSafe}{Correct}{}{Walltime}{Stdev}{153.0644747945908329271281715}%
\StoreBenchExecResult{Symbiotic}{SvcompHardSafe}{Correct}{True}{Count}{}{319}%
\StoreBenchExecResult{Symbiotic}{SvcompHardSafe}{Correct}{True}{Cputime}{}{26715.419290801}%
\StoreBenchExecResult{Symbiotic}{SvcompHardSafe}{Correct}{True}{Cputime}{Avg}{83.74739589592789968652037618}%
\StoreBenchExecResult{Symbiotic}{SvcompHardSafe}{Correct}{True}{Cputime}{Median}{0.489191758}%
\StoreBenchExecResult{Symbiotic}{SvcompHardSafe}{Correct}{True}{Cputime}{Min}{0.349282553}%
\StoreBenchExecResult{Symbiotic}{SvcompHardSafe}{Correct}{True}{Cputime}{Max}{687.00283368}%
\StoreBenchExecResult{Symbiotic}{SvcompHardSafe}{Correct}{True}{Cputime}{Stdev}{153.0256082401796875631039533}%
\StoreBenchExecResult{Symbiotic}{SvcompHardSafe}{Correct}{True}{Walltime}{}{26739.79898214130664644}%
\StoreBenchExecResult{Symbiotic}{SvcompHardSafe}{Correct}{True}{Walltime}{Avg}{83.82382126063105531799373041}%
\StoreBenchExecResult{Symbiotic}{SvcompHardSafe}{Correct}{True}{Walltime}{Median}{0.544163707876578}%
\StoreBenchExecResult{Symbiotic}{SvcompHardSafe}{Correct}{True}{Walltime}{Min}{0.36273218505084515}%
\StoreBenchExecResult{Symbiotic}{SvcompHardSafe}{Correct}{True}{Walltime}{Max}{686.6393246790394}%
\StoreBenchExecResult{Symbiotic}{SvcompHardSafe}{Correct}{True}{Walltime}{Stdev}{153.0644747945908329271281715}%

\StoreBenchExecResult{Symbiotic}{SvcompHardSafe}{Error}{}{Count}{}{1300}%
\StoreBenchExecResult{Symbiotic}{SvcompHardSafe}{Error}{}{Cputime}{}{933838.425403084}%
\StoreBenchExecResult{Symbiotic}{SvcompHardSafe}{Error}{}{Cputime}{Avg}{718.3372503100646153846153846}%
\StoreBenchExecResult{Symbiotic}{SvcompHardSafe}{Error}{}{Cputime}{Median}{960.9493953985}%
\StoreBenchExecResult{Symbiotic}{SvcompHardSafe}{Error}{}{Cputime}{Min}{0.807667001}%
\StoreBenchExecResult{Symbiotic}{SvcompHardSafe}{Error}{}{Cputime}{Max}{962.052878053}%
\StoreBenchExecResult{Symbiotic}{SvcompHardSafe}{Error}{}{Cputime}{Stdev}{320.2821666360694707707705306}%
\StoreBenchExecResult{Symbiotic}{SvcompHardSafe}{Error}{}{Walltime}{}{933767.5819996737396548}%
\StoreBenchExecResult{Symbiotic}{SvcompHardSafe}{Error}{}{Walltime}{Avg}{718.2827553843644151190769231}%
\StoreBenchExecResult{Symbiotic}{SvcompHardSafe}{Error}{}{Walltime}{Median}{960.6061742844759}%
\StoreBenchExecResult{Symbiotic}{SvcompHardSafe}{Error}{}{Walltime}{Min}{0.820166336139664}%
\StoreBenchExecResult{Symbiotic}{SvcompHardSafe}{Error}{}{Walltime}{Max}{961.8926863400266}%
\StoreBenchExecResult{Symbiotic}{SvcompHardSafe}{Error}{}{Walltime}{Stdev}{320.3854536702056101267696631}%
\StoreBenchExecResult{Symbiotic}{SvcompHardSafe}{Error}{Error}{Count}{}{266}%
\StoreBenchExecResult{Symbiotic}{SvcompHardSafe}{Error}{Error}{Cputime}{}{130816.304201232}%
\StoreBenchExecResult{Symbiotic}{SvcompHardSafe}{Error}{Error}{Cputime}{Avg}{491.7906172978646616541353383}%
\StoreBenchExecResult{Symbiotic}{SvcompHardSafe}{Error}{Error}{Cputime}{Median}{668.705378483}%
\StoreBenchExecResult{Symbiotic}{SvcompHardSafe}{Error}{Error}{Cputime}{Min}{0.807667001}%
\StoreBenchExecResult{Symbiotic}{SvcompHardSafe}{Error}{Error}{Cputime}{Max}{894.265954084}%
\StoreBenchExecResult{Symbiotic}{SvcompHardSafe}{Error}{Error}{Cputime}{Stdev}{313.2688165031149151950723355}%
\StoreBenchExecResult{Symbiotic}{SvcompHardSafe}{Error}{Error}{Walltime}{}{130885.1273988443430048}%
\StoreBenchExecResult{Symbiotic}{SvcompHardSafe}{Error}{Error}{Walltime}{Avg}{492.0493511234749737022556391}%
\StoreBenchExecResult{Symbiotic}{SvcompHardSafe}{Error}{Error}{Walltime}{Median}{669.03928822104355}%
\StoreBenchExecResult{Symbiotic}{SvcompHardSafe}{Error}{Error}{Walltime}{Min}{0.820166336139664}%
\StoreBenchExecResult{Symbiotic}{SvcompHardSafe}{Error}{Error}{Walltime}{Max}{894.7242210879922}%
\StoreBenchExecResult{Symbiotic}{SvcompHardSafe}{Error}{Error}{Walltime}{Stdev}{313.3843212766336761815411389}%
\StoreBenchExecResult{Symbiotic}{SvcompHardSafe}{Error}{OutOfMemory}{Count}{}{324}%
\StoreBenchExecResult{Symbiotic}{SvcompHardSafe}{Error}{OutOfMemory}{Cputime}{}{120998.833112840}%
\StoreBenchExecResult{Symbiotic}{SvcompHardSafe}{Error}{OutOfMemory}{Cputime}{Avg}{373.4531886198765432098765432}%
\StoreBenchExecResult{Symbiotic}{SvcompHardSafe}{Error}{OutOfMemory}{Cputime}{Median}{281.327814601}%
\StoreBenchExecResult{Symbiotic}{SvcompHardSafe}{Error}{OutOfMemory}{Cputime}{Min}{58.946201743}%
\StoreBenchExecResult{Symbiotic}{SvcompHardSafe}{Error}{OutOfMemory}{Cputime}{Max}{890.052412282}%
\StoreBenchExecResult{Symbiotic}{SvcompHardSafe}{Error}{OutOfMemory}{Cputime}{Stdev}{203.1917413217377600892243437}%
\StoreBenchExecResult{Symbiotic}{SvcompHardSafe}{Error}{OutOfMemory}{Walltime}{}{120856.68250632355905}%
\StoreBenchExecResult{Symbiotic}{SvcompHardSafe}{Error}{OutOfMemory}{Walltime}{Avg}{373.0144521800109847222222222}%
\StoreBenchExecResult{Symbiotic}{SvcompHardSafe}{Error}{OutOfMemory}{Walltime}{Median}{281.04090565862134}%
\StoreBenchExecResult{Symbiotic}{SvcompHardSafe}{Error}{OutOfMemory}{Walltime}{Min}{58.14370196079835}%
\StoreBenchExecResult{Symbiotic}{SvcompHardSafe}{Error}{OutOfMemory}{Walltime}{Max}{889.8073599201161}%
\StoreBenchExecResult{Symbiotic}{SvcompHardSafe}{Error}{OutOfMemory}{Walltime}{Stdev}{203.1838185136359036431161726}%
\StoreBenchExecResult{Symbiotic}{SvcompHardSafe}{Error}{Timeout}{Count}{}{710}%
\StoreBenchExecResult{Symbiotic}{SvcompHardSafe}{Error}{Timeout}{Cputime}{}{682023.288089012}%
\StoreBenchExecResult{Symbiotic}{SvcompHardSafe}{Error}{Timeout}{Cputime}{Avg}{960.5961804070591549295774648}%
\StoreBenchExecResult{Symbiotic}{SvcompHardSafe}{Error}{Timeout}{Cputime}{Median}{961.006041337}%
\StoreBenchExecResult{Symbiotic}{SvcompHardSafe}{Error}{Timeout}{Cputime}{Min}{905.121662672}%
\StoreBenchExecResult{Symbiotic}{SvcompHardSafe}{Error}{Timeout}{Cputime}{Max}{962.052878053}%
\StoreBenchExecResult{Symbiotic}{SvcompHardSafe}{Error}{Timeout}{Cputime}{Stdev}{4.394386905438701645853381896}%
\StoreBenchExecResult{Symbiotic}{SvcompHardSafe}{Error}{Timeout}{Walltime}{}{682025.7720945058376}%
\StoreBenchExecResult{Symbiotic}{SvcompHardSafe}{Error}{Timeout}{Walltime}{Avg}{960.5996790063462501408450704}%
\StoreBenchExecResult{Symbiotic}{SvcompHardSafe}{Error}{Timeout}{Walltime}{Median}{961.08516540995335}%
\StoreBenchExecResult{Symbiotic}{SvcompHardSafe}{Error}{Timeout}{Walltime}{Min}{905.8222145738546}%
\StoreBenchExecResult{Symbiotic}{SvcompHardSafe}{Error}{Timeout}{Walltime}{Max}{961.8926863400266}%
\StoreBenchExecResult{Symbiotic}{SvcompHardSafe}{Error}{Timeout}{Walltime}{Stdev}{4.340638810575657776203103392}%
\StoreBenchExecResult{Symbiotic}{SvcompHardSafe}{Unknown}{}{Count}{}{4}%
\StoreBenchExecResult{Symbiotic}{SvcompHardSafe}{Unknown}{}{Cputime}{}{2010.364930582}%
\StoreBenchExecResult{Symbiotic}{SvcompHardSafe}{Unknown}{}{Cputime}{Avg}{502.5912326455}%
\StoreBenchExecResult{Symbiotic}{SvcompHardSafe}{Unknown}{}{Cputime}{Median}{669.4533402145}%
\StoreBenchExecResult{Symbiotic}{SvcompHardSafe}{Unknown}{}{Cputime}{Min}{1.784415049}%
\StoreBenchExecResult{Symbiotic}{SvcompHardSafe}{Unknown}{}{Cputime}{Max}{669.673835104}%
\StoreBenchExecResult{Symbiotic}{SvcompHardSafe}{Unknown}{}{Cputime}{Stdev}{289.1409692234024215857950729}%
\StoreBenchExecResult{Symbiotic}{SvcompHardSafe}{Unknown}{}{Walltime}{}{2011.0610674598719663}%
\StoreBenchExecResult{Symbiotic}{SvcompHardSafe}{Unknown}{}{Walltime}{Avg}{502.765266864967991575}%
\StoreBenchExecResult{Symbiotic}{SvcompHardSafe}{Unknown}{}{Walltime}{Median}{669.6492024695035}%
\StoreBenchExecResult{Symbiotic}{SvcompHardSafe}{Unknown}{}{Walltime}{Min}{1.8459126139059663}%
\StoreBenchExecResult{Symbiotic}{SvcompHardSafe}{Unknown}{}{Walltime}{Max}{669.916749906959}%
\StoreBenchExecResult{Symbiotic}{SvcompHardSafe}{Unknown}{}{Walltime}{Stdev}{289.2059590884221939475769841}%
\StoreBenchExecResult{Symbiotic}{SvcompHardSafe}{Unknown}{Unknown}{Count}{}{4}%
\StoreBenchExecResult{Symbiotic}{SvcompHardSafe}{Unknown}{Unknown}{Cputime}{}{2010.364930582}%
\StoreBenchExecResult{Symbiotic}{SvcompHardSafe}{Unknown}{Unknown}{Cputime}{Avg}{502.5912326455}%
\StoreBenchExecResult{Symbiotic}{SvcompHardSafe}{Unknown}{Unknown}{Cputime}{Median}{669.4533402145}%
\StoreBenchExecResult{Symbiotic}{SvcompHardSafe}{Unknown}{Unknown}{Cputime}{Min}{1.784415049}%
\StoreBenchExecResult{Symbiotic}{SvcompHardSafe}{Unknown}{Unknown}{Cputime}{Max}{669.673835104}%
\StoreBenchExecResult{Symbiotic}{SvcompHardSafe}{Unknown}{Unknown}{Cputime}{Stdev}{289.1409692234024215857950729}%
\StoreBenchExecResult{Symbiotic}{SvcompHardSafe}{Unknown}{Unknown}{Walltime}{}{2011.0610674598719663}%
\StoreBenchExecResult{Symbiotic}{SvcompHardSafe}{Unknown}{Unknown}{Walltime}{Avg}{502.765266864967991575}%
\StoreBenchExecResult{Symbiotic}{SvcompHardSafe}{Unknown}{Unknown}{Walltime}{Median}{669.6492024695035}%
\StoreBenchExecResult{Symbiotic}{SvcompHardSafe}{Unknown}{Unknown}{Walltime}{Min}{1.8459126139059663}%
\StoreBenchExecResult{Symbiotic}{SvcompHardSafe}{Unknown}{Unknown}{Walltime}{Max}{669.916749906959}%
\StoreBenchExecResult{Symbiotic}{SvcompHardSafe}{Unknown}{Unknown}{Walltime}{Stdev}{289.2059590884221939475769841}%
\ifdefined\SymbioticSvcompHardSafeTotalCount\else\edef\SymbioticSvcompHardSafeTotalCount{0}\fi
\ifdefined\SymbioticSvcompHardSafeCorrectCount\else\edef\SymbioticSvcompHardSafeCorrectCount{0}\fi
\ifdefined\SymbioticSvcompHardSafeCorrectTrueCount\else\edef\SymbioticSvcompHardSafeCorrectTrueCount{0}\fi
\ifdefined\SymbioticSvcompHardSafeCorrectFalseCount\else\edef\SymbioticSvcompHardSafeCorrectFalseCount{0}\fi
\ifdefined\SymbioticSvcompHardSafeWrongTrueCount\else\edef\SymbioticSvcompHardSafeWrongTrueCount{0}\fi
\ifdefined\SymbioticSvcompHardSafeWrongFalseCount\else\edef\SymbioticSvcompHardSafeWrongFalseCount{0}\fi
\ifdefined\SymbioticSvcompHardSafeErrorTimeoutCount\else\edef\SymbioticSvcompHardSafeErrorTimeoutCount{0}\fi
\ifdefined\SymbioticSvcompHardSafeErrorOutOfMemoryCount\else\edef\SymbioticSvcompHardSafeErrorOutOfMemoryCount{0}\fi
\ifdefined\SymbioticSvcompHardSafeCorrectCputime\else\edef\SymbioticSvcompHardSafeCorrectCputime{0}\fi
\ifdefined\SymbioticSvcompHardSafeCorrectCputimeAvg\else\edef\SymbioticSvcompHardSafeCorrectCputimeAvg{None}\fi
\ifdefined\SymbioticSvcompHardSafeCorrectWalltime\else\edef\SymbioticSvcompHardSafeCorrectWalltime{0}\fi
\ifdefined\SymbioticSvcompHardSafeCorrectWalltimeAvg\else\edef\SymbioticSvcompHardSafeCorrectWalltimeAvg{None}\fi
\edef\SymbioticSvcompHardSafeErrorOtherInconclusiveCount{\the\numexpr \SymbioticSvcompHardSafeTotalCount - \SymbioticSvcompHardSafeCorrectCount - \SymbioticSvcompHardSafeWrongTrueCount - \SymbioticSvcompHardSafeWrongFalseCount - \SymbioticSvcompHardSafeErrorTimeoutCount - \SymbioticSvcompHardSafeErrorOutOfMemoryCount \relax}
\providecommand\StoreBenchExecResult[7]{\expandafter\newcommand\csname#1#2#3#4#5#6\endcsname{#7}}%
\StoreBenchExecResult{ReducedEvalSlDfSS}{ImcHardSafe}{Total}{}{Count}{}{870}%
\StoreBenchExecResult{ReducedEvalSlDfSS}{ImcHardSafe}{Total}{}{Cputime}{}{302948.913654700}%
\StoreBenchExecResult{ReducedEvalSlDfSS}{ImcHardSafe}{Total}{}{Cputime}{Avg}{348.2171421318390804597701149}%
\StoreBenchExecResult{ReducedEvalSlDfSS}{ImcHardSafe}{Total}{}{Cputime}{Median}{27.132602339}%
\StoreBenchExecResult{ReducedEvalSlDfSS}{ImcHardSafe}{Total}{}{Cputime}{Min}{4.200047668}%
\StoreBenchExecResult{ReducedEvalSlDfSS}{ImcHardSafe}{Total}{}{Cputime}{Max}{913.336971733}%
\StoreBenchExecResult{ReducedEvalSlDfSS}{ImcHardSafe}{Total}{}{Cputime}{Stdev}{416.8136934185865116386259746}%
\StoreBenchExecResult{ReducedEvalSlDfSS}{ImcHardSafe}{Total}{}{Walltime}{}{293740.8835312120357837}%
\StoreBenchExecResult{ReducedEvalSlDfSS}{ImcHardSafe}{Total}{}{Walltime}{Avg}{337.6331994611632595214942529}%
\StoreBenchExecResult{ReducedEvalSlDfSS}{ImcHardSafe}{Total}{}{Walltime}{Median}{14.9803082500147865}%
\StoreBenchExecResult{ReducedEvalSlDfSS}{ImcHardSafe}{Total}{}{Walltime}{Min}{1.642349715999444}%
\StoreBenchExecResult{ReducedEvalSlDfSS}{ImcHardSafe}{Total}{}{Walltime}{Max}{907.4551925710111}%
\StoreBenchExecResult{ReducedEvalSlDfSS}{ImcHardSafe}{Total}{}{Walltime}{Stdev}{412.6851360351360366936623188}%
\StoreBenchExecResult{ReducedEvalSlDfSS}{ImcHardSafe}{Correct}{}{Count}{}{530}%
\StoreBenchExecResult{ReducedEvalSlDfSS}{ImcHardSafe}{Correct}{}{Cputime}{}{29451.705478328}%
\StoreBenchExecResult{ReducedEvalSlDfSS}{ImcHardSafe}{Correct}{}{Cputime}{Avg}{55.56925561948679245283018868}%
\StoreBenchExecResult{ReducedEvalSlDfSS}{ImcHardSafe}{Correct}{}{Cputime}{Median}{11.1042503365}%
\StoreBenchExecResult{ReducedEvalSlDfSS}{ImcHardSafe}{Correct}{}{Cputime}{Min}{4.200047668}%
\StoreBenchExecResult{ReducedEvalSlDfSS}{ImcHardSafe}{Correct}{}{Cputime}{Max}{886.684116455}%
\StoreBenchExecResult{ReducedEvalSlDfSS}{ImcHardSafe}{Correct}{}{Cputime}{Stdev}{131.4976426228996454344169139}%
\StoreBenchExecResult{ReducedEvalSlDfSS}{ImcHardSafe}{Correct}{}{Walltime}{}{25049.8457871701102580}%
\StoreBenchExecResult{ReducedEvalSlDfSS}{ImcHardSafe}{Correct}{}{Walltime}{Avg}{47.26385997579266086415094340}%
\StoreBenchExecResult{ReducedEvalSlDfSS}{ImcHardSafe}{Correct}{}{Walltime}{Median}{4.672852226503892}%
\StoreBenchExecResult{ReducedEvalSlDfSS}{ImcHardSafe}{Correct}{}{Walltime}{Min}{1.642349715999444}%
\StoreBenchExecResult{ReducedEvalSlDfSS}{ImcHardSafe}{Correct}{}{Walltime}{Max}{855.7313299379894}%
\StoreBenchExecResult{ReducedEvalSlDfSS}{ImcHardSafe}{Correct}{}{Walltime}{Stdev}{127.8181859218133787172380298}%
\StoreBenchExecResult{ReducedEvalSlDfSS}{ImcHardSafe}{Correct}{True}{Count}{}{530}%
\StoreBenchExecResult{ReducedEvalSlDfSS}{ImcHardSafe}{Correct}{True}{Cputime}{}{29451.705478328}%
\StoreBenchExecResult{ReducedEvalSlDfSS}{ImcHardSafe}{Correct}{True}{Cputime}{Avg}{55.56925561948679245283018868}%
\StoreBenchExecResult{ReducedEvalSlDfSS}{ImcHardSafe}{Correct}{True}{Cputime}{Median}{11.1042503365}%
\StoreBenchExecResult{ReducedEvalSlDfSS}{ImcHardSafe}{Correct}{True}{Cputime}{Min}{4.200047668}%
\StoreBenchExecResult{ReducedEvalSlDfSS}{ImcHardSafe}{Correct}{True}{Cputime}{Max}{886.684116455}%
\StoreBenchExecResult{ReducedEvalSlDfSS}{ImcHardSafe}{Correct}{True}{Cputime}{Stdev}{131.4976426228996454344169139}%
\StoreBenchExecResult{ReducedEvalSlDfSS}{ImcHardSafe}{Correct}{True}{Walltime}{}{25049.8457871701102580}%
\StoreBenchExecResult{ReducedEvalSlDfSS}{ImcHardSafe}{Correct}{True}{Walltime}{Avg}{47.26385997579266086415094340}%
\StoreBenchExecResult{ReducedEvalSlDfSS}{ImcHardSafe}{Correct}{True}{Walltime}{Median}{4.672852226503892}%
\StoreBenchExecResult{ReducedEvalSlDfSS}{ImcHardSafe}{Correct}{True}{Walltime}{Min}{1.642349715999444}%
\StoreBenchExecResult{ReducedEvalSlDfSS}{ImcHardSafe}{Correct}{True}{Walltime}{Max}{855.7313299379894}%
\StoreBenchExecResult{ReducedEvalSlDfSS}{ImcHardSafe}{Correct}{True}{Walltime}{Stdev}{127.8181859218133787172380298}%

\StoreBenchExecResult{ReducedEvalSlDfSS}{ImcHardSafe}{Error}{}{Count}{}{340}%
\StoreBenchExecResult{ReducedEvalSlDfSS}{ImcHardSafe}{Error}{}{Cputime}{}{273497.208176372}%
\StoreBenchExecResult{ReducedEvalSlDfSS}{ImcHardSafe}{Error}{}{Cputime}{Avg}{804.4035534599176470588235294}%
\StoreBenchExecResult{ReducedEvalSlDfSS}{ImcHardSafe}{Error}{}{Cputime}{Median}{902.1410096465}%
\StoreBenchExecResult{ReducedEvalSlDfSS}{ImcHardSafe}{Error}{}{Cputime}{Min}{4.635623407}%
\StoreBenchExecResult{ReducedEvalSlDfSS}{ImcHardSafe}{Error}{}{Cputime}{Max}{913.336971733}%
\StoreBenchExecResult{ReducedEvalSlDfSS}{ImcHardSafe}{Error}{}{Cputime}{Stdev}{275.6648936263005232834765047}%
\StoreBenchExecResult{ReducedEvalSlDfSS}{ImcHardSafe}{Error}{}{Walltime}{}{268691.0377440419255257}%
\StoreBenchExecResult{ReducedEvalSlDfSS}{ImcHardSafe}{Error}{}{Walltime}{Avg}{790.2677580707115456638235294}%
\StoreBenchExecResult{ReducedEvalSlDfSS}{ImcHardSafe}{Error}{}{Walltime}{Median}{886.98942683799395}%
\StoreBenchExecResult{ReducedEvalSlDfSS}{ImcHardSafe}{Error}{}{Walltime}{Min}{1.8561149009910878}%
\StoreBenchExecResult{ReducedEvalSlDfSS}{ImcHardSafe}{Error}{}{Walltime}{Max}{907.4551925710111}%
\StoreBenchExecResult{ReducedEvalSlDfSS}{ImcHardSafe}{Error}{}{Walltime}{Stdev}{272.0555610794574634214431282}%
\StoreBenchExecResult{ReducedEvalSlDfSS}{ImcHardSafe}{Error}{Error}{Count}{}{38}%
\StoreBenchExecResult{ReducedEvalSlDfSS}{ImcHardSafe}{Error}{Error}{Cputime}{}{1410.752901000}%
\StoreBenchExecResult{ReducedEvalSlDfSS}{ImcHardSafe}{Error}{Error}{Cputime}{Avg}{37.12507634210526315789473684}%
\StoreBenchExecResult{ReducedEvalSlDfSS}{ImcHardSafe}{Error}{Error}{Cputime}{Median}{9.675710941}%
\StoreBenchExecResult{ReducedEvalSlDfSS}{ImcHardSafe}{Error}{Error}{Cputime}{Min}{4.635623407}%
\StoreBenchExecResult{ReducedEvalSlDfSS}{ImcHardSafe}{Error}{Error}{Cputime}{Max}{334.17184354}%
\StoreBenchExecResult{ReducedEvalSlDfSS}{ImcHardSafe}{Error}{Error}{Cputime}{Stdev}{71.62774899154493203216701385}%
\StoreBenchExecResult{ReducedEvalSlDfSS}{ImcHardSafe}{Error}{Error}{Walltime}{}{1267.3024649679718957}%
\StoreBenchExecResult{ReducedEvalSlDfSS}{ImcHardSafe}{Error}{Error}{Walltime}{Avg}{33.35006486757820778157894737}%
\StoreBenchExecResult{ReducedEvalSlDfSS}{ImcHardSafe}{Error}{Error}{Walltime}{Median}{6.2982278104973375}%
\StoreBenchExecResult{ReducedEvalSlDfSS}{ImcHardSafe}{Error}{Error}{Walltime}{Min}{1.8561149009910878}%
\StoreBenchExecResult{ReducedEvalSlDfSS}{ImcHardSafe}{Error}{Error}{Walltime}{Max}{330.41870098997606}%
\StoreBenchExecResult{ReducedEvalSlDfSS}{ImcHardSafe}{Error}{Error}{Walltime}{Stdev}{71.75963042913021155818250771}%
\StoreBenchExecResult{ReducedEvalSlDfSS}{ImcHardSafe}{Error}{OutOfMemory}{Count}{}{1}%
\StoreBenchExecResult{ReducedEvalSlDfSS}{ImcHardSafe}{Error}{OutOfMemory}{Cputime}{}{228.768928597}%
\StoreBenchExecResult{ReducedEvalSlDfSS}{ImcHardSafe}{Error}{OutOfMemory}{Cputime}{Avg}{228.768928597}%
\StoreBenchExecResult{ReducedEvalSlDfSS}{ImcHardSafe}{Error}{OutOfMemory}{Cputime}{Median}{228.768928597}%
\StoreBenchExecResult{ReducedEvalSlDfSS}{ImcHardSafe}{Error}{OutOfMemory}{Cputime}{Min}{228.768928597}%
\StoreBenchExecResult{ReducedEvalSlDfSS}{ImcHardSafe}{Error}{OutOfMemory}{Cputime}{Max}{228.768928597}%
\StoreBenchExecResult{ReducedEvalSlDfSS}{ImcHardSafe}{Error}{OutOfMemory}{Cputime}{Stdev}{0E-14}%
\StoreBenchExecResult{ReducedEvalSlDfSS}{ImcHardSafe}{Error}{OutOfMemory}{Walltime}{}{221.92137022598763}%
\StoreBenchExecResult{ReducedEvalSlDfSS}{ImcHardSafe}{Error}{OutOfMemory}{Walltime}{Avg}{221.92137022598763}%
\StoreBenchExecResult{ReducedEvalSlDfSS}{ImcHardSafe}{Error}{OutOfMemory}{Walltime}{Median}{221.92137022598763}%
\StoreBenchExecResult{ReducedEvalSlDfSS}{ImcHardSafe}{Error}{OutOfMemory}{Walltime}{Min}{221.92137022598763}%
\StoreBenchExecResult{ReducedEvalSlDfSS}{ImcHardSafe}{Error}{OutOfMemory}{Walltime}{Max}{221.92137022598763}%
\StoreBenchExecResult{ReducedEvalSlDfSS}{ImcHardSafe}{Error}{OutOfMemory}{Walltime}{Stdev}{0E-14}%
\StoreBenchExecResult{ReducedEvalSlDfSS}{ImcHardSafe}{Error}{Timeout}{Count}{}{301}%
\StoreBenchExecResult{ReducedEvalSlDfSS}{ImcHardSafe}{Error}{Timeout}{Cputime}{}{271857.686346775}%
\StoreBenchExecResult{ReducedEvalSlDfSS}{ImcHardSafe}{Error}{Timeout}{Cputime}{Avg}{903.1816822151993355481727575}%
\StoreBenchExecResult{ReducedEvalSlDfSS}{ImcHardSafe}{Error}{Timeout}{Cputime}{Median}{902.207494319}%
\StoreBenchExecResult{ReducedEvalSlDfSS}{ImcHardSafe}{Error}{Timeout}{Cputime}{Min}{901.340787547}%
\StoreBenchExecResult{ReducedEvalSlDfSS}{ImcHardSafe}{Error}{Timeout}{Cputime}{Max}{913.336971733}%
\StoreBenchExecResult{ReducedEvalSlDfSS}{ImcHardSafe}{Error}{Timeout}{Cputime}{Stdev}{2.915884515333338814582770271}%
\StoreBenchExecResult{ReducedEvalSlDfSS}{ImcHardSafe}{Error}{Timeout}{Walltime}{}{267201.8139088479660}%
\StoreBenchExecResult{ReducedEvalSlDfSS}{ImcHardSafe}{Error}{Timeout}{Walltime}{Avg}{887.7136674712557009966777409}%
\StoreBenchExecResult{ReducedEvalSlDfSS}{ImcHardSafe}{Error}{Timeout}{Walltime}{Median}{888.1546072549827}%
\StoreBenchExecResult{ReducedEvalSlDfSS}{ImcHardSafe}{Error}{Timeout}{Walltime}{Min}{868.4825293550093}%
\StoreBenchExecResult{ReducedEvalSlDfSS}{ImcHardSafe}{Error}{Timeout}{Walltime}{Max}{907.4551925710111}%
\StoreBenchExecResult{ReducedEvalSlDfSS}{ImcHardSafe}{Error}{Timeout}{Walltime}{Stdev}{7.472396889721180704275177030}%
\ifdefined\ReducedEvalSlDfSSImcHardSafeTotalCount\else\edef\ReducedEvalSlDfSSImcHardSafeTotalCount{0}\fi
\ifdefined\ReducedEvalSlDfSSImcHardSafeCorrectCount\else\edef\ReducedEvalSlDfSSImcHardSafeCorrectCount{0}\fi
\ifdefined\ReducedEvalSlDfSSImcHardSafeCorrectTrueCount\else\edef\ReducedEvalSlDfSSImcHardSafeCorrectTrueCount{0}\fi
\ifdefined\ReducedEvalSlDfSSImcHardSafeCorrectFalseCount\else\edef\ReducedEvalSlDfSSImcHardSafeCorrectFalseCount{0}\fi
\ifdefined\ReducedEvalSlDfSSImcHardSafeWrongTrueCount\else\edef\ReducedEvalSlDfSSImcHardSafeWrongTrueCount{0}\fi
\ifdefined\ReducedEvalSlDfSSImcHardSafeWrongFalseCount\else\edef\ReducedEvalSlDfSSImcHardSafeWrongFalseCount{0}\fi
\ifdefined\ReducedEvalSlDfSSImcHardSafeErrorTimeoutCount\else\edef\ReducedEvalSlDfSSImcHardSafeErrorTimeoutCount{0}\fi
\ifdefined\ReducedEvalSlDfSSImcHardSafeErrorOutOfMemoryCount\else\edef\ReducedEvalSlDfSSImcHardSafeErrorOutOfMemoryCount{0}\fi
\ifdefined\ReducedEvalSlDfSSImcHardSafeCorrectCputime\else\edef\ReducedEvalSlDfSSImcHardSafeCorrectCputime{0}\fi
\ifdefined\ReducedEvalSlDfSSImcHardSafeCorrectCputimeAvg\else\edef\ReducedEvalSlDfSSImcHardSafeCorrectCputimeAvg{None}\fi
\ifdefined\ReducedEvalSlDfSSImcHardSafeCorrectWalltime\else\edef\ReducedEvalSlDfSSImcHardSafeCorrectWalltime{0}\fi
\ifdefined\ReducedEvalSlDfSSImcHardSafeCorrectWalltimeAvg\else\edef\ReducedEvalSlDfSSImcHardSafeCorrectWalltimeAvg{None}\fi
\edef\ReducedEvalSlDfSSImcHardSafeErrorOtherInconclusiveCount{\the\numexpr \ReducedEvalSlDfSSImcHardSafeTotalCount - \ReducedEvalSlDfSSImcHardSafeCorrectCount - \ReducedEvalSlDfSSImcHardSafeWrongTrueCount - \ReducedEvalSlDfSSImcHardSafeWrongFalseCount - \ReducedEvalSlDfSSImcHardSafeErrorTimeoutCount - \ReducedEvalSlDfSSImcHardSafeErrorOutOfMemoryCount \relax}
\providecommand\StoreBenchExecResult[7]{\expandafter\newcommand\csname#1#2#3#4#5#6\endcsname{#7}}%
\StoreBenchExecResult{ReducedEvalSlDfSS}{ImcIgndfiHardSafe}{Total}{}{Count}{}{870}%
\StoreBenchExecResult{ReducedEvalSlDfSS}{ImcIgndfiHardSafe}{Total}{}{Cputime}{}{310043.770948392}%
\StoreBenchExecResult{ReducedEvalSlDfSS}{ImcIgndfiHardSafe}{Total}{}{Cputime}{Avg}{356.3721505153931034482758621}%
\StoreBenchExecResult{ReducedEvalSlDfSS}{ImcIgndfiHardSafe}{Total}{}{Cputime}{Median}{83.374383480}%
\StoreBenchExecResult{ReducedEvalSlDfSS}{ImcIgndfiHardSafe}{Total}{}{Cputime}{Min}{4.51306763}%
\StoreBenchExecResult{ReducedEvalSlDfSS}{ImcIgndfiHardSafe}{Total}{}{Cputime}{Max}{957.848831916}%
\StoreBenchExecResult{ReducedEvalSlDfSS}{ImcIgndfiHardSafe}{Total}{}{Cputime}{Stdev}{406.4338064196389823271583101}%
\StoreBenchExecResult{ReducedEvalSlDfSS}{ImcIgndfiHardSafe}{Total}{}{Walltime}{}{264936.8182710441470287}%
\StoreBenchExecResult{ReducedEvalSlDfSS}{ImcIgndfiHardSafe}{Total}{}{Walltime}{Avg}{304.5250784724645368145977011}%
\StoreBenchExecResult{ReducedEvalSlDfSS}{ImcIgndfiHardSafe}{Total}{}{Walltime}{Median}{30.1705278489971525}%
\StoreBenchExecResult{ReducedEvalSlDfSS}{ImcIgndfiHardSafe}{Total}{}{Walltime}{Min}{1.7615510629839264}%
\StoreBenchExecResult{ReducedEvalSlDfSS}{ImcIgndfiHardSafe}{Total}{}{Walltime}{Max}{895.3316802030022}%
\StoreBenchExecResult{ReducedEvalSlDfSS}{ImcIgndfiHardSafe}{Total}{}{Walltime}{Stdev}{370.4511561612247428520158603}%
\StoreBenchExecResult{ReducedEvalSlDfSS}{ImcIgndfiHardSafe}{Correct}{}{Count}{}{541}%
\StoreBenchExecResult{ReducedEvalSlDfSS}{ImcIgndfiHardSafe}{Correct}{}{Cputime}{}{41585.436629012}%
\StoreBenchExecResult{ReducedEvalSlDfSS}{ImcIgndfiHardSafe}{Correct}{}{Cputime}{Avg}{76.86772020150092421441774492}%
\StoreBenchExecResult{ReducedEvalSlDfSS}{ImcIgndfiHardSafe}{Correct}{}{Cputime}{Median}{16.096468534}%
\StoreBenchExecResult{ReducedEvalSlDfSS}{ImcIgndfiHardSafe}{Correct}{}{Cputime}{Min}{4.51306763}%
\StoreBenchExecResult{ReducedEvalSlDfSS}{ImcIgndfiHardSafe}{Correct}{}{Cputime}{Max}{895.354100396}%
\StoreBenchExecResult{ReducedEvalSlDfSS}{ImcIgndfiHardSafe}{Correct}{}{Cputime}{Stdev}{152.3745595216843019491785457}%
\StoreBenchExecResult{ReducedEvalSlDfSS}{ImcIgndfiHardSafe}{Correct}{}{Walltime}{}{25946.2683551012593467}%
\StoreBenchExecResult{ReducedEvalSlDfSS}{ImcIgndfiHardSafe}{Correct}{}{Walltime}{Avg}{47.95983060092654223049907579}%
\StoreBenchExecResult{ReducedEvalSlDfSS}{ImcIgndfiHardSafe}{Correct}{}{Walltime}{Median}{5.560596544994041}%
\StoreBenchExecResult{ReducedEvalSlDfSS}{ImcIgndfiHardSafe}{Correct}{}{Walltime}{Min}{1.7615510629839264}%
\StoreBenchExecResult{ReducedEvalSlDfSS}{ImcIgndfiHardSafe}{Correct}{}{Walltime}{Max}{787.5340112169797}%
\StoreBenchExecResult{ReducedEvalSlDfSS}{ImcIgndfiHardSafe}{Correct}{}{Walltime}{Stdev}{126.9781102904073725363789497}%
\StoreBenchExecResult{ReducedEvalSlDfSS}{ImcIgndfiHardSafe}{Correct}{True}{Count}{}{541}%
\StoreBenchExecResult{ReducedEvalSlDfSS}{ImcIgndfiHardSafe}{Correct}{True}{Cputime}{}{41585.436629012}%
\StoreBenchExecResult{ReducedEvalSlDfSS}{ImcIgndfiHardSafe}{Correct}{True}{Cputime}{Avg}{76.86772020150092421441774492}%
\StoreBenchExecResult{ReducedEvalSlDfSS}{ImcIgndfiHardSafe}{Correct}{True}{Cputime}{Median}{16.096468534}%
\StoreBenchExecResult{ReducedEvalSlDfSS}{ImcIgndfiHardSafe}{Correct}{True}{Cputime}{Min}{4.51306763}%
\StoreBenchExecResult{ReducedEvalSlDfSS}{ImcIgndfiHardSafe}{Correct}{True}{Cputime}{Max}{895.354100396}%
\StoreBenchExecResult{ReducedEvalSlDfSS}{ImcIgndfiHardSafe}{Correct}{True}{Cputime}{Stdev}{152.3745595216843019491785457}%
\StoreBenchExecResult{ReducedEvalSlDfSS}{ImcIgndfiHardSafe}{Correct}{True}{Walltime}{}{25946.2683551012593467}%
\StoreBenchExecResult{ReducedEvalSlDfSS}{ImcIgndfiHardSafe}{Correct}{True}{Walltime}{Avg}{47.95983060092654223049907579}%
\StoreBenchExecResult{ReducedEvalSlDfSS}{ImcIgndfiHardSafe}{Correct}{True}{Walltime}{Median}{5.560596544994041}%
\StoreBenchExecResult{ReducedEvalSlDfSS}{ImcIgndfiHardSafe}{Correct}{True}{Walltime}{Min}{1.7615510629839264}%
\StoreBenchExecResult{ReducedEvalSlDfSS}{ImcIgndfiHardSafe}{Correct}{True}{Walltime}{Max}{787.5340112169797}%
\StoreBenchExecResult{ReducedEvalSlDfSS}{ImcIgndfiHardSafe}{Correct}{True}{Walltime}{Stdev}{126.9781102904073725363789497}%

\StoreBenchExecResult{ReducedEvalSlDfSS}{ImcIgndfiHardSafe}{Error}{}{Count}{}{329}%
\StoreBenchExecResult{ReducedEvalSlDfSS}{ImcIgndfiHardSafe}{Error}{}{Cputime}{}{268458.334319380}%
\StoreBenchExecResult{ReducedEvalSlDfSS}{ImcIgndfiHardSafe}{Error}{}{Cputime}{Avg}{815.9827790862613981762917933}%
\StoreBenchExecResult{ReducedEvalSlDfSS}{ImcIgndfiHardSafe}{Error}{}{Cputime}{Median}{901.863149229}%
\StoreBenchExecResult{ReducedEvalSlDfSS}{ImcIgndfiHardSafe}{Error}{}{Cputime}{Min}{15.223739912}%
\StoreBenchExecResult{ReducedEvalSlDfSS}{ImcIgndfiHardSafe}{Error}{}{Cputime}{Max}{957.848831916}%
\StoreBenchExecResult{ReducedEvalSlDfSS}{ImcIgndfiHardSafe}{Error}{}{Cputime}{Stdev}{242.7677944589477214206769661}%
\StoreBenchExecResult{ReducedEvalSlDfSS}{ImcIgndfiHardSafe}{Error}{}{Walltime}{}{238990.549915942887682}%
\StoreBenchExecResult{ReducedEvalSlDfSS}{ImcIgndfiHardSafe}{Error}{}{Walltime}{Avg}{726.4150453372124245653495441}%
\StoreBenchExecResult{ReducedEvalSlDfSS}{ImcIgndfiHardSafe}{Error}{}{Walltime}{Median}{803.5965623650118}%
\StoreBenchExecResult{ReducedEvalSlDfSS}{ImcIgndfiHardSafe}{Error}{}{Walltime}{Min}{7.733986650011502}%
\StoreBenchExecResult{ReducedEvalSlDfSS}{ImcIgndfiHardSafe}{Error}{}{Walltime}{Max}{895.3316802030022}%
\StoreBenchExecResult{ReducedEvalSlDfSS}{ImcIgndfiHardSafe}{Error}{}{Walltime}{Stdev}{223.9466779932465639835854679}%
\StoreBenchExecResult{ReducedEvalSlDfSS}{ImcIgndfiHardSafe}{Error}{Error}{Count}{}{38}%
\StoreBenchExecResult{ReducedEvalSlDfSS}{ImcIgndfiHardSafe}{Error}{Error}{Cputime}{}{5649.498840732}%
\StoreBenchExecResult{ReducedEvalSlDfSS}{ImcIgndfiHardSafe}{Error}{Error}{Cputime}{Avg}{148.6710221245263157894736842}%
\StoreBenchExecResult{ReducedEvalSlDfSS}{ImcIgndfiHardSafe}{Error}{Error}{Cputime}{Median}{153.6023658035}%
\StoreBenchExecResult{ReducedEvalSlDfSS}{ImcIgndfiHardSafe}{Error}{Error}{Cputime}{Min}{15.223739912}%
\StoreBenchExecResult{ReducedEvalSlDfSS}{ImcIgndfiHardSafe}{Error}{Error}{Cputime}{Max}{466.98625936}%
\StoreBenchExecResult{ReducedEvalSlDfSS}{ImcIgndfiHardSafe}{Error}{Error}{Cputime}{Stdev}{81.68146584327738968656165356}%
\StoreBenchExecResult{ReducedEvalSlDfSS}{ImcIgndfiHardSafe}{Error}{Error}{Walltime}{}{4220.923634568956882}%
\StoreBenchExecResult{ReducedEvalSlDfSS}{ImcIgndfiHardSafe}{Error}{Error}{Walltime}{Avg}{111.0769377518146547894736842}%
\StoreBenchExecResult{ReducedEvalSlDfSS}{ImcIgndfiHardSafe}{Error}{Error}{Walltime}{Median}{124.73057947399502}%
\StoreBenchExecResult{ReducedEvalSlDfSS}{ImcIgndfiHardSafe}{Error}{Error}{Walltime}{Min}{7.733986650011502}%
\StoreBenchExecResult{ReducedEvalSlDfSS}{ImcIgndfiHardSafe}{Error}{Error}{Walltime}{Max}{380.61011422701995}%
\StoreBenchExecResult{ReducedEvalSlDfSS}{ImcIgndfiHardSafe}{Error}{Error}{Walltime}{Stdev}{69.00097628547769561660609892}%
\StoreBenchExecResult{ReducedEvalSlDfSS}{ImcIgndfiHardSafe}{Error}{Timeout}{Count}{}{291}%
\StoreBenchExecResult{ReducedEvalSlDfSS}{ImcIgndfiHardSafe}{Error}{Timeout}{Cputime}{}{262808.835478648}%
\StoreBenchExecResult{ReducedEvalSlDfSS}{ImcIgndfiHardSafe}{Error}{Timeout}{Cputime}{Avg}{903.1231459747353951890034364}%
\StoreBenchExecResult{ReducedEvalSlDfSS}{ImcIgndfiHardSafe}{Error}{Timeout}{Cputime}{Median}{901.969422507}%
\StoreBenchExecResult{ReducedEvalSlDfSS}{ImcIgndfiHardSafe}{Error}{Timeout}{Cputime}{Min}{901.171079033}%
\StoreBenchExecResult{ReducedEvalSlDfSS}{ImcIgndfiHardSafe}{Error}{Timeout}{Cputime}{Max}{957.848831916}%
\StoreBenchExecResult{ReducedEvalSlDfSS}{ImcIgndfiHardSafe}{Error}{Timeout}{Cputime}{Stdev}{4.226559842797812277838806423}%
\StoreBenchExecResult{ReducedEvalSlDfSS}{ImcIgndfiHardSafe}{Error}{Timeout}{Walltime}{}{234769.6262813739308}%
\StoreBenchExecResult{ReducedEvalSlDfSS}{ImcIgndfiHardSafe}{Error}{Timeout}{Walltime}{Avg}{806.7684751937248481099656357}%
\StoreBenchExecResult{ReducedEvalSlDfSS}{ImcIgndfiHardSafe}{Error}{Timeout}{Walltime}{Median}{804.242635371018}%
\StoreBenchExecResult{ReducedEvalSlDfSS}{ImcIgndfiHardSafe}{Error}{Timeout}{Walltime}{Min}{785.9986985789728}%
\StoreBenchExecResult{ReducedEvalSlDfSS}{ImcIgndfiHardSafe}{Error}{Timeout}{Walltime}{Max}{895.3316802030022}%
\StoreBenchExecResult{ReducedEvalSlDfSS}{ImcIgndfiHardSafe}{Error}{Timeout}{Walltime}{Stdev}{13.35132492057737508697746993}%
\ifdefined\ReducedEvalSlDfSSImcIgndfiHardSafeTotalCount\else\edef\ReducedEvalSlDfSSImcIgndfiHardSafeTotalCount{0}\fi
\ifdefined\ReducedEvalSlDfSSImcIgndfiHardSafeCorrectCount\else\edef\ReducedEvalSlDfSSImcIgndfiHardSafeCorrectCount{0}\fi
\ifdefined\ReducedEvalSlDfSSImcIgndfiHardSafeCorrectTrueCount\else\edef\ReducedEvalSlDfSSImcIgndfiHardSafeCorrectTrueCount{0}\fi
\ifdefined\ReducedEvalSlDfSSImcIgndfiHardSafeCorrectFalseCount\else\edef\ReducedEvalSlDfSSImcIgndfiHardSafeCorrectFalseCount{0}\fi
\ifdefined\ReducedEvalSlDfSSImcIgndfiHardSafeWrongTrueCount\else\edef\ReducedEvalSlDfSSImcIgndfiHardSafeWrongTrueCount{0}\fi
\ifdefined\ReducedEvalSlDfSSImcIgndfiHardSafeWrongFalseCount\else\edef\ReducedEvalSlDfSSImcIgndfiHardSafeWrongFalseCount{0}\fi
\ifdefined\ReducedEvalSlDfSSImcIgndfiHardSafeErrorTimeoutCount\else\edef\ReducedEvalSlDfSSImcIgndfiHardSafeErrorTimeoutCount{0}\fi
\ifdefined\ReducedEvalSlDfSSImcIgndfiHardSafeErrorOutOfMemoryCount\else\edef\ReducedEvalSlDfSSImcIgndfiHardSafeErrorOutOfMemoryCount{0}\fi
\ifdefined\ReducedEvalSlDfSSImcIgndfiHardSafeCorrectCputime\else\edef\ReducedEvalSlDfSSImcIgndfiHardSafeCorrectCputime{0}\fi
\ifdefined\ReducedEvalSlDfSSImcIgndfiHardSafeCorrectCputimeAvg\else\edef\ReducedEvalSlDfSSImcIgndfiHardSafeCorrectCputimeAvg{None}\fi
\ifdefined\ReducedEvalSlDfSSImcIgndfiHardSafeCorrectWalltime\else\edef\ReducedEvalSlDfSSImcIgndfiHardSafeCorrectWalltime{0}\fi
\ifdefined\ReducedEvalSlDfSSImcIgndfiHardSafeCorrectWalltimeAvg\else\edef\ReducedEvalSlDfSSImcIgndfiHardSafeCorrectWalltimeAvg{None}\fi
\edef\ReducedEvalSlDfSSImcIgndfiHardSafeErrorOtherInconclusiveCount{\the\numexpr \ReducedEvalSlDfSSImcIgndfiHardSafeTotalCount - \ReducedEvalSlDfSSImcIgndfiHardSafeCorrectCount - \ReducedEvalSlDfSSImcIgndfiHardSafeWrongTrueCount - \ReducedEvalSlDfSSImcIgndfiHardSafeWrongFalseCount - \ReducedEvalSlDfSSImcIgndfiHardSafeErrorTimeoutCount - \ReducedEvalSlDfSSImcIgndfiHardSafeErrorOutOfMemoryCount \relax}
\providecommand\StoreBenchExecResult[7]{\expandafter\newcommand\csname#1#2#3#4#5#6\endcsname{#7}}%
\StoreBenchExecResult{ReducedEvalSlDfSSo}{ImcHardSafe}{Total}{}{Count}{}{870}%
\StoreBenchExecResult{ReducedEvalSlDfSSo}{ImcHardSafe}{Total}{}{Cputime}{}{303683.845548166}%
\StoreBenchExecResult{ReducedEvalSlDfSSo}{ImcHardSafe}{Total}{}{Cputime}{Avg}{349.0618914346735632183908046}%
\StoreBenchExecResult{ReducedEvalSlDfSSo}{ImcHardSafe}{Total}{}{Cputime}{Median}{26.846311480}%
\StoreBenchExecResult{ReducedEvalSlDfSSo}{ImcHardSafe}{Total}{}{Cputime}{Min}{4.087006411}%
\StoreBenchExecResult{ReducedEvalSlDfSSo}{ImcHardSafe}{Total}{}{Cputime}{Max}{913.128687004}%
\StoreBenchExecResult{ReducedEvalSlDfSSo}{ImcHardSafe}{Total}{}{Cputime}{Stdev}{417.0799298982675518837273640}%
\StoreBenchExecResult{ReducedEvalSlDfSSo}{ImcHardSafe}{Total}{}{Walltime}{}{294518.7427456645828297}%
\StoreBenchExecResult{ReducedEvalSlDfSSo}{ImcHardSafe}{Total}{}{Walltime}{Avg}{338.5272905122581411835632184}%
\StoreBenchExecResult{ReducedEvalSlDfSSo}{ImcHardSafe}{Total}{}{Walltime}{Median}{13.636842309992062}%
\StoreBenchExecResult{ReducedEvalSlDfSSo}{ImcHardSafe}{Total}{}{Walltime}{Min}{1.6414567659958266}%
\StoreBenchExecResult{ReducedEvalSlDfSSo}{ImcHardSafe}{Total}{}{Walltime}{Max}{908.1039912319975}%
\StoreBenchExecResult{ReducedEvalSlDfSSo}{ImcHardSafe}{Total}{}{Walltime}{Stdev}{413.0815344827300468032675132}%
\StoreBenchExecResult{ReducedEvalSlDfSSo}{ImcHardSafe}{Correct}{}{Count}{}{532}%
\StoreBenchExecResult{ReducedEvalSlDfSSo}{ImcHardSafe}{Correct}{}{Cputime}{}{31220.413925447}%
\StoreBenchExecResult{ReducedEvalSlDfSSo}{ImcHardSafe}{Correct}{}{Cputime}{Avg}{58.68498858166729323308270677}%
\StoreBenchExecResult{ReducedEvalSlDfSSo}{ImcHardSafe}{Correct}{}{Cputime}{Median}{10.9839046375}%
\StoreBenchExecResult{ReducedEvalSlDfSSo}{ImcHardSafe}{Correct}{}{Cputime}{Min}{4.087006411}%
\StoreBenchExecResult{ReducedEvalSlDfSSo}{ImcHardSafe}{Correct}{}{Cputime}{Max}{857.619009866}%
\StoreBenchExecResult{ReducedEvalSlDfSSo}{ImcHardSafe}{Correct}{}{Cputime}{Stdev}{140.0547412291898305204687787}%
\StoreBenchExecResult{ReducedEvalSlDfSSo}{ImcHardSafe}{Correct}{}{Walltime}{}{26778.0916790796911361}%
\StoreBenchExecResult{ReducedEvalSlDfSSo}{ImcHardSafe}{Correct}{}{Walltime}{Avg}{50.33475879526257732349624060}%
\StoreBenchExecResult{ReducedEvalSlDfSSo}{ImcHardSafe}{Correct}{}{Walltime}{Median}{4.6757075320056175}%
\StoreBenchExecResult{ReducedEvalSlDfSSo}{ImcHardSafe}{Correct}{}{Walltime}{Min}{1.6414567659958266}%
\StoreBenchExecResult{ReducedEvalSlDfSSo}{ImcHardSafe}{Correct}{}{Walltime}{Max}{843.0681187960145}%
\StoreBenchExecResult{ReducedEvalSlDfSSo}{ImcHardSafe}{Correct}{}{Walltime}{Stdev}{136.5788548052185673600485407}%
\StoreBenchExecResult{ReducedEvalSlDfSSo}{ImcHardSafe}{Correct}{True}{Count}{}{532}%
\StoreBenchExecResult{ReducedEvalSlDfSSo}{ImcHardSafe}{Correct}{True}{Cputime}{}{31220.413925447}%
\StoreBenchExecResult{ReducedEvalSlDfSSo}{ImcHardSafe}{Correct}{True}{Cputime}{Avg}{58.68498858166729323308270677}%
\StoreBenchExecResult{ReducedEvalSlDfSSo}{ImcHardSafe}{Correct}{True}{Cputime}{Median}{10.9839046375}%
\StoreBenchExecResult{ReducedEvalSlDfSSo}{ImcHardSafe}{Correct}{True}{Cputime}{Min}{4.087006411}%
\StoreBenchExecResult{ReducedEvalSlDfSSo}{ImcHardSafe}{Correct}{True}{Cputime}{Max}{857.619009866}%
\StoreBenchExecResult{ReducedEvalSlDfSSo}{ImcHardSafe}{Correct}{True}{Cputime}{Stdev}{140.0547412291898305204687787}%
\StoreBenchExecResult{ReducedEvalSlDfSSo}{ImcHardSafe}{Correct}{True}{Walltime}{}{26778.0916790796911361}%
\StoreBenchExecResult{ReducedEvalSlDfSSo}{ImcHardSafe}{Correct}{True}{Walltime}{Avg}{50.33475879526257732349624060}%
\StoreBenchExecResult{ReducedEvalSlDfSSo}{ImcHardSafe}{Correct}{True}{Walltime}{Median}{4.6757075320056175}%
\StoreBenchExecResult{ReducedEvalSlDfSSo}{ImcHardSafe}{Correct}{True}{Walltime}{Min}{1.6414567659958266}%
\StoreBenchExecResult{ReducedEvalSlDfSSo}{ImcHardSafe}{Correct}{True}{Walltime}{Max}{843.0681187960145}%
\StoreBenchExecResult{ReducedEvalSlDfSSo}{ImcHardSafe}{Correct}{True}{Walltime}{Stdev}{136.5788548052185673600485407}%

\StoreBenchExecResult{ReducedEvalSlDfSSo}{ImcHardSafe}{Error}{}{Count}{}{338}%
\StoreBenchExecResult{ReducedEvalSlDfSSo}{ImcHardSafe}{Error}{}{Cputime}{}{272463.431622719}%
\StoreBenchExecResult{ReducedEvalSlDfSSo}{ImcHardSafe}{Error}{}{Cputime}{Avg}{806.1048272861508875739644970}%
\StoreBenchExecResult{ReducedEvalSlDfSSo}{ImcHardSafe}{Error}{}{Cputime}{Median}{902.1776709465}%
\StoreBenchExecResult{ReducedEvalSlDfSSo}{ImcHardSafe}{Error}{}{Cputime}{Min}{4.717716479}%
\StoreBenchExecResult{ReducedEvalSlDfSSo}{ImcHardSafe}{Error}{}{Cputime}{Max}{913.128687004}%
\StoreBenchExecResult{ReducedEvalSlDfSSo}{ImcHardSafe}{Error}{}{Cputime}{Stdev}{274.3700457982080290246166033}%
\StoreBenchExecResult{ReducedEvalSlDfSSo}{ImcHardSafe}{Error}{}{Walltime}{}{267740.6510665848916936}%
\StoreBenchExecResult{ReducedEvalSlDfSSo}{ImcHardSafe}{Error}{}{Walltime}{Avg}{792.1321037472925789751479290}%
\StoreBenchExecResult{ReducedEvalSlDfSSo}{ImcHardSafe}{Error}{}{Walltime}{Median}{887.30666817599555}%
\StoreBenchExecResult{ReducedEvalSlDfSSo}{ImcHardSafe}{Error}{}{Walltime}{Min}{1.8297864500200376}%
\StoreBenchExecResult{ReducedEvalSlDfSSo}{ImcHardSafe}{Error}{}{Walltime}{Max}{908.1039912319975}%
\StoreBenchExecResult{ReducedEvalSlDfSSo}{ImcHardSafe}{Error}{}{Walltime}{Stdev}{270.8665321358373373295028387}%
\StoreBenchExecResult{ReducedEvalSlDfSSo}{ImcHardSafe}{Error}{Error}{Count}{}{38}%
\StoreBenchExecResult{ReducedEvalSlDfSSo}{ImcHardSafe}{Error}{Error}{Cputime}{}{1463.388442053}%
\StoreBenchExecResult{ReducedEvalSlDfSSo}{ImcHardSafe}{Error}{Error}{Cputime}{Avg}{38.51022215928947368421052632}%
\StoreBenchExecResult{ReducedEvalSlDfSSo}{ImcHardSafe}{Error}{Error}{Cputime}{Median}{11.190677818}%
\StoreBenchExecResult{ReducedEvalSlDfSSo}{ImcHardSafe}{Error}{Error}{Cputime}{Min}{4.717716479}%
\StoreBenchExecResult{ReducedEvalSlDfSSo}{ImcHardSafe}{Error}{Error}{Cputime}{Max}{376.999135018}%
\StoreBenchExecResult{ReducedEvalSlDfSSo}{ImcHardSafe}{Error}{Error}{Cputime}{Stdev}{75.41633644460719492319217856}%
\StoreBenchExecResult{ReducedEvalSlDfSSo}{ImcHardSafe}{Error}{Error}{Walltime}{}{1317.8947572110045936}%
\StoreBenchExecResult{ReducedEvalSlDfSSo}{ImcHardSafe}{Error}{Error}{Walltime}{Avg}{34.68144097923696298947368421}%
\StoreBenchExecResult{ReducedEvalSlDfSSo}{ImcHardSafe}{Error}{Error}{Walltime}{Median}{6.5222940690000545}%
\StoreBenchExecResult{ReducedEvalSlDfSSo}{ImcHardSafe}{Error}{Error}{Walltime}{Min}{1.8297864500200376}%
\StoreBenchExecResult{ReducedEvalSlDfSSo}{ImcHardSafe}{Error}{Error}{Walltime}{Max}{373.6095939200022}%
\StoreBenchExecResult{ReducedEvalSlDfSSo}{ImcHardSafe}{Error}{Error}{Walltime}{Stdev}{75.59287219425154271057104670}%
\StoreBenchExecResult{ReducedEvalSlDfSSo}{ImcHardSafe}{Error}{Timeout}{Count}{}{300}%
\StoreBenchExecResult{ReducedEvalSlDfSSo}{ImcHardSafe}{Error}{Timeout}{Cputime}{}{271000.043180666}%
\StoreBenchExecResult{ReducedEvalSlDfSSo}{ImcHardSafe}{Error}{Timeout}{Cputime}{Avg}{903.3334772688866666666666667}%
\StoreBenchExecResult{ReducedEvalSlDfSSo}{ImcHardSafe}{Error}{Timeout}{Cputime}{Median}{902.324221153}%
\StoreBenchExecResult{ReducedEvalSlDfSSo}{ImcHardSafe}{Error}{Timeout}{Cputime}{Min}{900.501480853}%
\StoreBenchExecResult{ReducedEvalSlDfSSo}{ImcHardSafe}{Error}{Timeout}{Cputime}{Max}{913.128687004}%
\StoreBenchExecResult{ReducedEvalSlDfSSo}{ImcHardSafe}{Error}{Timeout}{Cputime}{Stdev}{2.867444672569619275700074623}%
\StoreBenchExecResult{ReducedEvalSlDfSSo}{ImcHardSafe}{Error}{Timeout}{Walltime}{}{266422.7563093738871}%
\StoreBenchExecResult{ReducedEvalSlDfSSo}{ImcHardSafe}{Error}{Timeout}{Walltime}{Avg}{888.0758543645796236666666667}%
\StoreBenchExecResult{ReducedEvalSlDfSSo}{ImcHardSafe}{Error}{Timeout}{Walltime}{Median}{888.14656609149825}%
\StoreBenchExecResult{ReducedEvalSlDfSSo}{ImcHardSafe}{Error}{Timeout}{Walltime}{Min}{869.2693176619941}%
\StoreBenchExecResult{ReducedEvalSlDfSSo}{ImcHardSafe}{Error}{Timeout}{Walltime}{Max}{908.1039912319975}%
\StoreBenchExecResult{ReducedEvalSlDfSSo}{ImcHardSafe}{Error}{Timeout}{Walltime}{Stdev}{7.770088150268849092263136910}%
\ifdefined\ReducedEvalSlDfSSoImcHardSafeTotalCount\else\edef\ReducedEvalSlDfSSoImcHardSafeTotalCount{0}\fi
\ifdefined\ReducedEvalSlDfSSoImcHardSafeCorrectCount\else\edef\ReducedEvalSlDfSSoImcHardSafeCorrectCount{0}\fi
\ifdefined\ReducedEvalSlDfSSoImcHardSafeCorrectTrueCount\else\edef\ReducedEvalSlDfSSoImcHardSafeCorrectTrueCount{0}\fi
\ifdefined\ReducedEvalSlDfSSoImcHardSafeCorrectFalseCount\else\edef\ReducedEvalSlDfSSoImcHardSafeCorrectFalseCount{0}\fi
\ifdefined\ReducedEvalSlDfSSoImcHardSafeWrongTrueCount\else\edef\ReducedEvalSlDfSSoImcHardSafeWrongTrueCount{0}\fi
\ifdefined\ReducedEvalSlDfSSoImcHardSafeWrongFalseCount\else\edef\ReducedEvalSlDfSSoImcHardSafeWrongFalseCount{0}\fi
\ifdefined\ReducedEvalSlDfSSoImcHardSafeErrorTimeoutCount\else\edef\ReducedEvalSlDfSSoImcHardSafeErrorTimeoutCount{0}\fi
\ifdefined\ReducedEvalSlDfSSoImcHardSafeErrorOutOfMemoryCount\else\edef\ReducedEvalSlDfSSoImcHardSafeErrorOutOfMemoryCount{0}\fi
\ifdefined\ReducedEvalSlDfSSoImcHardSafeCorrectCputime\else\edef\ReducedEvalSlDfSSoImcHardSafeCorrectCputime{0}\fi
\ifdefined\ReducedEvalSlDfSSoImcHardSafeCorrectCputimeAvg\else\edef\ReducedEvalSlDfSSoImcHardSafeCorrectCputimeAvg{None}\fi
\ifdefined\ReducedEvalSlDfSSoImcHardSafeCorrectWalltime\else\edef\ReducedEvalSlDfSSoImcHardSafeCorrectWalltime{0}\fi
\ifdefined\ReducedEvalSlDfSSoImcHardSafeCorrectWalltimeAvg\else\edef\ReducedEvalSlDfSSoImcHardSafeCorrectWalltimeAvg{None}\fi
\edef\ReducedEvalSlDfSSoImcHardSafeErrorOtherInconclusiveCount{\the\numexpr \ReducedEvalSlDfSSoImcHardSafeTotalCount - \ReducedEvalSlDfSSoImcHardSafeCorrectCount - \ReducedEvalSlDfSSoImcHardSafeWrongTrueCount - \ReducedEvalSlDfSSoImcHardSafeWrongFalseCount - \ReducedEvalSlDfSSoImcHardSafeErrorTimeoutCount - \ReducedEvalSlDfSSoImcHardSafeErrorOutOfMemoryCount \relax}
\providecommand\StoreBenchExecResult[7]{\expandafter\newcommand\csname#1#2#3#4#5#6\endcsname{#7}}%
\StoreBenchExecResult{ReducedEvalSlDfSSo}{ImcIgndfiHardSafe}{Total}{}{Count}{}{870}%
\StoreBenchExecResult{ReducedEvalSlDfSSo}{ImcIgndfiHardSafe}{Total}{}{Cputime}{}{312731.691417760}%
\StoreBenchExecResult{ReducedEvalSlDfSSo}{ImcIgndfiHardSafe}{Total}{}{Cputime}{Avg}{359.4617142732873563218390805}%
\StoreBenchExecResult{ReducedEvalSlDfSSo}{ImcIgndfiHardSafe}{Total}{}{Cputime}{Median}{83.6676897845}%
\StoreBenchExecResult{ReducedEvalSlDfSSo}{ImcIgndfiHardSafe}{Total}{}{Cputime}{Min}{4.446634201}%
\StoreBenchExecResult{ReducedEvalSlDfSSo}{ImcIgndfiHardSafe}{Total}{}{Cputime}{Max}{960.582682172}%
\StoreBenchExecResult{ReducedEvalSlDfSSo}{ImcIgndfiHardSafe}{Total}{}{Cputime}{Stdev}{407.4049890659628374174065749}%
\StoreBenchExecResult{ReducedEvalSlDfSSo}{ImcIgndfiHardSafe}{Total}{}{Walltime}{}{267829.3748844512158467}%
\StoreBenchExecResult{ReducedEvalSlDfSSo}{ImcIgndfiHardSafe}{Total}{}{Walltime}{Avg}{307.8498561890243860306896552}%
\StoreBenchExecResult{ReducedEvalSlDfSSo}{ImcIgndfiHardSafe}{Total}{}{Walltime}{Median}{28.9438241644966185}%
\StoreBenchExecResult{ReducedEvalSlDfSSo}{ImcIgndfiHardSafe}{Total}{}{Walltime}{Min}{1.7509042989986483}%
\StoreBenchExecResult{ReducedEvalSlDfSSo}{ImcIgndfiHardSafe}{Total}{}{Walltime}{Max}{906.3644544679846}%
\StoreBenchExecResult{ReducedEvalSlDfSSo}{ImcIgndfiHardSafe}{Total}{}{Walltime}{Stdev}{371.0314339349479780737381975}%
\StoreBenchExecResult{ReducedEvalSlDfSSo}{ImcIgndfiHardSafe}{Correct}{}{Count}{}{538}%
\StoreBenchExecResult{ReducedEvalSlDfSSo}{ImcIgndfiHardSafe}{Correct}{}{Cputime}{}{41574.639498322}%
\StoreBenchExecResult{ReducedEvalSlDfSSo}{ImcIgndfiHardSafe}{Correct}{}{Cputime}{Avg}{77.27628159539405204460966543}%
\StoreBenchExecResult{ReducedEvalSlDfSSo}{ImcIgndfiHardSafe}{Correct}{}{Cputime}{Median}{15.6193925925}%
\StoreBenchExecResult{ReducedEvalSlDfSSo}{ImcIgndfiHardSafe}{Correct}{}{Cputime}{Min}{4.446634201}%
\StoreBenchExecResult{ReducedEvalSlDfSSo}{ImcIgndfiHardSafe}{Correct}{}{Cputime}{Max}{882.44505151}%
\StoreBenchExecResult{ReducedEvalSlDfSSo}{ImcIgndfiHardSafe}{Correct}{}{Cputime}{Stdev}{153.5585677444348255763147959}%
\StoreBenchExecResult{ReducedEvalSlDfSSo}{ImcIgndfiHardSafe}{Correct}{}{Walltime}{}{26384.1773969015047557}%
\StoreBenchExecResult{ReducedEvalSlDfSSo}{ImcIgndfiHardSafe}{Correct}{}{Walltime}{Avg}{49.04122192732621701802973978}%
\StoreBenchExecResult{ReducedEvalSlDfSSo}{ImcIgndfiHardSafe}{Correct}{}{Walltime}{Median}{5.6449822834983935}%
\StoreBenchExecResult{ReducedEvalSlDfSSo}{ImcIgndfiHardSafe}{Correct}{}{Walltime}{Min}{1.7509042989986483}%
\StoreBenchExecResult{ReducedEvalSlDfSSo}{ImcIgndfiHardSafe}{Correct}{}{Walltime}{Max}{784.9636439690075}%
\StoreBenchExecResult{ReducedEvalSlDfSSo}{ImcIgndfiHardSafe}{Correct}{}{Walltime}{Stdev}{127.9864998532388843864722819}%
\StoreBenchExecResult{ReducedEvalSlDfSSo}{ImcIgndfiHardSafe}{Correct}{True}{Count}{}{538}%
\StoreBenchExecResult{ReducedEvalSlDfSSo}{ImcIgndfiHardSafe}{Correct}{True}{Cputime}{}{41574.639498322}%
\StoreBenchExecResult{ReducedEvalSlDfSSo}{ImcIgndfiHardSafe}{Correct}{True}{Cputime}{Avg}{77.27628159539405204460966543}%
\StoreBenchExecResult{ReducedEvalSlDfSSo}{ImcIgndfiHardSafe}{Correct}{True}{Cputime}{Median}{15.6193925925}%
\StoreBenchExecResult{ReducedEvalSlDfSSo}{ImcIgndfiHardSafe}{Correct}{True}{Cputime}{Min}{4.446634201}%
\StoreBenchExecResult{ReducedEvalSlDfSSo}{ImcIgndfiHardSafe}{Correct}{True}{Cputime}{Max}{882.44505151}%
\StoreBenchExecResult{ReducedEvalSlDfSSo}{ImcIgndfiHardSafe}{Correct}{True}{Cputime}{Stdev}{153.5585677444348255763147959}%
\StoreBenchExecResult{ReducedEvalSlDfSSo}{ImcIgndfiHardSafe}{Correct}{True}{Walltime}{}{26384.1773969015047557}%
\StoreBenchExecResult{ReducedEvalSlDfSSo}{ImcIgndfiHardSafe}{Correct}{True}{Walltime}{Avg}{49.04122192732621701802973978}%
\StoreBenchExecResult{ReducedEvalSlDfSSo}{ImcIgndfiHardSafe}{Correct}{True}{Walltime}{Median}{5.6449822834983935}%
\StoreBenchExecResult{ReducedEvalSlDfSSo}{ImcIgndfiHardSafe}{Correct}{True}{Walltime}{Min}{1.7509042989986483}%
\StoreBenchExecResult{ReducedEvalSlDfSSo}{ImcIgndfiHardSafe}{Correct}{True}{Walltime}{Max}{784.9636439690075}%
\StoreBenchExecResult{ReducedEvalSlDfSSo}{ImcIgndfiHardSafe}{Correct}{True}{Walltime}{Stdev}{127.9864998532388843864722819}%

\StoreBenchExecResult{ReducedEvalSlDfSSo}{ImcIgndfiHardSafe}{Error}{}{Count}{}{332}%
\StoreBenchExecResult{ReducedEvalSlDfSSo}{ImcIgndfiHardSafe}{Error}{}{Cputime}{}{271157.051919438}%
\StoreBenchExecResult{ReducedEvalSlDfSSo}{ImcIgndfiHardSafe}{Error}{}{Cputime}{Avg}{816.7381081910783132530120482}%
\StoreBenchExecResult{ReducedEvalSlDfSSo}{ImcIgndfiHardSafe}{Error}{}{Cputime}{Median}{901.852429783}%
\StoreBenchExecResult{ReducedEvalSlDfSSo}{ImcIgndfiHardSafe}{Error}{}{Cputime}{Min}{12.043165374}%
\StoreBenchExecResult{ReducedEvalSlDfSSo}{ImcIgndfiHardSafe}{Error}{}{Cputime}{Max}{960.582682172}%
\StoreBenchExecResult{ReducedEvalSlDfSSo}{ImcIgndfiHardSafe}{Error}{}{Cputime}{Stdev}{242.0634864786951756308022863}%
\StoreBenchExecResult{ReducedEvalSlDfSSo}{ImcIgndfiHardSafe}{Error}{}{Walltime}{}{241445.197487549711091}%
\StoreBenchExecResult{ReducedEvalSlDfSSo}{ImcIgndfiHardSafe}{Error}{}{Walltime}{Avg}{727.2445707456316599126506024}%
\StoreBenchExecResult{ReducedEvalSlDfSSo}{ImcIgndfiHardSafe}{Error}{}{Walltime}{Median}{803.5292788089864}%
\StoreBenchExecResult{ReducedEvalSlDfSSo}{ImcIgndfiHardSafe}{Error}{}{Walltime}{Min}{6.670381388015812}%
\StoreBenchExecResult{ReducedEvalSlDfSSo}{ImcIgndfiHardSafe}{Error}{}{Walltime}{Max}{906.3644544679846}%
\StoreBenchExecResult{ReducedEvalSlDfSSo}{ImcIgndfiHardSafe}{Error}{}{Walltime}{Stdev}{223.0865303580149251153741735}%
\StoreBenchExecResult{ReducedEvalSlDfSSo}{ImcIgndfiHardSafe}{Error}{Error}{Count}{}{38}%
\StoreBenchExecResult{ReducedEvalSlDfSSo}{ImcIgndfiHardSafe}{Error}{Error}{Cputime}{}{5577.734960056}%
\StoreBenchExecResult{ReducedEvalSlDfSSo}{ImcIgndfiHardSafe}{Error}{Error}{Cputime}{Avg}{146.7824989488421052631578947}%
\StoreBenchExecResult{ReducedEvalSlDfSSo}{ImcIgndfiHardSafe}{Error}{Error}{Cputime}{Median}{153.700114121}%
\StoreBenchExecResult{ReducedEvalSlDfSSo}{ImcIgndfiHardSafe}{Error}{Error}{Cputime}{Min}{12.043165374}%
\StoreBenchExecResult{ReducedEvalSlDfSSo}{ImcIgndfiHardSafe}{Error}{Error}{Cputime}{Max}{400.594954305}%
\StoreBenchExecResult{ReducedEvalSlDfSSo}{ImcIgndfiHardSafe}{Error}{Error}{Cputime}{Stdev}{70.10015482923224575198181365}%
\StoreBenchExecResult{ReducedEvalSlDfSSo}{ImcIgndfiHardSafe}{Error}{Error}{Walltime}{}{4189.756711122929091}%
\StoreBenchExecResult{ReducedEvalSlDfSSo}{ImcIgndfiHardSafe}{Error}{Error}{Walltime}{Avg}{110.2567555558665550263157895}%
\StoreBenchExecResult{ReducedEvalSlDfSSo}{ImcIgndfiHardSafe}{Error}{Error}{Walltime}{Median}{124.654941538989075}%
\StoreBenchExecResult{ReducedEvalSlDfSSo}{ImcIgndfiHardSafe}{Error}{Error}{Walltime}{Min}{6.670381388015812}%
\StoreBenchExecResult{ReducedEvalSlDfSSo}{ImcIgndfiHardSafe}{Error}{Error}{Walltime}{Max}{311.80091259797337}%
\StoreBenchExecResult{ReducedEvalSlDfSSo}{ImcIgndfiHardSafe}{Error}{Error}{Walltime}{Stdev}{59.53104763060035938337527091}%
\StoreBenchExecResult{ReducedEvalSlDfSSo}{ImcIgndfiHardSafe}{Error}{Timeout}{Count}{}{294}%
\StoreBenchExecResult{ReducedEvalSlDfSSo}{ImcIgndfiHardSafe}{Error}{Timeout}{Cputime}{}{265579.316959382}%
\StoreBenchExecResult{ReducedEvalSlDfSSo}{ImcIgndfiHardSafe}{Error}{Timeout}{Cputime}{Avg}{903.3310100659251700680272109}%
\StoreBenchExecResult{ReducedEvalSlDfSSo}{ImcIgndfiHardSafe}{Error}{Timeout}{Cputime}{Median}{901.9490228645}%
\StoreBenchExecResult{ReducedEvalSlDfSSo}{ImcIgndfiHardSafe}{Error}{Timeout}{Cputime}{Min}{901.166997803}%
\StoreBenchExecResult{ReducedEvalSlDfSSo}{ImcIgndfiHardSafe}{Error}{Timeout}{Cputime}{Max}{960.582682172}%
\StoreBenchExecResult{ReducedEvalSlDfSSo}{ImcIgndfiHardSafe}{Error}{Timeout}{Cputime}{Stdev}{4.617601071900769994411246870}%
\StoreBenchExecResult{ReducedEvalSlDfSSo}{ImcIgndfiHardSafe}{Error}{Timeout}{Walltime}{}{237255.4407764267820}%
\StoreBenchExecResult{ReducedEvalSlDfSSo}{ImcIgndfiHardSafe}{Error}{Timeout}{Walltime}{Avg}{806.9912951579142244897959184}%
\StoreBenchExecResult{ReducedEvalSlDfSSo}{ImcIgndfiHardSafe}{Error}{Timeout}{Walltime}{Median}{804.20998254149165}%
\StoreBenchExecResult{ReducedEvalSlDfSSo}{ImcIgndfiHardSafe}{Error}{Timeout}{Walltime}{Min}{793.3418347109982}%
\StoreBenchExecResult{ReducedEvalSlDfSSo}{ImcIgndfiHardSafe}{Error}{Timeout}{Walltime}{Max}{906.3644544679846}%
\StoreBenchExecResult{ReducedEvalSlDfSSo}{ImcIgndfiHardSafe}{Error}{Timeout}{Walltime}{Stdev}{13.40852904865733483371278200}%
\ifdefined\ReducedEvalSlDfSSoImcIgndfiHardSafeTotalCount\else\edef\ReducedEvalSlDfSSoImcIgndfiHardSafeTotalCount{0}\fi
\ifdefined\ReducedEvalSlDfSSoImcIgndfiHardSafeCorrectCount\else\edef\ReducedEvalSlDfSSoImcIgndfiHardSafeCorrectCount{0}\fi
\ifdefined\ReducedEvalSlDfSSoImcIgndfiHardSafeCorrectTrueCount\else\edef\ReducedEvalSlDfSSoImcIgndfiHardSafeCorrectTrueCount{0}\fi
\ifdefined\ReducedEvalSlDfSSoImcIgndfiHardSafeCorrectFalseCount\else\edef\ReducedEvalSlDfSSoImcIgndfiHardSafeCorrectFalseCount{0}\fi
\ifdefined\ReducedEvalSlDfSSoImcIgndfiHardSafeWrongTrueCount\else\edef\ReducedEvalSlDfSSoImcIgndfiHardSafeWrongTrueCount{0}\fi
\ifdefined\ReducedEvalSlDfSSoImcIgndfiHardSafeWrongFalseCount\else\edef\ReducedEvalSlDfSSoImcIgndfiHardSafeWrongFalseCount{0}\fi
\ifdefined\ReducedEvalSlDfSSoImcIgndfiHardSafeErrorTimeoutCount\else\edef\ReducedEvalSlDfSSoImcIgndfiHardSafeErrorTimeoutCount{0}\fi
\ifdefined\ReducedEvalSlDfSSoImcIgndfiHardSafeErrorOutOfMemoryCount\else\edef\ReducedEvalSlDfSSoImcIgndfiHardSafeErrorOutOfMemoryCount{0}\fi
\ifdefined\ReducedEvalSlDfSSoImcIgndfiHardSafeCorrectCputime\else\edef\ReducedEvalSlDfSSoImcIgndfiHardSafeCorrectCputime{0}\fi
\ifdefined\ReducedEvalSlDfSSoImcIgndfiHardSafeCorrectCputimeAvg\else\edef\ReducedEvalSlDfSSoImcIgndfiHardSafeCorrectCputimeAvg{None}\fi
\ifdefined\ReducedEvalSlDfSSoImcIgndfiHardSafeCorrectWalltime\else\edef\ReducedEvalSlDfSSoImcIgndfiHardSafeCorrectWalltime{0}\fi
\ifdefined\ReducedEvalSlDfSSoImcIgndfiHardSafeCorrectWalltimeAvg\else\edef\ReducedEvalSlDfSSoImcIgndfiHardSafeCorrectWalltimeAvg{None}\fi
\edef\ReducedEvalSlDfSSoImcIgndfiHardSafeErrorOtherInconclusiveCount{\the\numexpr \ReducedEvalSlDfSSoImcIgndfiHardSafeTotalCount - \ReducedEvalSlDfSSoImcIgndfiHardSafeCorrectCount - \ReducedEvalSlDfSSoImcIgndfiHardSafeWrongTrueCount - \ReducedEvalSlDfSSoImcIgndfiHardSafeWrongFalseCount - \ReducedEvalSlDfSSoImcIgndfiHardSafeErrorTimeoutCount - \ReducedEvalSlDfSSoImcIgndfiHardSafeErrorOutOfMemoryCount \relax}
\providecommand\StoreBenchExecResult[7]{\expandafter\newcommand\csname#1#2#3#4#5#6\endcsname{#7}}%
\StoreBenchExecResult{ReducedEvalSlDfSEn}{ImcHardSafe}{Total}{}{Count}{}{870}%
\StoreBenchExecResult{ReducedEvalSlDfSEn}{ImcHardSafe}{Total}{}{Cputime}{}{305086.682947756}%
\StoreBenchExecResult{ReducedEvalSlDfSEn}{ImcHardSafe}{Total}{}{Cputime}{Avg}{350.6743482158114942528735632}%
\StoreBenchExecResult{ReducedEvalSlDfSEn}{ImcHardSafe}{Total}{}{Cputime}{Median}{27.415579142}%
\StoreBenchExecResult{ReducedEvalSlDfSEn}{ImcHardSafe}{Total}{}{Cputime}{Min}{4.263151367}%
\StoreBenchExecResult{ReducedEvalSlDfSEn}{ImcHardSafe}{Total}{}{Cputime}{Max}{913.309152079}%
\StoreBenchExecResult{ReducedEvalSlDfSEn}{ImcHardSafe}{Total}{}{Cputime}{Stdev}{417.4433378790864560319291519}%
\StoreBenchExecResult{ReducedEvalSlDfSEn}{ImcHardSafe}{Total}{}{Walltime}{}{295915.3237369182466779}%
\StoreBenchExecResult{ReducedEvalSlDfSEn}{ImcHardSafe}{Total}{}{Walltime}{Avg}{340.1325560194462605493103448}%
\StoreBenchExecResult{ReducedEvalSlDfSEn}{ImcHardSafe}{Total}{}{Walltime}{Median}{14.9533766750100765}%
\StoreBenchExecResult{ReducedEvalSlDfSEn}{ImcHardSafe}{Total}{}{Walltime}{Min}{1.6599373899807688}%
\StoreBenchExecResult{ReducedEvalSlDfSEn}{ImcHardSafe}{Total}{}{Walltime}{Max}{905.8629780259798}%
\StoreBenchExecResult{ReducedEvalSlDfSEn}{ImcHardSafe}{Total}{}{Walltime}{Stdev}{413.3899396673756450557686617}%
\StoreBenchExecResult{ReducedEvalSlDfSEn}{ImcHardSafe}{Correct}{}{Count}{}{527}%
\StoreBenchExecResult{ReducedEvalSlDfSEn}{ImcHardSafe}{Correct}{}{Cputime}{}{28881.801138674}%
\StoreBenchExecResult{ReducedEvalSlDfSEn}{ImcHardSafe}{Correct}{}{Cputime}{Avg}{54.80417673372675521821631879}%
\StoreBenchExecResult{ReducedEvalSlDfSEn}{ImcHardSafe}{Correct}{}{Cputime}{Median}{11.147500049}%
\StoreBenchExecResult{ReducedEvalSlDfSEn}{ImcHardSafe}{Correct}{}{Cputime}{Min}{4.263151367}%
\StoreBenchExecResult{ReducedEvalSlDfSEn}{ImcHardSafe}{Correct}{}{Cputime}{Max}{890.564056695}%
\StoreBenchExecResult{ReducedEvalSlDfSEn}{ImcHardSafe}{Correct}{}{Cputime}{Stdev}{129.0267528508636155681593364}%
\StoreBenchExecResult{ReducedEvalSlDfSEn}{ImcHardSafe}{Correct}{}{Walltime}{}{24530.7590620871923636}%
\StoreBenchExecResult{ReducedEvalSlDfSEn}{ImcHardSafe}{Correct}{}{Walltime}{Avg}{46.54792990908385647741935484}%
\StoreBenchExecResult{ReducedEvalSlDfSEn}{ImcHardSafe}{Correct}{}{Walltime}{Median}{4.783207938016858}%
\StoreBenchExecResult{ReducedEvalSlDfSEn}{ImcHardSafe}{Correct}{}{Walltime}{Min}{1.6599373899807688}%
\StoreBenchExecResult{ReducedEvalSlDfSEn}{ImcHardSafe}{Correct}{}{Walltime}{Max}{863.0842669380072}%
\StoreBenchExecResult{ReducedEvalSlDfSEn}{ImcHardSafe}{Correct}{}{Walltime}{Stdev}{125.5701645557452421600522242}%
\StoreBenchExecResult{ReducedEvalSlDfSEn}{ImcHardSafe}{Correct}{True}{Count}{}{527}%
\StoreBenchExecResult{ReducedEvalSlDfSEn}{ImcHardSafe}{Correct}{True}{Cputime}{}{28881.801138674}%
\StoreBenchExecResult{ReducedEvalSlDfSEn}{ImcHardSafe}{Correct}{True}{Cputime}{Avg}{54.80417673372675521821631879}%
\StoreBenchExecResult{ReducedEvalSlDfSEn}{ImcHardSafe}{Correct}{True}{Cputime}{Median}{11.147500049}%
\StoreBenchExecResult{ReducedEvalSlDfSEn}{ImcHardSafe}{Correct}{True}{Cputime}{Min}{4.263151367}%
\StoreBenchExecResult{ReducedEvalSlDfSEn}{ImcHardSafe}{Correct}{True}{Cputime}{Max}{890.564056695}%
\StoreBenchExecResult{ReducedEvalSlDfSEn}{ImcHardSafe}{Correct}{True}{Cputime}{Stdev}{129.0267528508636155681593364}%
\StoreBenchExecResult{ReducedEvalSlDfSEn}{ImcHardSafe}{Correct}{True}{Walltime}{}{24530.7590620871923636}%
\StoreBenchExecResult{ReducedEvalSlDfSEn}{ImcHardSafe}{Correct}{True}{Walltime}{Avg}{46.54792990908385647741935484}%
\StoreBenchExecResult{ReducedEvalSlDfSEn}{ImcHardSafe}{Correct}{True}{Walltime}{Median}{4.783207938016858}%
\StoreBenchExecResult{ReducedEvalSlDfSEn}{ImcHardSafe}{Correct}{True}{Walltime}{Min}{1.6599373899807688}%
\StoreBenchExecResult{ReducedEvalSlDfSEn}{ImcHardSafe}{Correct}{True}{Walltime}{Max}{863.0842669380072}%
\StoreBenchExecResult{ReducedEvalSlDfSEn}{ImcHardSafe}{Correct}{True}{Walltime}{Stdev}{125.5701645557452421600522242}%

\StoreBenchExecResult{ReducedEvalSlDfSEn}{ImcHardSafe}{Error}{}{Count}{}{343}%
\StoreBenchExecResult{ReducedEvalSlDfSEn}{ImcHardSafe}{Error}{}{Cputime}{}{276204.881809082}%
\StoreBenchExecResult{ReducedEvalSlDfSEn}{ImcHardSafe}{Error}{}{Cputime}{Avg}{805.2620460906180758017492711}%
\StoreBenchExecResult{ReducedEvalSlDfSEn}{ImcHardSafe}{Error}{}{Cputime}{Median}{902.158502057}%
\StoreBenchExecResult{ReducedEvalSlDfSEn}{ImcHardSafe}{Error}{}{Cputime}{Min}{4.683089729}%
\StoreBenchExecResult{ReducedEvalSlDfSEn}{ImcHardSafe}{Error}{}{Cputime}{Max}{913.309152079}%
\StoreBenchExecResult{ReducedEvalSlDfSEn}{ImcHardSafe}{Error}{}{Cputime}{Stdev}{274.3545583905378982544812842}%
\StoreBenchExecResult{ReducedEvalSlDfSEn}{ImcHardSafe}{Error}{}{Walltime}{}{271384.5646748310543143}%
\StoreBenchExecResult{ReducedEvalSlDfSEn}{ImcHardSafe}{Error}{}{Walltime}{Avg}{791.2086433668543857559766764}%
\StoreBenchExecResult{ReducedEvalSlDfSEn}{ImcHardSafe}{Error}{}{Walltime}{Median}{887.760787103005}%
\StoreBenchExecResult{ReducedEvalSlDfSEn}{ImcHardSafe}{Error}{}{Walltime}{Min}{1.8850378100178204}%
\StoreBenchExecResult{ReducedEvalSlDfSEn}{ImcHardSafe}{Error}{}{Walltime}{Max}{905.8629780259798}%
\StoreBenchExecResult{ReducedEvalSlDfSEn}{ImcHardSafe}{Error}{}{Walltime}{Stdev}{270.7966939127032884761000868}%
\StoreBenchExecResult{ReducedEvalSlDfSEn}{ImcHardSafe}{Error}{Error}{Count}{}{38}%
\StoreBenchExecResult{ReducedEvalSlDfSEn}{ImcHardSafe}{Error}{Error}{Cputime}{}{1341.803556414}%
\StoreBenchExecResult{ReducedEvalSlDfSEn}{ImcHardSafe}{Error}{Error}{Cputime}{Avg}{35.31061990563157894736842105}%
\StoreBenchExecResult{ReducedEvalSlDfSEn}{ImcHardSafe}{Error}{Error}{Cputime}{Median}{10.0541020605}%
\StoreBenchExecResult{ReducedEvalSlDfSEn}{ImcHardSafe}{Error}{Error}{Cputime}{Min}{4.683089729}%
\StoreBenchExecResult{ReducedEvalSlDfSEn}{ImcHardSafe}{Error}{Error}{Cputime}{Max}{219.94205123}%
\StoreBenchExecResult{ReducedEvalSlDfSEn}{ImcHardSafe}{Error}{Error}{Cputime}{Stdev}{57.84437260366708345053927218}%
\StoreBenchExecResult{ReducedEvalSlDfSEn}{ImcHardSafe}{Error}{Error}{Walltime}{}{1197.2364536520326543}%
\StoreBenchExecResult{ReducedEvalSlDfSEn}{ImcHardSafe}{Error}{Error}{Walltime}{Avg}{31.50622246452717511315789474}%
\StoreBenchExecResult{ReducedEvalSlDfSEn}{ImcHardSafe}{Error}{Error}{Walltime}{Median}{6.106939566001529}%
\StoreBenchExecResult{ReducedEvalSlDfSEn}{ImcHardSafe}{Error}{Error}{Walltime}{Min}{1.8850378100178204}%
\StoreBenchExecResult{ReducedEvalSlDfSEn}{ImcHardSafe}{Error}{Error}{Walltime}{Max}{216.32514165499015}%
\StoreBenchExecResult{ReducedEvalSlDfSEn}{ImcHardSafe}{Error}{Error}{Walltime}{Stdev}{58.00335531239752111983743423}%
\StoreBenchExecResult{ReducedEvalSlDfSEn}{ImcHardSafe}{Error}{OutOfMemory}{Count}{}{1}%
\StoreBenchExecResult{ReducedEvalSlDfSEn}{ImcHardSafe}{Error}{OutOfMemory}{Cputime}{}{312.969245697}%
\StoreBenchExecResult{ReducedEvalSlDfSEn}{ImcHardSafe}{Error}{OutOfMemory}{Cputime}{Avg}{312.969245697}%
\StoreBenchExecResult{ReducedEvalSlDfSEn}{ImcHardSafe}{Error}{OutOfMemory}{Cputime}{Median}{312.969245697}%
\StoreBenchExecResult{ReducedEvalSlDfSEn}{ImcHardSafe}{Error}{OutOfMemory}{Cputime}{Min}{312.969245697}%
\StoreBenchExecResult{ReducedEvalSlDfSEn}{ImcHardSafe}{Error}{OutOfMemory}{Cputime}{Max}{312.969245697}%
\StoreBenchExecResult{ReducedEvalSlDfSEn}{ImcHardSafe}{Error}{OutOfMemory}{Cputime}{Stdev}{0E-14}%
\StoreBenchExecResult{ReducedEvalSlDfSEn}{ImcHardSafe}{Error}{OutOfMemory}{Walltime}{}{305.21781224399456}%
\StoreBenchExecResult{ReducedEvalSlDfSEn}{ImcHardSafe}{Error}{OutOfMemory}{Walltime}{Avg}{305.21781224399456}%
\StoreBenchExecResult{ReducedEvalSlDfSEn}{ImcHardSafe}{Error}{OutOfMemory}{Walltime}{Median}{305.21781224399456}%
\StoreBenchExecResult{ReducedEvalSlDfSEn}{ImcHardSafe}{Error}{OutOfMemory}{Walltime}{Min}{305.21781224399456}%
\StoreBenchExecResult{ReducedEvalSlDfSEn}{ImcHardSafe}{Error}{OutOfMemory}{Walltime}{Max}{305.21781224399456}%
\StoreBenchExecResult{ReducedEvalSlDfSEn}{ImcHardSafe}{Error}{OutOfMemory}{Walltime}{Stdev}{0E-14}%
\StoreBenchExecResult{ReducedEvalSlDfSEn}{ImcHardSafe}{Error}{SegmentationFault}{Count}{}{1}%
\StoreBenchExecResult{ReducedEvalSlDfSEn}{ImcHardSafe}{Error}{SegmentationFault}{Cputime}{}{821.59432435}%
\StoreBenchExecResult{ReducedEvalSlDfSEn}{ImcHardSafe}{Error}{SegmentationFault}{Cputime}{Avg}{821.59432435}%
\StoreBenchExecResult{ReducedEvalSlDfSEn}{ImcHardSafe}{Error}{SegmentationFault}{Cputime}{Median}{821.59432435}%
\StoreBenchExecResult{ReducedEvalSlDfSEn}{ImcHardSafe}{Error}{SegmentationFault}{Cputime}{Min}{821.59432435}%
\StoreBenchExecResult{ReducedEvalSlDfSEn}{ImcHardSafe}{Error}{SegmentationFault}{Cputime}{Max}{821.59432435}%
\StoreBenchExecResult{ReducedEvalSlDfSEn}{ImcHardSafe}{Error}{SegmentationFault}{Cputime}{Stdev}{0E-14}%
\StoreBenchExecResult{ReducedEvalSlDfSEn}{ImcHardSafe}{Error}{SegmentationFault}{Walltime}{}{814.4603031660081}%
\StoreBenchExecResult{ReducedEvalSlDfSEn}{ImcHardSafe}{Error}{SegmentationFault}{Walltime}{Avg}{814.4603031660081}%
\StoreBenchExecResult{ReducedEvalSlDfSEn}{ImcHardSafe}{Error}{SegmentationFault}{Walltime}{Median}{814.4603031660081}%
\StoreBenchExecResult{ReducedEvalSlDfSEn}{ImcHardSafe}{Error}{SegmentationFault}{Walltime}{Min}{814.4603031660081}%
\StoreBenchExecResult{ReducedEvalSlDfSEn}{ImcHardSafe}{Error}{SegmentationFault}{Walltime}{Max}{814.4603031660081}%
\StoreBenchExecResult{ReducedEvalSlDfSEn}{ImcHardSafe}{Error}{SegmentationFault}{Walltime}{Stdev}{0E-14}%
\StoreBenchExecResult{ReducedEvalSlDfSEn}{ImcHardSafe}{Error}{Timeout}{Count}{}{303}%
\StoreBenchExecResult{ReducedEvalSlDfSEn}{ImcHardSafe}{Error}{Timeout}{Cputime}{}{273728.514682621}%
\StoreBenchExecResult{ReducedEvalSlDfSEn}{ImcHardSafe}{Error}{Timeout}{Cputime}{Avg}{903.3944378964389438943894389}%
\StoreBenchExecResult{ReducedEvalSlDfSEn}{ImcHardSafe}{Error}{Timeout}{Cputime}{Median}{902.289314001}%
\StoreBenchExecResult{ReducedEvalSlDfSEn}{ImcHardSafe}{Error}{Timeout}{Cputime}{Min}{901.341123723}%
\StoreBenchExecResult{ReducedEvalSlDfSEn}{ImcHardSafe}{Error}{Timeout}{Cputime}{Max}{913.309152079}%
\StoreBenchExecResult{ReducedEvalSlDfSEn}{ImcHardSafe}{Error}{Timeout}{Cputime}{Stdev}{3.013789848750642943355695897}%
\StoreBenchExecResult{ReducedEvalSlDfSEn}{ImcHardSafe}{Error}{Timeout}{Walltime}{}{269067.6501057690190}%
\StoreBenchExecResult{ReducedEvalSlDfSEn}{ImcHardSafe}{Error}{Timeout}{Walltime}{Avg}{888.0120465536931320132013201}%
\StoreBenchExecResult{ReducedEvalSlDfSEn}{ImcHardSafe}{Error}{Timeout}{Walltime}{Median}{888.4079600760015}%
\StoreBenchExecResult{ReducedEvalSlDfSEn}{ImcHardSafe}{Error}{Timeout}{Walltime}{Min}{870.7320083079976}%
\StoreBenchExecResult{ReducedEvalSlDfSEn}{ImcHardSafe}{Error}{Timeout}{Walltime}{Max}{905.8629780259798}%
\StoreBenchExecResult{ReducedEvalSlDfSEn}{ImcHardSafe}{Error}{Timeout}{Walltime}{Stdev}{7.470355958661885613721876739}%
\ifdefined\ReducedEvalSlDfSEnImcHardSafeTotalCount\else\edef\ReducedEvalSlDfSEnImcHardSafeTotalCount{0}\fi
\ifdefined\ReducedEvalSlDfSEnImcHardSafeCorrectCount\else\edef\ReducedEvalSlDfSEnImcHardSafeCorrectCount{0}\fi
\ifdefined\ReducedEvalSlDfSEnImcHardSafeCorrectTrueCount\else\edef\ReducedEvalSlDfSEnImcHardSafeCorrectTrueCount{0}\fi
\ifdefined\ReducedEvalSlDfSEnImcHardSafeCorrectFalseCount\else\edef\ReducedEvalSlDfSEnImcHardSafeCorrectFalseCount{0}\fi
\ifdefined\ReducedEvalSlDfSEnImcHardSafeWrongTrueCount\else\edef\ReducedEvalSlDfSEnImcHardSafeWrongTrueCount{0}\fi
\ifdefined\ReducedEvalSlDfSEnImcHardSafeWrongFalseCount\else\edef\ReducedEvalSlDfSEnImcHardSafeWrongFalseCount{0}\fi
\ifdefined\ReducedEvalSlDfSEnImcHardSafeErrorTimeoutCount\else\edef\ReducedEvalSlDfSEnImcHardSafeErrorTimeoutCount{0}\fi
\ifdefined\ReducedEvalSlDfSEnImcHardSafeErrorOutOfMemoryCount\else\edef\ReducedEvalSlDfSEnImcHardSafeErrorOutOfMemoryCount{0}\fi
\ifdefined\ReducedEvalSlDfSEnImcHardSafeCorrectCputime\else\edef\ReducedEvalSlDfSEnImcHardSafeCorrectCputime{0}\fi
\ifdefined\ReducedEvalSlDfSEnImcHardSafeCorrectCputimeAvg\else\edef\ReducedEvalSlDfSEnImcHardSafeCorrectCputimeAvg{None}\fi
\ifdefined\ReducedEvalSlDfSEnImcHardSafeCorrectWalltime\else\edef\ReducedEvalSlDfSEnImcHardSafeCorrectWalltime{0}\fi
\ifdefined\ReducedEvalSlDfSEnImcHardSafeCorrectWalltimeAvg\else\edef\ReducedEvalSlDfSEnImcHardSafeCorrectWalltimeAvg{None}\fi
\edef\ReducedEvalSlDfSEnImcHardSafeErrorOtherInconclusiveCount{\the\numexpr \ReducedEvalSlDfSEnImcHardSafeTotalCount - \ReducedEvalSlDfSEnImcHardSafeCorrectCount - \ReducedEvalSlDfSEnImcHardSafeWrongTrueCount - \ReducedEvalSlDfSEnImcHardSafeWrongFalseCount - \ReducedEvalSlDfSEnImcHardSafeErrorTimeoutCount - \ReducedEvalSlDfSEnImcHardSafeErrorOutOfMemoryCount \relax}
\providecommand\StoreBenchExecResult[7]{\expandafter\newcommand\csname#1#2#3#4#5#6\endcsname{#7}}%
\StoreBenchExecResult{ReducedEvalSlDfSEn}{ImcIgndfiHardSafe}{Total}{}{Count}{}{870}%
\StoreBenchExecResult{ReducedEvalSlDfSEn}{ImcIgndfiHardSafe}{Total}{}{Cputime}{}{314108.966067504}%
\StoreBenchExecResult{ReducedEvalSlDfSEn}{ImcIgndfiHardSafe}{Total}{}{Cputime}{Avg}{361.0447885833379310344827586}%
\StoreBenchExecResult{ReducedEvalSlDfSEn}{ImcIgndfiHardSafe}{Total}{}{Cputime}{Median}{85.175027013}%
\StoreBenchExecResult{ReducedEvalSlDfSEn}{ImcIgndfiHardSafe}{Total}{}{Cputime}{Min}{4.628164513}%
\StoreBenchExecResult{ReducedEvalSlDfSEn}{ImcIgndfiHardSafe}{Total}{}{Cputime}{Max}{913.168474268}%
\StoreBenchExecResult{ReducedEvalSlDfSEn}{ImcIgndfiHardSafe}{Total}{}{Cputime}{Stdev}{406.1017143526813864378334963}%
\StoreBenchExecResult{ReducedEvalSlDfSEn}{ImcIgndfiHardSafe}{Total}{}{Walltime}{}{268829.0228041830119774}%
\StoreBenchExecResult{ReducedEvalSlDfSEn}{ImcIgndfiHardSafe}{Total}{}{Walltime}{Avg}{308.9988767864172551464367816}%
\StoreBenchExecResult{ReducedEvalSlDfSEn}{ImcIgndfiHardSafe}{Total}{}{Walltime}{Median}{29.9961842964985405}%
\StoreBenchExecResult{ReducedEvalSlDfSEn}{ImcIgndfiHardSafe}{Total}{}{Walltime}{Min}{1.750724600016838}%
\StoreBenchExecResult{ReducedEvalSlDfSEn}{ImcIgndfiHardSafe}{Total}{}{Walltime}{Max}{900.3924184349889}%
\StoreBenchExecResult{ReducedEvalSlDfSEn}{ImcIgndfiHardSafe}{Total}{}{Walltime}{Stdev}{369.6006022395599640058329352}%
\StoreBenchExecResult{ReducedEvalSlDfSEn}{ImcIgndfiHardSafe}{Correct}{}{Count}{}{543}%
\StoreBenchExecResult{ReducedEvalSlDfSEn}{ImcIgndfiHardSafe}{Correct}{}{Cputime}{}{47434.903179116}%
\StoreBenchExecResult{ReducedEvalSlDfSEn}{ImcIgndfiHardSafe}{Correct}{}{Cputime}{Avg}{87.35709609413627992633517495}%
\StoreBenchExecResult{ReducedEvalSlDfSEn}{ImcIgndfiHardSafe}{Correct}{}{Cputime}{Median}{16.186721544}%
\StoreBenchExecResult{ReducedEvalSlDfSEn}{ImcIgndfiHardSafe}{Correct}{}{Cputime}{Min}{4.628164513}%
\StoreBenchExecResult{ReducedEvalSlDfSEn}{ImcIgndfiHardSafe}{Correct}{}{Cputime}{Max}{892.288265732}%
\StoreBenchExecResult{ReducedEvalSlDfSEn}{ImcIgndfiHardSafe}{Correct}{}{Cputime}{Stdev}{171.5421310179800711925327235}%
\StoreBenchExecResult{ReducedEvalSlDfSEn}{ImcIgndfiHardSafe}{Correct}{}{Walltime}{}{31481.7229264999332354}%
\StoreBenchExecResult{ReducedEvalSlDfSEn}{ImcIgndfiHardSafe}{Correct}{}{Walltime}{Avg}{57.97739028821350503756906077}%
\StoreBenchExecResult{ReducedEvalSlDfSEn}{ImcIgndfiHardSafe}{Correct}{}{Walltime}{Median}{5.61692165301065}%
\StoreBenchExecResult{ReducedEvalSlDfSEn}{ImcIgndfiHardSafe}{Correct}{}{Walltime}{Min}{1.750724600016838}%
\StoreBenchExecResult{ReducedEvalSlDfSEn}{ImcIgndfiHardSafe}{Correct}{}{Walltime}{Max}{790.1960601290048}%
\StoreBenchExecResult{ReducedEvalSlDfSEn}{ImcIgndfiHardSafe}{Correct}{}{Walltime}{Stdev}{144.7892950029202447691943423}%
\StoreBenchExecResult{ReducedEvalSlDfSEn}{ImcIgndfiHardSafe}{Correct}{True}{Count}{}{543}%
\StoreBenchExecResult{ReducedEvalSlDfSEn}{ImcIgndfiHardSafe}{Correct}{True}{Cputime}{}{47434.903179116}%
\StoreBenchExecResult{ReducedEvalSlDfSEn}{ImcIgndfiHardSafe}{Correct}{True}{Cputime}{Avg}{87.35709609413627992633517495}%
\StoreBenchExecResult{ReducedEvalSlDfSEn}{ImcIgndfiHardSafe}{Correct}{True}{Cputime}{Median}{16.186721544}%
\StoreBenchExecResult{ReducedEvalSlDfSEn}{ImcIgndfiHardSafe}{Correct}{True}{Cputime}{Min}{4.628164513}%
\StoreBenchExecResult{ReducedEvalSlDfSEn}{ImcIgndfiHardSafe}{Correct}{True}{Cputime}{Max}{892.288265732}%
\StoreBenchExecResult{ReducedEvalSlDfSEn}{ImcIgndfiHardSafe}{Correct}{True}{Cputime}{Stdev}{171.5421310179800711925327235}%
\StoreBenchExecResult{ReducedEvalSlDfSEn}{ImcIgndfiHardSafe}{Correct}{True}{Walltime}{}{31481.7229264999332354}%
\StoreBenchExecResult{ReducedEvalSlDfSEn}{ImcIgndfiHardSafe}{Correct}{True}{Walltime}{Avg}{57.97739028821350503756906077}%
\StoreBenchExecResult{ReducedEvalSlDfSEn}{ImcIgndfiHardSafe}{Correct}{True}{Walltime}{Median}{5.61692165301065}%
\StoreBenchExecResult{ReducedEvalSlDfSEn}{ImcIgndfiHardSafe}{Correct}{True}{Walltime}{Min}{1.750724600016838}%
\StoreBenchExecResult{ReducedEvalSlDfSEn}{ImcIgndfiHardSafe}{Correct}{True}{Walltime}{Max}{790.1960601290048}%
\StoreBenchExecResult{ReducedEvalSlDfSEn}{ImcIgndfiHardSafe}{Correct}{True}{Walltime}{Stdev}{144.7892950029202447691943423}%

\StoreBenchExecResult{ReducedEvalSlDfSEn}{ImcIgndfiHardSafe}{Error}{}{Count}{}{327}%
\StoreBenchExecResult{ReducedEvalSlDfSEn}{ImcIgndfiHardSafe}{Error}{}{Cputime}{}{266674.062888388}%
\StoreBenchExecResult{ReducedEvalSlDfSEn}{ImcIgndfiHardSafe}{Error}{}{Cputime}{Avg}{815.5170118910948012232415902}%
\StoreBenchExecResult{ReducedEvalSlDfSEn}{ImcIgndfiHardSafe}{Error}{}{Cputime}{Median}{901.895586757}%
\StoreBenchExecResult{ReducedEvalSlDfSEn}{ImcIgndfiHardSafe}{Error}{}{Cputime}{Min}{11.649928607}%
\StoreBenchExecResult{ReducedEvalSlDfSEn}{ImcIgndfiHardSafe}{Error}{}{Cputime}{Max}{913.168474268}%
\StoreBenchExecResult{ReducedEvalSlDfSEn}{ImcIgndfiHardSafe}{Error}{}{Cputime}{Stdev}{242.8606240628231470157362970}%
\StoreBenchExecResult{ReducedEvalSlDfSEn}{ImcIgndfiHardSafe}{Error}{}{Walltime}{}{237347.299877683078742}%
\StoreBenchExecResult{ReducedEvalSlDfSEn}{ImcIgndfiHardSafe}{Error}{}{Walltime}{Avg}{725.8327213384803631253822630}%
\StoreBenchExecResult{ReducedEvalSlDfSEn}{ImcIgndfiHardSafe}{Error}{}{Walltime}{Median}{803.6048432399984}%
\StoreBenchExecResult{ReducedEvalSlDfSEn}{ImcIgndfiHardSafe}{Error}{}{Walltime}{Min}{5.812912791006966}%
\StoreBenchExecResult{ReducedEvalSlDfSEn}{ImcIgndfiHardSafe}{Error}{}{Walltime}{Max}{900.3924184349889}%
\StoreBenchExecResult{ReducedEvalSlDfSEn}{ImcIgndfiHardSafe}{Error}{}{Walltime}{Stdev}{224.1585591066572176194846897}%
\StoreBenchExecResult{ReducedEvalSlDfSEn}{ImcIgndfiHardSafe}{Error}{Error}{Count}{}{38}%
\StoreBenchExecResult{ReducedEvalSlDfSEn}{ImcIgndfiHardSafe}{Error}{Error}{Cputime}{}{5641.429582570}%
\StoreBenchExecResult{ReducedEvalSlDfSEn}{ImcIgndfiHardSafe}{Error}{Error}{Cputime}{Avg}{148.4586732255263157894736842}%
\StoreBenchExecResult{ReducedEvalSlDfSEn}{ImcIgndfiHardSafe}{Error}{Error}{Cputime}{Median}{153.642233866}%
\StoreBenchExecResult{ReducedEvalSlDfSEn}{ImcIgndfiHardSafe}{Error}{Error}{Cputime}{Min}{11.649928607}%
\StoreBenchExecResult{ReducedEvalSlDfSEn}{ImcIgndfiHardSafe}{Error}{Error}{Cputime}{Max}{332.056478855}%
\StoreBenchExecResult{ReducedEvalSlDfSEn}{ImcIgndfiHardSafe}{Error}{Error}{Cputime}{Stdev}{63.22244329513165229132759963}%
\StoreBenchExecResult{ReducedEvalSlDfSEn}{ImcIgndfiHardSafe}{Error}{Error}{Walltime}{}{4194.203521342133142}%
\StoreBenchExecResult{ReducedEvalSlDfSEn}{ImcIgndfiHardSafe}{Error}{Error}{Walltime}{Avg}{110.3737768774245563684210526}%
\StoreBenchExecResult{ReducedEvalSlDfSEn}{ImcIgndfiHardSafe}{Error}{Error}{Walltime}{Median}{125.459962096501835}%
\StoreBenchExecResult{ReducedEvalSlDfSEn}{ImcIgndfiHardSafe}{Error}{Error}{Walltime}{Min}{5.812912791006966}%
\StoreBenchExecResult{ReducedEvalSlDfSEn}{ImcIgndfiHardSafe}{Error}{Error}{Walltime}{Max}{243.47325310698943}%
\StoreBenchExecResult{ReducedEvalSlDfSEn}{ImcIgndfiHardSafe}{Error}{Error}{Walltime}{Stdev}{50.54950050443627221481018657}%
\StoreBenchExecResult{ReducedEvalSlDfSEn}{ImcIgndfiHardSafe}{Error}{Timeout}{Count}{}{289}%
\StoreBenchExecResult{ReducedEvalSlDfSEn}{ImcIgndfiHardSafe}{Error}{Timeout}{Cputime}{}{261032.633305818}%
\StoreBenchExecResult{ReducedEvalSlDfSEn}{ImcIgndfiHardSafe}{Error}{Timeout}{Cputime}{Avg}{903.2271048644221453287197232}%
\StoreBenchExecResult{ReducedEvalSlDfSEn}{ImcIgndfiHardSafe}{Error}{Timeout}{Cputime}{Median}{901.98877378}%
\StoreBenchExecResult{ReducedEvalSlDfSEn}{ImcIgndfiHardSafe}{Error}{Timeout}{Cputime}{Min}{901.187588489}%
\StoreBenchExecResult{ReducedEvalSlDfSEn}{ImcIgndfiHardSafe}{Error}{Timeout}{Cputime}{Max}{913.168474268}%
\StoreBenchExecResult{ReducedEvalSlDfSEn}{ImcIgndfiHardSafe}{Error}{Timeout}{Cputime}{Stdev}{3.198517956156121619777765131}%
\StoreBenchExecResult{ReducedEvalSlDfSEn}{ImcIgndfiHardSafe}{Error}{Timeout}{Walltime}{}{233153.0963563409456}%
\StoreBenchExecResult{ReducedEvalSlDfSEn}{ImcIgndfiHardSafe}{Error}{Timeout}{Walltime}{Avg}{806.7581188800724761245674740}%
\StoreBenchExecResult{ReducedEvalSlDfSEn}{ImcIgndfiHardSafe}{Error}{Timeout}{Walltime}{Median}{804.2591822970135}%
\StoreBenchExecResult{ReducedEvalSlDfSEn}{ImcIgndfiHardSafe}{Error}{Timeout}{Walltime}{Min}{791.9621774319967}%
\StoreBenchExecResult{ReducedEvalSlDfSEn}{ImcIgndfiHardSafe}{Error}{Timeout}{Walltime}{Max}{900.3924184349889}%
\StoreBenchExecResult{ReducedEvalSlDfSEn}{ImcIgndfiHardSafe}{Error}{Timeout}{Walltime}{Stdev}{12.75833147361737321053673230}%
\ifdefined\ReducedEvalSlDfSEnImcIgndfiHardSafeTotalCount\else\edef\ReducedEvalSlDfSEnImcIgndfiHardSafeTotalCount{0}\fi
\ifdefined\ReducedEvalSlDfSEnImcIgndfiHardSafeCorrectCount\else\edef\ReducedEvalSlDfSEnImcIgndfiHardSafeCorrectCount{0}\fi
\ifdefined\ReducedEvalSlDfSEnImcIgndfiHardSafeCorrectTrueCount\else\edef\ReducedEvalSlDfSEnImcIgndfiHardSafeCorrectTrueCount{0}\fi
\ifdefined\ReducedEvalSlDfSEnImcIgndfiHardSafeCorrectFalseCount\else\edef\ReducedEvalSlDfSEnImcIgndfiHardSafeCorrectFalseCount{0}\fi
\ifdefined\ReducedEvalSlDfSEnImcIgndfiHardSafeWrongTrueCount\else\edef\ReducedEvalSlDfSEnImcIgndfiHardSafeWrongTrueCount{0}\fi
\ifdefined\ReducedEvalSlDfSEnImcIgndfiHardSafeWrongFalseCount\else\edef\ReducedEvalSlDfSEnImcIgndfiHardSafeWrongFalseCount{0}\fi
\ifdefined\ReducedEvalSlDfSEnImcIgndfiHardSafeErrorTimeoutCount\else\edef\ReducedEvalSlDfSEnImcIgndfiHardSafeErrorTimeoutCount{0}\fi
\ifdefined\ReducedEvalSlDfSEnImcIgndfiHardSafeErrorOutOfMemoryCount\else\edef\ReducedEvalSlDfSEnImcIgndfiHardSafeErrorOutOfMemoryCount{0}\fi
\ifdefined\ReducedEvalSlDfSEnImcIgndfiHardSafeCorrectCputime\else\edef\ReducedEvalSlDfSEnImcIgndfiHardSafeCorrectCputime{0}\fi
\ifdefined\ReducedEvalSlDfSEnImcIgndfiHardSafeCorrectCputimeAvg\else\edef\ReducedEvalSlDfSEnImcIgndfiHardSafeCorrectCputimeAvg{None}\fi
\ifdefined\ReducedEvalSlDfSEnImcIgndfiHardSafeCorrectWalltime\else\edef\ReducedEvalSlDfSEnImcIgndfiHardSafeCorrectWalltime{0}\fi
\ifdefined\ReducedEvalSlDfSEnImcIgndfiHardSafeCorrectWalltimeAvg\else\edef\ReducedEvalSlDfSEnImcIgndfiHardSafeCorrectWalltimeAvg{None}\fi
\edef\ReducedEvalSlDfSEnImcIgndfiHardSafeErrorOtherInconclusiveCount{\the\numexpr \ReducedEvalSlDfSEnImcIgndfiHardSafeTotalCount - \ReducedEvalSlDfSEnImcIgndfiHardSafeCorrectCount - \ReducedEvalSlDfSEnImcIgndfiHardSafeWrongTrueCount - \ReducedEvalSlDfSEnImcIgndfiHardSafeWrongFalseCount - \ReducedEvalSlDfSEnImcIgndfiHardSafeErrorTimeoutCount - \ReducedEvalSlDfSEnImcIgndfiHardSafeErrorOutOfMemoryCount \relax}
\providecommand\StoreBenchExecResult[7]{\expandafter\newcommand\csname#1#2#3#4#5#6\endcsname{#7}}%
\StoreBenchExecResult{ReducedEvalSlDfSOsf}{ImcHardSafe}{Total}{}{Count}{}{870}%
\StoreBenchExecResult{ReducedEvalSlDfSOsf}{ImcHardSafe}{Total}{}{Cputime}{}{305532.394167176}%
\StoreBenchExecResult{ReducedEvalSlDfSOsf}{ImcHardSafe}{Total}{}{Cputime}{Avg}{351.1866599622712643678160920}%
\StoreBenchExecResult{ReducedEvalSlDfSOsf}{ImcHardSafe}{Total}{}{Cputime}{Median}{28.751897005}%
\StoreBenchExecResult{ReducedEvalSlDfSOsf}{ImcHardSafe}{Total}{}{Cputime}{Min}{4.24575107}%
\StoreBenchExecResult{ReducedEvalSlDfSOsf}{ImcHardSafe}{Total}{}{Cputime}{Max}{913.428313018}%
\StoreBenchExecResult{ReducedEvalSlDfSOsf}{ImcHardSafe}{Total}{}{Cputime}{Stdev}{418.9128791943050501696813042}%
\StoreBenchExecResult{ReducedEvalSlDfSOsf}{ImcHardSafe}{Total}{}{Walltime}{}{296376.6881494697017713}%
\StoreBenchExecResult{ReducedEvalSlDfSOsf}{ImcHardSafe}{Total}{}{Walltime}{Avg}{340.6628599419191974382758621}%
\StoreBenchExecResult{ReducedEvalSlDfSOsf}{ImcHardSafe}{Total}{}{Walltime}{Median}{17.324335808007163}%
\StoreBenchExecResult{ReducedEvalSlDfSOsf}{ImcHardSafe}{Total}{}{Walltime}{Min}{1.702810141025111}%
\StoreBenchExecResult{ReducedEvalSlDfSOsf}{ImcHardSafe}{Total}{}{Walltime}{Max}{907.5938753430091}%
\StoreBenchExecResult{ReducedEvalSlDfSOsf}{ImcHardSafe}{Total}{}{Walltime}{Stdev}{414.8860362724548308925823591}%
\StoreBenchExecResult{ReducedEvalSlDfSOsf}{ImcHardSafe}{Correct}{}{Count}{}{525}%
\StoreBenchExecResult{ReducedEvalSlDfSOsf}{ImcHardSafe}{Correct}{}{Cputime}{}{26923.512741373}%
\StoreBenchExecResult{ReducedEvalSlDfSOsf}{ImcHardSafe}{Correct}{}{Cputime}{Avg}{51.28288141213904761904761905}%
\StoreBenchExecResult{ReducedEvalSlDfSOsf}{ImcHardSafe}{Correct}{}{Cputime}{Median}{11.169113507}%
\StoreBenchExecResult{ReducedEvalSlDfSOsf}{ImcHardSafe}{Correct}{}{Cputime}{Min}{4.24575107}%
\StoreBenchExecResult{ReducedEvalSlDfSOsf}{ImcHardSafe}{Correct}{}{Cputime}{Max}{897.502118471}%
\StoreBenchExecResult{ReducedEvalSlDfSOsf}{ImcHardSafe}{Correct}{}{Cputime}{Stdev}{122.5291494916102588823422500}%
\StoreBenchExecResult{ReducedEvalSlDfSOsf}{ImcHardSafe}{Correct}{}{Walltime}{}{22596.9011857656006936}%
\StoreBenchExecResult{ReducedEvalSlDfSOsf}{ImcHardSafe}{Correct}{}{Walltime}{Avg}{43.04171654431542989257142857}%
\StoreBenchExecResult{ReducedEvalSlDfSOsf}{ImcHardSafe}{Correct}{}{Walltime}{Median}{4.618105987989111}%
\StoreBenchExecResult{ReducedEvalSlDfSOsf}{ImcHardSafe}{Correct}{}{Walltime}{Min}{1.702810141025111}%
\StoreBenchExecResult{ReducedEvalSlDfSOsf}{ImcHardSafe}{Correct}{}{Walltime}{Max}{869.1426772929844}%
\StoreBenchExecResult{ReducedEvalSlDfSOsf}{ImcHardSafe}{Correct}{}{Walltime}{Stdev}{118.9365133443496695548240206}%
\StoreBenchExecResult{ReducedEvalSlDfSOsf}{ImcHardSafe}{Correct}{True}{Count}{}{525}%
\StoreBenchExecResult{ReducedEvalSlDfSOsf}{ImcHardSafe}{Correct}{True}{Cputime}{}{26923.512741373}%
\StoreBenchExecResult{ReducedEvalSlDfSOsf}{ImcHardSafe}{Correct}{True}{Cputime}{Avg}{51.28288141213904761904761905}%
\StoreBenchExecResult{ReducedEvalSlDfSOsf}{ImcHardSafe}{Correct}{True}{Cputime}{Median}{11.169113507}%
\StoreBenchExecResult{ReducedEvalSlDfSOsf}{ImcHardSafe}{Correct}{True}{Cputime}{Min}{4.24575107}%
\StoreBenchExecResult{ReducedEvalSlDfSOsf}{ImcHardSafe}{Correct}{True}{Cputime}{Max}{897.502118471}%
\StoreBenchExecResult{ReducedEvalSlDfSOsf}{ImcHardSafe}{Correct}{True}{Cputime}{Stdev}{122.5291494916102588823422500}%
\StoreBenchExecResult{ReducedEvalSlDfSOsf}{ImcHardSafe}{Correct}{True}{Walltime}{}{22596.9011857656006936}%
\StoreBenchExecResult{ReducedEvalSlDfSOsf}{ImcHardSafe}{Correct}{True}{Walltime}{Avg}{43.04171654431542989257142857}%
\StoreBenchExecResult{ReducedEvalSlDfSOsf}{ImcHardSafe}{Correct}{True}{Walltime}{Median}{4.618105987989111}%
\StoreBenchExecResult{ReducedEvalSlDfSOsf}{ImcHardSafe}{Correct}{True}{Walltime}{Min}{1.702810141025111}%
\StoreBenchExecResult{ReducedEvalSlDfSOsf}{ImcHardSafe}{Correct}{True}{Walltime}{Max}{869.1426772929844}%
\StoreBenchExecResult{ReducedEvalSlDfSOsf}{ImcHardSafe}{Correct}{True}{Walltime}{Stdev}{118.9365133443496695548240206}%

\StoreBenchExecResult{ReducedEvalSlDfSOsf}{ImcHardSafe}{Error}{}{Count}{}{345}%
\StoreBenchExecResult{ReducedEvalSlDfSOsf}{ImcHardSafe}{Error}{}{Cputime}{}{278608.881425803}%
\StoreBenchExecResult{ReducedEvalSlDfSOsf}{ImcHardSafe}{Error}{}{Cputime}{Avg}{807.5619751472550724637681159}%
\StoreBenchExecResult{ReducedEvalSlDfSOsf}{ImcHardSafe}{Error}{}{Cputime}{Median}{902.189764358}%
\StoreBenchExecResult{ReducedEvalSlDfSOsf}{ImcHardSafe}{Error}{}{Cputime}{Min}{4.763845204}%
\StoreBenchExecResult{ReducedEvalSlDfSOsf}{ImcHardSafe}{Error}{}{Cputime}{Max}{913.428313018}%
\StoreBenchExecResult{ReducedEvalSlDfSOsf}{ImcHardSafe}{Error}{}{Cputime}{Stdev}{273.0226745811424856765711598}%
\StoreBenchExecResult{ReducedEvalSlDfSOsf}{ImcHardSafe}{Error}{}{Walltime}{}{273779.7869637041010777}%
\StoreBenchExecResult{ReducedEvalSlDfSOsf}{ImcHardSafe}{Error}{}{Walltime}{Avg}{793.5645998947944958773913043}%
\StoreBenchExecResult{ReducedEvalSlDfSOsf}{ImcHardSafe}{Error}{}{Walltime}{Median}{887.3124313660082}%
\StoreBenchExecResult{ReducedEvalSlDfSOsf}{ImcHardSafe}{Error}{}{Walltime}{Min}{1.8929744130000472}%
\StoreBenchExecResult{ReducedEvalSlDfSOsf}{ImcHardSafe}{Error}{}{Walltime}{Max}{907.5938753430091}%
\StoreBenchExecResult{ReducedEvalSlDfSOsf}{ImcHardSafe}{Error}{}{Walltime}{Stdev}{269.4968329334964355231461144}%
\StoreBenchExecResult{ReducedEvalSlDfSOsf}{ImcHardSafe}{Error}{Error}{Count}{}{38}%
\StoreBenchExecResult{ReducedEvalSlDfSOsf}{ImcHardSafe}{Error}{Error}{Cputime}{}{1279.540862945}%
\StoreBenchExecResult{ReducedEvalSlDfSOsf}{ImcHardSafe}{Error}{Error}{Cputime}{Avg}{33.67212797223684210526315789}%
\StoreBenchExecResult{ReducedEvalSlDfSOsf}{ImcHardSafe}{Error}{Error}{Cputime}{Median}{9.2803889315}%
\StoreBenchExecResult{ReducedEvalSlDfSOsf}{ImcHardSafe}{Error}{Error}{Cputime}{Min}{4.763845204}%
\StoreBenchExecResult{ReducedEvalSlDfSOsf}{ImcHardSafe}{Error}{Error}{Cputime}{Max}{259.092632682}%
\StoreBenchExecResult{ReducedEvalSlDfSOsf}{ImcHardSafe}{Error}{Error}{Cputime}{Stdev}{60.39581288324114791190272923}%
\StoreBenchExecResult{ReducedEvalSlDfSOsf}{ImcHardSafe}{Error}{Error}{Walltime}{}{1138.4715097860607777}%
\StoreBenchExecResult{ReducedEvalSlDfSOsf}{ImcHardSafe}{Error}{Error}{Walltime}{Avg}{29.95977657331738888684210526}%
\StoreBenchExecResult{ReducedEvalSlDfSOsf}{ImcHardSafe}{Error}{Error}{Walltime}{Median}{5.8181837829906725}%
\StoreBenchExecResult{ReducedEvalSlDfSOsf}{ImcHardSafe}{Error}{Error}{Walltime}{Min}{1.8929744130000472}%
\StoreBenchExecResult{ReducedEvalSlDfSOsf}{ImcHardSafe}{Error}{Error}{Walltime}{Max}{255.84466692100978}%
\StoreBenchExecResult{ReducedEvalSlDfSOsf}{ImcHardSafe}{Error}{Error}{Walltime}{Stdev}{60.58649337159613658643428024}%
\StoreBenchExecResult{ReducedEvalSlDfSOsf}{ImcHardSafe}{Error}{Timeout}{Count}{}{307}%
\StoreBenchExecResult{ReducedEvalSlDfSOsf}{ImcHardSafe}{Error}{Timeout}{Cputime}{}{277329.340562858}%
\StoreBenchExecResult{ReducedEvalSlDfSOsf}{ImcHardSafe}{Error}{Timeout}{Cputime}{Avg}{903.3529008562149837133550489}%
\StoreBenchExecResult{ReducedEvalSlDfSOsf}{ImcHardSafe}{Error}{Timeout}{Cputime}{Median}{902.289691275}%
\StoreBenchExecResult{ReducedEvalSlDfSOsf}{ImcHardSafe}{Error}{Timeout}{Cputime}{Min}{901.335064061}%
\StoreBenchExecResult{ReducedEvalSlDfSOsf}{ImcHardSafe}{Error}{Timeout}{Cputime}{Max}{913.428313018}%
\StoreBenchExecResult{ReducedEvalSlDfSOsf}{ImcHardSafe}{Error}{Timeout}{Cputime}{Stdev}{2.995627763233228882699055673}%
\StoreBenchExecResult{ReducedEvalSlDfSOsf}{ImcHardSafe}{Error}{Timeout}{Walltime}{}{272641.3154539180403}%
\StoreBenchExecResult{ReducedEvalSlDfSOsf}{ImcHardSafe}{Error}{Timeout}{Walltime}{Avg}{888.0824607619480140065146580}%
\StoreBenchExecResult{ReducedEvalSlDfSOsf}{ImcHardSafe}{Error}{Timeout}{Walltime}{Median}{888.0719826789864}%
\StoreBenchExecResult{ReducedEvalSlDfSOsf}{ImcHardSafe}{Error}{Timeout}{Walltime}{Min}{870.6730530850182}%
\StoreBenchExecResult{ReducedEvalSlDfSOsf}{ImcHardSafe}{Error}{Timeout}{Walltime}{Max}{907.5938753430091}%
\StoreBenchExecResult{ReducedEvalSlDfSOsf}{ImcHardSafe}{Error}{Timeout}{Walltime}{Stdev}{7.491201392679265582989766228}%
\ifdefined\ReducedEvalSlDfSOsfImcHardSafeTotalCount\else\edef\ReducedEvalSlDfSOsfImcHardSafeTotalCount{0}\fi
\ifdefined\ReducedEvalSlDfSOsfImcHardSafeCorrectCount\else\edef\ReducedEvalSlDfSOsfImcHardSafeCorrectCount{0}\fi
\ifdefined\ReducedEvalSlDfSOsfImcHardSafeCorrectTrueCount\else\edef\ReducedEvalSlDfSOsfImcHardSafeCorrectTrueCount{0}\fi
\ifdefined\ReducedEvalSlDfSOsfImcHardSafeCorrectFalseCount\else\edef\ReducedEvalSlDfSOsfImcHardSafeCorrectFalseCount{0}\fi
\ifdefined\ReducedEvalSlDfSOsfImcHardSafeWrongTrueCount\else\edef\ReducedEvalSlDfSOsfImcHardSafeWrongTrueCount{0}\fi
\ifdefined\ReducedEvalSlDfSOsfImcHardSafeWrongFalseCount\else\edef\ReducedEvalSlDfSOsfImcHardSafeWrongFalseCount{0}\fi
\ifdefined\ReducedEvalSlDfSOsfImcHardSafeErrorTimeoutCount\else\edef\ReducedEvalSlDfSOsfImcHardSafeErrorTimeoutCount{0}\fi
\ifdefined\ReducedEvalSlDfSOsfImcHardSafeErrorOutOfMemoryCount\else\edef\ReducedEvalSlDfSOsfImcHardSafeErrorOutOfMemoryCount{0}\fi
\ifdefined\ReducedEvalSlDfSOsfImcHardSafeCorrectCputime\else\edef\ReducedEvalSlDfSOsfImcHardSafeCorrectCputime{0}\fi
\ifdefined\ReducedEvalSlDfSOsfImcHardSafeCorrectCputimeAvg\else\edef\ReducedEvalSlDfSOsfImcHardSafeCorrectCputimeAvg{None}\fi
\ifdefined\ReducedEvalSlDfSOsfImcHardSafeCorrectWalltime\else\edef\ReducedEvalSlDfSOsfImcHardSafeCorrectWalltime{0}\fi
\ifdefined\ReducedEvalSlDfSOsfImcHardSafeCorrectWalltimeAvg\else\edef\ReducedEvalSlDfSOsfImcHardSafeCorrectWalltimeAvg{None}\fi
\edef\ReducedEvalSlDfSOsfImcHardSafeErrorOtherInconclusiveCount{\the\numexpr \ReducedEvalSlDfSOsfImcHardSafeTotalCount - \ReducedEvalSlDfSOsfImcHardSafeCorrectCount - \ReducedEvalSlDfSOsfImcHardSafeWrongTrueCount - \ReducedEvalSlDfSOsfImcHardSafeWrongFalseCount - \ReducedEvalSlDfSOsfImcHardSafeErrorTimeoutCount - \ReducedEvalSlDfSOsfImcHardSafeErrorOutOfMemoryCount \relax}
\providecommand\StoreBenchExecResult[7]{\expandafter\newcommand\csname#1#2#3#4#5#6\endcsname{#7}}%
\StoreBenchExecResult{ReducedEvalSlDfSOsf}{ImcIgndfiHardSafe}{Total}{}{Count}{}{870}%
\StoreBenchExecResult{ReducedEvalSlDfSOsf}{ImcIgndfiHardSafe}{Total}{}{Cputime}{}{311992.686511905}%
\StoreBenchExecResult{ReducedEvalSlDfSOsf}{ImcIgndfiHardSafe}{Total}{}{Cputime}{Avg}{358.6122833470172413793103448}%
\StoreBenchExecResult{ReducedEvalSlDfSOsf}{ImcIgndfiHardSafe}{Total}{}{Cputime}{Median}{87.3663342285}%
\StoreBenchExecResult{ReducedEvalSlDfSOsf}{ImcIgndfiHardSafe}{Total}{}{Cputime}{Min}{4.509444045}%
\StoreBenchExecResult{ReducedEvalSlDfSOsf}{ImcIgndfiHardSafe}{Total}{}{Cputime}{Max}{913.139820507}%
\StoreBenchExecResult{ReducedEvalSlDfSOsf}{ImcIgndfiHardSafe}{Total}{}{Cputime}{Stdev}{406.4028511462078241937204152}%
\StoreBenchExecResult{ReducedEvalSlDfSOsf}{ImcIgndfiHardSafe}{Total}{}{Walltime}{}{267020.7790330193823276}%
\StoreBenchExecResult{ReducedEvalSlDfSOsf}{ImcIgndfiHardSafe}{Total}{}{Walltime}{Avg}{306.9204356701372210662068966}%
\StoreBenchExecResult{ReducedEvalSlDfSOsf}{ImcIgndfiHardSafe}{Total}{}{Walltime}{Median}{30.78812884000945}%
\StoreBenchExecResult{ReducedEvalSlDfSOsf}{ImcIgndfiHardSafe}{Total}{}{Walltime}{Min}{1.7349930929776747}%
\StoreBenchExecResult{ReducedEvalSlDfSOsf}{ImcIgndfiHardSafe}{Total}{}{Walltime}{Max}{905.944337975001}%
\StoreBenchExecResult{ReducedEvalSlDfSOsf}{ImcIgndfiHardSafe}{Total}{}{Walltime}{Stdev}{370.1048769271596679732308641}%
\StoreBenchExecResult{ReducedEvalSlDfSOsf}{ImcIgndfiHardSafe}{Correct}{}{Count}{}{539}%
\StoreBenchExecResult{ReducedEvalSlDfSOsf}{ImcIgndfiHardSafe}{Correct}{}{Cputime}{}{42845.098652973}%
\StoreBenchExecResult{ReducedEvalSlDfSOsf}{ImcIgndfiHardSafe}{Correct}{}{Cputime}{Avg}{79.48997894800185528756957328}%
\StoreBenchExecResult{ReducedEvalSlDfSOsf}{ImcIgndfiHardSafe}{Correct}{}{Cputime}{Median}{15.836609339}%
\StoreBenchExecResult{ReducedEvalSlDfSOsf}{ImcIgndfiHardSafe}{Correct}{}{Cputime}{Min}{4.509444045}%
\StoreBenchExecResult{ReducedEvalSlDfSOsf}{ImcIgndfiHardSafe}{Correct}{}{Cputime}{Max}{876.825633696}%
\StoreBenchExecResult{ReducedEvalSlDfSOsf}{ImcIgndfiHardSafe}{Correct}{}{Cputime}{Stdev}{156.4033929641206571502026317}%
\StoreBenchExecResult{ReducedEvalSlDfSOsf}{ImcIgndfiHardSafe}{Correct}{}{Walltime}{}{27410.5304155150661766}%
\StoreBenchExecResult{ReducedEvalSlDfSOsf}{ImcIgndfiHardSafe}{Correct}{}{Walltime}{Avg}{50.85441635531552166345083488}%
\StoreBenchExecResult{ReducedEvalSlDfSOsf}{ImcIgndfiHardSafe}{Correct}{}{Walltime}{Median}{5.574438512005145}%
\StoreBenchExecResult{ReducedEvalSlDfSOsf}{ImcIgndfiHardSafe}{Correct}{}{Walltime}{Min}{1.7349930929776747}%
\StoreBenchExecResult{ReducedEvalSlDfSOsf}{ImcIgndfiHardSafe}{Correct}{}{Walltime}{Max}{779.235800880997}%
\StoreBenchExecResult{ReducedEvalSlDfSOsf}{ImcIgndfiHardSafe}{Correct}{}{Walltime}{Stdev}{130.6963718590760943435005548}%
\StoreBenchExecResult{ReducedEvalSlDfSOsf}{ImcIgndfiHardSafe}{Correct}{True}{Count}{}{539}%
\StoreBenchExecResult{ReducedEvalSlDfSOsf}{ImcIgndfiHardSafe}{Correct}{True}{Cputime}{}{42845.098652973}%
\StoreBenchExecResult{ReducedEvalSlDfSOsf}{ImcIgndfiHardSafe}{Correct}{True}{Cputime}{Avg}{79.48997894800185528756957328}%
\StoreBenchExecResult{ReducedEvalSlDfSOsf}{ImcIgndfiHardSafe}{Correct}{True}{Cputime}{Median}{15.836609339}%
\StoreBenchExecResult{ReducedEvalSlDfSOsf}{ImcIgndfiHardSafe}{Correct}{True}{Cputime}{Min}{4.509444045}%
\StoreBenchExecResult{ReducedEvalSlDfSOsf}{ImcIgndfiHardSafe}{Correct}{True}{Cputime}{Max}{876.825633696}%
\StoreBenchExecResult{ReducedEvalSlDfSOsf}{ImcIgndfiHardSafe}{Correct}{True}{Cputime}{Stdev}{156.4033929641206571502026317}%
\StoreBenchExecResult{ReducedEvalSlDfSOsf}{ImcIgndfiHardSafe}{Correct}{True}{Walltime}{}{27410.5304155150661766}%
\StoreBenchExecResult{ReducedEvalSlDfSOsf}{ImcIgndfiHardSafe}{Correct}{True}{Walltime}{Avg}{50.85441635531552166345083488}%
\StoreBenchExecResult{ReducedEvalSlDfSOsf}{ImcIgndfiHardSafe}{Correct}{True}{Walltime}{Median}{5.574438512005145}%
\StoreBenchExecResult{ReducedEvalSlDfSOsf}{ImcIgndfiHardSafe}{Correct}{True}{Walltime}{Min}{1.7349930929776747}%
\StoreBenchExecResult{ReducedEvalSlDfSOsf}{ImcIgndfiHardSafe}{Correct}{True}{Walltime}{Max}{779.235800880997}%
\StoreBenchExecResult{ReducedEvalSlDfSOsf}{ImcIgndfiHardSafe}{Correct}{True}{Walltime}{Stdev}{130.6963718590760943435005548}%

\StoreBenchExecResult{ReducedEvalSlDfSOsf}{ImcIgndfiHardSafe}{Error}{}{Count}{}{331}%
\StoreBenchExecResult{ReducedEvalSlDfSOsf}{ImcIgndfiHardSafe}{Error}{}{Cputime}{}{269147.587858932}%
\StoreBenchExecResult{ReducedEvalSlDfSOsf}{ImcIgndfiHardSafe}{Error}{}{Cputime}{Avg}{813.1347065224531722054380665}%
\StoreBenchExecResult{ReducedEvalSlDfSOsf}{ImcIgndfiHardSafe}{Error}{}{Cputime}{Median}{901.823261368}%
\StoreBenchExecResult{ReducedEvalSlDfSOsf}{ImcIgndfiHardSafe}{Error}{}{Cputime}{Min}{9.26089801}%
\StoreBenchExecResult{ReducedEvalSlDfSOsf}{ImcIgndfiHardSafe}{Error}{}{Cputime}{Max}{913.139820507}%
\StoreBenchExecResult{ReducedEvalSlDfSOsf}{ImcIgndfiHardSafe}{Error}{}{Cputime}{Stdev}{246.6233027847716293733012746}%
\StoreBenchExecResult{ReducedEvalSlDfSOsf}{ImcIgndfiHardSafe}{Error}{}{Walltime}{}{239610.248617504316151}%
\StoreBenchExecResult{ReducedEvalSlDfSOsf}{ImcIgndfiHardSafe}{Error}{}{Walltime}{Avg}{723.8980320770523146555891239}%
\StoreBenchExecResult{ReducedEvalSlDfSOsf}{ImcIgndfiHardSafe}{Error}{}{Walltime}{Median}{803.5542384049913}%
\StoreBenchExecResult{ReducedEvalSlDfSOsf}{ImcIgndfiHardSafe}{Error}{}{Walltime}{Min}{3.172922376979841}%
\StoreBenchExecResult{ReducedEvalSlDfSOsf}{ImcIgndfiHardSafe}{Error}{}{Walltime}{Max}{905.944337975001}%
\StoreBenchExecResult{ReducedEvalSlDfSOsf}{ImcIgndfiHardSafe}{Error}{}{Walltime}{Stdev}{227.0950016888455851406018262}%
\StoreBenchExecResult{ReducedEvalSlDfSOsf}{ImcIgndfiHardSafe}{Error}{Error}{Count}{}{38}%
\StoreBenchExecResult{ReducedEvalSlDfSOsf}{ImcIgndfiHardSafe}{Error}{Error}{Cputime}{}{5488.821504534}%
\StoreBenchExecResult{ReducedEvalSlDfSOsf}{ImcIgndfiHardSafe}{Error}{Error}{Cputime}{Avg}{144.4426711719473684210526316}%
\StoreBenchExecResult{ReducedEvalSlDfSOsf}{ImcIgndfiHardSafe}{Error}{Error}{Cputime}{Median}{153.698045410}%
\StoreBenchExecResult{ReducedEvalSlDfSOsf}{ImcIgndfiHardSafe}{Error}{Error}{Cputime}{Min}{13.529566944}%
\StoreBenchExecResult{ReducedEvalSlDfSOsf}{ImcIgndfiHardSafe}{Error}{Error}{Cputime}{Max}{332.89663563}%
\StoreBenchExecResult{ReducedEvalSlDfSOsf}{ImcIgndfiHardSafe}{Error}{Error}{Cputime}{Stdev}{60.44447963596844941740948003}%
\StoreBenchExecResult{ReducedEvalSlDfSOsf}{ImcIgndfiHardSafe}{Error}{Error}{Walltime}{}{4104.267507130978610}%
\StoreBenchExecResult{ReducedEvalSlDfSOsf}{ImcIgndfiHardSafe}{Error}{Error}{Walltime}{Avg}{108.0070396613415423684210526}%
\StoreBenchExecResult{ReducedEvalSlDfSOsf}{ImcIgndfiHardSafe}{Error}{Error}{Walltime}{Median}{125.52169833300286}%
\StoreBenchExecResult{ReducedEvalSlDfSOsf}{ImcIgndfiHardSafe}{Error}{Error}{Walltime}{Min}{7.008002563990885}%
\StoreBenchExecResult{ReducedEvalSlDfSOsf}{ImcIgndfiHardSafe}{Error}{Error}{Walltime}{Max}{246.89015440101502}%
\StoreBenchExecResult{ReducedEvalSlDfSOsf}{ImcIgndfiHardSafe}{Error}{Error}{Walltime}{Stdev}{49.99067453601426611576735593}%
\StoreBenchExecResult{ReducedEvalSlDfSOsf}{ImcIgndfiHardSafe}{Error}{Exception}{Count}{}{1}%
\StoreBenchExecResult{ReducedEvalSlDfSOsf}{ImcIgndfiHardSafe}{Error}{Exception}{Cputime}{}{9.26089801}%
\StoreBenchExecResult{ReducedEvalSlDfSOsf}{ImcIgndfiHardSafe}{Error}{Exception}{Cputime}{Avg}{9.26089801}%
\StoreBenchExecResult{ReducedEvalSlDfSOsf}{ImcIgndfiHardSafe}{Error}{Exception}{Cputime}{Median}{9.26089801}%
\StoreBenchExecResult{ReducedEvalSlDfSOsf}{ImcIgndfiHardSafe}{Error}{Exception}{Cputime}{Min}{9.26089801}%
\StoreBenchExecResult{ReducedEvalSlDfSOsf}{ImcIgndfiHardSafe}{Error}{Exception}{Cputime}{Max}{9.26089801}%
\StoreBenchExecResult{ReducedEvalSlDfSOsf}{ImcIgndfiHardSafe}{Error}{Exception}{Cputime}{Stdev}{0E-14}%
\StoreBenchExecResult{ReducedEvalSlDfSOsf}{ImcIgndfiHardSafe}{Error}{Exception}{Walltime}{}{3.172922376979841}%
\StoreBenchExecResult{ReducedEvalSlDfSOsf}{ImcIgndfiHardSafe}{Error}{Exception}{Walltime}{Avg}{3.172922376979841}%
\StoreBenchExecResult{ReducedEvalSlDfSOsf}{ImcIgndfiHardSafe}{Error}{Exception}{Walltime}{Median}{3.172922376979841}%
\StoreBenchExecResult{ReducedEvalSlDfSOsf}{ImcIgndfiHardSafe}{Error}{Exception}{Walltime}{Min}{3.172922376979841}%
\StoreBenchExecResult{ReducedEvalSlDfSOsf}{ImcIgndfiHardSafe}{Error}{Exception}{Walltime}{Max}{3.172922376979841}%
\StoreBenchExecResult{ReducedEvalSlDfSOsf}{ImcIgndfiHardSafe}{Error}{Exception}{Walltime}{Stdev}{0E-15}%
\StoreBenchExecResult{ReducedEvalSlDfSOsf}{ImcIgndfiHardSafe}{Error}{Timeout}{Count}{}{292}%
\StoreBenchExecResult{ReducedEvalSlDfSOsf}{ImcIgndfiHardSafe}{Error}{Timeout}{Cputime}{}{263649.505456388}%
\StoreBenchExecResult{ReducedEvalSlDfSOsf}{ImcIgndfiHardSafe}{Error}{Timeout}{Cputime}{Avg}{902.9092652616027397260273973}%
\StoreBenchExecResult{ReducedEvalSlDfSOsf}{ImcIgndfiHardSafe}{Error}{Timeout}{Cputime}{Median}{901.880735609}%
\StoreBenchExecResult{ReducedEvalSlDfSOsf}{ImcIgndfiHardSafe}{Error}{Timeout}{Cputime}{Min}{901.163204668}%
\StoreBenchExecResult{ReducedEvalSlDfSOsf}{ImcIgndfiHardSafe}{Error}{Timeout}{Cputime}{Max}{913.139820507}%
\StoreBenchExecResult{ReducedEvalSlDfSOsf}{ImcIgndfiHardSafe}{Error}{Timeout}{Cputime}{Stdev}{2.839405303512905261038246829}%
\StoreBenchExecResult{ReducedEvalSlDfSOsf}{ImcIgndfiHardSafe}{Error}{Timeout}{Walltime}{}{235502.8081879963577}%
\StoreBenchExecResult{ReducedEvalSlDfSOsf}{ImcIgndfiHardSafe}{Error}{Timeout}{Walltime}{Avg}{806.5164663972478003424657534}%
\StoreBenchExecResult{ReducedEvalSlDfSOsf}{ImcIgndfiHardSafe}{Error}{Timeout}{Walltime}{Median}{803.9419942974928}%
\StoreBenchExecResult{ReducedEvalSlDfSOsf}{ImcIgndfiHardSafe}{Error}{Timeout}{Walltime}{Min}{792.2311651179916}%
\StoreBenchExecResult{ReducedEvalSlDfSOsf}{ImcIgndfiHardSafe}{Error}{Timeout}{Walltime}{Max}{905.944337975001}%
\StoreBenchExecResult{ReducedEvalSlDfSOsf}{ImcIgndfiHardSafe}{Error}{Timeout}{Walltime}{Stdev}{12.90224137138459236842677915}%
\ifdefined\ReducedEvalSlDfSOsfImcIgndfiHardSafeTotalCount\else\edef\ReducedEvalSlDfSOsfImcIgndfiHardSafeTotalCount{0}\fi
\ifdefined\ReducedEvalSlDfSOsfImcIgndfiHardSafeCorrectCount\else\edef\ReducedEvalSlDfSOsfImcIgndfiHardSafeCorrectCount{0}\fi
\ifdefined\ReducedEvalSlDfSOsfImcIgndfiHardSafeCorrectTrueCount\else\edef\ReducedEvalSlDfSOsfImcIgndfiHardSafeCorrectTrueCount{0}\fi
\ifdefined\ReducedEvalSlDfSOsfImcIgndfiHardSafeCorrectFalseCount\else\edef\ReducedEvalSlDfSOsfImcIgndfiHardSafeCorrectFalseCount{0}\fi
\ifdefined\ReducedEvalSlDfSOsfImcIgndfiHardSafeWrongTrueCount\else\edef\ReducedEvalSlDfSOsfImcIgndfiHardSafeWrongTrueCount{0}\fi
\ifdefined\ReducedEvalSlDfSOsfImcIgndfiHardSafeWrongFalseCount\else\edef\ReducedEvalSlDfSOsfImcIgndfiHardSafeWrongFalseCount{0}\fi
\ifdefined\ReducedEvalSlDfSOsfImcIgndfiHardSafeErrorTimeoutCount\else\edef\ReducedEvalSlDfSOsfImcIgndfiHardSafeErrorTimeoutCount{0}\fi
\ifdefined\ReducedEvalSlDfSOsfImcIgndfiHardSafeErrorOutOfMemoryCount\else\edef\ReducedEvalSlDfSOsfImcIgndfiHardSafeErrorOutOfMemoryCount{0}\fi
\ifdefined\ReducedEvalSlDfSOsfImcIgndfiHardSafeCorrectCputime\else\edef\ReducedEvalSlDfSOsfImcIgndfiHardSafeCorrectCputime{0}\fi
\ifdefined\ReducedEvalSlDfSOsfImcIgndfiHardSafeCorrectCputimeAvg\else\edef\ReducedEvalSlDfSOsfImcIgndfiHardSafeCorrectCputimeAvg{None}\fi
\ifdefined\ReducedEvalSlDfSOsfImcIgndfiHardSafeCorrectWalltime\else\edef\ReducedEvalSlDfSOsfImcIgndfiHardSafeCorrectWalltime{0}\fi
\ifdefined\ReducedEvalSlDfSOsfImcIgndfiHardSafeCorrectWalltimeAvg\else\edef\ReducedEvalSlDfSOsfImcIgndfiHardSafeCorrectWalltimeAvg{None}\fi
\edef\ReducedEvalSlDfSOsfImcIgndfiHardSafeErrorOtherInconclusiveCount{\the\numexpr \ReducedEvalSlDfSOsfImcIgndfiHardSafeTotalCount - \ReducedEvalSlDfSOsfImcIgndfiHardSafeCorrectCount - \ReducedEvalSlDfSOsfImcIgndfiHardSafeWrongTrueCount - \ReducedEvalSlDfSOsfImcIgndfiHardSafeWrongFalseCount - \ReducedEvalSlDfSOsfImcIgndfiHardSafeErrorTimeoutCount - \ReducedEvalSlDfSOsfImcIgndfiHardSafeErrorOutOfMemoryCount \relax}
\edef\CpacheckerRev{42901}
\edef\ImcIgndffMinusOverImcReducedCount{4}
\edef\ImcIgndffNetOverImcReducedCount{2}
\edef\ImcIgndffPlusOverImcReducedCount{6}
\edef\ImcIgndfiMinusOverImcReducedCount{13}
\edef\ImcIgndfiNetOverImcReducedCount{10}
\edef\ImcIgndfiPlusOverImcReducedCount{23}
\edef\ImcIgndfiPlusOverImpactCount{88}
\edef\ImcIgndfiPlusOverImpactReducedCount{76}
\edef\ImcIgndfiPlusOverKiIgndfCount{324}
\edef\ImcIgndfiPlusOverKiIgndfReducedCount{275}
\edef\ImcIgndfiPlusOverLsCount{423}
\edef\ImcIgndfiPlusOverLsReducedCount{309}
\edef\ImcIgndfiPlusOverPredabsCount{143}
\edef\ImcIgndfiPlusOverPredabsReducedCount{79}
\edef\ImcIgndfiPlusOverSymbioticCount{624}
\edef\ImcIgndfiPlusOverSymbioticReducedCount{331}
\edef\NonTrivInvCount{888}
\edef\NonTrivNonIndInvCount{18}
\edef\UniqSolveCpaSmtCount{10}
\edef\UniqSolveCpaSmtReducedCount{4}
\edef\UniqSolveOtherToolsCount{342}
\edef\UniqSolveOtherToolsReducedCount{246}
\edef\UniqSolveOverallCount{3}
\edef\UniqSolveOverallReducedCount{3}
\edef\nTasksFewerItpCall{56}
\edef\nTasksFewerK{35}
\edef\nTasksLessItpTime{490}
\edef\nTasksLessWalltime{426}
\edef\nTasksMoreItpCall{21}
\edef\nTasksMoreItpTime{288}
\edef\nTasksMoreK{4}
\edef\nTasksMoreWalltime{444}
\edef\nTasksTenXLessItpTime{20}
\edef\nTasksTenXLessWalltime{10}
\edef\nTasksTenXMoreItpTime{4}
\edef\nTasksTenXMoreWalltime{24}

%% file: abstract.tex
Software model checking is a challenging problem,
and generating relevant invariants is a key factor
in proving the safety properties of a program.
Program invariants can be obtained by various approaches,
including lightweight procedures based on data-flow analysis
and intensive techniques using Craig interpolation.
Although data-flow analysis runs efficiently,
it often produces invariants that are too weak to prove the properties.
By contrast, interpolation-based approaches
build strong invariants from interpolants,
but they might not scale well due to expensive interpolation procedures.
Invariants can also be injected into
model-checking algorithms to assist the analysis.
Invariant injection has been studied for many well-known approaches,
including \kinduction, predicate abstraction, and symbolic execution.
We propose an augmented interpolation-based verification algorithm that
injects external invariants into \emph{interpolation-based model checking}
(\href{https://doi.org/10.1007/978-3-540-45069-6_1}{McMillan, 2003}),
a hardware model-checking algorithm recently adopted for software verification.
The auxiliary invariants help prune unreachable states in Craig interpolants
and confine the analysis to the reachable parts of a program.
We implemented the proposed technique in the verification framework~\cpachecker
and evaluated it against mature SMT-based methods in~\cpachecker
as well as other state-of-the-art software verifiers.
We found that injecting invariants
reduces the number of interpolation queries needed to prove safety properties and
improves the run-time efficiency.
Consequently, the proposed invariant-injection approach verified difficult tasks
that none of its plain version (i.e., without invariants), the invariant generator,
or any compared tools could solve.

%% file: introduction.tex
\section{Introduction}
\label{sect:introduction}

Assuring that programs execute correctly with respect to their specifications
is fundamental for deploying them in our daily lives.
In the software industry,
testing~\cite{ArtOfSoftwareTesting} is the most popular methodology
to validate the quality of programs.
However, software testing can only spot the presence of bugs in programs
but not guarantee their absence.
To prove the correctness of programs with mathematical rigorousness,
automatic software model checking~\cite{SoftwareModelChecking} has been extensively studied.
One of the greatest challenges to formally verify a program is to deduce
suitable \emph{program invariants} that can be used to prove the program's safety properties.
A program invariant is a logical formula over some program variables
that must hold at a certain program location for all feasible program paths.

There are numerous approaches to invariant generation,
with different performance characteristics and
strengths of the produced invariants.
Data-flow analysis~\cite{Kildall,Kam76,SharirPnueli,Kennedy81,JonesMuchnick,Ryder83,RHS95}
is a category of methods that sacrifice path sensitivity for scalability.
To reduce the size of the abstract reachability graph,
data-flow analysis merges
abstract program states arising from different execution paths
when the control flow meets.
Since merging abstract states usually loses information,
the resulting invariants might not be strong
enough to prove the safety properties.
In contrast to data-flow analysis,
techniques~\cite{McMillanCraig,AbstractionsFromProofs,IMPACT,VizelFMCAD09,LazyAnnotation,SoftwareIC3}
based on Craig interpolation~\cite{Craig57}
iteratively derive interpolants%
\footnote{An interpolant~$\itp$ for an implication $A \implies B$
    is a formula that uses only the common variables of formulas~$A$ and~$B$
    such that $A \implies \itp$ and $\itp \implies B$ are valid.}
to construct strong invariants.
Although these approaches are capable of generating useful invariants,
they might take too many refinement iterations and fail to converge.

To leverage the information carried by invariants,
researchers also use invariant generators as an auxiliary component
to aid various model-checking techniques.
Algorithms that have been augmented by \emph{invariant injection}
include bounded model checking (BMC)~\cite{InvBMC,Ganai06,Cheng08},
\kinduction~\cite{AuxiliaryInvariants,kInduction,kikiKroening,ESBMCDEPTHK,PDR-kInduction},
predicate abstraction~\cite{PredicatedLattice,InvariantsForPredicateAbstraction},
symbolic execution~\cite{Pasareanu04}, and
IC3/PDR~\cite{SEAHORN}.
The idea is to confine the scope of a verification algorithm
to the reachable state space of a program,
as safety properties often do not hold in the unreachable parts~\cite{kInduction}.

\subsection{Our Research Goal}
Despite the extensive study of injecting invariants to
various software-verification methods,
the possibilities of leveraging auxiliary invariants to assist
\emph{interpolation-based model checking}~\cite{McMillanCraig},
a hardware model-checking algorithm published by McMillan in 2003
and adopted to software recently~\cite{IMC-JAR},
have not been explored yet.
Although this algorithm was invented two decades ago,
it still remains state-of-the-art for safety verification.
In this paper, our goal is to find out
\emph{whether invariant injection can enhance
    McMillan's interpolation-based model-checking algorithm from 2003 for software verification}.
In particular,
we aim to reduce the number of interpolation queries
required to prove the safety property of a program,
since interpolation is usually the most time-consuming step
in interpolation-based algorithms.
To avoid confusion between McMillan's algorithm from 2003~\cite{McMillanCraig}
and other interpolation-based verification approaches,
we refer to McMillan's approach from 2003 as IMC from now on.
IMC has been combined with expensive SAT-based invariant generators
for hardware model checking~\cite{Case07,Cabodi09},
but its characteristics when assisted by
lightweight invariant generators~\cite{CPA-DF} based on data-flow analysis
remain unknown for software verification.

To motivate how auxiliary invariants could assist IMC,
we briefly introduce the algorithm first.
IMC extends BMC via constructing inductive invariants by Craig interpolants~\cite{Craig57}
arising from unsatisfiable BMC queries.
Whenever a BMC query starting from initial states is unsatisfiable,
IMC derives an interpolant from the query
and replaces the initial states in the query with the interpolant.
If the new query is unsatisfiable again,
another interpolant can be obtained.
The process is repeated until either
(1)~the newest interpolant is contained in the union of initial states and previous interpolants,
or (2)~the query starting from the newest interpolant becomes satisfiable.
In case~(1), an inductive invariant (i.e., a fixed point) is found and the property is proven.
In case~(2), the satisfied query might be a spurious counterexample,
so IMC will continue to unfold the program.

Interpolants tend to abstract away irrelevant information
and may intersect with the unreachable state space.
Therefore, conjoining them with auxiliary invariants
prunes away some unreachable states.
In the IMC algorithm outlined above,
if the newest interpolant is strengthened by an auxiliary invariant,
it is more likely to be contained
in the union of initial states and previous interpolants,
and the query starting from a strengthened interpolant
is less likely to be satisfiable.
In other words,
a fixed point might be found with fewer interpolation queries in case~(1),
and a spurious alarm plus extra program unrollings could be avoided in case~(2).
An example will be shown in~\cref{sect:motivating-example}
to illustrate these benefits.

\subsection{Our Contributions}

\subsubsection{Augmenting IMC with Auxiliary Invariants}
We devise two methods to augment IMC with auxiliary invariants.
The first method confines the containment check between
the newest interpolant and the union of initial states and previous interpolants
with auxiliary invariants.
The strengthened check is more likely to succeed,
so a fixed point can be found with fewer unrollings and interpolation calls.
The second method conjoins derived interpolants with auxiliary invariants.
The BMC query starting from a strengthened interpolant
is more likely to be unsatisfiable,
and more interpolants can be derived to form a fixed point
before IMC further unrolls the program.
We rigorously prove the correctness of the proposed techniques.
The proposed augmenting approaches are \textbf{novel}
because they are the first invariant-injection techniques for IMC applied to program analysis.
Moreover, the theoretical results in this paper are applicable to IMC for hardware verification
even though we focus on software verification.

\subsubsection{Open-Source Implementation and Extensive Evaluation}
We implemented the proposed approaches in the open-source framework~\cpachecker~\cite{CPACHECKER}
and conducted an extensive evaluation
on more than \num{1600} difficult verification tasks of C~programs
from the 2022 Intl. Competition on Software Verification~\cite{SVCOMP22}.
In our experiments,
the \textit{plain} IMC algorithm (i.e., without invariant injection),
three other mature SMT-based algorithms in~\cpachecker,
and two other software verifiers
were used as references to evaluate our implementation.
Our experimental results show that invariant injection
(1)~effectively reduces the numbers of program unrollings and interpolation queries
needed by plain IMC to prove safety properties,
(2)~reduces the wall-time usage,
(3)~proves 16 tasks that neither plain IMC nor the data-flow analyzer used to generate invariants could solve, and
(4)~outperformed other well-established software verifiers.
These observations are \textbf{significant}
because they enhance the knowledge about the effect of auxiliary invariants on IMC.
Our open-source implementation of the proposed approaches also helps
other researchers to understand the details of the algorithms better
and provides a solid baseline for future studies.

\subsection{Motivating Example}
\label{sect:motivating-example}

\newsavebox{\exampleCode}
\begin{lrbox}{\exampleCode}
    \begin{minipage}[b]{0.4\textwidth}
        \centering
\begin{lstlisting}[
            style=C,
            basicstyle=\ttfamilywithbold,
            numberstyle=\scriptsize,
            lineskip=2pt,
            aboveskip=0pt,
            belowskip=0pt,
        ]
extern int nondet();
int main(void) {
  unsigned int x = 0;
  unsigned int i = 0;
  while (nondet()) {
    x += 2;
    if (i == 3) x++;
    i++;
    if (i == 2) i = 0;
  }
  if (x % 2) {
    ERROR: return 1;
  }
  return 0;
}
\end{lstlisting}
    \end{minipage}
\end{lrbox}

\begin{wrapfigure}{R}{0.36\textwidth}
    \vspace{-7mm}
    \centering
    \scalebox{0.8}{\usebox{\exampleCode}}
    \caption{An example C program to motivate how auxiliary invariants help IMC}
    \label{fig:even-code}
    \vspace{-7.5mm}
\end{wrapfigure}

We use the example C~program in~\cref{fig:even-code}
to explain why invariant injection
can reduce the numbers of program unrollings and interpolation queries for IMC.
In the program,
variables~\texttt{x} and~\texttt{i} are both initialized to $0$.
The loop will be executed nondeterministically many times
depending on the values returned by function~\texttt{nondet()}.
The values of~\texttt{x} and~\texttt{i} are modified in each loop iteration:
\texttt{x} will be incremented by $2$;
if~\texttt{i} equals $3$,
\texttt{x} will be further incremented by $1$.
Variable~\texttt{i} will be incremented by $1$
and reset to $0$ if its value equals $2$ after being incremented.
An error occurs if~\texttt{x} is odd
when the control flow exits the loop.

IMC first checks if the error location line~\texttt{12}
is reachable by skipping the loop
(i.e., assuming \texttt{nondet()} returns $0$).
As the conjunction of the initial states $x=0 \land i=0$
and the guard to the error location $x\%2 \neq 0$ is unsatisfiable,
namely, line~\texttt{12} is unreachable if the loop is not entered,
IMC will try to build an inductive invariant at the loop head line~\texttt{5}
to prove that all paths entering the loop cannot reach the error location.
IMC unrolls the program and builds a BMC query%
\footnote{
    The query is
    $(x=0 \land i=0) \land (i=3 \implies x'=x+3) \land (i\neq3 \implies x'=x+2) \land (i+1=2 \implies i'=0) \land (i+1\neq2 \implies i'=i+1) \land (x'\%2\neq 0)$,
    where the prime symbols indicate variables after a loop iteration,
    and the returned values of function \texttt{nondet()} are omitted for simplicity.
}
starting from the initial location line~\texttt{3},
going through the loop once (i.e., visiting line~\texttt{5} twice),
and ending at the error location line~\texttt{12}.
This BMC query is unsatisfiable,
so an interpolant at the loop head line~\texttt{5}
can be derived to overapproximate the program states reachable after one loop iteration.
Assume the interpolant is $x\%2=0$ (instantiated with variables of the initial states).
IMC will replace the initial states $x=0 \land i=0$ in the BMC query by $x\%2=0$
and pose another query to derive the next interpolant.
Unfortunately, the new query is satisfiable,
indicating that there exists a feasible path
from line~\texttt{5} to the error location,
which assumes~\texttt{x} to be even at the beginning of the path
and goes through the loop body once.
Indeed, a solution to the new query is $(x,i)=(0,3)$,
which leads to $x=3$ after exiting the loop.
To decide whether this solution is a spurious counterexample or not,
IMC has to further unroll the program
and fails to converge to a fixed point under the current unrolling.
The reason behind this situation is that the interpolant
$x\%2=0$ contains unreachable program states
allowing the infeasible assignment $i=3$.
In fact, the safety property of the program is fulfilled
as variable~$i$ never grows beyond $2$,
and the problematic statement \texttt{x++;} at line~\texttt{7} is unreachable.

By contrast, it is easy for data-flow analysis,
e.g., one based on the abstract domain of intervals~\cite{CPA-DF},
to identify that $0 \leq i \leq 1$ is an invariant at line~\texttt{5} of the program.
If this invariant is injected to strengthen the interpolant $x\%2=0$,
IMC reaches a fixed point $0 \leq i \leq 1 \land x\%2=0$ immediately,
without further unrolling the program.
This reasoning is confirmed by our implementation in~\cpachecker,
which injects auxiliary invariants produced by
a continuously-refining interval analysis~\cite{CPA-DF}
into the plain IMC algorithm~\cite{IMC-JAR}.
The table below summarizes the first interpolants obtained at line~$\texttt{5}$
and the numbers of unrollings and interpolation queries
needed to prove the program in~\cref{fig:even-code}
for the plain and augmented IMC implementations in~\cpachecker.

\begin{table*}[h]
    \centering
    \begin{tabular}{l|c|c|c|c}
        Algorithm               & First interpolant at line~\texttt{5} & Fixed point & \#Unrolling & \#Itp-queries \\
        \hline
        Plain IMC~\cite{IMC-JAR} & $x\%2=0$                             & No          & 3           & 7             \\
        Augmented IMC           & $0 \leq i \leq 1 \land x\%2=0$       & Yes         & 1           & 2             \\
    \end{tabular}
\end{table*}

%% file: related-work.tex
\section{Related Work}
\label{sect:related-work}
Our paper is primarily related to invariant generation
and verification approaches aided by auxiliary invariants.

\subsection{Invariant Generation}
Various approaches exist for invariant generation,
ranging from lightweight ones based on data-flow analysis
to computationally expensive ones based on
SAT/SMT solving,
predicate abstraction,
or Craig interpolation.

Data-flow analysis has been extensively studied~\cite{Kam76,SharirPnueli,Kennedy81,JonesMuchnick,Ryder83,RHS95}.
Classic approaches usually consider program variables over an abstract domain
structured as a semi-lattice.
A standard fixed-point procedure is used to iteratively explore a program,
and abstract states reached by different program traces are merged.
When the procedure finishes,
the merged abstract states at each program location contain the corresponding program invariants.
Common choices of abstract domains include a value domain~\cite{Rosen88,BodikAnik98}
or an interval domain~\cite{Interval}.


Interpolation-based approaches construct invariants by
either predicates collected from interpolants~\cite{AbstractionsFromProofs}
or interpolants themselves~\cite{McMillanCraig,IMPACT,VizelFMCAD09,LazyAnnotation}.
Although they are able to build strong invariants,
the expensive interpolation queries might hinder their scalability.
Iterative SAT solving is used to prove the inductiveness of
candidate invariants extracted from simulation data~\cite{Case07}.
Invariant generation can also be guided by
the safety property~\cite{PropertyDirectedInvariants,IC3}
or the syntax of the program~\cite{Fedyukovich18}.

\subsection{Invariant-Aided Verification}
Injecting invariants to enhance model-checking algorithms is a popular technique.
There are techniques to restrict the state space in BMC queries by
high-level design information~\cite{Ganai06} or data mining~\cite{Cheng08}.
It is well known that the induction hypothesis of \kinduction
is often too weak on its own and has to be strengthened by auxiliary invariants
from static analysis~\cite{AuxiliaryInvariants,kikiKroening}
or property-directed reachability~\cite{PDR-kInduction,PDR}.
Alternatively, \kinduction can be used to prove candidate invariants
instantiated from a template,
and the confirmed invariants can be used in an induction proof~\cite{PKind}.
Recently, an interval-based analysis that produces continuously-refined invariants
is successfully applied to boost the performance of \kinduction~\cite{kInduction}.
Transition relations can be strengthened by invariants from the octagon abstract domain
to reduce the number of refinement loops in predicate abstraction~\cite{InvariantsForPredicateAbstraction}.
Loop invariants can be used as annotations in symbolic execution~\cite{Pasareanu04}.
Predicate abstraction is also employed to improve candidate invariants
formed by Craig interpolants to reduce the number of refinements~\cite{UFO}.
In hardware model checking,
IMC is also combined with inductive invariants to avoid spurious counterexamples~\cite{Cabodi09}.
Different from continuously-refined invariants commonly used in software verification,
this approach computes a single invariant by SAT solving and
uses it repeatedly throughout the entire verification process.
\spacer~\cite{SPACER},
the backend engine of the verification framework \seahorn~\cite{SEAHORN},
uses invariants produced by the abstract-interpretation tool \ikos~\cite{IKOS}
to speed up the analysis.

%% file: background.tex
\section{Background}
\label{sect:background}

In this section,
we provide necessary background knowledge for the rest of the paper.
All used logical formulas are quantifier-free and
belong to the first-order theory of equality with
uninterpreted functions, arrays, bit-vectors, and floats.
We consider the satisfiability and validity of a logical formula with respect to this theory.
A first-order predicate over state variables is interpreted interchangeably
as a set of system states that satisfy the predicate.

\subsection{Model Checking}
\label{sect:background-model-checking}
First, we recap the problem formulation of model checking.
To simplify the presentation in~\cref{sect:imc-invariants},
we describe model checking with the notation of state-transition systems
and view a program as a state-transition system.
The implementation of the proposed methods,
discussed in~\cref{sect:implementation},
represents a program as a control-flow automaton
and verifies C programs.

\subsubsection{Describing State-Transition Systems}
A state-transition system $M$ is characterized
by two predicates $I(s)$ and $T(s,s')$ over state variables:
$I(s)$ evaluates to true if state~$s$ is an initial state of $M$;
$T(s,s')$ evaluates to true if~$M$ can transit from state~$s$ to state~$s'$.
In the following,
a state-transition system is represented by $M=(I(s),T(s,s'))$ and
state variables after a transition are denoted with a prime.
A primed variable is used for a state after transition.
We write $R(s)$ to represent the set of all reachable states of $M$.

\subsubsection{Verifying Safety Properties}
Model checking concerns if a state-transition system fulfills a safety property.
A safety property can be described as a predicate $P(s)$
that evaluates to true if state~$s$ satisfies the property.
Given a state-transition system $M=(I(s),T(s,s'))$ and a safety property $P(s)$,
$M$ fulfills $P(s)$ if $R(s) \implies P(s)$ is valid,
namely, the property is satisfied by all reachable states of~$M$.
Model-checking algorithms receive $M=(I(s),T(s,s'))$ and $P(s)$ as input
and aim at proving or disproving $R(s) \implies P(s)$.

We discuss two criteria widely used by model-checking algorithms
to establish $R(s) \implies P(s)$.
First, some approaches construct an \emph{inductive} set $F(s)$ that implies $P(s)$.
That is, $F(s)$ must conform to the constraint below:
\begin{align*}
    (I(s) \implies F(s)) \land (F(s) \land T(s,s') \implies F(s')) \land (F(s) \implies P(s)).
\end{align*}
Since $R(s)$ is the smallest inductive set of~$M$,
we know that $R(s) \implies F(s)$ holds.
The implication $R(s) \implies P(s)$ follows from
$R(s) \implies F(s)$ and $F(s) \implies P(s)$.
Interpolation-based approaches such as IMC~\cite{McMillanCraig}
and interpolation-sequence-based model checking~\cite{VizelFMCAD09}
fall into this category
because they try to build an inductive state set $F(s)$ from Craig interpolants~\cite{Craig57}.

The second criterion takes advantage of \emph{invariants}.
An invariant $\inv(s)$ of~$M$ is a predicate that holds for all reachable states of $M$,
namely, $R(s) \implies \inv(s)$ is valid.
Given an invariant $\inv(s)$ of~$M$,
some methods produce a set $G(s)$ that
is \emph{relatively inductive} to $\inv(s)$
and implies $P(s)$.
In other words, $G(s)$ must fulfill:
\begin{align}\label{eq:relative-induction}
    (I(s) \implies G(s)) \land (\inv(s) \land G(s) \land T(s,s') \implies G(s')) \land (G(s) \implies P(s)).
\end{align}
Because $\inv(s)$ holds for every reachable state
and $G(s)$ is relatively inductive to $\inv(s)$,
we have $R(s) \implies G(s)$.
The implication $R(s) \implies P(s)$ is thus concluded from
$R(s) \implies G(s)$ and $G(s) \implies P(s)$.
IC3/PDR~\cite{IC3} is a prominent example based on this criterion.
It chooses $G(s)$ to be the safety property $P(s)$
and generates clauses to form an invariant to which $P(s)$ is relatively inductive.
As will be seen in~\cref{sect:imc-invariants},
instead of producing invariants internally during model checking,
we leverage external invariant generators and augment IMC
to compute a relatively inductive set to the auxiliary invariants.

\subsection{Interpolation-Based Model Checking}
IMC is an algorithm proposed by McMillan in 2003~\cite{McMillanCraig}.
Originally designed for hardware model checking,
it has been adopted to verify software programs recently~\cite{IMC-JAR}.
IMC extends BMC by constructing inductive invariants (i.e., fixed points)
from \emph{Craig interpolants}~\cite{Craig57}.

\subsubsection{Craig Interpolation}
Given two formulas~$A$ and~$B$,
if $A \implies B$,
Craig's interpolation theorem~\cite{Craig57} assures the existence of
a formula~$\itp$ such that
$A \implies \itp$ and $\itp \implies B$ are valid,
and $\itp$ only involves variables appearing in both~$A$ and~$B$.
$\itp$ is called an \emph{interpolant} of~$A$ and~$B$
as it is logically between~$A$ and~$B$.
In the model-checking community,
Craig's interpolation theorem is usually stated in the equivalent form below.

\begin{theorem}\label{thm:Craig}
    Given an unsatisfiable formula $A \land B$,
    there exists an interpolant~$\itp$ of this formula such that
    (1)~$A \implies \itp$ is valid,
    (2)~$\itp \land B$ is unsatisfiable, and
    (3)~$\itp$ only refers to the common variables of~$A$ and~$B$.
\end{theorem}

\subsubsection{Computational Stages in IMC}
IMC~\cite{McMillanCraig} has two nested stages in its computation:
The outer stage unrolls a state-transition system and poses BMC queries to satisfiability solvers;
the inner stage derives Craig interpolants and constructs fixed points.
In the following,
we name the outer stage \emph{BMC stage} and
the inner stage \emph{interpolation stage}.

In the BMC stage,
a state-transition system is unfolded into several copies,
which is controlled by an unrolling counter.
Suppose the value of the counter is~$k$.
A BMC query depicting all possible paths from an initial state
to a property-violating state via at most $k$~transitions is posed:
\begin{align}
    \underbrace{I(s_0)T(s_0,s_1)}_{\text{$A(s_0,s_1)$}}
    \underbrace{T(s_1,s_2)\ldots T(s_{k-1},s_k)(\neg P(s_1) \lor\ldots\lor \neg P(s_k))}_{\text{$B(s_1,s_2,\ldots,s_k)$}}.
    \label{eq:BMC}
\end{align}
In the above formula,
variable~$s_i$ denotes the state variable after the $i$-th transition.
If~\cref{eq:BMC} is satisfiable,
a feasible counterexample to the safety property is found.
Otherwise, IMC enters the interpolation stage.

In the interpolation stage,
a BMC query such as~\cref{eq:BMC} is partitioned into two formulas $A$ and $B$.
According to~\cref{thm:Craig},
an interpolant $\itp_1(s_1)$ for~\cref{eq:BMC} exists such that:
\begin{align*}
     & I(s_0)T(s_0,s_1) \rightarrow \itp_1(s_1)\text{~is valid and}                                                  \\
     & \itp_1(s_1)\land\bigwedge_{i=1}^{k-1}T(s_i,s_{i+1})\land\bigvee_{i=1}^{k}\neg P(s_i)\text{ is unsatisfiable.}
\end{align*}
Conceptually, $\itp_1(s_1)$ summarizes the reason why~\cref{eq:BMC} is unsatisfiable:
It overapproximates the set of states that are
(1)~reachable via one transition from an initial state and
(2)~do not violate the safety property within $(k-1)$ transitions.

A fixed point of the state-transition system
may be constructed by computing such interpolants iteratively.
Substituting variable~$s_0$ into $\itp_1$
and replacing the initial states $I(s_0)$ in~\cref{eq:BMC} by~$\itp_1(s_0)$,
the interpolation stage of IMC poses another BMC query
from the first interpolant~$\itp_1$:
\begin{align*}
    \underbrace{\itp_1(s_0)T(s_0,s_1)}_{\text{$A'(s_0,s_1)$}}
    \underbrace{T(s_1,s_2)\ldots T(s_{k-1},s_k)(\neg P(s_1) \lor\ldots\lor \neg P(s_k))}_{\text{$B'(s_1,s_2,\ldots,s_k)$}}
\end{align*}
If the formula is still unsatisfiable,
a second interpolant $\itp_2(s_1)$ can be derived,
which overapproximates the set of states reachable from an initial state via two transitions.
Replacing~$\itp_1$ with~$\itp_2$,
one can derive the next interpolant~$\itp_3$ if
the BMC query starting from~$\itp_2$ is again unsatisfiable.
Suppose the above process is repeated~$n$ times,
and a list of interpolants $\itp_1,\itp_2,\ldots,\itp_n$ is derived.
Upon the derivation of the newest interpolant~$\itp_n$,
IMC performs a \emph{fixed-point check}
to decide whether a fixed point has been reached:
If $\itp_n$ is contained in the union of the initial states and previous interpolants,
namely, $\itp_n \implies I \lor \bigvee_{j=1}^{n-1}\itp_j$ holds,
the set $I \lor \bigvee_{j=1}^{n-1}\itp_j$ is an inductive invariant of the system~\cite{McMillanCraig}.
The inductive invariant is also safe
because every interpolant does not violate the safety property
according to the second condition of~\cref{thm:Craig}.
As a result, IMC concludes the system preserves the safety property.

If a BMC query starting from an interpolant is satisfiable,
the safety property is not definitely violated.
The corresponding error path might be spurious
because the interpolant could involve unreachable states.
In this situation,
IMC will return to the BMC stage and increment the unrolling counter
to confirm the existence of a feasible error path.

%% file: approach.tex
\section{Augmenting IMC with Auxiliary Invariants}
\label{sect:imc-invariants}

We propose two methods to augment IMC with auxiliary invariants.
The first one strengthens the fixed-point checks,
and the second one strengthens the interpolants
derived in the interpolation stage.
Given a state-transition system $M=(I(s),T(s,s'))$,
we further assume that auxiliary invariants are inductive.
In other words, an invariant $\inv(s)$ satisfies
$I(s) \implies \inv(s)$ and $\inv(s) \land T(s,s') \implies \inv(s')$.
This assumption is reasonable in practice because
invariant generators,
such as those based on data-flow analysis,
usually perform a fixed-point iteration and produce inductive invariants.

\subsection{Approach 1: Strengthening Fixed-Point Checks}
\label{sect:imc-strengthen-fixed-point}
The first approach strengthens fixed-point checks
by restricting the newest interpolant with an auxiliary invariant.
Let $\itp_1,\ldots,\itp_n$ be a list of interpolants
derived in the interpolation stage.
Instead of checking $\itp_n \implies I \lor \bigvee_{j=1}^{n-1}\itp_j$,
we strengthen the check to $\inv \land \itp_n \implies I \lor \bigvee_{j=1}^{n-1}\itp_j$.
The intuition is to confine the scope of fixed-point checks
mainly within reachable states of the system such that
the union of the initial states and previous interpolants
is more likely to contain the newest interpolant.
The correctness of this approach is stated in~\cref{thm:strengthen-fixed-point}.

\begin{theorem}\label{thm:strengthen-fixed-point}
    Given a state-transition system $M=(I(s),T(s,s'))$ and a safety property~$P(s)$,
    let $\itp_1,\ldots,\itp_n$ be a list of interpolants
    derived in the interpolation stage of IMC.
    If the strengthened fixed-point check
    $\inv \land \itp_n \implies I \lor \bigvee_{j=1}^{n-1}\itp_j$
    holds for some auxiliary invariant $\inv$,
    $M$ fulfills the safety property~$P(s)$.
\end{theorem}
\begin{proof}
    We rely on the criterion described by~\cref{eq:relative-induction}
    to show that~$M$ fulfills~$P$.
    Defining $G$ to be $I \lor \bigvee_{j=1}^{n-1}\itp_j$,
    we will prove that $G$ satisfies~\cref{eq:relative-induction}.

    Proving $I \implies G$ is trivial.
    Moreover, \cref{thm:Craig} guarantees that
    $\itp_1,\ldots,\itp_n$ do not violate the safety property,
    which assures that $G \implies P$ holds.

    To show $\inv(s) \land G(s) \land T(s,s') \implies G(s')$ is valid,
    recall that both $I(s) \land T(s,s') \implies \itp_1(s')$ and
    $\itp_j(s) \land T(s,s') \implies \itp_{j+1}(s')$ for $j=1,\ldots,n-1$
    hold according to~\cref{thm:Craig}.
    Combining these conditions and the inductiveness of the auxiliary invariant $\inv$,
    we simplify the implication to
    \begin{align*}
        \inv(s) \land (I(s) \lor \bigvee_{j=1}^{n-1}\itp_j(s)) \land T(s,s') \implies
        \inv(s') \land (\bigvee_{j=1}^{n-1}\itp_j(s') \lor \itp_n(s')).
    \end{align*}
    Since the strengthened fixed-point check
    $\inv \land \itp_n \implies I \lor \bigvee_{j=1}^{n-1}\itp_j$ holds,
    the right-hand side of the above implication further implies
    $(\inv(s') \land \bigvee_{j=1}^{n-1}\itp_j(s')) \lor (I(s') \lor \bigvee_{j=1}^{n-1}\itp_j(s'))$,
    which equals $G(s')$.

    Therefore, we proved that $G$ satisfies~\cref{eq:relative-induction},
    and hence~$M$ fulfills~$P$.
\end{proof}

\subsection{Approach 2: Strengthening Interpolants}
\label{sect:imc-strengthen-interpolant}

The second approach strengthens interpolants
by conjoining them with an auxiliary invariant.
Unlike the first approach,
which only restricts the newest interpolant
with an auxiliary invariant in a fixed-point check,
the second approach replaces the original interpolants
returned by the interpolation procedure with the strengthened ones.
That is, given a BMC query starting from the initial states $I$
or a previously strengthened interpolant $\itp_j \land \inv$,
the interpolant $\itp_{j+1}$ is replaced by $\itp_{j+1} \land \inv$.
Note that if interpolants are strengthened,
fixed-point checks are effectively strengthened as well.
\cref{thm:strengthen-interpolant} states that $\itp_{j+1} \land \inv$
is also an interpolant for the BMC query,
and hence the correctness of the approach follows from the plain IMC algorithm.

\begin{theorem}\label{thm:strengthen-interpolant}
    Given a transition system $M=(I(s),T(s,s'))$ and a safety property~$P(s)$,
    consider an unsatisfiable BMC query
    $\start(s_0) \land T(s_0,s_1) \land \ldots \land T(s_{k-1},s_k) \land \bigvee_{i=1}^{k}\neg P(s_i)$
    posed in the interpolation stage of IMC,
    where $\start(s_0)$ is either $I(s_0)$ or
    a previously strengthened interpolant $\itp_j(s_0) \land \inv(s_0)$.
    Suppose $\itp_{j+1}(s_1)$ is an interpolant of the BMC query
    with $\start(s_0) \land T(s_0,s_1)$ labeled as formula~$A$ and
    $T(s_1,s_2) \land \ldots \land T(s_{k-1},s_k) \land \bigvee_{i=1}^{k}\neg P(s_i)$ labeled as formula~$B$.
    Then $\itp_{j+1}(s_1) \land \inv(s_1)$ is also an interpolant of the BMC query.
\end{theorem}
\begin{proof}
    To prove that $\itp_{j+1}(s_1) \land \inv(s_1)$ is also an interpolant,
    we have to show that it satisfies the three conditions in~\cref{thm:Craig}.
    Namely,
    (1)~$\start(s_0) \land T(s_0,s_1) \implies \itp_{j+1}(s_1) \land \inv(s_1)$ is valid;
    (2)~$\itp_{j+1}(s_1) \land \inv(s_1) \land T(s_1,s_2) \land \ldots
        \land T(s_{k-1},s_k) \land \bigvee_{i=1}^{k}\neg P(s_i)$ is unsatisfiable; and
    (3)~$\itp_{j+1}(s_1) \land \inv(s_1)$ only refers to the common variables of formulas~$A$ and~$B$.
    Condition~(2) holds because
    $\itp_{j+1}(s_1) \land T(s_1,s_2) \land \ldots
        \land T(s_{k-1},s_k) \land \bigvee_{i=1}^{k}\neg P(s_i)$ is already unsatisfiable
    according to~\cref{thm:Craig}.
    Condition~(3) holds because $\itp_{j+1}(s_1) \land \inv(s_1)$ only uses the common variable~$s_1$.
    In the following, we prove condition~(1) by splitting into two cases.

    When $\start(s_0)=I(s_0)$,
    our proof goal is $I(s_0) \land T(s_0,s_1) \implies \itp_{j+1}(s_1) \land \inv(s_1)$.
    Applying~\cref{thm:Craig} to $\itp_{j+1}$,
    we have $I(s_0) \land T(s_0,s_1) \implies \itp_{j+1}(s_1)$.
    Since $\inv$ is an invariant and hence contains all reachable states,
    we also have $I(s_0) \land T(s_0,s_1) \implies \inv(s_1)$.
    Therefore, our proof goal is achieved.

    When $\start(s_0)=\itp_j(s_0)\land\inv(s_0)$,
    our proof goal is $\itp_j(s_0)\land\inv(s_0) \land T(s_0,s_1) \implies
        \itp_{j+1}(s_1) \land \inv(s_1)$.
    Applying~\cref{thm:Craig} to $\itp_{j+1}$,
    we have $\itp_j(s_0)\land\inv(s_0) \land T(s_0,s_1) \implies \itp_{j+1}(s_1)$.
    Since $\inv$ is inductive,
    we also have $\inv(s_0) \land T(s_0,s_1) \implies \inv(s_1)$.
    Our goal follows from the two implications.

    As both cases are proved,
    we conclude that $\itp_{j+1}(s_1) \land \inv(s_1)$
    is also an interpolant of the BMC query
    and can be used in the interpolation stage of IMC.
\end{proof}

\subsection{Comparison between the Two Approaches}
\label{sect:imc-strengthen-discussion}

The design spirit behind the two approaches is to make IMC
capable of proving the safety of a system
with fewer iterations in the BMC and interpolation stages.
We attain this objective with the help of auxiliary invariants.
In the first approach,
instead of an inductive invariant,
IMC generates a relatively inductive set of states,
which might require fewer interpolation calls to produce.
In the second approach,
in addition to strengthening fixed-point checks,
strengthened interpolants also make BMC queries
in the interpolation stage more likely to be unsatisfiable.
That is,
IMC will remain in the interpolation stage more often,
searching for a proof of the safety property.
The second approach is more ``aggressive'' than the first one because
strengthened interpolants will change the interpolants obtained later,
and IMC will follow a different computational footprint.
By contrast,
the interpolants encountered in the first approach
are identical to the plain IMC algorithm
if the interpolation procedure is the same.
Moreover, the first approach can in principle be adopted by other verification algorithms
with similar fixed-point checks,
whereas the second approach is specifically tailored towards IMC.
In our evaluation, we also experimented with a basic injection approach
that conjoins the safety property with auxiliary invariants.
However, this approach performed roughly the same as the plain IMC algorithm.

\subsection{Augmented IMC Procedure}

\newcommand{\strf}{\mathit{strengthen\_fpc}}
\newcommand{\stri}{\mathit{strengthen\_itp}}
\begin{algorithm}[t]
    \caption{Augmenting IMC~\cite{IMC-JAR} with Auxiliary Invariants}
    \label{alg:strengthened-imc}
    \begin{algorithmic}[1]
        \REQUIRE
        a state-transition system~$M=(I(s),T(s,s'))$, a safety property~$P(s)$,\\
        an upper bound $k_{max}$ for the unrolling counter $k$,\\
        a Boolean flag~$\strf$ to enable strengthening fixed-point checks, and\\
        a Boolean flag~$\stri$ to enable strengthening interpolants
        \ENSURE
        \TRUE{} if~$M$ fulfills~$P$;
        \FALSE{} if a feasible error path is found;
        \textbf{unknown} otherwise
        \VARDECL the unrolling counter $k$,\\
        the auxiliary invariant computed by an external invariant generator $\inv(s)$,\\
        the candidate (relatively) inductive set $G(s)$,
        the computed interpolant $\itp$, and\\
        the two formulas $A(s_0, s_1)$ and $B(s_1, \ldots, s_k)$ for interpolation

        \IF{$\mathsf{sat}$($I(s_0) \land \lnot P(s_0)$)}
        \RETURN \FALSE{}
        \ENDIF
        \STATE $k := 1$
        \STATE $\inv(s) := \top$
        \WHILE{$k \leq k_{max}$}
        \STATE $G(s) := I(s)$
        \STATE $A(s_0, s_1) := I(s_0) \land T(s_0,s_1)$
        \STATE $B(s_1, \ldots, s_k) := \bigwedge_{i=1}^{k-1} T(s_i,s_{i+1}) \land (\bigvee_{i=1}^{k} \lnot P(s_i))$
        \IF{$\mathsf{sat}$($A(s_0, s_1) \land B(s_1, \ldots, s_k)$)}
        \RETURN \FALSE{} \hfill \COMMENT{BMC finds a feasible error path}
        \ENDIF
        \IF{$\strf \lor \stri$}
        \STATE $\inv(s) := \mathsf{get\_auxiliary\_invariant}()$
        \label{alg:get-auxiliary-invariants}
        \ENDIF
        \WHILE{$\lnot\mathsf{sat}$($A(s_0, s_1) \land B(s_1, \ldots, s_k)$)}
        \STATE $\itp(s_1) := \mathsf{get\_interpolant}(A(s_0, s_1), B(s_1, \ldots, s_k))$\label{alg:get-interpolant}
        \IF{$\stri$}
        \STATE $\itp(s) := \inv(s)\land\itp(s)$
        \label{alg:strengthen-itp}
        \ENDIF
        \IF{$\strf$}
        \IF{$\lnot\mathsf{sat}$($\inv(s)\land\itp(s)\land\lnot G(s)$)}
        \label{alg:strengthen-fpc}
        \RETURN \TRUE{} \hfill \COMMENT{$G$ is relatively inductive to $\inv$}
        \ENDIF
        \ELSE
        \IF{$\lnot\mathsf{sat}$($\itp(s)\land\lnot G(s)$)}
        \label{alg:imc-fpc}
        \RETURN \TRUE{} \hfill \COMMENT{$G$ is an inductive invariant}
        \ENDIF
        \ENDIF
        \STATE $G(s) := G(s) \lor \itp(s)$ \hfill \COMMENT{Add new states to $G$}
        \STATE $A(s_0, s_1) := \itp(s_0) \land T(s_0,s_1)$\hfill \COMMENT{Start next BMC query from new states}
        \ENDWHILE
        \STATE $k := k+1$
        \ENDWHILE
        \RETURN \textbf{unknown}
    \end{algorithmic}
\end{algorithm}

The proposed procedure to augment IMC with auxiliary invariants
is summarized in~\cref{alg:strengthened-imc}.
It has two arguments $\strf$ and $\stri$ to enable the discussed strengthening
of fixed-point checks and interpolants, respectively,
on top of the original IMC algorithm.
An auxiliary invariant~$\inv(s)$ is injected into IMC at~\cref{alg:get-auxiliary-invariants}
if the current BMC query is unsatisfiable.
If~$\strf$ is true,
the first approach to strengthen fixed-point checks is performed at~\cref{alg:strengthen-fpc}.
If~$\stri$ is true,
the second approach to strengthen interpolants is performed at~\cref{alg:strengthen-itp}.
Observe that fixed-point checks are also strengthened if interpolants are strengthened:
\cref{alg:strengthen-fpc} and \cref{alg:imc-fpc} are equivalent
if the derived interpolant~$\itp$ at~\cref{alg:get-interpolant}
is replaced by $\inv\land\itp$.

%% file: implementation.tex
\section{Implementation}
\label{sect:implementation}

To investigate whether invariant injection is helpful for IMC,
we apply the augmented IMC procedure in~\cref{alg:strengthened-imc} to verify programs.
The proposed algorithm was implemented in~\cpachecker~\cite{CPACHECKER},
a software-verification framework for the programming language C.
We chose~\cpachecker because it provides
an implementation~\cite{IMC-JAR} of the plain IMC algorithm.
The IMC implementation extracts transition relations from programs
with \emph{large-block encoding}~\cite{LBE}.
Error labels in a program are used to specify the negation of safety properties.
In addition to an IMC implementation,
\cpachecker also has a continuously-refining invariant generator based on intervals~\cite{CPA-DF},
which is used to enhance \kinduction~\cite{kInduction}.
Before continuing the discussion,
we emphasize that the proposed approaches discussed in~\cref{sect:imc-invariants} are
independent of the implementation framework and not limited to specific invariant generators.

\cpachecker is based on \emph{configurable program analysis} (CPA)~\cite{CPA},
which defines an abstract domain used in a program analysis.
The plain IMC algorithm is implemented using several CPAs,
with one of them tracking path formulas
between program locations in its abstract states~\cite{ABE}.
This CPA is configured to perform large-block encoding,
which constructs formulas for complete loop unrollings as transition relations~\cite{IMC-JAR}.
Thanks to its flexibility,
it underpins the implementations of other SMT-based algorithms
in \cpachecker~\cite{AlgorithmComparison-JAR},
including predicate analysis, \impact, and \kinduction.
We will compare the proposed algorithm to these techniques in~\cref{sect:evaluation}.

The continuously-refining invariant generator in~\cpachecker uses a CPA
with an abstract domain based on expressions over intervals~\cite{CPA-DF}.
Compared to methods concerning single intervals only,
it can represent complex ranges like the disjunction and conjunction of intervals,
e.g., $(x<5\lor{}x>7)\land{}(1\leq{}y\leq{}8)$.
Its precision includes a set of important variables and a maximum depth of expressions.
The precision can be dynamically adjusted during the analysis~\cite{CPAplus}
to produce more refined invariants.

Given a CPA and an initial abstract state,
\cpachecker constructs a set of reachable abstract states
by iteratively processing states and computing their abstract successors.
To perform a specific algorithm,
required information can be collected from the abstract states.
For example, IMC collects the path formulas to assemble BMC queries
and derive interpolants~\cite{IMC-JAR}.
The interval-based invariant generator outputs
the union of expressions as invariants~\cite{CPA-DF}.
In our implementation,
we run~\cref{alg:strengthened-imc} and
the invariant generator in parallel
and inject auxiliary invariants to augment IMC.

%% file: evaluation.tex
\newcommand\plotpath{evaluation/tex}
\input{\plotpath/plot-defs}
\newcommand{\ntasks}{\num{\EvalSlDfSImcHardSafeTotalCount}}
\newcommand{\nreducedtasks}{\ReducedEvalSlDfSImcHardSafeTotalCount}

\section{Evaluation}
\label{sect:evaluation}

To understand the effects of the proposed invariant-injection approaches
on the proof-finding ability of IMC,
we pose the following research questions:

\begin{itemize}
  \item Part 1: augmented IMC vs. plain IMC
        \begin{itemize}
          \item \textbf{RQ1}: Can auxiliary invariants reduce the numbers of program unrollings and interpolation queries?
          \item \textbf{RQ2}: Can augmented IMC prove the correctness of additional programs?
          \item \textbf{RQ3}: Can invariant injection improve the run-time efficiency?
        \end{itemize}
  \item Part 2: augmented IMC vs. other approaches and tools
        \begin{itemize}
          \item \textbf{RQ4}: Can augmented IMC find more proofs than other SMT-based software-verification algorithms in~\cpachecker?
          \item \textbf{RQ5}: Can augmented IMC find more proofs than other state-of-the-art software verifiers?
        \end{itemize}
\end{itemize}

\subsection{Evaluated Approaches and Tools}
\label{sect:evaluation-approaches}

To answer the above research questions,
we evaluated the proposed invariant-injection methods for IMC
against (1)~its plain version plus three other SMT-based algorithms
in the same verification framework \cpachecker~\cite{CPACHECKER} and
(2)~two other state-of-the-art software verifiers.

The plain IMC algorithm was recently adopted to verify programs~\cite{IMC-JAR}.
The three other approaches in \cpachecker are
predicate abstraction~\cite{AbstractionsFromProofs,GrafSaidi97},
\impact~\cite{IMPACT},
and \kinduction boosted by continuously-refined invariants~\cite{kInduction},
whose characteristics were compared in a recent article on
SMT-based software verification~\cite{AlgorithmComparison-JAR}.
All proposed methods and compared algorithms above are
implemented in the same framework,
so the confounding variables are kept to a minimum
(identical frontend, program encodings, SMT solvers, etc.)
to facilitate the comparison of algorithmic differences.
For invariant generation,
we used the continuously-refining data-flow analysis (DF)~\cite{CPA-DF} described in~\cref{sect:implementation}.
The invariant generator can prove the safety properties of some programs on its own.
In our experiments, we ignored the answers computed by DF
because our goal is to study the effects of auxiliary invariants
on the main analyses, namely, IMC and \kinduction.
In the following,
we denote the injection of continuously-refined invariants
generated by DF into \kinduction as \kidf{},
and the augmented IMC algorithms with
fixed-point checks and interpolants strengthened by auxiliary invariants
as IMC\injecttof{}DF and IMC\injecttoi{}DF, respectively.

To reflect the state of the art on software verification and invariant injection,
we further compared our approaches to
\twols~\cite{kikiKroening},
which has a mature implementation of \kinduction boosted by auxiliary invariants,
and \symbiotic~\cite{SymbioticApproach},
which is based on symbolic execution
and performs well in the Intl. Competitions on Software Verification~\cite{SVCOMP22}
(the overall winner in 2022).

\subsection{Benchmark Set}
\label{sect:experimental-benchmark}

Since our goal is to investigate whether auxiliary invariants
can improve IMC's capability of delivering correctness proofs,
we selected verification tasks without known violation
to their reachability-safety properties
from the 2022 Intl. Competition on Software Verification (SV-COMP\,'22)~\cite{SVCOMP22}.
Furthermore, to concentrate the evaluation on hard verification problems,
we excluded the trivial tasks solvable by BMC of~\cpachecker.
To keep the program encodings consistent across all evaluated approaches in \cpachecker,
we only considered single-loop programs in our experiments
because the IMC implementation in~\cpachecker needs to transform
a multi-loop program into a single-loop program as preprocessing~\cite{IMC-JAR}.
However, this is not a limitation of the proposed augmenting methods,
since they can work on multi-loop programs as well
if single-loop transformation~\cite{DragonBook,kIndForDMARaces} is applied.
In total, we collected \ntasks~verification tasks,
each of which contains a single-loop program and a challenging safety property.

\subsection{Experimental Setup}
\label{sect:experimental-setup}

All experiments were conducted on machines
using a 3.4\,GHz CPU (Intel Xeon E3-1230~v5)
with 8~processing units and 33\,GB of RAM.
The operating system was Ubuntu 22.04 (64 bit),
using Linux~5.15 and OpenJDK~17.0.5.
Each verification task was limited to 4~CPU cores,
\SI{15}{min} of CPU time,
and \SI{15}{GB} of RAM.
In our evaluation, we utilized the benchmarking framework
\benchexec~\cite{Benchmarking-STTT}
to ensure reproducibility of our results,
and \cpachecker at revision
\href{https://svn.sosy-lab.org/software/cpachecker/branches/imc-with-invariants@\CpacheckerRev}{\texttt{\CpacheckerRev}}
of branch \texttt{imc-with-invariants}.
All SMT and interpolation queries in~\cpachecker were handled by~\mathsat~\cite{MATHSAT5}.
Since we want to observe IMC's behavior in the presence of auxiliary invariants,
we limited the CPU time allocated to the invariant generator to \SI{2.5}{min}
such that IMC had enough time to perform its analysis.
For the comparison to other software verifiers,
we downloaded \twols and \symbiotic from the tool archives of SV-COMP\,'22~\cite{SVCOMP22-TOOLS-artifact}.

\subsection{Results}
\label{sect:experimental-results}


\subsubsection{RQ1: Reduction of program unrollings and interpolation queries}
To understand whether the invariant-injection techniques
could reduce the numbers of program unrollings and interpolation queries
required by plain IMC to find a proof,
we conducted a case study on IMC\injecttoi{}DF.
As discussed in~\cref{sect:imc-strengthen-discussion},
IMC\injecttoi{}DF derives different interpolants from plain IMC
and is supposed to exhibit distinct computational behavior.
We further identified \nreducedtasks{}~tasks for which DF was able to generate
non-trivial and inductive invariants%
\footnote{The trivial invariant $\top$ represents the entire state space.
}
because trivial invariants provide no additional information to help IMC.
The scatter plots~\cref{fig:evaluation:scatter-k} and~\cref{fig:evaluation:scatter-itpcall}
show the comparison of IMC\injecttoi{}DF and IMC on the \nreducedtasks{}~tasks
on the numbers of program unrollings and interpolation queries, respectively.
A data point ($x$, $y$) in the plots means that there is a task
solvable by both IMC and IMC\injecttoi{}DF,
and IMC (resp. IMC\injecttoi{}DF) requires $x$ (resp. $y$) times of the indicated operation.
The color of a data point shows the number of tasks falling into this coordinate.
Data points under the diagonal represent the tasks for which
IMC\injecttoi{}DF needed fewer operations than plain IMC.
Observe that IMC\injecttoi{}DF often required
fewer program unrollings and interpolation calls.
For programs correctly proved by both,
our augmenting approach reduced
the number of program unrollings in \nTasksFewerK{}~tasks,
and the number of interpolation calls in \nTasksFewerItpCall{}~tasks.

\newsavebox{\scatterK}
\begin{lrbox}{\scatterK}
  \begin{minipage}[b]{0.5\textwidth}
    \centering
    \scalebox{1.4}{\input{\plotpath/scatter-k}}
  \end{minipage}
\end{lrbox}
\newsavebox{\scatterItpCall}
\begin{lrbox}{\scatterItpCall}
  \begin{minipage}[b]{0.5\textwidth}
    \centering
    \scalebox{1.4}{\input{\plotpath/scatter-itpcall}}
  \end{minipage}
\end{lrbox}
\begin{figure*}[t]
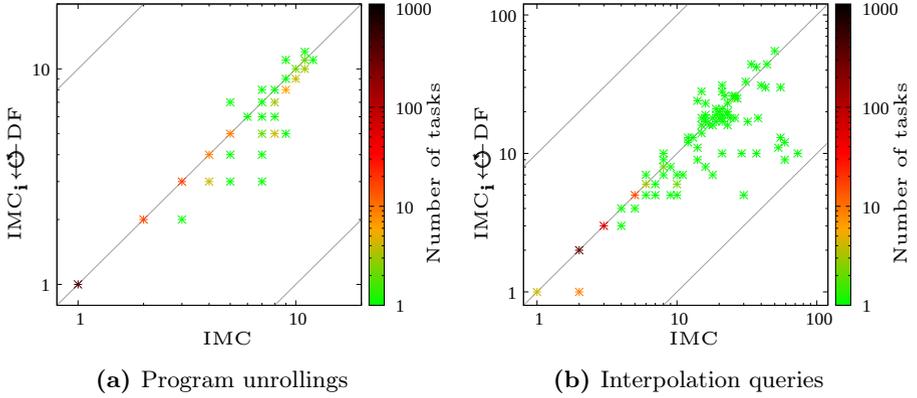

  \centering
  \subfloat[Program unrollings]{\usebox{\scatterK}\label{fig:evaluation:scatter-k}}
  \subfloat[Interpolation queries]{\usebox{\scatterItpCall}\label{fig:evaluation:scatter-itpcall}}\\
  \caption{Comparison of the numbers of program unrollings and interpolation queries}
  \label{fig:evaluation:scatter-discrete}
\end{figure*}

\Cref{tab:top-results} shows the tasks
for which augmented IMC exhibited significant performance improvement.
Upon these tasks,
plain IMC either ran into timeouts or
required considerably more run-time,
whereas the augmented versions were able to deliver proofs efficiently.
Under each algorithm, the table lists
the verification result~(solved or timeout),
the numbers of program unrollings and interpolation queries,
wall-time,
and interpolation time~(the time spent on all interpolation queries).
Note that since plain IMC suffered from timeouts in most of these tasks,
the reported numbers of unrollings and interpolation queries are lower bounds of the actual numbers
required to compute proofs, which could be much higher.

\begin{table*}[t]
  \smaller
  \centering
  \caption{Statistics of tasks with significant improvement
    (time unit: \second{} with two significant digits;
    \solved for solved, and \timeout for timeout)
  }
  \label{tab:top-results}
  \newcommand\precnum[1]{\tablenum[table-format=3.0]{#1}}
  \newcommand\rndnum[1]{\tablenum[round-mode=figures,round-precision=2,table-format=2.0]{#1}}
  \scalebox{0.94}{\input{\plotpath/imc-ignDfi-top-results}}
  \\[1mm]
  \scalebox{0.94}{\input{\plotpath/imc-ignDff-top-results}}
\end{table*}


\subsubsection{RQ2: Effectiveness of augmented IMC vs. plain IMC}
We compared the number of tasks solved by augmented IMC versus that by plain IMC
to evaluate the effectiveness of the proposed invariant-injection approaches.
\cref{tab:plain-vs-aug} summarizes the results
on the \nreducedtasks{}~tasks with non-trivial invariants.
Observe that
IMC\injecttof{}DF and IMC\injecttoi{}DF solved more tasks than plain IMC.
Specifically,
IMC\injecttoi{}DF (resp. IMC\injecttof{}DF) found proofs for
\ImcIgndfiPlusOverImcReducedCount{} (resp. \ImcIgndffPlusOverImcReducedCount) tasks
where plain IMC ran into timeouts.
However, it was not able to solve
\ImcIgndfiMinusOverImcReducedCount{} (resp. \ImcIgndffMinusOverImcReducedCount) tasks
proven by plain IMC,
partially due to the extra time spent on invariant generation
(see~\cref{sect:experimental-discussion} for more discussion).
Overall, IMC\injecttoi{}DF (resp. IMC\injecttof{}DF) proved
\ImcIgndfiNetOverImcReducedCount{} (resp. \ImcIgndffNetOverImcReducedCount)
additional tasks compared to plain IMC,
and there were 16 tasks solvable by IMC\injecttoi{}DF
but not by plain IMC or DF.
Although not extraordinary,
the increase shows invariant injection can improve the effectiveness of plain IMC.

\begin{table}[t]
  \centering
  \caption{Summary of the results for plain vs. augmented IMC on~\nreducedtasks{} tasks with non-trivial auxiliary invariants}
  \label{tab:plain-vs-aug}
  \newcommand\precnum[1]{\tablenum[table-format=4]{#1}}
  \begin{tabular}{l@{\hspace{2mm}}|@{\hspace{2mm}}c@{\hspace{2mm}}c@{\hspace{2mm}}c@{\hspace{2mm}}}
    \toprule
    Algorithm          & IMC                                                                     & IMC\injecttof{}DF & IMC\injecttoi{}DF \\
    \midrule
    Proofs             & \precnum{\ReducedEvalSlDfSImcHardSafeCorrectCount}
                       & \precnum{\ReducedEvalSlDfSImcIgndffHardSafeCorrectCount}
                       & \textbf{\precnum{\ReducedEvalSlDfSImcIgndfiHardSafeCorrectCount}}
    \\
    Timeouts           & \precnum{\ReducedEvalSlDfSImcHardSafeErrorTimeoutCount}
                       & \precnum{\ReducedEvalSlDfSImcIgndffHardSafeErrorTimeoutCount}
                       & \precnum{\ReducedEvalSlDfSImcIgndfiHardSafeErrorTimeoutCount}
    \\
    Out of memory      & \precnum{\ReducedEvalSlDfSImcHardSafeErrorOutOfMemoryCount}
                       & \precnum{\ReducedEvalSlDfSImcIgndffHardSafeErrorOutOfMemoryCount}
                       & \precnum{\ReducedEvalSlDfSImcIgndfiHardSafeErrorOutOfMemoryCount}
    \\
    Other inconclusive & \precnum{\ReducedEvalSlDfSImcHardSafeErrorOtherInconclusiveCount}
                       & \precnum{\ReducedEvalSlDfSImcIgndffHardSafeErrorOtherInconclusiveCount}
                       & \precnum{\ReducedEvalSlDfSImcIgndfiHardSafeErrorOtherInconclusiveCount}
    \\
    \bottomrule
  \end{tabular}
\end{table}

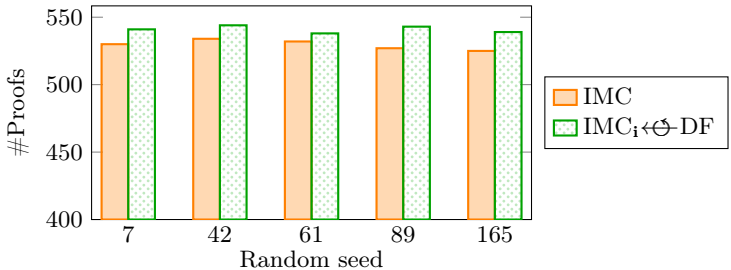
\begin{figure}[t]
  \centering
  \scalebox{1}{\input{\plotpath/bar-rand-seed-2d}}
  \caption{Numbers of proofs found by IMC and IMC\injecttoi{}DF using different random seeds for SMT solving (default seed: 42)}
  \label{fig:evaluation:bar-rand-seed}
\end{figure}

To account for the randomness in SMT solving and interpolation,
we repeated the experiment with five different random seeds for the underlying SMT solver.
The results are shown in~\cref{fig:evaluation:bar-rand-seed}.
Observe that IMC\injecttoi{}DF consistently found more proofs than plain IMC
with all five random seeds,
which demonstrates the robustness of the proposed invariant-injection approaches.

\newsavebox{\quantileCoopCpu}
\begin{lrbox}{\quantileCoopCpu}
  \begin{minipage}[b]{0.5\textwidth}
    \centering
    \scalebox{0.8}{\input{\plotpath/quantile-coop-cputime}}
  \end{minipage}
\end{lrbox}
\newsavebox{\quantileCoopWall}
\begin{lrbox}{\quantileCoopWall}
  \begin{minipage}[b]{0.5\textwidth}
    \centering
    \scalebox{0.8}{\input{\plotpath/quantile-coop-walltime}}
  \end{minipage}
\end{lrbox}
\begin{figure*}[t]
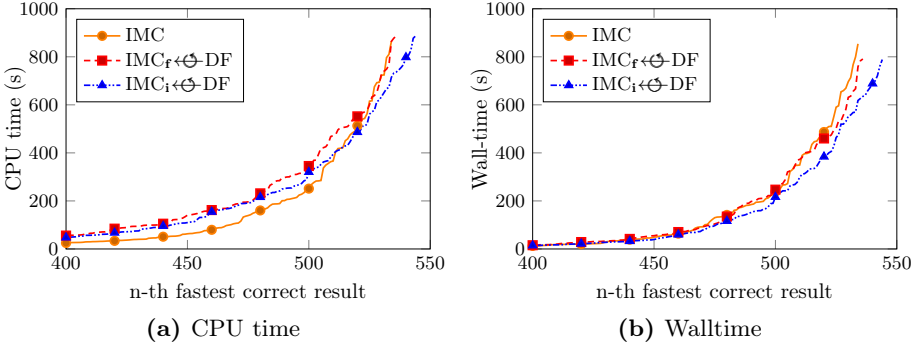

  \centering
  \subfloat[CPU time]{\usebox{\quantileCoopCpu}\label{fig:evaluation:quantile-coop-cpu}}
  \subfloat[Walltime]{\usebox{\quantileCoopWall}\label{fig:evaluation:quantile-coop-wall}}
  \caption{Quantile plots comparing the run-time of plain and augmented IMC}
  \label{fig:evaluation:quantile-coop}
\end{figure*}

\newsavebox{\scatterItpTime}
\begin{lrbox}{\scatterItpTime}
  \begin{minipage}[b]{0.5\textwidth}
    \centering
    \scalebox{0.7}{\input{\plotpath/scatter-itptime}}
  \end{minipage}
\end{lrbox}
\newsavebox{\scatterWalltime}
\begin{lrbox}{\scatterWalltime}
  \begin{minipage}[b]{0.5\textwidth}
    \centering
    \scalebox{0.7}{\input{\plotpath/scatter-walltime}}
  \end{minipage}
\end{lrbox}
\begin{figure*}[t]
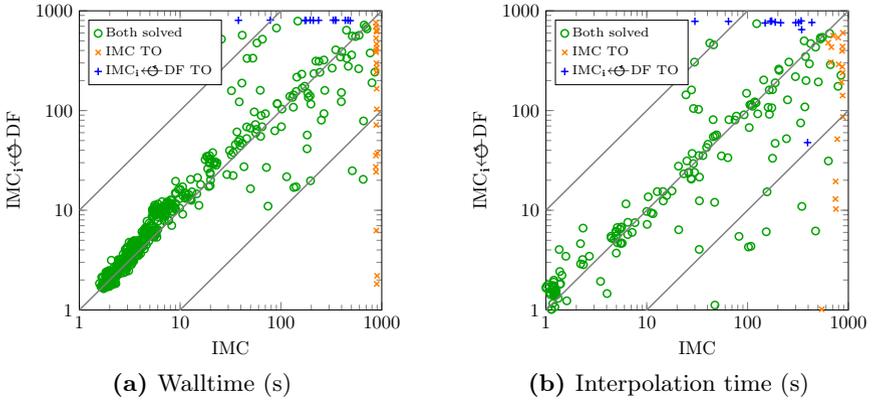

  \centering
  \subfloat[Walltime (\second)]{\usebox{\scatterWalltime}\label{fig:evaluation:scatter-walltime}}
  \subfloat[Interpolation time (\second)]{\usebox{\scatterItpTime}\label{fig:evaluation:scatter-itptime}}
  \caption{Comparison of wall-time and interpolation time (TO stands for timeout)}
  \label{fig:evaluation:scatter-continuous}
\end{figure*}

\subsubsection{RQ3: Run-time efficiency of augmented IMC vs. plain IMC}
Quantile plots comparing the CPU time and wall-time usage
of IMC, IMC\injecttof{}DF, and IMC\injecttoi{}DF
on the tasks with non-trivial invariants
are shown in~\cref{fig:evaluation:quantile-coop-cpu}
and~\cref{fig:evaluation:quantile-coop-wall}, respectively.
A data point ($x$, $y$) in the plots indicates
that $x$ tasks were correctly solved by the respective algorithm within
a time bound of $y$ seconds.
To focus on the performance differences for difficult tasks,
we crop the first \num{400} tasks from the figures
because they can be solved within \SI{1}{min}.
From~\cref{fig:evaluation:quantile-coop-cpu},
observe that IMC dominated its two augmented variants
when the CPU time was below \SI{400}{s}
because it did not have a parallel invariant generator that consumed extra time.
However, as the elapsed CPU time increased,
IMC\injecttoi{}DF took over and became the best among the three.
By contrast,
\cref{fig:evaluation:quantile-coop-wall} shows a clear advantage of IMC\injecttoi{}DF
regarding the wall-time consumption.
It was able to solve the most tasks when the elapsed wall-time was beyond \SI{200}{s}.
In addition, although IMC\injecttof{}DF did not not perform
as well as IMC\injecttoi{}DF on these tasks,
it still showed some wall-time improvement over plain IMC.
The results indicate that our proposed augmenting approaches
can not only prove the correctness of additional programs
but also reduce the wall-time for solving difficult tasks.

The scatter plots in~\cref{fig:evaluation:scatter-continuous} compare
the wall-time and interpolation time
consumed by IMC and IMC\injecttoi{}DF
on the tasks with non-trivial invariants.
A data point ($x$, $y$) in the plots indicates a task
that can be solved by IMC or IMC\injecttoi{}DF,
for which the former took $x$ seconds of wall-time (resp. interpolation time),
and the latter took $y$ seconds of wall-time (resp. interpolation time).
Observe that invariant injection helps improve plain IMC's wall-time efficiency
for time-consuming tasks
because more data points are below the diagonal
in the top-right quarter of~\cref{fig:evaluation:scatter-walltime}.
For interpolation time,
invariant injection generally lowered the summation of time spent on all queries,
as can be seen from~\cref{fig:evaluation:scatter-itptime}.
Moreover, there were \nTasksTenXLessItpTime{}~tasks
showing reduction by an order of magnitude.


\subsubsection{RQ4: Comparison with other SMT-based methods in \cpachecker}

\begin{table}[t]
  \centering
  \caption{Summary of the results for IMC\injecttoi{}DF vs. other SMT-based algorithms in \cpachecker, \twols, and \symbiotic, on all~\ntasks{} tasks}
  \label{tab:other-smt}
  \newcommand\precnum[1]{\tablenum[table-format=4]{#1}}
  \begin{tabular}{l@{\hspace{2mm}}|@{\hspace{2mm}}c@{\hspace{2mm}}c@{\hspace{2mm}}c@{\hspace{2mm}}c@{\hspace{2mm}}|@{\hspace{2mm}}c@{\hspace{2mm}}c}
    \toprule
    Tool      & \multicolumn{4}{c@{\hspace{2mm}}|@{\hspace{2mm}}}{\cpachecker}   & \twols & \symbiotic                   \\
    Algorithm & IMC\injecttoi{}DF                                                & \kidf  & \impact    & PredAbs & - & - \\
    \midrule
    Proofs
              & \textbf{\precnum{\EvalSlDfSImcIgndfiHardSafeCorrectCount}}
              & \precnum{\EvalSlDfSKiIgndfHardSafeCorrectCount}
              & \precnum{\EvalSlDfSImpactHardSafeCorrectCount}
              & \precnum{\EvalSlDfSPredabsHardSafeCorrectCount}
              & \precnum{\LsDefaultHardSafeCorrectCount}
              & \precnum{\SymbioticSvcompHardSafeCorrectCount}
    \\
    Timeouts
              & \precnum{\EvalSlDfSImcIgndfiHardSafeErrorTimeoutCount}
              & \precnum{\EvalSlDfSKiIgndfHardSafeErrorTimeoutCount}
              & \precnum{\EvalSlDfSImpactHardSafeErrorTimeoutCount}
              & \precnum{\EvalSlDfSPredabsHardSafeErrorTimeoutCount}
              & \precnum{\LsDefaultHardSafeErrorTimeoutCount}
              & \precnum{\SymbioticSvcompHardSafeErrorTimeoutCount}
    \\
    Out of memory
              & \precnum{\EvalSlDfSImcIgndfiHardSafeErrorOutOfMemoryCount}
              & \precnum{\EvalSlDfSKiIgndfHardSafeErrorOutOfMemoryCount}
              & \precnum{\EvalSlDfSImpactHardSafeErrorOutOfMemoryCount}
              & \precnum{\EvalSlDfSPredabsHardSafeErrorOutOfMemoryCount}
              & \precnum{\LsDefaultHardSafeErrorOutOfMemoryCount}
              & \precnum{\SymbioticSvcompHardSafeErrorOutOfMemoryCount}
    \\
    Other inconclusive
              & \precnum{\EvalSlDfSImcIgndfiHardSafeErrorOtherInconclusiveCount}
              & \precnum{\EvalSlDfSKiIgndfHardSafeErrorOtherInconclusiveCount}
              & \precnum{\EvalSlDfSImpactHardSafeErrorOtherInconclusiveCount}
              & \precnum{\EvalSlDfSPredabsHardSafeErrorOtherInconclusiveCount}
              & \precnum{\LsDefaultHardSafeErrorOtherInconclusiveCount}
              & \precnum{\SymbioticSvcompHardSafeErrorOtherInconclusiveCount}
    \\
    \bottomrule
  \end{tabular}
\end{table}

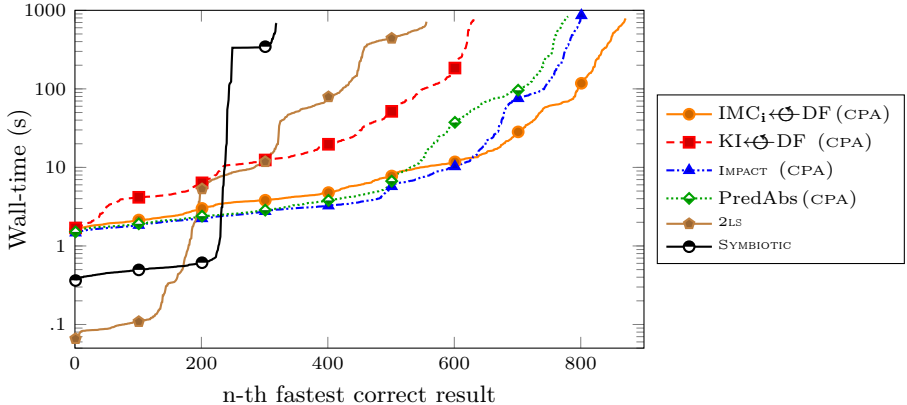
\begin{figure}[t]
  \centering
  \scalebox{1}{\input{\plotpath/quantile-others-walltime}}
  \caption{Quantile plot comparing IMC\injecttoi{}DF with other approaches and tools}
  \label{fig:evaluation:quantile-others}
\end{figure}

To study whether the the proposed invariant-injection approaches for IMC
could deliver more proofs than other SMT-based methods in~\cpachecker,
we evaluated IMC\injecttoi{}DF against
\kidf~\cite{kInduction},
\impact~\cite{IMPACT},
and predicate abstraction (PredAbs)~\cite{GrafSaidi97,AbstractionsFromProofs}.%
\footnote{Plain IMC has been compared with other approaches in a technical report~\cite{IMC-JAR}.}
The results are summarized in ~\cref{tab:other-smt},
where IMC\injecttoi{}DF solved the most tasks among the four compared algorithms.
\Cref{fig:evaluation:quantile-others} shows the quantile plot
of the wall-time usage for each compared algorithm.
IMC\injecttoi{}DF outperformed other polished SMT-based approaches in \cpachecker
in terms of both the number of solved tasks and the run-time efficiency.
There were
\ImcIgndfiPlusOverKiIgndfCount,
\ImcIgndfiPlusOverImpactCount,
and \ImcIgndfiPlusOverPredabsCount{}~additional tasks
solvable by IMC\injecttoi{}DF,
but not by \kidf, \impact, and predicate abstraction, respectively.
In total, IMC\injecttoi{}DF uniquely solved \UniqSolveCpaSmtCount{}~tasks
among the compared algorithms in \cpachecker,
demonstrating our contribution to SMT-based software verification.

\subsubsection{RQ5: Comparison with other software verifiers}
\Cref{tab:other-smt} and~\cref{fig:evaluation:quantile-others} also report
the results for \twols~\cite{kikiKroening} and \symbiotic~\cite{SymbioticApproach}
(executed with the configurations used in SV-COMP\,'22),
and our proposed method IMC\injecttoi{}DF
found significantly more proofs than both of them.
In particular,
\ImcIgndfiPlusOverLsCount{} and \ImcIgndfiPlusOverSymbioticCount{}~additional
tasks were solved by IMC\injecttoi{}DF,
but not by \twols and \symbiotic, respectively,
and a total of \UniqSolveOtherToolsCount{}~tasks were uniquely solved by IMC\injecttoi{}DF.
The results demonstrate the value of our invariant-injection methods for IMC
and justify our contribution to the state of the art of software model checking.

\subsection{Discussion}
\label{sect:experimental-discussion}
In general, strengthened interpolants help to decrease
Our goal is to study whether invariant injection is helpful for IMC,
and the experimental results reported above confirm its usefulness.
In our experiments, strengthened interpolants helped decrease
the number of interpolation queries (\cref{fig:evaluation:scatter-itpcall})
and the total interpolation time (\cref{fig:evaluation:scatter-itptime}),
but the effect came at a price: SMT and interpolation queries may become more challenging.
There were several tasks solvable by plain IMC in a minute
but stuck at one difficult SMT query with a strengthened interpolant for hundreds of seconds.
This phenomenon is an overhead against the performance of augmented IMC.
In principle,
one could limit the time spent on a call for SMT solving or interpolation
and fall back to the original interpolant
if the strengthened one makes the query too time-consuming.
Nevertheless, it is not straightforward to reuse an SMT solver stack
after it receives a timeout signal
because most solvers do not specify their behavior in this situation.
To summarize, the effects of invariant injection on IMC are mixed:
On one hand, it reduces the numbers of unrollings and interpolation calls;
on the other hand, each SMT or interpolation query might consume more time.
A similar trade-off between fewer refinement steps
and extra computation effort to achieve the reduction was
reported for the verifier~\ufo~\cite{UFO}.
In our evaluation,
we observed a clear decrease in the numbers of unrollings and interpolation calls,
but the number of solved tasks did not increase remarkably.
Similar observations that
the run-time does not necessarily benefit from auxiliary invariants
and that the number of solved tasks might not increase significantly
have also been reported in previous publications~\cite{Cabodi09,kikiKroening} on invariant-aided verification.

%% file: evaluation/tex/plot-defs.tex
\pgfplotsset{quantile plot/.style={
    width=9cm,
    height=4cm,
    scale only axis,
    /pgfplots/table/x expr={\coordindex+1},
    /pgfplots/table/y index=3,
    /pgfplots/table/header=false,
    ylabel shift=-1em,
    ticklabel style={font={\smaller}},
    xmin=0,
    ymin=1,
    ymax=1000,
    legend cell align={left},
    legend style={at={(0,1)}, anchor=north west, outer xsep=5pt, outer ysep=5pt, fill=none, font={\smaller}},
    legend columns=3,
    cycle multiindex list={
      orange, red, blue, green, brown, black\nextlist
      mark list*\nextlist
      solid, densely dashed, densely dashdotdotted, densely dotted},
  },
  every axis plot/.append style={thick}
}

\onlyifstandalone{
  \pgfplotsset{quantile plot/.append style={
      ylabel=CPU time (\second),
    },
  }
}

\newcommand\addgraph[2]{{
  \newcommand\csvfile{\plotpath/\detokenize{#2}}
  \IfFileExists\csvfile{
    \addplot+ table {\csvfile}; \addlegendentry{#1}
  }{
    \addplot coordinates {};
  }
}}

\pgfkeys{/tikz/.cd,
cube top color/.store in=\CubeTopColor,
cube top color=blue!60,
cube front color/.store in=\CubeFrontColor,
cube front color=blue!30,
cube side color/.store in=\CubeSideColor,
cube side color=blue!40,
3d cube color/.code={\colorlet{mycolor}{#1}%
\tikzset{cube top color=mycolor!60,cube front color=mycolor!30,%
cube side color=mycolor!40,draw=mycolor}}
}
\makeatletter
\pgfdeclareplotmark{half cube*}
                {%
                        \pgfplots@cube@gethalf@x
                        \let\pgfplots@cube@halfx=\pgfmathresult
                        \pgfplots@cube@gethalf@y
                        \let\pgfplots@cube@halfy=\pgfmathresult
                        \pgfplots@cube@gethalf@z
                        \let\pgfplots@cube@halfz=\pgfmathresult
                        \pgfmathparse{0*\pgfplots@cube@halfz}%
                        \let\pgfplots@cube@topz=\pgfmathresult
                        \pgfmathparse{-1*\pgfplots@cube@halfz}%
                        \let\pgfplots@cube@bottomz=\pgfmathresult
                        \pgfplotsifaxissurfaceisforeground{0vv}{%
                                \pgfsetfillcolor{\CubeFrontColor}
                                \pgfpathmoveto{\pgfplotsqpointxyz{-\pgfplots@cube@halfx}{-\pgfplots@cube@halfy}{\pgfplots@cube@bottomz}}%
                                \pgfpathlineto{\pgfplotsqpointxyz{-\pgfplots@cube@halfx}{-\pgfplots@cube@halfy}{\pgfplots@cube@topz}}%
                                \pgfpathlineto{\pgfplotsqpointxyz{-\pgfplots@cube@halfx}{ \pgfplots@cube@halfy}{\pgfplots@cube@topz}}%
                                \pgfpathlineto{\pgfplotsqpointxyz{-\pgfplots@cube@halfx}{ \pgfplots@cube@halfy}{\pgfplots@cube@bottomz}}%
                                \pgfpathclose
                                \pgfusepathqfillstroke
                        }{%
                                \pgfsetfillcolor{\CubeFrontColor}
                                \pgfpathmoveto{\pgfplotsqpointxyz{ \pgfplots@cube@halfx}{-\pgfplots@cube@halfy}{\pgfplots@cube@bottomz}}%
                                \pgfpathlineto{\pgfplotsqpointxyz{ \pgfplots@cube@halfx}{-\pgfplots@cube@halfy}{\pgfplots@cube@topz}}%
                                \pgfpathlineto{\pgfplotsqpointxyz{ \pgfplots@cube@halfx}{ \pgfplots@cube@halfy}{\pgfplots@cube@topz}}%
                                \pgfpathlineto{\pgfplotsqpointxyz{ \pgfplots@cube@halfx}{ \pgfplots@cube@halfy}{\pgfplots@cube@bottomz}}%
                                \pgfpathclose
                                \pgfusepathqfillstroke
                        }%
                        \pgfplotsifaxissurfaceisforeground{v0v}{%
                                \pgfsetfillcolor{\CubeSideColor}
                                \pgfpathmoveto{\pgfplotsqpointxyz{-\pgfplots@cube@halfx}{-\pgfplots@cube@halfy}{\pgfplots@cube@bottomz}}%
                                \pgfpathlineto{\pgfplotsqpointxyz{-\pgfplots@cube@halfx}{-\pgfplots@cube@halfy}{\pgfplots@cube@topz}}%
                                \pgfpathlineto{\pgfplotsqpointxyz{ \pgfplots@cube@halfx}{-\pgfplots@cube@halfy}{\pgfplots@cube@topz}}%
                                \pgfpathlineto{\pgfplotsqpointxyz{ \pgfplots@cube@halfx}{-\pgfplots@cube@halfy}{\pgfplots@cube@bottomz}}%
                                \pgfpathclose
                                \pgfusepathqfillstroke
                        }{%
                                \pgfsetfillcolor{\CubeSideColor}
                                \pgfpathmoveto{\pgfplotsqpointxyz{-\pgfplots@cube@halfx}{ \pgfplots@cube@halfy}{\pgfplots@cube@bottomz}}%
                                \pgfpathlineto{\pgfplotsqpointxyz{-\pgfplots@cube@halfx}{ \pgfplots@cube@halfy}{\pgfplots@cube@topz}}%
                                \pgfpathlineto{\pgfplotsqpointxyz{ \pgfplots@cube@halfx}{ \pgfplots@cube@halfy}{\pgfplots@cube@topz}}%
                                \pgfpathlineto{\pgfplotsqpointxyz{ \pgfplots@cube@halfx}{ \pgfplots@cube@halfy}{\pgfplots@cube@bottomz}}%
                                \pgfpathclose
                                \pgfusepathqfillstroke
                        }%
                        \pgfplotsifaxissurfaceisforeground{vv0}{%
                                \pgfsetfillcolor{\CubeTopColor}
                                \pgfpathmoveto{\pgfplotsqpointxyz{-\pgfplots@cube@halfx}{-\pgfplots@cube@halfy}{\pgfplots@cube@bottomz}}%
                                \pgfpathlineto{\pgfplotsqpointxyz{-\pgfplots@cube@halfx}{ \pgfplots@cube@halfy}{\pgfplots@cube@bottomz}}%
                                \pgfpathlineto{\pgfplotsqpointxyz{ \pgfplots@cube@halfx}{ \pgfplots@cube@halfy}{\pgfplots@cube@bottomz}}%
                                \pgfpathlineto{\pgfplotsqpointxyz{ \pgfplots@cube@halfx}{-\pgfplots@cube@halfy}{\pgfplots@cube@bottomz}}%
                                \pgfpathclose
                                \pgfusepathqfillstroke
                        }{%
                                \pgfsetfillcolor{\CubeTopColor}
                                \pgfpathmoveto{\pgfplotsqpointxyz{-\pgfplots@cube@halfx}{-\pgfplots@cube@halfy}{\pgfplots@cube@topz}}%
                                \pgfpathlineto{\pgfplotsqpointxyz{-\pgfplots@cube@halfx}{ \pgfplots@cube@halfy}{\pgfplots@cube@topz}}%
                                \pgfpathlineto{\pgfplotsqpointxyz{ \pgfplots@cube@halfx}{ \pgfplots@cube@halfy}{\pgfplots@cube@topz}}%
                                \pgfpathlineto{\pgfplotsqpointxyz{ \pgfplots@cube@halfx}{-\pgfplots@cube@halfy}{\pgfplots@cube@topz}}%
                                \pgfpathclose
                                \pgfusepathqfillstroke
                        }%
            }
\makeatother

%% file: evaluation/tex/scatter-k.tex
\begin{tikzpicture}
    \node[inner sep=0] (image) at (0,0) {\includegraphics[width=4cm]{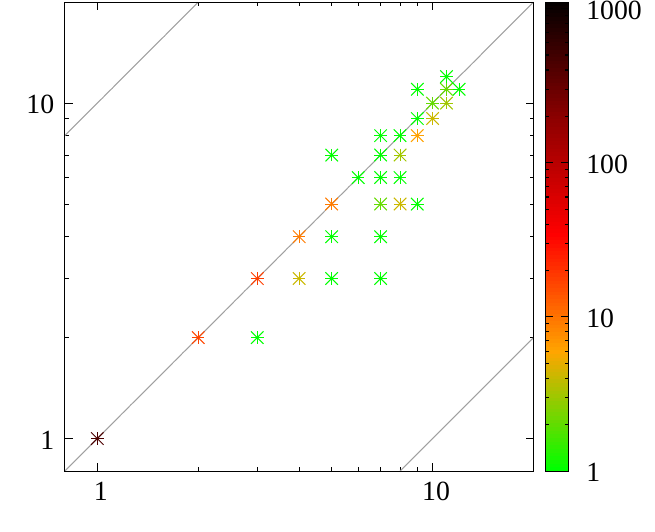}};
    \node[left=-2.5mm of image, rotate=90, anchor=south, font=\tiny] {IMC\injecttoi{}DF};
    \node[below=-2.5mm of image, font=\tiny] {IMC};
    \node[right=-2.5mm of image, rotate=90, anchor=north, font=\tiny] {Number of tasks};
\end{tikzpicture}

%% file: evaluation/tex/scatter-itpcall.tex
\begin{tikzpicture}
    \node[inner sep=0] (image) at (0,0) {\includegraphics[width=4cm]{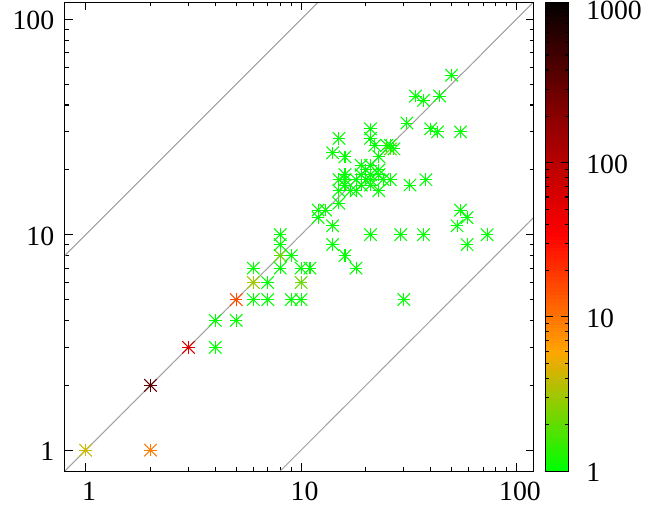}};
    \node[left=-2.5mm of image, rotate=90, anchor=south, font=\tiny] {IMC\injecttoi{}DF};
    \node[below=-2.5mm of image, font=\tiny] {IMC};
    \node[right=-2.5mm of image, rotate=90, anchor=north, font=\tiny] {Number of tasks};
\end{tikzpicture}

%% file: evaluation/tex/imc-ignDfi-top-results.tex

\begin{tabular}{@{\hspace{2mm}}l@{\hspace{1mm}}|@{\hspace{1mm}}c@{\hspace{1mm}}c@{\hspace{1mm}}c@{\hspace{1mm}}c@{\hspace{1mm}}c@{\hspace{1mm}}|@{\hspace{1mm}}c@{\hspace{1mm}}c@{\hspace{1mm}}c@{\hspace{1mm}}c@{\hspace{1mm}}c@{\hspace{3.5mm}}}
    \toprule
           & \multicolumn{5}{@{\hspace{1mm}}c|@{\hspace{1mm}}}{IMC}  & \multicolumn{5}{@{\hspace{1mm}}c@{\hspace{1mm}}}{IMC\injecttoi{}DF}    \\
    Task   & result & \#unroll & \#itp & wall-time & itp-time        & result & \#unroll & \#itp & wall-time & itp-time                        \\
    \midrule

\href{https://gitlab.com/sosy-lab/benchmarking/sv-benchmarks/-/blob/svcomp22/c/eca-rers2012/Problem03_label03.c}{Problem03_label03} & \timeout & \precnum{8} & \precnum{50} & \rndnum{872.22865271708} & \rndnum{744.979} & \solved & \precnum{5} & \precnum{11} & \rndnum{27.238232209114358} & \rndnum{12.961} \\
\href{https://gitlab.com/sosy-lab/benchmarking/sv-benchmarks/-/blob/svcomp22/c/eca-rers2012/Problem03_label15.c}{Problem03_label15} & \timeout & \precnum{7} & \precnum{54} & \rndnum{876.379350814037} & \rndnum{747.834} & \solved & \precnum{5} & \precnum{11} & \rndnum{35.356361616868526} & \rndnum{19.451} \\
\href{https://gitlab.com/sosy-lab/benchmarking/sv-benchmarks/-/blob/svcomp22/c/eca-rers2012/Problem03_label51.c}{Problem03_label51} & \timeout & \precnum{10} & \precnum{57} & \rndnum{878.3521001820918} & \rndnum{753.434} & \solved & \precnum{5} & \precnum{11} & \rndnum{24.287363136885688} & \rndnum{10.283} \\
\href{https://gitlab.com/sosy-lab/benchmarking/sv-benchmarks/-/blob/svcomp22/c/loop-zilu/benchmark37_conjunctive.c}{benchmark37_conj} & \timeout & \precnum{317} & \precnum{316} & \rndnum{892.0208312040195} & \rndnum{851.141} & \solved & \precnum{1} & \precnum{2} & \rndnum{1.8257458130829036} & \rndnum{0.008} \\
\href{https://gitlab.com/sosy-lab/benchmarking/sv-benchmarks/-/blob/svcomp22/c/seq-mthreaded-reduced/pals_lcr-var-start-time.4.ufo.UNBOUNDED.pals.c.v+sep-reducer.c}{pals_lcr-var-start} & \timeout & \precnum{11} & \precnum{16} & \rndnum{875.1322450151201} & \rndnum{632.583} & \solved & \precnum{10} & \precnum{24} & \rndnum{760.8478177771904} & \rndnum{479.382} \\
\href{https://gitlab.com/sosy-lab/benchmarking/sv-benchmarks/-/blob/svcomp22/c/seq-mthreaded-reduced/pals_lcr.4.ufo.UNBOUNDED.pals.c.v+lhb-reducer.c}{pals_lcr.4.ufo.UNB} & \solved & \precnum{11} & \precnum{23} & \rndnum{198.25573511817493} & \rndnum{164.821} & \solved & \precnum{10} & \precnum{16} & \rndnum{116.0566888100002} & \rndnum{91.702} \\
\href{https://gitlab.com/sosy-lab/benchmarking/sv-benchmarks/-/blob/svcomp22/c/loop-acceleration/phases_2-2.c}{phases_2-2} & \timeout & \precnum{1} & \precnum{1} & \rndnum{898.6676160399802} & \rndnum{897.066} & \solved & \precnum{1} & \precnum{2} & \rndnum{2.201416672905907} & \rndnum{0.057} \\
\href{https://gitlab.com/sosy-lab/benchmarking/sv-benchmarks/-/blob/svcomp22/c/bitvector/s3_srvr_1a.BV.c.cil.c}{s3_srvr_1a.BV.c.cil} & \timeout & \precnum{64} & \precnum{441} & \rndnum{885.0645597530529} & \rndnum{544.280} & \solved & \precnum{5} & \precnum{13} & \rndnum{6.2615407118573785} & \rndnum{1.016} \\
\href{https://gitlab.com/sosy-lab/benchmarking/sv-benchmarks/-/blob/svcomp22/c/bitvector/s3_srvr_2a.BV.c.cil.c}{s3_srvr_2a.BV.c.cil} & \timeout & \precnum{39} & \precnum{290} & \rndnum{886.5472344069276} & \rndnum{793.493} & \solved & \precnum{35} & \precnum{252} & \rndnum{635.0260343221016} & \rndnum{538.770} \\
\href{https://gitlab.com/sosy-lab/benchmarking/sv-benchmarks/-/blob/svcomp22/c/bitvector/s3_srvr_2a_alt.BV.c.cil.c}{s3_srvr_2a_alt.BV} & \timeout & \precnum{37} & \precnum{278} & \rndnum{882.8717780259904} & \rndnum{776.454} & \solved & \precnum{19} & \precnum{125} & \rndnum{71.68477753107436} & \rndnum{51.658} \\

    \bottomrule
\end{tabular}

%% file: evaluation/tex/imc-ignDff-top-results.tex

\begin{tabular}{@{\hspace{2mm}}l@{\hspace{1mm}}|@{\hspace{1mm}}c@{\hspace{1mm}}c@{\hspace{1mm}}c@{\hspace{1mm}}c@{\hspace{1mm}}c@{\hspace{1mm}}|@{\hspace{1mm}}c@{\hspace{1mm}}c@{\hspace{1mm}}c@{\hspace{1mm}}c@{\hspace{1mm}}c@{\hspace{2.5mm}}}
    \toprule
           & \multicolumn{5}{@{\hspace{1mm}}c|@{\hspace{1mm}}}{IMC}  & \multicolumn{5}{@{\hspace{1mm}}c@{\hspace{1mm}}}{IMC\injecttof{}DF}    \\
    Task   & result & \#unroll & \#itp & wall-time & itp-time        & result & \#unroll & \#itp & wall-time & itp-time                        \\
    \midrule

\href{https://gitlab.com/sosy-lab/benchmarking/sv-benchmarks/-/blob/svcomp22/c/eca-rers2012/Problem03_label01.c}{Problem03_label01} & \solved & \precnum{9} & \precnum{73} & \rndnum{654.156850613188} & \rndnum{472.896} & \solved & \precnum{6} & \precnum{13} & \rndnum{37.46677428903058} & \rndnum{17.998} \\
\href{https://gitlab.com/sosy-lab/benchmarking/sv-benchmarks/-/blob/svcomp22/c/eca-rers2012/Problem03_label03.c}{Problem03_label03} & \timeout & \precnum{8} & \precnum{50} & \rndnum{872.22865271708} & \rndnum{744.979} & \solved & \precnum{6} & \precnum{15} & \rndnum{85.67790243797936} & \rndnum{61.715} \\
\href{https://gitlab.com/sosy-lab/benchmarking/sv-benchmarks/-/blob/svcomp22/c/eca-rers2012/Problem03_label04.c}{Problem03_label04} & \solved & \precnum{7} & \precnum{53} & \rndnum{141.33076911582612} & \rndnum{102.077} & \solved & \precnum{5} & \precnum{11} & \rndnum{17.948978120926768} & \rndnum{4.849} \\
\href{https://gitlab.com/sosy-lab/benchmarking/sv-benchmarks/-/blob/svcomp22/c/eca-rers2012/Problem03_label11.c}{Problem03_label11} & \solved & \precnum{7} & \precnum{55} & \rndnum{113.87224058085121} & \rndnum{82.201} & \solved & \precnum{5} & \precnum{9} & \rndnum{17.427703568944708} & \rndnum{4.054} \\
\href{https://gitlab.com/sosy-lab/benchmarking/sv-benchmarks/-/blob/svcomp22/c/eca-rers2012/Problem03_label12.c}{Problem03_label12} & \solved & \precnum{8} & \precnum{59} & \rndnum{509.70322883711196} & \rndnum{345.975} & \solved & \precnum{8} & \precnum{18} & \rndnum{80.8643268099986} & \rndnum{59.340} \\
\href{https://gitlab.com/sosy-lab/benchmarking/sv-benchmarks/-/blob/svcomp22/c/eca-rers2012/Problem03_label15.c}{Problem03_label15} & \timeout & \precnum{7} & \precnum{54} & \rndnum{876.379350814037} & \rndnum{747.834} & \solved & \precnum{6} & \precnum{18} & \rndnum{107.30843193409964} & \rndnum{80.402} \\
\href{https://gitlab.com/sosy-lab/benchmarking/sv-benchmarks/-/blob/svcomp22/c/eca-rers2012/Problem03_label19.c}{Problem03_label19} & \solved & \precnum{8} & \precnum{59} & \rndnum{134.35687817796133} & \rndnum{107.946} & \solved & \precnum{5} & \precnum{9} & \rndnum{22.799386984901503} & \rndnum{6.634} \\
\href{https://gitlab.com/sosy-lab/benchmarking/sv-benchmarks/-/blob/svcomp22/c/eca-rers2012/Problem03_label51.c}{Problem03_label51} & \timeout & \precnum{10} & \precnum{57} & \rndnum{878.3521001820918} & \rndnum{753.434} & \solved & \precnum{7} & \precnum{16} & \rndnum{49.77535207592882} & \rndnum{32.453} \\
\href{https://gitlab.com/sosy-lab/benchmarking/sv-benchmarks/-/blob/svcomp22/c/bitvector/s3_srvr_1a.BV.c.cil.c}{s3_srvr_1a.BV.c.cil} & \timeout & \precnum{64} & \precnum{441} & \rndnum{885.0645597530529} & \rndnum{544.280} & \solved & \precnum{61} & \precnum{422} & \rndnum{656.1564103758428} & \rndnum{391.957} \\
\href{https://gitlab.com/sosy-lab/benchmarking/sv-benchmarks/-/blob/svcomp22/c/bitvector/s3_srvr_2a.BV.c.cil.c}{s3_srvr_2a.BV.c.cil} & \timeout & \precnum{39} & \precnum{290} & \rndnum{886.5472344069276} & \rndnum{793.493} & \solved & \precnum{19} & \precnum{123} & \rndnum{77.1594064598903} & \rndnum{56.020} \\

    \bottomrule
\end{tabular}

%% file: evaluation/tex/bar-rand-seed-2d.tex
\begin{tikzpicture}
  \begin{axis} [
    ybar=0,
    width=.6\linewidth,
    height=.36\linewidth,
    bar width=10pt,
    xtick={0,1,2,3,4},
    xticklabels={7,42,61,89,165},
    ytick={400,450,...,550},
    ymin=400,
    xlabel={Random seed},
    ylabel={\#Proofs},
    xtick style={draw=none},
    xticklabel style={yshift=5pt},
    xlabel style={yshift=3pt},
    legend image code/.code={
        \draw [#1] (0cm,-0.1cm) rectangle (0.3cm,0.1cm);
    },
    legend style={
      at={(1,0.5)},
      anchor=west,
      outer xsep=5pt,
      outer ysep=5pt,
      fill=none,
    },
    legend cell align={left},
    /pgf/number format/1000 sep={},
  ]

    \addplot [
        draw = orange,
        fill=orange,
        fill opacity=0.3,
    ] coordinates {
        (0,\ReducedEvalSlDfSSImcHardSafeCorrectCount)
        (1,\ReducedEvalSlDfSImcHardSafeCorrectCount)
        (2,\ReducedEvalSlDfSSoImcHardSafeCorrectCount)
        (3,\ReducedEvalSlDfSEnImcHardSafeCorrectCount)
        (4,\ReducedEvalSlDfSOsfImcHardSafeCorrectCount)
    };

    \addplot [draw = green,
        fill=green,
        fill opacity=0.5,
        pattern = crosshatch dots,
        pattern color = green,
    ] coordinates {
        (0,\ReducedEvalSlDfSSImcIgndfiHardSafeCorrectCount)
        (1,\ReducedEvalSlDfSImcIgndfiHardSafeCorrectCount)
        (2,\ReducedEvalSlDfSSoImcIgndfiHardSafeCorrectCount)
        (3,\ReducedEvalSlDfSEnImcIgndfiHardSafeCorrectCount)
        (4,\ReducedEvalSlDfSOsfImcIgndfiHardSafeCorrectCount)
    };

    \legend {IMC, IMC\injecttoi{}DF};
  \end{axis}
\end{tikzpicture}

%% file: evaluation/tex/quantile-coop-cputime.tex
\begin{tikzpicture}
\begin{axis}[
    quantile plot,
    xlabel=n-th fastest correct result,
    ylabel=CPU time (\second),
    width=6cm,
    mark repeat=20,
    /pgf/number format/1000 sep={},
    legend columns=1,
    xmax=550,
    xmin=400,
    ymin=0
    ]
    \addgraph{IMC}{../csv/reduced.imc.quantile-cputime.csv}
    \addgraph{IMC\injecttof{}DF}{../csv/reduced.imc-ignDff.quantile-cputime.csv}
    \addgraph{IMC\injecttoi{}DF}{../csv/reduced.imc-ignDfi.quantile-cputime.csv}
\end{axis}
\end{tikzpicture}

%% file: evaluation/tex/quantile-coop-walltime.tex
\begin{tikzpicture}
\begin{axis}[
    quantile plot,
    xlabel=n-th fastest correct result,
    ylabel=Wall-time (\second),
    width=6cm,
    /pgfplots/table/y index=4,
    mark repeat=20,
    /pgf/number format/1000 sep={},
    legend columns=1,
    xmax=550,
    xmin=400,
    ymin=0
    ]
    \addgraph{IMC}{../csv/reduced.imc.quantile-walltime.csv}
    \addgraph{IMC\injecttof{}DF}{../csv/reduced.imc-ignDff.quantile-walltime.csv}
    \addgraph{IMC\injecttoi{}DF}{../csv/reduced.imc-ignDfi.quantile-walltime.csv}
\end{axis}
\end{tikzpicture}

%% file: evaluation/tex/scatter-itptime.tex
\begin{tikzpicture}
\begin{loglogaxis}[
    xlabel=IMC,
    ylabel=IMC\injecttoi{}DF,
    xmin=1,
    xmax=1000,
    ymin=1,
    ymax=1000,
    domain=1:1001,
    clip mode=individual,
    axis equal image,
    legend pos=north west,
    legend style={font=\scriptsize},
    legend cell align={left},
    /pgf/number format/1000 sep={},
    ]
    \addplot+[green, mark=o, only marks]
         table[
             header=false,
             skip first n=3, 
             x index=2, 
             y index=4  
             ] {evaluation/csv/reduced.imc-ignDfi.scatter-itptime.table.csv};
    \addlegendentry{Both solved}
    \addplot+[orange, mark=x, only marks]
        table[
            header=false,
            skip first n=3, 
            x index=2, 
            y index=4  
            ] {evaluation/csv/reduced.imc-ignDfi.scatter-itptime-timeout-x.table.csv};
    \addlegendentry{IMC TO}
    \addplot+[blue, mark=+, only marks]
    table[
        header=false,
        skip first n=3, 
        x index=2, 
        y index=4  
        ] {evaluation/csv/reduced.imc-ignDfi.scatter-itptime-timeout-y.table.csv};
    \addlegendentry{IMC\injecttoi{}DF TO}
    \addplot[gray] {x};
    \addplot[gray] {10*x};
    \addplot[gray] {x/10};
\end{loglogaxis}
\end{tikzpicture}

%% file: evaluation/tex/scatter-walltime.tex
\begin{tikzpicture}
\begin{loglogaxis}[
    xlabel=IMC,
    ylabel=IMC\injecttoi{}DF,
    xmin=1,
    xmax=1000,
    ymin=1,
    ymax=1000,
    domain=1:1001,
    clip mode=individual,
    axis equal image,
    legend pos=north west,
    legend style={font=\scriptsize},
    legend cell align={left},
    /pgf/number format/1000 sep={},
    ]
    \addplot+[green, mark=o, only marks]
         table[
             header=false,
             skip first n=3, 
             x index=2, 
             y index=4  
             ] {evaluation/csv/reduced.imc-ignDfi.scatter-walltime.table.csv};
    \addlegendentry{Both solved}
    \addplot+[orange, mark=x, only marks]
         table[
             header=false,
             skip first n=3, 
             x index=2, 
             y index=4  
             ] {evaluation/csv/reduced.imc-ignDfi.scatter-walltime-timeout-x.table.csv};
    \addlegendentry{IMC TO}
    \addplot+[blue, mark=+, only marks]
         table[
             header=false,
             skip first n=3, 
             x index=2, 
             y index=4  
             ] {evaluation/csv/reduced.imc-ignDfi.scatter-walltime-timeout-y.table.csv};
    \addlegendentry{IMC\injecttoi{}DF TO}
    \addplot[gray] {x};
    \addplot[gray] {10*x};
    \addplot[gray] {x/10};
\end{loglogaxis}
\end{tikzpicture}

%% file: evaluation/tex/quantile-others-walltime.tex
\begin{tikzpicture}
\begin{semilogyaxis}[
    quantile plot,
    width=7.5cm,
    height=4.5cm,
    xlabel=n-th fastest correct result,
    ylabel=Wall-time (\second),
    /pgfplots/table/y index=4,
    ymin=0.05,
    mark repeat=100,
    /pgf/number format/1000 sep={},
    legend style={at={(1,0.5)}, anchor=west, outer xsep=5pt, outer ysep=5pt, fill=none},
    legend columns=1,
    xmax=900
    ]
    \addgraph{IMC\injecttoi{}DF\,(\cpacheckerabbrv)}{../csv/imc-ignDfi.quantile-walltime.csv}
    \addgraph{\kidf\,(\cpacheckerabbrv)}{../csv/ki-ignDf.quantile-walltime.csv}
    \addgraph{\impact\,(\cpacheckerabbrv)}{../csv/impact.quantile-walltime.csv}
    \addgraph{{PredAbs}\,(\cpacheckerabbrv)}{../csv/predAbs.quantile-walltime.csv}
    \addgraph{\twols}{../csv/2ls.quantile-walltime.csv}
    \addgraph{\symbiotic}{../csv/symbiotic.quantile-walltime.csv}
\end{semilogyaxis}
\end{tikzpicture}

%% file: conclusion.tex
\section{Conclusion}
We augmented IMC,
an interpolation-based model-checking algorithm
published by McMillan in 2003~\cite{McMillanCraig},
via injecting auxiliary invariants to reduce
the numbers of program unrollings and interpolation queries
needed to prove the correctness of programs.
Invariants are used to strengthen
(1)~the checks for determining whether a fixed point is reached, or
(2)~the interpolants derived during the procedure.
We rigorously proved the correctness of the proposed approaches
and implemented both techniques in the verification tool~\cpachecker.
We empirically evaluated our implementations
against four SMT-based verification approaches in~\cpachecker
and two state-of-the-art software verifiers
over C~programs whose safety properties are hard to prove.
Our experiments show that the proposed techniques
effectively reduced the numbers of program unrollings and interpolation calls
and therefore reduced the wall-time usage compared to plain IMC.
Furthermore, the proposed augmentation helped IMC
deliver more proofs than the compared SMT-based algorithms in \cpachecker
and solve \UniqSolveOtherToolsCount{}~tasks unsolvable by the compared tools.
For future work,
as the strengthened interpolants might lead to extra time spent on SMT solving or interpolation,
we plan to devise a strategy to selectively inject invariants into IMC
only when they are likely to be helpful,
in order to further improve the performance of the proposed methods.